\documentclass{ptephy_v1}%%%%where ptephy_v1 is the template name
\usepackage{graphics} %for dealing with figures.
\usepackage{dcolumn}% Align table columns on decimal point
\usepackage{bm}% bold math
\usepackage{amsfonts}
\usepackage{latexsym}
\usepackage{amssymb}
\usepackage{verbatim}
\usepackage{amsthm}
\usepackage{amsmath}
\usepackage{mathrsfs}
\usepackage{wrapft}
\newcommand{\FF}{{{\mathbb F}}}

\newcommand{\HF}{{{\mathbb H}}}

\newcommand{\RF}{{{\mathbb R}}}

\newcommand{\ScrF}{{{\mathscr F}}}
\newcommand{\ScrG}{{{\mathscr G}}}
\newcommand{\ScrH}{{{\mathscr H}}}

\newcommand{\ScrK}{{{\mathscr K}}}

\newcommand{\ScrM}{{{\mathscr M}}}
\newcommand{\ScrN}{{{\mathscr N}}}

\newcommand{\ScrQ}{{{\mathscr Q}}}

\newcommand{\ScrT}{{{\mathscr T}}}

\newcommand{\ScrX}{{{\mathscr X}}}
\newcommand{\ScrY}{{{\mathscr Y}}}

\newtheorem{theorem}{Theorem}[section]

\newtheorem{conjecture}{Conjecture}[section]
\newtheorem{proposal}{Proposal}[section]
\begin{document}

%\draft

%%%%%%%%%%%%%%%%%%%%%%%%%%%%%%%%%%%%%%%%%%%%%%%%
%%%%%%%%%%%%%%%%%%%%%%%%%%%%%%%%%%%%%%%%%%%%%%%%
%%% the equation numbering as sectionwise
%%%%%%%%%%%%%%%%%%%%%%%%%%%%%%%%%%%%%%%%%%%%%%%%
\numberwithin{equation}{section}
%%%%%%%%%%%%%%%%%%%%%%%%%%%%%%%%%%%%%%%%%%%%%%%%
%%%%%%%%%%%%%%%%%%%%%%%%%%%%%%%%%%%%%%%%%%%%%%%%

%%%%%%%%%%%%%%%%%%%%%%%%%%%%%%%%%%%%%%%%%%%%%%%%%%%%%%
%%%%%%%%%%%%%%%%%%%%%%%%%%%%%%%%%%%%%%%%%%%%%%%%%%%%%%
%%%%%%%%%%%%%%%%%%%%%%%%%%%%%%%%%%%%%%%%%%%%%%%%%%%%%%
%%%%%%%%%%%%%%%%%%%%%%%%%%%%%%%%%%%%%%%%%%%%%%%%%%%%%%
\title{
  Gauge-invariant perturbation theory on the Schwarzschild
  background spacetime Part I : \\
  --- Formulation and odd-mode perturbations ---
}
%%%%%%%%%%%%%%%%%%%%%%%%%%%%%%%%%%%%%%%%%%%%%%%%%%%%%%
%%%%%%%%%%%%%%%%%%%%%%%%%%%%%%%%%%%%%%%%%%%%%%%%%%%%%%
%%%%%%%%%%%%%%%%%%%%%%%%%%%%%%%%%%%%%%%%%%%%%%%%%%%%%%
%%%%%%%%%%%%%%%%%%%%%%%%%%%%%%%%%%%%%%%%%%%%%%%%%%%%%%
\author{
  Kouji Nakamura
  \footnote{E-mail address: dr.kouji.nakamura@gmail.com}
}
%%%%%%%%%%%%%%%%%%%%%%%%%%%%%%%%%%%%%%%%%%%%%%%%%%%%%%
%%%%%%%%%%%%%%%%%%%%%%%%%%%%%%%%%%%%%%%%%%%%%%%%%%%%%%
%%%%%%%%%%%%%%%%%%%%%%%%%%%%%%%%%%%%%%%%%%%%%%%%%%%%%%
%%%%%%%%%%%%%%%%%%%%%%%%%%%%%%%%%%%%%%%%%%%%%%%%%%%%%%
\address{
  Gravitational-Wave Science Project,
  National Astronomical Observatory of Japan,\\
  2-21-1, Osawa, Mitaka, Tokyo 181-8588, Japan
}
%\ead{kouji.nakamura@nao.ac.jp}
%%%%%%%%%%%%%%%%%%%%%%%%%%%%%%%%%%%%%%%%%%%%%%%%%%%%%%
%%%%%%%%%%%%%%%%%%%%%%%%%%%%%%%%%%%%%%%%%%%%%%%%%%%%%%
%%%%%%%%%%%%%%%%%%%%%%%%%%%%%%%%%%%%%%%%%%%%%%%%%%%%%%
%%%%%%%%%%%%%%%%%%%%%%%%%%%%%%%%%%%%%%%%%%%%%%%%%%%%%%
\date{\today}
%%%%%%%%%%%%%%%%%%%%%%%%%%%%%%%%%%%%%%%%%%%%%%%%%%%%%%
%%%%%%%%%%%%%%%%%%%%%%%%%%%%%%%%%%%%%%%%%%%%%%%%%%%%%%
%%%%%%%%%%%%%%%%%%%%%%%%%%%%%%%%%%%%%%%%%%%%%%%%%%%%%%
%%%%%%%%%%%%%%%%%%%%%%%%%%%%%%%%%%%%%%%%%%%%%%%%%%%%%%
\begin{abstract}
  This is the Part I paper of our series of full papers on a
  gauge-invariant {\it linear} perturbation theory on the
  Schwarzschild background  spacetime which was briefly reported in
  our short papers [K.~Nakamura, Class. Quantum Grav. {\bf 38} (2021),
  145010; K.~Nakamura, Letters in High Energy Physics {\bf 2021}
  (2021), 215.].
  We first review our general framework of the gauge-invariant
  perturbation theory, which can be easily extended to the
  {\it higher-order} perturbation theory.
  When we apply this general framework to perturbations on the
  Schwarzschild background spacetime, gauge-invariant treatments
  of $l=0,1$ mode perturbations are required.
  On the other hand, in the current consensus on the perturbations of
  the Schwarzschild spacetime, gauge-invariant treatments for $l=0,1$
  modes are difficult if we keep the reconstruction of the original
  metric perturbations in our mind.
  Due to this situation, we propose a strategy of a gauge-invariant
  treatment of $l=0,1$ mode perturbations through the decomposition
  of the metric perturbations by singular harmonic functions at once
  and the regularization of these singularities through the imposition
  of the boundary conditions to the Einstein equations.
  Following this proposal, we derive the linearized Einstein equations
  for any modes of $l\geq 0$ in a gauge-invariant manner.
  We discuss the solutions to the odd-mode perturbation equations in
  the linearized Einstein equations and show that these perturbations
  include the Kerr parameter perturbation in these odd-mode
  perturbation, which is physically reasonable.
  In the Part II and Part III papers [K.~Nakamura, arXiv:2110.13512
  [gr-qc]; arXiv:2110.13519 [gr-qc].] of this series of papers, we
  will show that the even-mode solutions to the linearized Einstein
  equations obtained through our proposal are also physically
  reasonable.
  Then, we conclude that our proposal of a gauge-invariant treatment
  for $l=0,1$-mode perturbations is also physically reasonable.
\end{abstract}

\maketitle

%\pacs{04.20.-q, 04.20.Cv, 04.50.+h, 98.80.Jk}
%\submitto{\CQG}
%\maketitle

%%%%%%%%%%%%%%%%%%%%%%%%%%%%%%%%%%%%%%%%%%%%%%%%%%%%%%
%%%%%%%%%%%%%%%%%%%%%%%%%%%%%%%%%%%%%%%%%%%%%%%%%%%%%%
%%%%%%%%%%%%%%%%%%%%%%%%%%%%%%%%%%%%%%%%%%%%%%%%%%%%%%
%%%%%%%%%%%%%%%%%%%%%%%%%%%%%%%%%%%%%%%%%%%%%%%%%%%%%%
\section{Introduction}
\label{sec:introduction}
%%%%%%%%%%%%%%%%%%%%%%%%%%%%%%%%%%%%%%%%%%%%%%%%%%%%%%
%%%%%%%%%%%%%%%%%%%%%%%%%%%%%%%%%%%%%%%%%%%%%%%%%%%%%%
%%%%%%%%%%%%%%%%%%%%%%%%%%%%%%%%%%%%%%%%%%%%%%%%%%%%%%
%%%%%%%%%%%%%%%%%%%%%%%%%%%%%%%%%%%%%%%%%%%%%%%%%%%%%%

%*************************************************************

Gravitational-wave astronomy has begun from the first event GW150914
of the direct observation of gravitational waves in
2015~\cite{B.P.Abbot-et-al-2016-GW150914}.
This event was also the beginning of the multi-messenger astronomy
including gravitational waves~\cite{LIGO-home-page}.
We are now on the stage where we can directly measure gravitational
waves and we can carry out scientific research through these
gravitational-wave events.
We can also expect that one future direction of gravitational-wave
astronomy is the development as a precise science by the detailed
studies of source science, the tests of general-relativity, and the
developments of the global network of gravitational-wave
detectors~\cite{LIGO-home-page,Virgo-home-page,KAGRA-home-page,LIGO-INDIA-home-page}.
In addition to the current network of ground-based detectors, as
future ground-based gravitational-wave detectors, the projects of
Einstein Telescope~\cite{ET-home-page} and Cosmic
Explorer~\cite{CosmicExplorer-home-page} are also progressing to
achieve more sensitive detections.

%*************************************************************

Besides these ground-based detectors, some projects of space
gravitational-wave antenna are also
progressing~\cite{LISA-home-page,DECIGO-PTEP-2021,TianQin-PTEP-2021,Taiji-PTEP-2021}.
Among them, the Extreme-Mass-Ratio-Inspiral (EMRI), which is a source
of gravitational waves from the motion of a stellar mass object around
a supermassive black hole, is a promising target of the Laser
Interferometer Space Antenna~\cite{LISA-home-page}.
To describe the gravitational wave from EMRIs, black hole
perturbations are used~\cite{L.Barack-A.Pound-2019}.
Furthermore, the sophistication of higher-order black hole
perturbation theories is required to support these gravitational-wave
physics as a precise science.
Very recently, the backaction effect of mass and angular momentum
accretion on the Schwarzschild black hole due to the Blandford-Znajek
process~\cite{R.D.Blandford-R.L.Znajek-1977} was also
discussed~\cite{M.Kimura-T.Harada-A.Naruko-K.Toma-2021}, which are
higher-order effects of two-parameter
perturbations~\cite{K.Nakamura-2003,K.Nakamura-2005}.
The motivation of this paper is in the theoretical sophistication of
black hole perturbation theories toward higher-order perturbations for
very wide physical situations including the topic in
Ref.~\cite{M.Kimura-T.Harada-A.Naruko-K.Toma-2021}.

%*************************************************************

In the current situation of black hole perturbation theories, we may
say that further sophistications are possible even in perturbation
theories on the Schwarzschild background spacetime, although realistic
black holes have their angular momentum and we have to consider the
perturbation theory of a Kerr black hole for direct applications to EMRI.
From the pioneering works by Regge and
Wheeler~\cite{T.Regge-J.A.Wheeler-1957} and
Zerilli~\cite{F.Zerilli-1970-PRL,F.Zerilli-1970-PRD,H.Nakano-2019},
there have been many studies on the perturbations in the Schwarzschild
background
spacetime~\cite{H.Nakano-2019,V.Moncrief-1974a,V.Moncrief-1974b,C.T.Cunningham-R.H.Price-V.Moncrief-1978,Chandrasekhar-1983,Gerlach_Sengupta-1979a,Gerlach_Sengupta-1979b,Gerlach_Sengupta-1979c,Gerlach_Sengupta-1980,T.Nakamura-K.Oohara-Y.Kojima-1987,Gundlach-Martine-Garcia-2000,Gundlach-Martine-Garcia-2001,A.Nagar-L.Rezzolla-2005-2006,K.Martel-E.Poisson-2005}.
They usually decompose the perturbations on the Schwarzschild
spacetime using the spherical harmonics $Y_{lm}$ and classify them
into odd- and even-modes based on their parity, because the
Schwarzschild spacetime has the spherical symmetry.
However, in the current situations, $l=0$ and $l=1$ modes should be
separately treated through a gauge-fixing
procedure~\cite{Gundlach-Martine-Garcia-2000,Gundlach-Martine-Garcia-2001,A.Nagar-L.Rezzolla-2005-2006,K.Martel-E.Poisson-2005}.
From the arguments in
Refs.~\cite{Gundlach-Martine-Garcia-2000,Gundlach-Martine-Garcia-2001,A.Nagar-L.Rezzolla-2005-2006,K.Martel-E.Poisson-2005},
it is the current consensus that the constructions of ``{\it
  gauge-invariant}'' variables for $l=0,1$ mode perturbations are
difficult if we keep the reconstruction of the original metric
perturbations in our mind.

%*************************************************************

On the other hand, toward unambiguous sophisticated {\it nonlinear}
general-relativistic perturbation theories, we have been developing
the general formulation of a higher-order gauge-invariant perturbation
theory on a generic background
spacetime~\cite{K.Nakamura-2003,K.Nakamura-2005,K.Nakamura-2011,K.Nakamura-IJMPD-2012,K.Nakamura-2013,K.Nakamura-2014}
and have been applying it to cosmological perturbations~\cite{kouchan-cosmo-second-2006,kouchan-cosmo-second-2007,K.Nakamura-2008,K.Nakamura-2009a,K.Nakamura-2009b,K.Nakamura-2010,A.J.Christopherson-K.A.Malik-D.R.Matravers-K.Nakamura-2011,K.Nakamura-2020}.
We review our framework of the linear gauge-invariant perturbation
theory on generic background
spacetime~\cite{K.Nakamura-2003,K.Nakamura-2005} in Sec.~\ref{sec:review-of-general-framework-GI-perturbation-theroy} of
this paper.
This framework can be easily extended to {\it higher-order}
perturbations, since the reconstruction of the original metric is trivial.
This framework starts from the distinction of the notions of the first-
and the second-kind gauges.
These two notions of gauges in perturbations are different from each
other and this distinction of the first- and second-kind gauges is
quite important to understand the development of perturbation theory
in this series of our papers.
We point out the fact that we often use the first-kind gauge
transformation when we predict or interpret the measurement results of
observations or experiments.
Since actual measurement results includes the information of the
detector directivity and the relative motion of the detector and
observational targets, we exclude these information using the
first-kind gauge transformation when we predict or interpret the
experimental results.
On the other hand, the second-kind gauge have nothing to do with the
nature of physical spacetime and the second-kind gauge should be
regarded as unphysical modes.
More details are described in
Sec.~\ref{sec:review-of-general-framework-GI-perturbation-theroy}.

%*************************************************************

The general framework of gauge-invariant perturbation theories
developed in Refs.~\cite{K.Nakamura-2003,K.Nakamura-2005,K.Nakamura-2011,K.Nakamura-IJMPD-2012,K.Nakamura-2013,K.Nakamura-2014}
is based on a conjecture
(Conjecture~\ref{conjecture:decomposition-conjecture} below), which
roughly states that {\it we already know the procedure to find
  gauge-invariant variables for linear-order metric perturbations.}
Throughout this series of papers and in Refs.~\cite{K.Nakamura-2003,K.Nakamura-2005,K.Nakamura-2011,K.Nakamura-IJMPD-2012,K.Nakamura-2013,K.Nakamura-2014},
we use the terminology ``gauge-invariant variables'' as the
variables in which the gauge-degree of freedom of the ``{\it second
  kind}'' are completely excluded, if there is no possibility of any
confusions.
Owing to Conjecture~\ref{conjecture:decomposition-conjecture}, the
reconstruction of the original metric from the gauge-invariant
variables is trivial.
A proof of Conjecture~\ref{conjecture:decomposition-conjecture} was
already discussed in
Ref.~\cite{K.Nakamura-2011,K.Nakamura-IJMPD-2012,K.Nakamura-2013}.
In this proof, we had to introduce some Green functions for some
elliptic derivative operators and ignored the kernel modes of these
elliptic derivative operators due to a technical reason.
We called these kernel modes ``{\it zero modes},'' and the treatment
of these zero modes remained unclear.
We also called the problem to find a gauge-invariant treatment of
these zero modes as the ``{\it zero-mode problem}.''
This zero-mode problem is the serious problem to be resolved when we
develop higher-order gauge-invariant perturbation theory, since
mode-coupling effects including the above ``zero modes'' occur in
higher-order perturbations.

%*************************************************************

In the case of the perturbations on the Schwarzschild background
spacetime, as we will see in Sec.~\ref{sec:spherical_background_case},
these ``{\it zero modes}'' correspond to the above $l=0,1$ modes.
The above conventional special treatments of $l=0,1$ modes in many
literature correspond to a partial gauge-fixing procedure.
If arguments are completed within the linear perturbations on a single
patch of the spacetime, this partial gauge-fixing procedure will be
harmless, because there is no mode-coupling in the linear perturbation
level.
However, from the viewpoint of the application of our
{\it higher-order} perturbation theory, the above special treatments of
these modes become an obstacle when we develop nonlinear perturbation
theory because the mode-couplings owing to the nonlinear effects
make the couplings between linear-order $l=0,1$ modes and other modes,
as mentioned above.
Actually, higher-order $l=0,1$ modes are also created due to the
mode-coupling owing to the nonlinear effects of Einstein equations~\cite{D.Brizuela-J.M.Martin-Garcia-G.A.Mena-Marugan-2007}.
Due to this mode-coupling, the special treatments by gauge-fixing for
the linear $l=0,1$ modes in many literature make the ``gauge
covariance'' of the higher-order perturbations unclear.
Moreover, in the EMRI case, we separate the whole spacetime of the
system into some regions and derive the perturbative solutions
including $l=0,1$ mode in each region at once, then we construct
global solutions through some matching method such as the matched
asymptotic expansion.
To exclude ``gauge-ambiguity'' in these matching, we have to carry out
these matching procedure under the ``same gauge.''
To guarantee that the matching procedure is under ``same gauge'', it
is convenient to discuss the perturbation theory in which ``gauge
covariance'' is manifest.
Since this ``gauge covariance'' is already manifest for $l\geq 2$
modes of the perturbations on the Schwarzschild spacetime in the
gauge-invariant perturbation theory, it is natural to hope that there
is a gauge-invariant treatment for $l=0,1$-modes perturbations in
spite of the current consensus mentioned above.
Thus, the finding of a gauge-invariant treatment of $l=0,1$ modes in
the perturbations on Schwarzschild background spacetime is not only a
resolution of the above technical zero-mode problem in a specific
background spacetime but also is quite physically crucial in the
arguments of EMRI.

%*************************************************************

This paper is the Part I paper of the series of full papers on the
application of our gauge-invariant perturbation theory on generic
background spacetime to that on the Schwarzschild background
spacetime, which is already reported in our short
papers~\cite{K.Nakamura-2021a,K.Nakamura-2021b}.
This series of papers is the full paper version of our short
paper~\cite{K.Nakamura-2021a}.
In this Part I paper, we propose a gauge-invariant treatment of the
$l=0,1$-mode perturbations on the Schwarzschild background spacetime
and show that Conjecture~\ref{conjecture:decomposition-conjecture} is
true even for these modes if we accept our proposal.
If we consider the mode decompositions for $l=0,1$ modes by the
spherical harmonic functions $Y_{lm}$, the vector and tensor harmonics
vanish for $l=0$ mode and the tensor harmonics vanish for $l=1$ mode.
This is the essential reason why we have to treat $l=0,1$ modes
separately in the conventional approaches as explained in
Sec.~\ref{sec:2+2-formulation-Perturbation-decomposition}.
The mode decomposition based on the conventional spherical harmonic
function $Y_{lm}$ corresponds to the imposition of the boundary
condition due to the restriction of the functions to $L^{2}$-space at
the starting point.
Due to this regular boundary condition at the starting point, vector
and tensor harmonics for $l=0$ modes and tensor harmonics for $l=1$
mode vanishes.
This requires the special treatments of $l=0,1$ modes in the
conventional approaches.
In Sec.~\ref{sec:gauge-inv-pertur-sphe-sym-spacetime}, we also
explained the explicit reason for the difficulties of the construction
of a gauge-invariant variables for $l=0,1$ modes through the
gauge-transformation rules of the metric perturbations.

%*************************************************************

In contrast with this conventional approaches, in our proposal, we
introduce singular harmonic functions at once to prepare the
nonvanishing vector and tensor harmonics for $l=0,1$ mode.
Owing to this introduction of the singular harmonic functions, we can
treat $l=0,1$ modes of perturbations in the similar manner to the
treatment of $l\geq 2$ modes in which the gauge-degree-of-freedom of
the second kind is completely excluded.
We can also construct the gauge-invariant variables for $l=0,1$-mode
perturbations in the similar manner to those of $l\geq 2$-modes
perturbations in which the reconstruction of the original metric from
the gauge-invariant variables is trivial.
This unified construction of gauge-invariant variables including
$l=0,1$ modes enable us to define gauge-invariant variables for
perturbations of any tensor fields of any-order in our higher-order
gauge-invariant perturbation theory~\cite{K.Nakamura-2003,K.Nakamura-2005,K.Nakamura-2011,K.Nakamura-IJMPD-2012,K.Nakamura-2013,K.Nakamura-2014,kouchan-cosmo-second-2006,kouchan-cosmo-second-2007,K.Nakamura-2008,K.Nakamura-2009a,K.Nakamura-2009b,K.Nakamura-2010,A.J.Christopherson-K.A.Malik-D.R.Matravers-K.Nakamura-2011,K.Nakamura-2020},
in which mode-couplings between $l=0,1$ modes and the other modes are
naturally included.
After the derivation of the linear-order Einstein equations in terms
of these gauge-invariant variables, we eliminate the introduced
singular harmonics by imposing the regularity of perturbations as the
boundary conditions.
This is the main scenario of our proposal in this paper.

%*************************************************************

In this paper, we show that we can resolve the above ``{\it zero-mode
  problem}'' if we accept the above proposal.
This resolution will be an important step of the development of the
higher-order gauge-invariant perturbation theory on the Schwarzschild
background spacetime which includes the analyses of EMRI.
In addition to the perturbation theory on a specific background
spacetime, this resolution will become a clue to the perturbation
theory on a generic background spacetime.
We note that we do not intend to insist that this proposal is the unique
resolution of the above ``{\it zero-mode problem}.''
However, in the series of our papers, we derive the solutions to the
linearized Einstein equation through our proposal and point out that
these solutions are physically reasonable.
In this Part I paper, we derive the odd-mode perturbative solutions
which are physically reasonable.
In the Part II paper~\cite{K.Nakamura-2021c}, we will discuss the
strategy to solve the even-mode perturbations following our
Proposal~\ref{proposal:treatment-proposal-on-pert-on-spherical-BG} and
derived their $l=0,1$-mode solutions.
Then, we show these solutions are physically reasonable.
Furthermore, in the Part III paper~\cite{K.Nakamura-2021d}, we will
discuss the realization of two exact solutions in terms of the linear
perturbations on the Schwarzschild background spacetime.
Owing to these supports, we may say that our proposal in this paper is
also physically reasonable.
A brief discussion on the extension to the higher-order perturbations
are already given in Ref.~\cite{K.Nakamura-2021b}.

%*************************************************************

The organization of this Part I paper is as follows.
In Sec.~\ref{sec:review-of-general-framework-GI-perturbation-theroy},
we briefly review the framework of the general-relativistic
gauge-invariant perturbation theory within the linear perturbation
theory, as mentioned above.
This framework can be easily extended to higher-order
perturbations~\cite{K.Nakamura-2003,K.Nakamura-2005,K.Nakamura-2011,K.Nakamura-IJMPD-2012,K.Nakamura-2013,K.Nakamura-2014},
since the reconstruction of the original metric is trivial through the
Conjecture~\ref{conjecture:decomposition-conjecture}.
In this
Sec.~\ref{sec:review-of-general-framework-GI-perturbation-theroy}, we
emphasize that the distinction of the first-kind gauge and the
second-kind gauge is an important premise of our gauge-invariant
perturbation theory.
In Sec.~\ref{sec:spherical_background_case}, we explain the situation
in many studies why the special treatments of $l=0,1$ modes are
required.
Then, we propose a strategy for gauge-invariant treatments of $l=0,1$
modes.
In Sec.~\ref{sec:gauge-inv-pertur-sphe-sym-spacetime}, we construct
gauge-invariant variables including $l=0,1$ modes through the proposal
described in Sec.~\ref{sec:spherical_background_case}.
This is a proof of
Conjecture~\ref{conjecture:decomposition-conjecture} for all modes
of perturbations, $l\geq 0$, on the background spacetimes with
spherical symmetry.
In Sec.~\ref{sec:Einstein_equations}, we derive the Einstein equations
for any mode perturbations following the proposal in
Sec.~\ref{sec:spherical_background_case}.
In
Sec.~\ref{sec:Schwarzschild_Background-non-vaccum-odd-Nakano-treatment},
we show the strategy to solve the odd-mode perturbations and derive
the explicit solutions for $l=0,1$ mode perturbations through the
component treatment of gauge-invariant variables in the Einstein
equations derived in Sec.~\ref{sec:Einstein_equations}.
The final section~\ref{sec:summary_and_discussions} is devoted to the
summary and discussions within this Part I paper.

%*************************************************************

Throughout this paper, we use the unit $G=c=1$, where $G$ is Newton's
constant of gravitation, and $c$ is the velocity of light.

%*************************************************************

%%%%%%%%%%%%%%%%%%%%%%%%%%%%%%%%%%%%%%%%%%%%%%%%%%%%%%
%%%%%%%%%%%%%%%%%%%%%%%%%%%%%%%%%%%%%%%%%%%%%%%%%%%%%%
%%%%%%%%%%%%%%%%%%%%%%%%%%%%%%%%%%%%%%%%%%%%%%%%%%%%%%
%%%%%%%%%%%%%%%%%%%%%%%%%%%%%%%%%%%%%%%%%%%%%%%%%%%%%%
\section{Review of our general-relativistic gauge-invariant perturbation theory}
\label{sec:review-of-general-framework-GI-perturbation-theroy}
%%%%%%%%%%%%%%%%%%%%%%%%%%%%%%%%%%%%%%%%%%%%%%%%%%%%%%
%%%%%%%%%%%%%%%%%%%%%%%%%%%%%%%%%%%%%%%%%%%%%%%%%%%%%%
%%%%%%%%%%%%%%%%%%%%%%%%%%%%%%%%%%%%%%%%%%%%%%%%%%%%%%
%%%%%%%%%%%%%%%%%%%%%%%%%%%%%%%%%%%%%%%%%%%%%%%%%%%%%%

%*************************************************************

In this section, we briefly review our general framework of the
gauge-invariant perturbation
theory~\cite{K.Nakamura-2003,K.Nakamura-2005}.
Although the main purpose of the framework of the gauge-invariant
perturbation theory developed in
Refs.~\cite{K.Nakamura-2003,K.Nakamura-2005} is the extension to the
higher-order perturbation theory, in this review, we concentrate only
on the linear perturbations.
This is because we treat only the linear perturbations within this
paper.
Since we want to explain the gauge-invariant perturbation theory in
general relativity, first of all, we have to explain the notions of
``gauges'' in general
relativity~\cite{K.Nakamura-2008,K.Nakamura-2010,K.Nakamura-2020}.

%****************************************************************

General relativity is a theory with general covariance.
This general covariance intuitively states that there is no preferred
coordinate system in nature.
This general covariance also introduces the notion of ``gauge'' in the
theory.
In the theory with general covariance, these ``gauges'' give rise to
the unphysical degree of freedom and we have to fix the ``gauges'' or
to extract some invariant quantities to obtain physical results.
Therefore, treatments of ``gauges'' are crucial in general relativity
and this situation becomes more delicate in general relativistic
perturbation theories.

%****************************************************************

In 1964, Sachs~\cite{R.K.Sachs-1964} pointed out that there are two
kinds of ``gauges'' in general relativity.
Sachs called these two ``gauges'' as the first- and the second-kind
gauges, respectively.
Here, we review these concepts of ``gauge,'' which are different from
each other.
Furthermore, the distinction of these ``gauges'' is important to
understand the results of this paper and
papers~\cite{K.Nakamura-2021c,K.Nakamura-2021d}.

%****************************************************************

In Sec.~\ref{sec:The-first-kind-gauge}, we first explain the notion of
the first kind gauge.
Second, we explain the notion of the second-kind gauge in
Sec.~\ref{sec:The-second-kind-gauge}.
We expect that the reader can distinguish these two different notions
of gauges in general relativity through these explanations.
Then, we review our general framework of the general-relativistic
gauge-invariant perturbation theory on generic background spacetimes
in Sec.~\ref{sec:The_general-relativistic_gauge-invariant_linear_perturbation_theory}.
We have to emphasize that the aim of our general formulation of
general-relativistic gauge-invariant perturbation theory is to exclude
the degree of freedom of the second-kind gauge, completely.

%****************************************************************

%%%%%%%%%%%%%%%%%%%%%%%%%%%%%%%%%%%%%%%%%%%%%%%%%%%%%%
%%%%%%%%%%%%%%%%%%%%%%%%%%%%%%%%%%%%%%%%%%%%%%%%%%%%%%
%%%%%%%%%%%%%%%%%%%%%%%%%%%%%%%%%%%%%%%%%%%%%%%%%%%%%%
\subsection{First kind gauge}
\label{sec:The-first-kind-gauge}
%%%%%%%%%%%%%%%%%%%%%%%%%%%%%%%%%%%%%%%%%%%%%%%%%%%%%%
%%%%%%%%%%%%%%%%%%%%%%%%%%%%%%%%%%%%%%%%%%%%%%%%%%%%%%
%%%%%%%%%%%%%%%%%%%%%%%%%%%%%%%%%%%%%%%%%%%%%%%%%%%%%%

%****************************************************************

{\it The first kind gauge} is a coordinate system on a single manifold
$\ScrM$.
This first kind gauge is not the ``gauge'' of our ``gauge-invariant
perturbation theory.''
However, we have to explain this first kind gauge to distinguish the
notions of the first-kind gauge and the second-kind gauge, as
emphasized above.

%****************************************************************

\begin{figure}
  \begin{center}
    \includegraphics[width=0.98\textwidth]{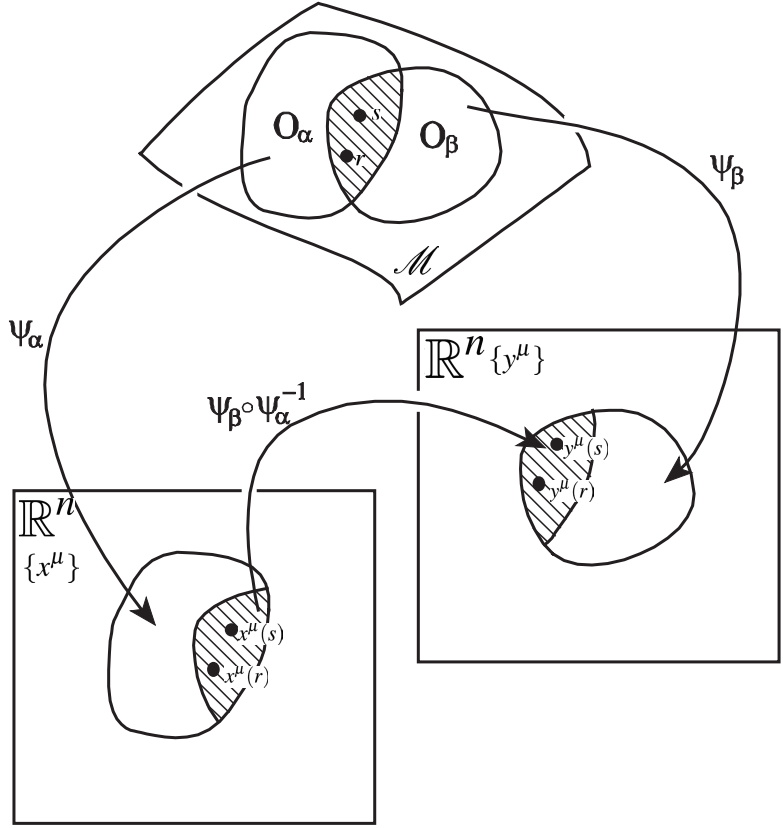}
  \end{center}
  \caption{
    The first kind gauge is a coordinate system of a single manifold.
    The points $r$ and $s$ and its coordinates
    $\{x^{\mu}(s),x^{\mu}(r)\}$ and $\{y^{\mu}(s),y^{\mu}(r)\}$ are
    used in the explanations at the paragraph of
    Eq.~(\ref{eq:tensor-pull-back-first-kind-gauge-def}).
  }
  \label{fig:first-kind-gauge}
\end{figure}

%****************************************************************

In standard textbooks of manifolds (for example,
see~\cite{Kobayashi-Nomizu-I-1996}), the following property of a
manifold is written, ``On a manifold, we can always introduce a
coordinate system as a diffeomorphism $\psi_{\alpha}$ from an open set
$O_{\alpha}\subset\ScrM$ to an open set
$\psi_{\alpha}(O_{\alpha})\subset\RF^{n}$ ($n=\dim\ScrM$).''
This diffeomorphism $\psi_{\alpha}$, i.e., coordinate system of the
open set $O_{\alpha}$, is called {\it gauge choice} (of the first
kind).
If we consider another open set in $O_{\beta}\subset\ScrM$, we have
another gauge choice
$\psi_{\beta}:O_{\beta}\mapsto\psi_{\beta}(O_{\beta})\subset\RF^{n}$
for $O_{\beta}$.
If these two open sets $O_{\alpha}$ and $O_{\beta}$ have the
intersection $O_{\alpha}\cap O_{\beta}\neq\emptyset$, we can consider
the diffeomorphism $\psi_{\beta}\circ\psi_{\alpha}^{-1}$.
This diffeomorphism $\psi_{\beta}\circ\psi_{\alpha}^{-1}$ is just a
coordinate transformation: $\psi_{\alpha}(O_{\alpha}\cap
O_{\beta})\subset\RF^{n}\mapsto \psi_{\beta}(O_{\alpha}\cap
O_{\beta})\subset\RF^{n}$, which is called {\it gauge transformation}
(of the first kind) in general relativity.
These are depicted in Fig.~\ref{fig:first-kind-gauge} which is a
famous figure in many textbooks of the theory of manifolds.

%****************************************************************

According to the theory of manifolds, coordinate systems are not on a
manifold itself, but we can always introduce a coordinate system
as a map from an open set on the manifold $\ScrM$ to an open set of
$\RF^{n}$.
Furthermore, we may choose an different coordinate system through the
different map from an open set in the manifold $\ScrM$ to an open
set of $\RF^{n}$.
We can always change the coordinate system as we want.
This is a realization of the statement of the general covariance that
``there is no preferred coordinate system in nature.''
For this reason, general covariance in general relativity is
automatically included in the premise that our spacetime is regarded
as a single manifold.
The first kind gauge does arise due to this general covariance.
The gauge issue of the first kind is usually represented by the question,
``Which coordinate system is convenient?''
The answer to this question depends on the problem which we are
addressing, i.e., what we want to clarify.
In some cases, this gauge issue of the first kind is important.
On the other hand, in many cases, this gauge issue becomes harmless if
we apply a covariant theory on the manifold.

%****************************************************************

We also note that the fact that we often use this first-kind gauge
transformation when we predict or interpret the measurement results in
observations and experiments as mentioned in
Sec.~\ref{sec:introduction}.
In general, directly measured results in observations or experiments
include the information of the detector directivity and the relative
motion of the detector and observational targets.
When we predict or interpret the results of these directly-measured
results, we have to take into account of these information of our
detectors.

%****************************************************************

One of typical examples is the dipole mode in the fluctuations of the
cosmic microwave background (CMB).
It is well-known that the dipole mode of CMB is actually detected by
the detectors.
Usually, this detected dipole mode is interpreted as the relative
motion of the detector against the last scattering surface of the
universe.
Then, this detected dipole mode is regarded as unimportant detected
data when we want to discuss the primordial fluctuations in CMB which
are generated in the early history of universe.
Regarding the reason of the detection of these dipole fluctuations in
CMB is the proper motion of the detector against the last scattering
surface, we use the coordinate transformation to eliminate our
relative motion of the detector against the last scattering surface so
that the dipole fluctuations disappear.
This coordinate transformation is a typical example of the first-kind
gauge transformation.
We can also give the inclination of rotating star or a binary system
and the antenna pattern function of interferometric gravitational-wave
detectors as examples of the first-kind gauges.

%****************************************************************

The final example of the first-kind gauge transformation is the most
important one for general relativistic perturbation theories.
This is the identification of the actual replacement of points within
the single manifold $\ScrM$ with an infinitesimal coordinate
transformation~\cite{Landau:1975pou}.
To explain this, we consider the replacement of a points
$r\in\ScrM$ to the other point $s\in\ScrM$ in a neighborhood $r$.
This replacement $r\mapsto s$ is represented by a diffeomorphism
$\Psi_{\lambda}:\ScrM\rightarrow\ScrM$ as
$s=\Psi_{\lambda}(r)$, where $\lambda$ is an infinitesimal parameter
satisfying $\Psi_{\lambda=0}(r)=r$.
The pullback $\Psi_{\lambda}^{*}$ of any tensor field $Q$ on
$\ScrM$ is given by
\begin{eqnarray}
  \label{eq:tensor-pull-back-first-kind-gauge-def}
  Q(s) = (\Psi_{\lambda}^{*}Q)(r) = Q(r) +
  \lambda\left.{\pounds}_{\xi}Q \right|_{\lambda=0} + O(\lambda^{2}),
\end{eqnarray}
where $\xi^{a}$ is the generator of the pull-back
$\Psi_{\lambda}^{*}$ and a vector field on the tangent space of
$\ScrM$.
We consider this expression
(\ref{eq:tensor-pull-back-first-kind-gauge-def}) by a coordinate
transformation.
To see this, we introduce the coordinate system
$\{O_{\alpha},\psi_{\alpha}\}$ on $\ScrM$ as above and assume that
$r,s\in O_{\alpha}\cap O_{\beta}\neq\emptyset$ as in
Fig.~\ref{fig:first-kind-gauge}.
Here, we denote the coordinates $\psi_{\alpha}$ $:$
$O_{\alpha}\subset \ScrM$ $\mapsto$ $\RF^{n}(\{x^{\mu}\})$ and
$\psi_{\beta}$ $:$ $O_{\beta}\subset \ScrM$ $\mapsto$
$\RF^{n}(\{y^{\mu}\})$.
Through these coordinate systems, we can assign the coordinate labels
$(x^{\mu}(r),x^{\mu}(s))$ $\in$ $\RF^{n}(\{x^{\mu}\})$ and
$(y^{\mu}(r),y^{\mu}(s))$ $\in$ $\RF^{n}(\{y^{\mu}\})$ for the points
$r$ and $s$ as in Fig.~\ref{fig:first-kind-gauge}.
When the variable $Q$ is the coordinate function $x^{\mu}$ associated
with the chart $\psi_{\alpha}$, we obtain $x^{\mu}(s)$ $=$
$x^{\mu}(r)$ $+$ $\lambda \xi^{\mu}(r)$ $+$ $O(\lambda^{2})$.
Now, we consider the coordinate transformation
$\psi_{\beta}\circ\psi_{\alpha}^{-1}$ so that
$y^{\mu}(s) := x^{\mu}(s)$ and
we have the relation between the different coordinates as
\begin{eqnarray}
  \label{eq:infinitesimal-coordinate-trans-first-kind}
  y^{\mu}(s) := x^{\mu}(r) + \lambda \xi^{\mu}(r) + O(\lambda^{2}).
\end{eqnarray}
As an example of tensor field, we consider the metric $g_{ab}$ on
$\ScrM$.
Under the infinitesimal coordinate transformation
(\ref{eq:infinitesimal-coordinate-trans-first-kind}), the metric at
the point $s$ is given by
\begin{eqnarray}
  g_{ab}(s)
  &=&
      \left.g_{\mu\nu}(x(s)) (dx^{\mu})_{a} (dx^{\nu})_{b} \right|_{s}
      =
      \left.g_{\mu\nu}(y(s)) (dy^{\mu})_{a} (dy^{\nu})_{b} \right|_{s}
      \nonumber\\
  &=&
      \left.g_{\mu\nu}\left(x(r)+\lambda\xi(r)+O(\lambda^{2})\right)
      \frac{\partial y^{\mu}}{\partial x^{\rho}}
      \frac{\partial y^{\nu}}{\partial x^{\sigma}}
      (dx^{\rho})_{a}
      (dx^{\sigma})_{b}
      \right|_{r}
      \nonumber\\
  &=&
      g_{ab}(r)
      +
      \lambda
      \left.
      \left(
      \xi^{\tau}
      \partial_{\tau}g_{\rho\sigma}
      +
      g_{\mu\sigma}
      \partial_{\rho}\xi^{\mu}
      +
      g_{\rho\nu}
      \partial_{\sigma}\xi^{\mu}
      \right)
      (dx^{\rho})_{a}
      (dx^{\sigma})_{b}
      \right|_{r}
      +
      O(\lambda^{2})
      \nonumber\\
  &=&
      g_{ab}(r)
      +
      \lambda
      \left.
      {\pounds}_{\xi}g_{ab}
      \right|_{r}
      +
      O(\lambda^{2})
      .
      \label{eq:infinitesimal-coordinate-trans-first-kind-metric}
\end{eqnarray}
Because of $g_{ab}(s)=\Psi_{\lambda}^{*}g_{ab}(r)$,
Eq.~(\ref{eq:infinitesimal-coordinate-trans-first-kind-metric}) is
usually written as
\begin{eqnarray}
  (\Psi_{\lambda}^{*}g_{ab})(r)
  =
  g_{ab}(r)
  +
  \lambda
  \left.
  {\pounds}_{\xi}g_{ab}
  \right|_{r}
  +
  O(\lambda^{2})
  .
  \label{eq:infinitesimal-coordinate-trans-first-kind-metric-Lie-def}
\end{eqnarray}
This is just the definition of the Lie derivative and the realization of
Eq.~(\ref{eq:tensor-pull-back-first-kind-gauge-def}) itself~\footnote{
  In the derivation of the Lie derivative in \S~94 of
  Ref.~\cite{Landau:1975pou}, the coordinate transformation
  ${x'}^{i}=x^{i}+\xi^{i}$ is performed, at first, and the comparison
  inverse metrices ${g'}^{ik}({x'}^{l})$ and $g^{ik}(x^{l})$ at
  the ``same coordinate value'' $x^{l}$ is carried out.
  The comparison at the ``same coordinate value'' $x^{l}$ under the
  coordinate transformation ${x'}^{i}=x^{i}+\xi^{i}$ means the
  comparison the inverse metrics at the ``different points'' on the
  same manifold as shown in
  Eq.~(\ref{eq:infinitesimal-coordinate-trans-first-kind-metric}).
}.
From the action of the coordinate transformation
(\ref{eq:infinitesimal-coordinate-trans-first-kind}), the coordinate
transformation should be regarded as the action of the diffeomorphism
\begin{eqnarray}
  \label{eq:psibeta-Psizeta-psialphainv}
  \psi_{\beta}\circ\Psi_{\lambda}\circ\psi_{\alpha}^{-1}
\end{eqnarray}
rather than the simple coordinate transformation
$\psi_{\beta}\circ\psi_{\alpha}^{-1}$.
However, in our perturbation theory, we also regard the infinitesimal
coordinate transformation
(\ref{eq:infinitesimal-coordinate-trans-first-kind}) is the first-kind
gauge transformation, since the above arguments are restricted within
a single manifold $\ScrM$.
Namely, the Taylor expansion through the infinitesimal parameter
$\lambda$ is to the tangential direction within the manifold
$\ScrM$.

%****************************************************************

We may write the metric $g_{ab}$ as $g_{ab}$ $=$ ${}^{(0)}\!g_{ab}$ $+$
$\lambda$ $h_{ab}$ $+$ $O(\lambda^{2})$ within $\ScrM$.
We emphasize that the direction of this Taylor expansion through the
infinitesimal parameter $\lambda$ is still ``tangential'' to  $\ScrM$.
In this case,
Eq.~(\ref{eq:infinitesimal-coordinate-trans-first-kind-metric}) yields
\begin{eqnarray}
  {}^{(0)}\!g_{ab}(s)
  +
  \lambda
  h_{ab}(s)
  =
  {}^{(0)}\!g_{ab}(r)
  +
  \lambda
  \left(
  h_{ab}(r)
  +
  \left.
  {\pounds}_{\xi}{}^{(0)}\!g_{ab}
  \right|_{r}
  \right)
  +
  O(\lambda^{2})
  .
  \label{eq:infinitesimal-coordinate-trans-first-kind-metric-pert}
\end{eqnarray}
In many literature, arguments start from the infinitesimal coordinate
transformation (\ref{eq:infinitesimal-coordinate-trans-first-kind})
and reach to the conclusion
(\ref{eq:infinitesimal-coordinate-trans-first-kind-metric-pert}).
For this reason, the term of Lie derivative of the background metric
in the right-hand side in
Eq.~(\ref{eq:infinitesimal-coordinate-trans-first-kind-metric}) is
understood as the ``degree of freedom of coordinate transformations''
and it is ``unphysical degree of freedom'', in many literature.
However, the appearance of the Lie derivative of the background metric
in Eq.~(\ref{eq:infinitesimal-coordinate-trans-first-kind-metric-pert})
is just due the change of the reference point within the single
manifold $\ScrM$ and this situation is same as the above example of
CMB dipole measurement.
For this reason, we regard this example as the appearance of the
first-kind gauge.
This example appears when we interpret our results in
Sec.~\ref{sec:Schwarzschild_Background-non-vaccum-odd-Nakano-treatment}
of this paper.

%****************************************************************

We will be able to find many other examples of the first-kind gauges.
All of these are interpreted as the changes of reference point within
the single manifold.
In some case, these change of reference point within the single
manifold included in the measurement results in observations and
experiments in some case.
For this reason, we do not regard this above first-kind gauge is
``unphysical degree of freedom''.
On the other hand, the second-kind gauge which is explained in
Sec.~\ref{sec:The-second-kind-gauge} have nothing to do with our
physical spacetime but are included in perturbative variables as
explained below.
We have to emphasize that the this second-kind gauge is the
``unphysical degree of freedom'' which should be excluded  in general
relativistic perturbation theory.

%****************************************************************

%%%%%%%%%%%%%%%%%%%%%%%%%%%%%%%%%%%%%%%%%%%%%%%%%%%%%%
%%%%%%%%%%%%%%%%%%%%%%%%%%%%%%%%%%%%%%%%%%%%%%%%%%%%%%
%%%%%%%%%%%%%%%%%%%%%%%%%%%%%%%%%%%%%%%%%%%%%%%%%%%%%%
\subsection{Second kind gauge}
\label{sec:The-second-kind-gauge}
%%%%%%%%%%%%%%%%%%%%%%%%%%%%%%%%%%%%%%%%%%%%%%%%%%%%%%
%%%%%%%%%%%%%%%%%%%%%%%%%%%%%%%%%%%%%%%%%%%%%%%%%%%%%%
%%%%%%%%%%%%%%%%%%%%%%%%%%%%%%%%%%%%%%%%%%%%%%%%%%%%%%

%****************************************************************

{\it The second kind gauge} appears in perturbation theories in a
theory with general covariance.
To explain this, we have to remind what we are doing in perturbation
theories.

%****************************************************************

First, in any perturbation theories, we always treat two spacetime
manifolds.
One is the {\it physical spacetime} $\ScrM_{{\rm ph}}$.
We want to describe the properties of this physical spacetime
$\ScrM_{{\rm ph}}$ through perturbative analyses.
This physical spacetime $\ScrM_{{\rm ph}}$ is usually identified
with our nature itself.
The other is the {\it background spacetime} $\ScrM$.
This background spacetime has nothing to do with our nature and
is a fictitious manifold which is introduced as a reference to carry
out perturbative analyses by us.
We emphasize that these two spacetime manifolds $\ScrM_{{\rm ph}}$
and $\ScrM$ are distinct.
Let us denote the physical spacetime by
$(\ScrM_{{\rm ph}},\bar{g}_{ab})$ and the background spacetime by
$(\ScrM,g_{ab})$, where $\bar{g}_{ab}$ is the metric on the
physical spacetime manifold, $\ScrM_{{\rm ph}}$, and $g_{ab}$ is
the metric on the background spacetime manifold, $\ScrM$.
Further, we formally denote the spacetime metric and the other
physical tensor fields on $\ScrM_{{\rm ph}}$ by $Q$ and its
background value on $\ScrM$ by $Q_{0}$.

%****************************************************************

\begin{figure}
  \begin{center}
    \includegraphics[width=0.9\textwidth]{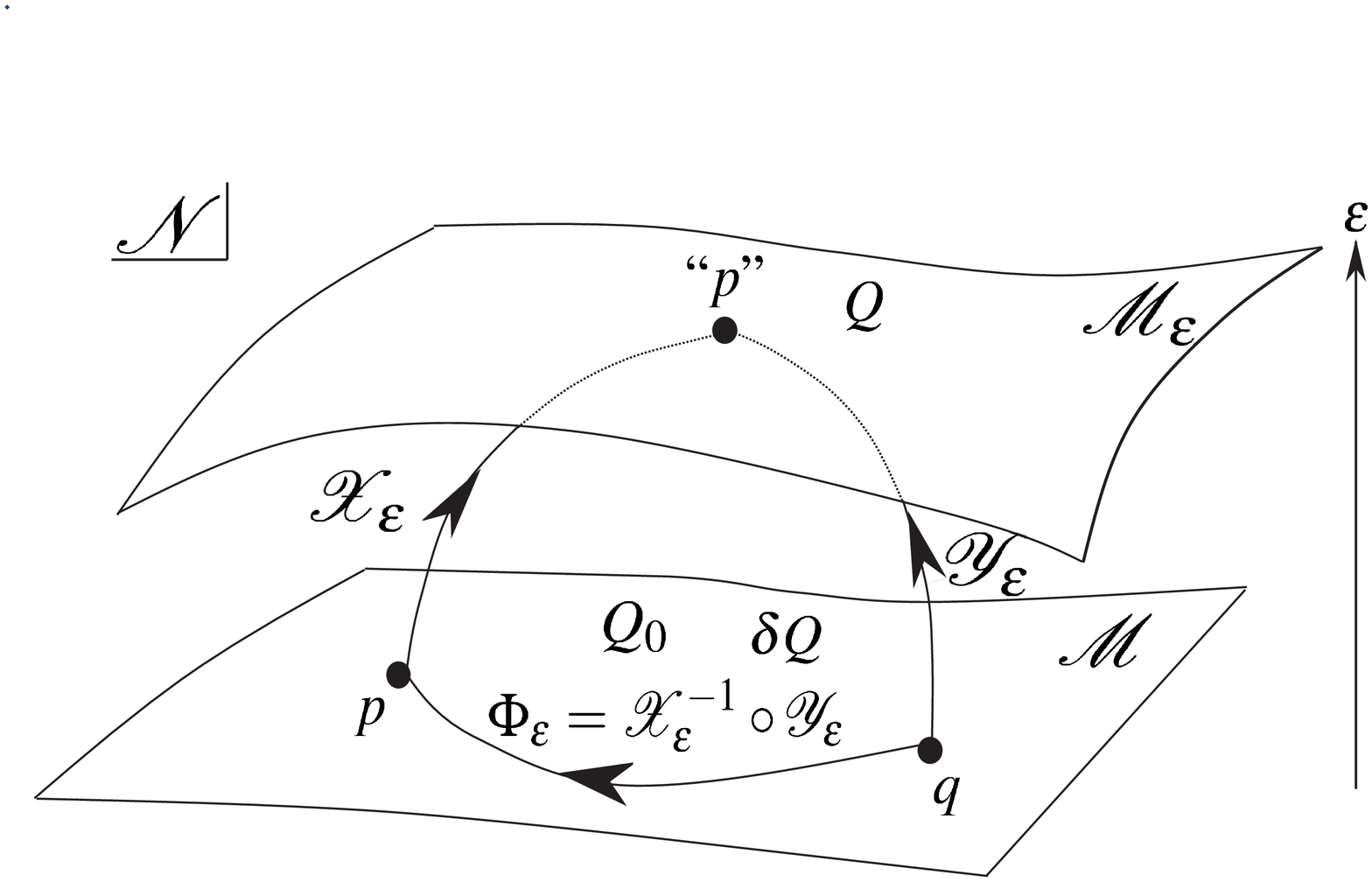}
  \end{center}
  \caption{
    The second kind gauge is a point-identification between the
    physical spacetime $\ScrM_{{\rm ph}}=\ScrM_{\epsilon}$ and
    the background spacetime $\ScrM$ on the extended manifold
    $\ScrN$.
    Through Eq.~(\ref{eq:variable-symbolic-perturbation}), we
    implicitly assume the existence of a point-identification map
    between $\ScrM_{\epsilon}$ and $\ScrM$.
    However, this point-identification is not unique by virtue of the
    general covariance in the theory.
    We may choose the gauge of the second kind so that
    $p\in\ScrM$ and ``$p$''$\in\ScrM_{\epsilon}$ is same
    ($\ScrX_{\epsilon}$).
    We may also choose the gauge so that $q\in\ScrM_{0}$ and
    ``$p$''$\in\ScrM_{\epsilon}$ is same ($\ScrY_{\epsilon}$).
    These are different gauge choices.
    The gauge transformation
    $\ScrX_{\epsilon}\rightarrow\ScrY_{\epsilon}$ is given
    by the diffeomorphism
    $\Phi_{\epsilon}=\ScrX_{\epsilon}^{-1}\circ\ScrY_{\epsilon}$.
  }
  \label{fig:gauge-choice-is-point-identification}
\end{figure}

%****************************************************************

Second, in any perturbation theory, we always write equations for the
perturbation of the variable $Q$ as follows:
\begin{equation}
  \label{eq:variable-symbolic-perturbation}
  Q(``p\mbox{''}) = Q_{0}(p) + \delta Q(p).
\end{equation}
Equation (\ref{eq:variable-symbolic-perturbation}) gives a
relation between variables on different manifolds.
Actually, $Q(``p\mbox{''})$ in
Eq.~(\ref{eq:variable-symbolic-perturbation}) is a variable on
$\ScrM_{\rm ph}$, whereas $Q_{0}(p)$ and $\delta Q(p)$ are variables
on $\ScrM$.
Because we regard Eq.~(\ref{eq:variable-symbolic-perturbation}) as
a field equation, Eq.~(\ref{eq:variable-symbolic-perturbation})
includes an implicit assumption of the existence of a point
identification map $\ScrM\rightarrow\ScrM_{\rm ph}$ $:$
$p\in\ScrM\mapsto ``p\mbox{''}\in\ScrM_{\rm ph}$.
This identification map is a {\it gauge choice} in
general-relativistic perturbation theories (see
Fig.~\ref{fig:gauge-choice-is-point-identification}).
This is the notion of the {\it second-kind gauge} pointed out by
Sachs~\cite{R.K.Sachs-1964}.
Note that this second-kind gauge is a different notion from the
degree of freedom of the coordinate transformation on the single
manifold which is explained in Sec.~\ref{sec:The-first-kind-gauge}.

%****************************************************************

To develop this understanding of the ``gauge of the second kind,'' we
introduce an infinitesimal parameter $\epsilon$ for perturbations and
$4+1$-dimensional manifold $\ScrN=\ScrM_{{\rm ph}}\times\RF$
($4=\dim\ScrM$) such that
$\ScrM=\left.\ScrN\right|_{\epsilon=0}$ and
$\ScrM_{{\rm ph}}=\ScrM_{\epsilon}=\left.\ScrN\right|_{\RF=\epsilon}$.
On $\ScrN$, the point-identification choice is regarded as a
diffeomorphism $\ScrX_{\epsilon}:\ScrN\rightarrow\ScrN$ such
that $\ScrX_{\epsilon}:\ScrM\rightarrow\ScrM_{\epsilon}$.
This point-identification is a gauge choice of the second
kind~\cite{R.K.Sachs-1964,J.M.Stewart-M.Walker-1974,J.M.Stewart-M.Walker-1990,J.M.Stewart-1990,K.Nakamura-2008,K.Nakamura-2010,K.Nakamura-2020}.
Furthermore, we introduce a gauge choice $\ScrX_{\epsilon}$ as an
exponential map with a generator ${}^{\ScrX}\!\eta^{a}$,
which is chosen such that its integral curve in $\ScrN$ is
transverse to each $\ScrM_{\epsilon}$ everywhere on
$\ScrN$.
Points lying on the same integral curve are regarded as the
``same point'' by the gauge choice $\ScrX_{\epsilon}$.
Note that the action of $\ScrX_{\epsilon}$ is transverse to each
$\ScrM_{\epsilon}$.

%****************************************************************

The first-order perturbation of the variable $Q$ on
$\ScrM_{\epsilon}$ is defined as the pulled-back
$\ScrX_{\epsilon}^{*}Q$ on $\ScrM$, which is induced by
$\ScrX_{\epsilon}$, and is expanded as
\begin{eqnarray}
  \ScrX_{\epsilon}^{*}Q
  =
  Q_{0}
  + \epsilon \left.{\pounds}_{{}^{\ScrX}\!\eta}Q\right|_{\ScrM}
  + O(\epsilon^{2}),
  \label{eq:perturbative-expansion-of-Q-def}
\end{eqnarray}
where $Q_{0}=\left.Q\right|_{\ScrM}$ is the background value of
$Q$ and all terms in Eq.~(\ref{eq:perturbative-expansion-of-Q-def})
are evaluated on the background spacetime $\ScrM$.
Because Eq.~(\ref{eq:perturbative-expansion-of-Q-def}) is the
perturbative expansion of $\ScrX^{*}_{\epsilon}Q_{\epsilon}$,
the first-order perturbation of $Q$ is given by
${}^{(1)}_{\ScrX}\!Q:=\left.{\pounds}_{{}^{\ScrX}\!\eta}Q\right|_{\ScrM}$.

%****************************************************************

When we have two gauge choices $\ScrX_{\epsilon}$ and
$\ScrY_{\epsilon}$ with the generators ${}^{\ScrX}\!\eta^{a}$
and ${}^{\ScrY}\!\eta^{a}$, respectively, and when these
generators have different tangential components
to each $\ScrM_{\epsilon}$, $\ScrX_{\epsilon}$ and
$\ScrY_{\epsilon}$ are regarded as {\it different gauge choices}.
A {\it gauge-transformation} is regarded as the change of the
point-identification
$\ScrX_{\epsilon}\rightarrow\ScrY_{\epsilon}$, which is given by
the diffeomorphism $\Phi_{\epsilon}$ $:=$
$\left(\ScrX_{\epsilon}\right)^{-1}\circ\ScrY_{\epsilon}$
$:$ $\ScrM$ $\rightarrow$ $\ScrM$.
The diffeomorphism $\Phi_{\epsilon}$ does change the
point-identification.
Here, $\Phi_{\epsilon}$ induces a pull-back from the representation
$\ScrX_{\epsilon}^{*}\!Q_{\epsilon}$ to the representation
$\ScrY_{\epsilon}^{*}\!Q_{\epsilon}$ as~\footnote{
  As depicted in Fig.~\ref{fig:gauge-choice-is-point-identification},
  the action of the diffeomorphism $\Phi_{\epsilon}$ $:=$
  $\ScrX_{\epsilon}^{-1}\circ\ScrY_{\epsilon}$ is the
  replacement of $\Phi_{\epsilon}(q)$ $=$ $p$.
  However, the evaluations of the both-side of
  Eq.~(\ref{eq:Bruni-46-one-pre}) are carry out
  at the same point on the background spacetime $\ScrM$ and
  Eq.~(\ref{eq:Bruni-47-one}) is also evaluated at the same point on
  the background spacetime $\ScrM$ as the result, while
  Eq.~(\ref{eq:infinitesimal-coordinate-trans-first-kind-metric-pert})
  represents the difference between the tensor field at different
  points on the same manifold.
  To explain this, we consider the points
  $``p\mbox{''}\in\ScrM_{ph}$, $``q\mbox{''}\in\ScrM_{ph}$
  ($``p\mbox{''}\neq``q\mbox{''}$),  and $q\in\ScrM$ and the action
  of the diffeomorphisms $\ScrY_{\epsilon}$, and $\ScrX_{\epsilon}$ so that
  $``p\mbox{''}=\ScrY_{\epsilon}(q)$ and
  $``q\mbox{''}=\ScrX_{\epsilon}(q)$.
  Through this setup, Eq.~(\ref{eq:Bruni-46-one-pre}) derived as
  \begin{eqnarray}
    Q(``p\mbox{''})
    &=&
        Q(\ScrY_{\epsilon}(q))
        \nonumber\\
    &=&
        \ScrY_{\epsilon}^{*}Q(q)
        =
        \ScrY_{\epsilon}^{*}Q(
        \ScrX_{\epsilon}^{-1}(``q\mbox{''})
        )
        =
        \ScrY_{\epsilon}^{*}\circ
        (\ScrX_{\epsilon}^{-1})^{*}Q(``q\mbox{''})
        =
        \ScrY_{\epsilon}^{*}\circ
        (\ScrX_{\epsilon}^{-1})^{*}Q(\ScrX_{\epsilon}(q))
        \nonumber\\
    &=&
        \ScrY_{\epsilon}^{*}\circ
        (\ScrX_{\epsilon}^{-1})^{*}\circ
        \ScrX_{\epsilon}^{*}Q(q)
        =
        \left(
        \ScrX_{\epsilon}^{-1}
        \circ
        \ScrY_{\epsilon}
        \right)^{*}\circ
        \ScrX_{\epsilon}^{*}Q(q)
        \nonumber\\
    &=&
        \Phi_{\epsilon}^{*}
        \ScrX_{\epsilon}^{*}Q(q)
        .
        \label{eq:Bruni-46-one-pre-derivation}
  \end{eqnarray}
  Then, through Eqs.~(\ref{eq:perturbative-expansion-of-Q-def}) and
  (\ref{eq:Bruni-46-one}), we reach to the gauge-transformation rule
  (\ref{eq:Bruni-47-one}) at the same point, which should be regarded
  as ${}^{(1)}_{\;\ScrY}\!Q(q)$ $-$ ${}^{(1)}_{\;\ScrX}\!Q(q)$
  $=$ ${\pounds}_{\xi_{(1)}}Q_{0}(q)$.
}
\begin{eqnarray}
  \label{eq:Bruni-46-one-pre}
  \ScrY_{\epsilon}^{*}\!Q_{\epsilon} = \Phi_{\epsilon}^{*}\ScrX_{\epsilon}^{*}\!Q_{\epsilon}.
\end{eqnarray}
From general arguments of the Taylor
expansion~\cite{M.Bruni-S.Matarrese-S.Mollerach-S.Sonego-1997}, the
pull-back $\Phi_{\epsilon}^{*}$ is expanded as
\begin{eqnarray}
  \ScrY_{\epsilon}^{*}\!Q_{\epsilon}
  =
  \ScrX_{\epsilon}^{*}\!Q_{\epsilon}
  + \epsilon {\pounds}_{\xi_{(1)}} \ScrX_{\epsilon}^{*}\!Q_{\epsilon}
  + O(\epsilon^{2}),
  \label{eq:Bruni-46-one}
\end{eqnarray}
where $\xi_{(1)}^{a}$ is the generator of $\Phi_{\epsilon}$.
From Eqs.~(\ref{eq:perturbative-expansion-of-Q-def}) and
(\ref{eq:Bruni-46-one}), the linear-order gauge-transformation is
given as
\begin{eqnarray}
  \label{eq:Bruni-47-one}
  {}^{(1)}_{\;\ScrY}\!Q - {}^{(1)}_{\;\ScrX}\!Q &=&
  {\pounds}_{\xi_{(1)}}Q_{0}.
\end{eqnarray}
We also employ the {\it order by order gauge invariance} (of the
second kind) as a concept of gauge
invariance~\cite{K.Nakamura-2009a}.
We call the $k$th-order perturbation ${}^{(k)}_{\ScrX}\!Q$ as
gauge invariant (of the second-kind) if and only if
\begin{eqnarray}
  \label{eq:order-by-order-gauge-inv-notion-def}
  {}^{(k)}_{\;\ScrX}\!Q = {}^{(k)}_{\;\ScrY}\!Q
\end{eqnarray}
for any gauge choice $\ScrX_{\epsilon}$ and $\ScrY_{\epsilon}$.

%****************************************************************

Here, we have to emphasize the importance of the gauge invariance of
the second kind.
As explained above, the second kind gauge have nothing to do with the
properties of the physical spacetime.
The physical spacetime is usually identified with our nature itself.
We are living not on the background spacetime but on the physical
spacetime.
Any experiment and observation are carried out within the physical
spacetime through the physical process within the physical spacetime.
Therefore, measurement results of experiments and observations should
have nothing to do with the background spacetime nor the gauge-degree
of freedom of the second kind.
For this reason, measurement results of experiments and observations
should be gauge invariant in sense of the second kind.
Keeping in our mind these premise, the gauge-transformation rule
(\ref{eq:Bruni-47-one}) indicates that the first-order perturbation
${}^{(1)}\!Q$ for an arbitrary tensor field $Q$ is transformed through
the gauge-transformation, i.e., the change of the point identification
of the points of the physical spacetime and the background spacetime,
in general.
This implies that the first-order perturbation ${}^{(1)}\!Q$ includes
the unphysical degree of freedom, i.e., the gauge degree of freedom in
the second kind, in general.
Thus, order-by-order gauge-invariant variables defined by
Eq.~(\ref{eq:order-by-order-gauge-inv-notion-def}) does not include
the gauge degree of freedom in the second kind and is quite important
for perturbation theories in general relativity.

%****************************************************************

Finally, we comment on the difference between the notion of this
second-kind gauge and the first-kind gauge especially the example in
the paragraph which contains
Eq.~(\ref{eq:tensor-pull-back-first-kind-gauge-def}) and in the next
paragraph.
First, we point out that the Taylor expansion through the
infinitesimal parameter $\lambda$ in
Eqs.~(\ref{eq:tensor-pull-back-first-kind-gauge-def}) to
(\ref{eq:infinitesimal-coordinate-trans-first-kind-metric-pert}) is
the expansion within the single manifold $\ScrM$.
Therefore, even if we includes higher-order perturbations of the
infinitesimal parameter $\lambda$, this Taylor expansion is still
within the single manifold.
On the other hand, the direction of the Taylor expansion
(\ref{eq:perturbative-expansion-of-Q-def}) for the perturbative
variable $\ScrX_{\epsilon}^{*}Q$ is the transverse direction from
the background spacetime $\ScrM$ to the physical spacetime
$\ScrM_{ph}$ in the extended manifold $\ScrN$.
Although the action of the diffeomorphism $\Phi_{\epsilon}^{*}$ is
within the background spacetime, the Taylor expansion of
$\ScrY^{*}Q_{\epsilon}$ and $\ScrX_{\epsilon}Q_{\epsilon}$
through the infinitesimal parameter $\epsilon$ is the transverse
direction to each manifolds $\ScrM_{\epsilon}$ in the extended
manifold $\ScrN$.
Therefore, the metric perturbation in
Eq.~(\ref{eq:infinitesimal-coordinate-trans-first-kind-metric-pert})
cannot direct to the physical spacetime $\ScrM_{ph}$, but the
perturbation in Eq.~(\ref{eq:perturbative-expansion-of-Q-def})
actually direct to the physical spacetime $\ScrM_{ph}$.
Therefore, the perturbation of $h_{ab}$ in
Eq.~(\ref{eq:infinitesimal-coordinate-trans-first-kind-metric-pert})
does not have any information of $\ScrM_{ph}$ if the manifold
$\ScrM$ for Eq.~(\ref{eq:tensor-pull-back-first-kind-gauge-def}) is
the background spacetime of perturbation, but ${}^{(1)}_{\;\ScrX}\!Q$
in Eq.~(\ref{eq:perturbative-expansion-of-Q-def}) should have the
information of $\ScrM_{ph}$.

%****************************************************************

However, as shown in Eq.~(\ref{eq:Bruni-47-one}) indicates the
variables ${}^{(1)}_{\;\cal X}\!Q$ includes the information of the
second-kind gauge and we have to excludes this second-kind gauge
completely.
This is accomplished by the construction of gauge-invariant variables
(in the sense of the second-kind).
The general-relativistic gauge-invariant perturbation theory explained
below (in
Sec.~\ref{sec:The_general-relativistic_gauge-invariant_linear_perturbation_theory})
automatically treats only gauge-invariant variables in the sense of
the second-kind defined by
Eq.~(\ref{eq:order-by-order-gauge-inv-notion-def}).
Thus, the development of our gauge-invariant perturbation theory is
crucially important in physics.
Here, we emphasize the important fact that the gauge-degree of freedom
in perturbations to be excluded by the gauge-invariant perturbation
theory is not the above first-kind gauge but the second-kind gauge as
explained below.

%****************************************************************

%%%%%%%%%%%%%%%%%%%%%%%%%%%%%%%%%%%%%%%%%%%%%%%%%%%%%%
%%%%%%%%%%%%%%%%%%%%%%%%%%%%%%%%%%%%%%%%%%%%%%%%%%%%%%
%%%%%%%%%%%%%%%%%%%%%%%%%%%%%%%%%%%%%%%%%%%%%%%%%%%%%%
\subsection{The general-relativistic gauge-invariant linear perturbation theory}
\label{sec:The_general-relativistic_gauge-invariant_linear_perturbation_theory}
%%%%%%%%%%%%%%%%%%%%%%%%%%%%%%%%%%%%%%%%%%%%%%%%%%%%%%
%%%%%%%%%%%%%%%%%%%%%%%%%%%%%%%%%%%%%%%%%%%%%%%%%%%%%%
%%%%%%%%%%%%%%%%%%%%%%%%%%%%%%%%%%%%%%%%%%%%%%%%%%%%%%

%****************************************************************

Based on the above setup, we proposed a procedure to construct
gauge-invariant variables of higher-order
perturbations~\cite{K.Nakamura-2003,K.Nakamura-2005}.
In this paper, we concentrate only on the explanations of the linear
perturbations.
First, we expand the metric on the physical spacetime
$\ScrM_{\epsilon}$, which was pulled back to the background
spacetime $\ScrM$ through a gauge choice $\ScrX_{\epsilon}$ as
\begin{eqnarray}
  \ScrX^{*}_{\epsilon}\bar{g}_{ab}
  &=&
  g_{ab} + \epsilon {}_{\ScrX}\!h_{ab}
  + O(\epsilon^{2}).
  \label{eq:metric-expansion}
\end{eqnarray}
Although the expression (\ref{eq:metric-expansion}) depends
entirely on the gauge choice $\ScrX_{\epsilon}$, henceforth,
we do not explicitly express the index of the gauge choice
$\ScrX_{\epsilon}$ in the expression if there is no
possibility of confusion.
The important premise of our proposal was the following
conjecture~\cite{K.Nakamura-2003,K.Nakamura-2005} for the linear
metric perturbation $h_{ab}$:
\begin{conjecture}
  \label{conjecture:decomposition-conjecture}
  If the gauge-transformation rule for a perturbative pulled-back
  tensor field $h_{ab}$ to the background spacetime $\ScrM$ is
  given by ${}_{\ScrY}\!h_{ab}$ $-$ ${}_{\ScrX}\!h_{ab}$ $=$
  ${\pounds}_{\xi_{(1)}}g_{ab}$ with the background metric $g_{ab}$,
  there then exist a tensor field $\ScrF_{ab}$ and a vector field
  $Y^{a}$ such that $h_{ab}$ is decomposed as $h_{ab}$ $=:$
  $\ScrF_{ab}$ $+$ ${\pounds}_{Y}g_{ab}$, where $\ScrF_{ab}$ and
  $Y^{a}$ are transformed as ${}_{\ScrY}\!\ScrF_{ab}$ $-$
  ${}_{\ScrX}\!\ScrF_{ab}$ $=$ $0$ and ${}_{\ScrY}\!Y^{a}$
  $-$ ${}_{\ScrX}\!Y^{a}$ $=$ $\xi^{a}_{(1)}$ under the gauge
  transformation, respectively.
\end{conjecture}
We call $\ScrF_{ab}$ and $Y^{a}$ as the
{\it gauge-invariant} and {\it gauge-variant} parts
of $h_{ab}$, respectively.
In our higher-order gauge-invariant perturbation
theory~\cite{K.Nakamura-2003,K.Nakamura-2005,K.Nakamura-2011,K.Nakamura-IJMPD-2012,K.Nakamura-2013,K.Nakamura-2014,kouchan-cosmo-second-2006,kouchan-cosmo-second-2007,K.Nakamura-2008,K.Nakamura-2009a,K.Nakamura-2009b,K.Nakamura-2010,A.J.Christopherson-K.A.Malik-D.R.Matravers-K.Nakamura-2011,K.Nakamura-2020},
Conjecture~\ref{conjecture:decomposition-conjecture} play an essential
role in the derivation of the formula for the decomposition of any
variables of higher-order perturbations into their gauge-invariant and
gauge-variant variables.

%*************************************************************

The proof of Conjecture~\ref{conjecture:decomposition-conjecture} is
highly nontrivial~\cite{K.Nakamura-2011,K.Nakamura-2013}, and it was
found that gauge-invariant variables are essentially non-local.
Despite this non-triviality, once we accept
Conjecture~\ref{conjecture:decomposition-conjecture},
we can construct gauge-invariant variables for the linear perturbation
of an arbitrary tensor field ${}_{\ScrX}^{(1)}\!Q$, whose
gauge-transformation is given by Eq.~(\ref{eq:Bruni-47-one}), through
the gauge-variant part of the metric perturbation $Y_{a}$ in
Conjecture~\ref{conjecture:decomposition-conjecture} as
\begin{eqnarray}
  \label{eq:gauge-inv-Q-def}
  {}^{(1)}\!\ScrQ := {}_{\ScrX}^{(1)}\!Q - {\pounds}_{{}_{\ScrX}\!Y}Q_{0}.
\end{eqnarray}
This definition implies that the linear perturbation
${}_{\ScrX}^{(1)}\!Q$ of an arbitrary tensor field
$\ScrX_{\epsilon}^{*}Q$ is always decomposed into its gauge-invariant
part ${}^{(1)}\!\ScrQ$ and gauge-variant part
${\pounds}_{{}_{\ScrX}\!Y}Q_{0}$ as
\begin{eqnarray}
  \label{eq:arbitrary-Q-decomp}
  {}_{\ScrX}^{(1)}\!Q = {}^{(1)}\!\ScrQ + {\pounds}_{{}_{\ScrX}\!Y}Q_{0}.
\end{eqnarray}
As examples, the linearized Einstein tensor
${}_{\ScrX}^{(1)}\!G_{a}^{\;\;b}$ and the linear perturbation of the
energy-momentum tensor ${}_{\ScrX}^{(1)}\!T_{a}^{\;\;b}$ are also
decomposed as
\begin{eqnarray}
  \label{eq:Gab-Tab-decomp}
  {}_{\ScrX}^{(1)}\!G_{a}^{\;\;b}
  =
  {}^{(1)}\!\ScrG_{a}^{\;\;b}\left[\ScrF\right] + {\pounds}_{{}_{\ScrX}\!Y}G_{a}^{\;\;b}
  ,
  \quad
  {}_{\ScrX}^{(1)}\!T_{a}^{\;\;b}
  =
  {}^{(1)}\!\ScrT_{a}^{\;\;b}\left[\ScrF,\phi\right] + {\pounds}_{{}_{\ScrX}\!Y}T_{a}^{\;\;b}
  ,
\end{eqnarray}
where $G_{ab}$ and $T_{ab}$ are the background values of the Einstein
tensor and the energy-momentum tensor, respectively, and $\phi$ in the
gauge-invariant variable
${}^{(1)}\!\ScrT_{a}^{\;\;b}\left[\ScrF,\phi\right]$
symbolically represents the matter degree of freedom.
The gauge-invariant part ${}^{(1)}\!\ScrG_{a}^{\;\;b}$ of the
linear-order perturbation of the Einstein tensor is given by
\begin{eqnarray}
  \label{eq:linear-Einstein-AIA2010-2}
  \!\!\!\!\!\!\!\!\!\!\!\!\!\!\!\!
  &&
     {}^{(1)}\!\ScrG_{a}^{\;\;b}\left[A\right]
     :=
     {}^{(1)}\!\Sigma_{a}^{\;\;b}\left[A\right]
     - \frac{1}{2} \delta_{a}^{\;\;b} {}^{(1)}\!\Sigma_{c}^{\;\;c}\left[A\right]
     ,
  \\
  \label{eq:(1)Sigma-def-linear}
  \!\!\!\!\!\!\!\!\!\!\!\!\!\!\!\!
  &&
     {}^{(1)}\!\Sigma_{a}^{\;\;b}\left[A\right]
     :=
     - 2 \nabla_{[a}^{}H_{d]}^{\;\;\;bd}\left[A\right]
     - A^{cb} R_{ac}
     , \quad
     H_{ba}^{\;\;\;\;c}\left[A\right]
     :=
     \nabla_{(a}A_{b)}^{\;\;c} - \frac{1}{2} \nabla^{c}A_{ab}
     .
\end{eqnarray}
Then, using the background Einstein equation
$G_{a}^{\;\;b}$ $=$ $8\pi T_{a}^{\;\;b}$, the linearized Einstein equation
${}_{\ScrX}^{(1)}\!G_{a}^{\;\;b}$ $=$
$8\pi{}_{\ScrX}^{(1)}\!T_{a}^{\;\;b}$ is automatically given in the
gauge-invariant form
\begin{eqnarray}
  \label{eq:einstein-equation-gauge-inv}
  {}^{(1)}\!\ScrG_{a}^{\;\;b}\left[\ScrF\right]
  =
  8 \pi {}^{(1)}\!\ScrT_{a}^{\;\;b}\left[\ScrF,\phi\right]
\end{eqnarray}
even if the background Einstein equation is nontrivial.
We also note that, in the case of a vacuum background case, i.e.,
$G_{a}^{\;\;b} =8\pi T_{a}^{\;\;b}=0$, Eq.~(\ref{eq:Gab-Tab-decomp})
shows that the linear perturbations of the Einstein tensor and the
energy-momentum tensor are automatically gauge-invariant.

%*************************************************************

We can also derive the perturbation of the divergence of
$\bar{\nabla}_{a}\bar{T}_{b}^{\;\;a}$ of the second-rank tensor
$\bar{T}_{b}^{\;\;a}$ on $(\ScrM_{\rm ph},\bar{g}_{ab})$.
Through the gauge choice $\ScrX_{\epsilon}$, the tensor
$\bar{T}_{b}^{\;\;a}$ is pulled back to
$\ScrX_{\epsilon}^{*}\bar{T}_{b}^{\;\;a}$ on the background
spacetime $(\ScrM,g_{ab})$, and the covariant derivative operator
$\bar{\nabla}_{a}$ on $(\ScrM_{\rm ph},\bar{g}_{ab})$ is pulled
back to a derivative operator
\begin{eqnarray}
  \label{eq:pulled-back-covariant-derive-from-Me-to-M0}
  \bar{\nabla}_{a} := \ScrX_{\epsilon}^{*}\bar{\nabla}_{a}(\ScrX_{\epsilon}^{-1})^{*}
\end{eqnarray}
on $(\ScrM,g_{ab})$.
Note that the derivative $\bar{\nabla}_{a}$ is the covariant
derivative associated with the metric
$\ScrX_{\epsilon}\bar{g}_{ab}$, whereas the derivative $\nabla_{a}$ on
the background spacetime $(\ScrM,g_{ab})$ is the covariant derivative
associated with the background metric $g_{ab}$.
Bearing in mind the difference in these derivatives, the first-order
perturbation of $\bar{\nabla}_{a}\bar{T}_{b}^{\;\;a}$ is given by
\begin{eqnarray}
  \label{eq:linear-perturbation-of-div-Tab}
  {}^{(1)}\!\left(\bar{\nabla}_{a}\bar{T}_{b}^{\;\;a}\right)
  =
  \nabla_{a}{}^{(1)}\!\ScrT_{b}^{\;\;a}\left[\ScrF,\phi\right]
  +
  H_{ca}^{\;\;\;\;a}\left[\ScrF\right] T_{b}^{\;\;c}
  -
  H_{ba}^{\;\;\;\;c}\left[\ScrF\right] T_{c}^{\;\;a}
  +
  {\pounds}_{Y}\nabla_{a}T_{b}^{\;\;a}
  .
\end{eqnarray}
The derivation of the formula
(\ref{eq:linear-perturbation-of-div-Tab}) is given in
Ref.~\cite{K.Nakamura-2005}.
If the tensor field $\bar{T}_{b}^{\;\;a}$ is the Einstein tensor
$\bar{G}_{a}^{\;\;b}$, Eq.~(\ref{eq:linear-perturbation-of-div-Tab})
yields the linear-order perturbation of the Bianchi identity
\begin{eqnarray}
  \label{eq:linear-perturbation-of-div-Gab}
  \nabla_{a}{}^{(1)}\!\ScrG_{b}^{\;\;a}\left[\ScrF\right]
  +
  H_{ca}^{\;\;\;\;a}\left[\ScrF\right] G_{b}^{\;\;c}
  -
  H_{ba}^{\;\;\;\;c}\left[\ScrF\right] G_{c}^{\;\;a}
  =
  0
  .
\end{eqnarray}
Furthermore, if the background Einstein tensor vanishes
$G_{a}^{\;\;b}=0$, we obtain the identity
\begin{eqnarray}
  \label{eq:linear-perturbation-of-div-Gab-vacuum}
  \nabla_{a}{}^{(1)}\!\ScrG_{b}^{\;\;a}\left[\ScrF\right]
  =
  0.
\end{eqnarray}
By contrast, if the tensor field $\bar{T}_{b}^{\;\;a}$ is the
energy-momentum tensor, Eq.~(\ref{eq:linear-perturbation-of-div-Tab})
yields the continuity equation of the energy-momentum tensor
\begin{eqnarray}
  \label{eq:linear-perturbation-of-div-Tab-ene-mon}
  \nabla_{a}{}^{(1)}\!\ScrT_{b}^{\;\;a}\left[\ScrF,\phi\right]
  +
  H_{ca}^{\;\;\;\;a}\left[\ScrF\right] T_{b}^{\;\;c}
  -
  H_{ba}^{\;\;\;\;c}\left[\ScrF\right] T_{c}^{\;\;a}
  =
  0
  ,
\end{eqnarray}
where we used the background continuity equation
$\nabla_{a}T_{b}^{\;\;a}=0$.
If the background spacetime is vacuum $T_{ab}=0$,
Eq.~(\ref{eq:linear-perturbation-of-div-Tab-ene-mon}) yields
a linear perturbation of the energy-momentum tensor given by
\begin{eqnarray}
  \label{eq:divergence-barTab-linear-vac-back-u}
  \nabla_{a}{}^{(1)}\!\ScrT_{b}^{\;\;a}\left[\phi\right]
  =
  0
  .
\end{eqnarray}
Thus, starting from the
Conjecture~\ref{conjecture:decomposition-conjecture}, we can develop
the gauge-invariant perturbation theory through the above framework.
Furthermore, this formulation can be extended to any order
perturbations~\cite{K.Nakamura-2003,K.Nakamura-2005,K.Nakamura-2011,K.Nakamura-2014}
from Conjecture~\ref{conjecture:decomposition-conjecture}.
In this sense, the proof of the
Conjecture~\ref{conjecture:decomposition-conjecture} is crucial to
this framework.

%*************************************************************

We should also note that the decomposition of the metric perturbation
$h_{ab}$ into its gauge-invariant part $\ScrF_{ab}$ and
into its gauge-variant part $Y^{a}$ is not
unique~\cite{K.Nakamura-2009a,K.Nakamura-2010,K.Nakamura-2020}.
For example, the gauge-invariant part $\ScrF_{ab}$ has six
components and we can create the gauge-invariant vector field $Z^{a}$
through the component $\ScrF_{ab}$ such that the
gauge-transformation of the vector field $Z^{a}$ is given by
${}_{\ScrY}\!Z^{a}$ $-$ ${}_{\ScrX}\!Z^{a}$ $=$ $0$.
Using this gauge-invariant vector field $Z^{a}$, the original metric
perturbation can be expressed as follows:
\begin{eqnarray}
  \label{eq:gauge-inv-nonunique}
  h_{ab}
  =
  \ScrF_{ab} - {\pounds}_{Z}g_{ab}
  + {\pounds}_{Z+Y}g_{ab}
  =:
  \ScrH_{ab} + {\pounds}_{X}g_{ab}
  .
\end{eqnarray}
The tensor field $\ScrH_{ab}:=\ScrF_{ab} - {\pounds}_{Z}g_{ab}$
is also regarded as the gauge-invariant part of the perturbation
$h_{ab}$ because ${}_{\ScrY}\!\ScrH_{ab}$ $-$
${}_{\ScrX}\!\ScrH_{ab}$ $=$ $0$.
Similarly, the vector field $X^{a}:=Z^{a}+Y^{a}$ is also regarded as
the gauge-variant part of the perturbation $h_{ab}$ because
${}_{\ScrY}\!X^{a}$ $-$ ${}_{\ScrX}\!X^{a}$ $=$ $\xi^{a}_{(1)}$.

%****************************************************************

Equation (\ref{eq:gauge-inv-nonunique}) does show that the definition
of the gauge-invariant variable $\ScrF_{ab}$ is not unique.
At the same time, this non-uniqueness of the definition of the
gauge-invariant variable $\ScrF_{ab}$ implies the symmetry of the
linearized Einstein equation (\ref{eq:einstein-equation-gauge-inv}).
Through the same derivation of the formulae (\ref{eq:Gab-Tab-decomp}),
we can also derive the linearized Einstein tensor
${}_{\ScrX}^{(1)}\!G_{a}^{\;\;b}$ and the linear perturbation of the
energy-momentum tensor ${}_{\ScrX}^{(1)}\!T_{a}^{\;\;b}$ as
\begin{eqnarray}
  \label{eq:Gab-Tab-decomp-calH}
  {}_{\ScrX}^{(1)}\!G_{a}^{\;\;b}
  =
  {}^{(1)}\!\ScrG_{a}^{\;\;b}\left[\ScrH\right]
  +
  {\pounds}_{{}_{\ScrX}\!X}G_{a}^{\;\;b}
  ,
  \quad
  {}_{\ScrX}^{(1)}\!T_{a}^{\;\;b}
  =
  {}^{(1)}\!\ScrT_{a}^{\;\;b}\left[\ScrH,\phi\right]
  +
  {\pounds}_{{}_{\ScrX}\!X}T_{a}^{\;\;b}
  .
\end{eqnarray}
Then, through the same logic for the derivation of
Eq.~(\ref{eq:einstein-equation-gauge-inv}), we reach to the conclusion
\begin{eqnarray}
  \label{eq:einstein-equation-gauge-inv-calH}
  {}^{(1)}\!\ScrG_{a}^{\;\;b}\left[\ScrH\right]
  =
  8 \pi {}^{(1)}\!\ScrT_{a}^{\;\;b}\left[\ScrH,\phi\right]
  .
\end{eqnarray}
Equations (\ref{eq:einstein-equation-gauge-inv}) and
(\ref{eq:einstein-equation-gauge-inv-calH}) indicate the symmetry of
the linearized Einstein equation.
Namely, if the gauge-invariant metric perturbation $\ScrF_{ab}$ is
a solution to the linearized Einstein equation
(\ref{eq:einstein-equation-gauge-inv}), the gauge-invariant metric
perturbation $\ScrH_{ab}$ $:=$ $\ScrF_{ab}$ $-$
${\pounds}_{Z}g_{ab}$ is also a solution to the linearized Einstein
equation.
This symmetry of the linearized Einstein equation implies that
solutions to the linearized Einstein equation may includes the term
${\pounds}_{Z}g_{ab}$ as a gauge-invariant arbitrary degree of
freedom.
Actually, we will see the fact that the gauge-invariant term
${\pounds}_{Z}g_{ab}$ appears in the solutions derived in
Sec.~\ref{sec:Schwarzschild_Background-non-vaccum-odd-Nakano-treatment}.

%****************************************************************

\begin{figure}
  \begin{center}
    \includegraphics[width=0.958\textwidth]{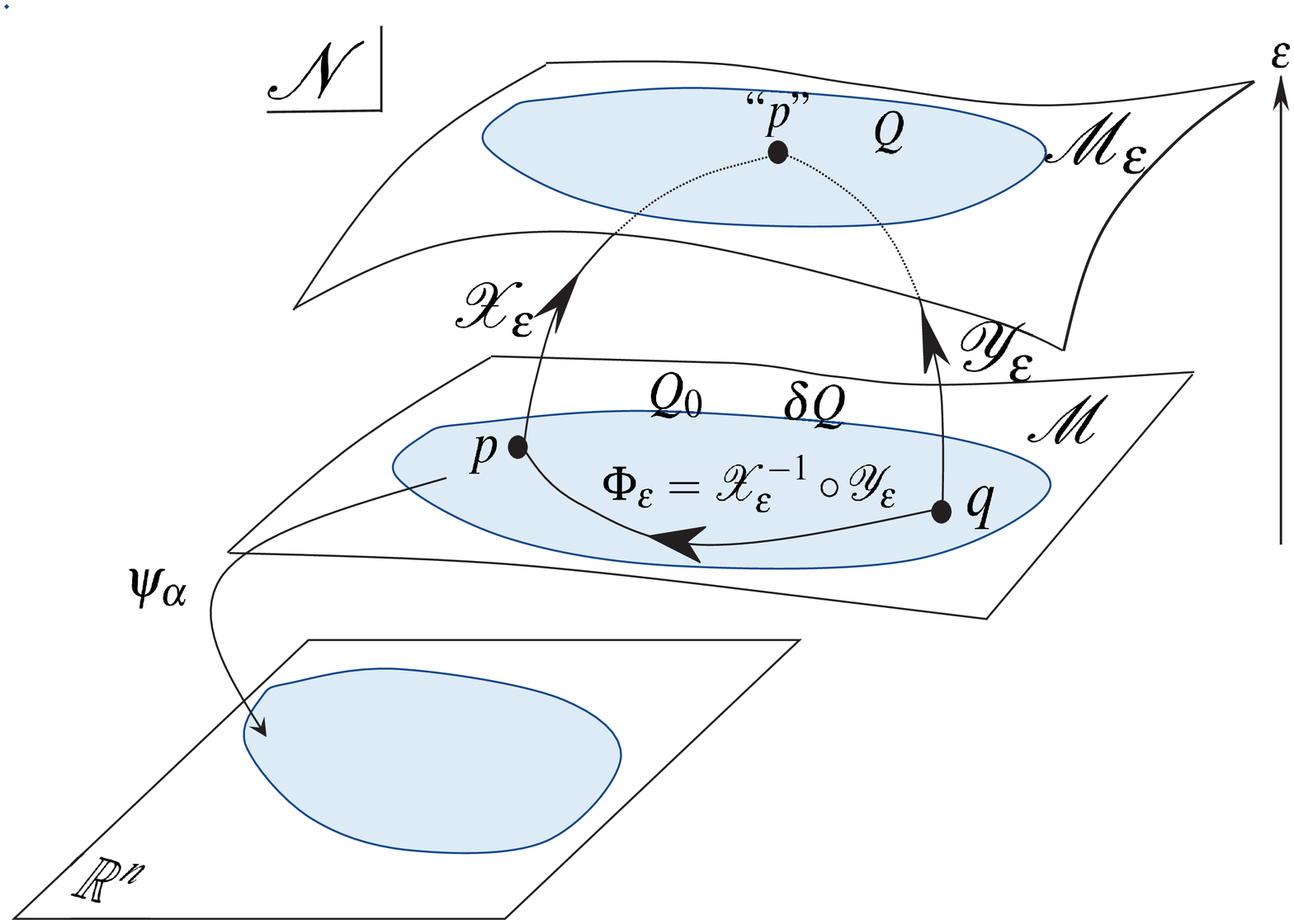}
  \end{center}
  \caption{
    A second kind gauge transformation induces a coordinate transformation.
    The diffeomorphism $\psi_{\alpha}\circ\ScrX_{\epsilon}^{-1}$
    maps the open set
    $\ScrX_{\epsilon}(O_{\alpha})\subset\ScrM_{{\rm ph}}$ to a
    open set on $\RF^{4}$.
    If we change the gauge choice from $\ScrX_{\epsilon}$ to
    $\ScrY_{\epsilon}$, this change induces the coordinate
    transformation $\psi_{\alpha}\circ\ScrX_{\epsilon}^{-1}$ to
    $\psi_{\alpha}\circ\ScrY_{\epsilon}^{-1}$.
  }
  \label{fig:SecondKindGaugeInduceCoordianteTans}
\end{figure}

%****************************************************************

Finally, we comment on the relation between the gauge-transformation
$\Phi_{\epsilon}$ and the coordinate
transformation~\cite{M.Bruni-S.Sonego-CQG1999,S.Sonego-M.Bruni-1998,K.Nakamura-2010,K.Nakamura-2020}.
As mentioned above, the notion of the second-kind gauges above is
different from the notion of the degree of freedom of the coordinate
transformation on a single manifold which is called first-kind gauge.
However, the gauge-transformation $\Phi_{\epsilon}$ of the second kind
induces the coordinate transformations.
To see this, we introduce the coordinate system
$\{O_{\alpha},\psi_{\alpha}\}$ on the background spacetime $\ScrM$,
where $O_{\alpha}$ are open sets on the background spacetime and
$\psi_{\alpha}$ are diffeomorphisms from $O_{\alpha}$ to $\RF^{4}$
($4=\dim\ScrM$) as depicted in
Fig.~\ref{fig:SecondKindGaugeInduceCoordianteTans}.
The coordinate system $\{O_{\alpha},\psi_{\alpha}\}$ is the set of
collections of the pair of open sets $O_{\alpha}$ and diffeomorphism
$O_{\alpha}\mapsto\RF^{4}$.
If we employ a gauge choice $\ScrX_{\epsilon}$ of the second kind,
we have the correspondence of the physical spacetime
$\ScrM_{\epsilon}=\ScrM_{{\rm ph}}$ and the background spacetime
$\ScrM$.
Together with the coordinate system $\psi_{\alpha}$ on $\ScrM$,
this correspondence between $\ScrM_{\epsilon}$ and $\ScrM$
induces the coordinate system on $\ScrM_{\epsilon}$.
Actually, $\ScrX_{\epsilon}(O_{\alpha})$ for each $\alpha$ is an
open set of $\ScrM_{\epsilon}$.
Then, $\psi_{\alpha}\circ\ScrX_{\epsilon}^{-1}$ becomes a
diffeomorphism from an open set
$\ScrX_{\epsilon}(O_{\alpha})\subset \ScrM_{\epsilon}$
to $\RF^{4}(\{x^{\mu}\})$.
This diffeomorphism $\psi_{\alpha}\circ\ScrX_{\epsilon}^{-1}$
induces a coordinate system of an open set on $\ScrM_{\epsilon}$.
When we have two different gauge choices $\ScrX_{\epsilon}$ and
$\ScrY_{\epsilon}$ of the second kind,
$\psi_{\alpha}\circ\ScrX_{\epsilon}^{-1}$ $\mapsto$
$\RF^{4}(\{x^{\mu}\})$ and
$\psi_{\alpha}\circ\ScrY_{\epsilon}^{-1}$ $\mapsto$
$\RF^{4}(\{y^{\mu}\})$ become different coordinate systems on
$\ScrM_{\epsilon}$.
We can also consider the coordinate transformation from the coordinate
system $\psi_{\alpha}\circ\ScrX_{\epsilon}^{-1}$ to another
coordinate system $\psi_{\alpha}\circ\ScrY_{\epsilon}^{-1}$.
Because the gauge transformation
$\ScrX_{\epsilon}\rightarrow\ScrY_{\epsilon}$ is induced by the
diffeomorphism $\Phi_{\epsilon}$ $:=$
$\left(\ScrX_{\epsilon}\right)^{-1}\circ\ScrY_{\epsilon}$, this
diffeomorphism $\Phi_{\epsilon}$ induces the coordinate transformation
as
\begin{eqnarray}
  \label{eq:induced-coordinate-trans}
  y^{\mu}(q) := x^{\mu}(p) = \left((\Phi_{\epsilon}^{-1})^{*}x^{\mu}\right)(q)
\end{eqnarray}
in the passive point of
view~\cite{K.Nakamura-2003,M.Bruni-S.Matarrese-S.Mollerach-S.Sonego-1997},
where $p\in \ScrM$, $\ScrX_{\epsilon}(p)$ $=$ $``p"\in \ScrM_{ph}$
and $q \in \ScrM$, $\ScrY_{\epsilon}(q)$ $=$ $``p"\in \ScrM_{ph}$.
If we represent this coordinate transformation in terms of the Taylor
expansion (\ref{eq:Bruni-46-one}), we have the coordinate transformation
\begin{eqnarray}
  \label{eq:infinitesimal-coordinate-trans-explicit}
  y^{\mu}(q) = x^{\mu}(q) - \epsilon \xi^{\mu}_{(1)}(q) + O(\epsilon^{2}).
\end{eqnarray}
We should emphasize that the coordinate transformation
(\ref{eq:infinitesimal-coordinate-trans-explicit}) is not the starting
point of the gauge-transformation but a result of the above framework.
Because our above framework of the gauge-invariant perturbation theory
is constructed without a coordinate transformation
(\ref{eq:infinitesimal-coordinate-trans-explicit}), we {\it do not}
use the coordinate transformation
(\ref{eq:infinitesimal-coordinate-trans-explicit}) in our formulation.

%****************************************************************

%%%%%%%%%%%%%%%%%%%%%%%%%%%%%%%%%%%%%%%%%%%
%%%%%%%%%%%%%%%%%%%%%%%%%%%%%%%%%%%%%%%%%%%
%%%%%%%%%%%%%%%%%%%%%%%%%%%%%%%%%%%%%%%%%%%
\section{Linear perturbations on spherically symmetric background}
\label{sec:spherical_background_case}
%%%%%%%%%%%%%%%%%%%%%%%%%%%%%%%%%%%%%%%%%%%
%%%%%%%%%%%%%%%%%%%%%%%%%%%%%%%%%%%%%%%%%%%
%%%%%%%%%%%%%%%%%%%%%%%%%%%%%%%%%%%%%%%%%%%

%*********************************************************************

Here, we consider the 2+2 formulation of perturbations of a
spherically symmetric background spacetime, which originally proposed
by Gerlach and Sengupta~\cite{Gerlach_Sengupta-1979a,Gerlach_Sengupta-1979b,Gerlach_Sengupta-1979c,Gerlach_Sengupta-1980}.
In this formulation, we pay attention to the symmetry of the
background spacetime.
Spherically symmetric spacetimes are characterized by the direct
product $\ScrM=\ScrM_{1}\times S^{2}$ and the metric on this
spacetime is given by
\begin{eqnarray}
  \label{eq:background-metric-2+2}
  g_{ab}
  &=&
  y_{ab} + r^{2}\gamma_{ab}
  , \\
  y_{ab} &=& y_{AB} (dx^{A})_{a}(dx^{B})_{b}, \quad
             \gamma_{ab} = \gamma_{pq} (dx^{p})_{a} (dx^{q})_{b},
\end{eqnarray}
where $x^{A} = (t,r)$, $x^{p}=(\theta,\phi)$, and $\gamma_{pq}$ is the
metric on the unit sphere.
In the case of the Schwarzschild spacetime, the metric
(\ref{eq:background-metric-2+2}) is given by
\begin{eqnarray}
  \label{eq:background-metric-2+2-y-comp-Schwarzschild}
  y_{ab}
  &=&
      - f (dt)_{a}(dt)_{b}
      +
      f^{-1} (dr)_{a}(dr)_{b}
      ,
      \quad
      f := 1 - \frac{2M}{r}
  ,\\
  \label{eq:background-metric-2+2-gamma-comp-Schwarzschild}
  \gamma_{ab}
  &=&
      (d\theta)_{a}(d\theta)_{b}
      +
      \sin^{2}\theta(d\phi)_{a}(d\phi)_{b}
      =
      \theta_{a}\theta_{b} + \phi_{a}\phi_{b}
      ,
  \\
  \label{eq:S2-unit-basis-def}
  \theta_{a}
  &=&
      (d\theta)_{a}, \quad
      \phi_{a}
      =
      \sin\theta (d\phi)_{a}
      .
\end{eqnarray}

%*********************************************************************

In Sec.~\ref{sec:2+2-formulation-Perturbation-decomposition}, we
review the conventional decomposition of the metric perturbation and
its inverse relation and show that the conventional decomposition is
essentially non-local and the two Green functions for the derivative
operators are necessary to derive its inverse relation.
The kernel modes of these derivative operators are $l=0,1$ modes.
This is the reason why $l=0,1$ modes in the perturbations on the
spherically symmetric background spacetime should be treated,
separately.
In Sec.~\ref{sec:2+2-formulation-Addition-of-the-kernel-modes-1}, we
discuss a treatment in which the special treatments of these kernel
modes are not necessary.
To develop such treatment, we use the different scalar harmonic
functions from the conventional spherical harmonic functions.
We also summarize the conditions for the harmonic functions should be
satisfied.
In Sec.~\ref{sec:Explict_form_of_the_mode_functions}, we derive the
explicit form of the mode functions.
In Sec.~\ref{sec:Proposal}, we propose a treatment of $l=0,1$ modes in
perturbations on spherically symmetric background spacetime.

%*********************************************************************

%%%%%%%%%%%%%%%%%%%%%%%%%%%%%%%%%%%%%%%%%%%
%%%%%%%%%%%%%%%%%%%%%%%%%%%%%%%%%%%%%%%%%%%
\subsection{Conventional perturbation decomposition and its inverse relation}
\label{sec:2+2-formulation-Perturbation-decomposition}
%%%%%%%%%%%%%%%%%%%%%%%%%%%%%%%%%%%%%%%%%%%
%%%%%%%%%%%%%%%%%%%%%%%%%%%%%%%%%%%%%%%%%%%

%*********************************************************************

On the above background spacetime $(\ScrM,g_{ab})$, the components
of the metric perturbation are given by
\begin{eqnarray}
  \label{eq:metric-perturbation-components}
  h_{ab}
  =
  h_{AB} (dx^{A})_{a}(dx^{B})_{b}
  +
  2 h_{Ap} (dx^{A})_{(a}(dx^{p})_{b)}
  +
  h_{pq} (dx^{p})_{a}(dx^{q})_{b}
  .
\end{eqnarray}
Here, we note that the components $h_{AB}$, $h_{Ap}$, and
$h_{pq}$ are regarded as components of scalar, vector, and
tensor on $S^{2}$, respectively.
In many literatures, these components are decomposed through the
decomposition~\cite{J.W.York-1973,J.W.York-1974,S.Deser-1967} using
the spherical harmonics $S=Y_{lm}$ as follows:
\begin{eqnarray}
  \label{eq:hAB-fourier}
  h_{AB}
  &=&
      \sum_{l,m} \tilde{h}_{AB} S
      ,
  \\
  \label{eq:hAp-fourier}
  h_{Ap}
  &=&
      r \sum_{l,m} \left[
      \tilde{h}_{(e1)A} \hat{D}_{p}S
      +
      \tilde{h}_{(o1)A} \epsilon_{pq} \hat{D}^{q}S
      \right]
      ,
  \\
  \label{eq:hpq-fourier}
  h_{pq}
  &=&
      r^{2} \sum_{l,m} \left[
      \frac{1}{2} \gamma_{pq} \tilde{h}_{(e0)} S
      +
      \tilde{h}_{(e2)} \left(
      \hat{D}_{p}\hat{D}_{q} - \frac{1}{2} \gamma_{pq} \hat{D}^{r}\hat{D}_{r}
      \right) S
%      \right.
%      \nonumber\\
%  && \quad\quad\quad
%     \left.
      +
      2 \tilde{h}_{(o2)} \epsilon_{r(p} \hat{D}_{q)}\hat{D}^{r} S
      \right]
      ,
\end{eqnarray}
where $\hat{D}_{p}$ is the covariant derivative associated with
the metric $\gamma_{pq}$ on $S^{2}$,
$\hat{D}^{p}=\gamma^{pq}\hat{D}_{q}$,
$\epsilon_{pq}=\epsilon_{[pq]}=2\theta_{[p}\phi_{q]}$ is the totally
antisymmetric tensor on $S^{2}$.
Here, we note that the covariant derivatives of the basis $\theta_{p}$
and $\phi_{p}$ on $S^{2}$ are given by
\begin{eqnarray}
  \hat{D}_{p}\theta_{q}
  =
  \cot\theta \phi_{p}\phi_{q}
  , \quad
  \hat{D}_{p}\phi_{q}
  =
  -
  \cot\theta \phi_{p}\theta_{q}
  .
\end{eqnarray}
Through these formulae, we can check $\hat{D}_{r}\epsilon_{pq} = 0$.
We also note that the curvature tensors ${}^{(2)}R_{pqrs}$ and
${}^{(2)}R_{pr}$ associated with the metric $\gamma_{pq}$ are given by
\begin{eqnarray}
  \label{eq:unit-sphere-curvatures}
  {}^{(2)}R_{pqrs} = 2 \gamma_{p[r}\gamma_{s]q}, \quad
  {}^{(2)}R_{pr} = \gamma_{pr}.
\end{eqnarray}

%*********************************************************************

Although the matrix representations of the independent harmonic
functions are used in the pioneer
papers~\cite{T.Regge-J.A.Wheeler-1957,F.Zerilli-1970-PRL,F.Zerilli-1970-PRD,H.Nakano-2019},
these are equivalent to the covariant form
(\ref{eq:hAB-fourier})--(\ref{eq:hpq-fourier}) with the choice $S=Y_{lm}$.
The choice $S=Y_{lm}$ is the starting point of the original 2+2
formulation proposed by Gerlach and
Sengupta~\cite{Gerlach_Sengupta-1979a,Gerlach_Sengupta-1979b,Gerlach_Sengupta-1979c,Gerlach_Sengupta-1980}.
They showed the constructions of gauge-invariant variables for $l\geq
2$ modes and derived Einstein equations.
If we apply the decomposition
(\ref{eq:hAB-fourier})--(\ref{eq:hpq-fourier}) with $S=Y_{lm}$ to the
metric perturbation $h_{ab}$, special treatments for $l=0,1$ modes are
required~\cite{T.Regge-J.A.Wheeler-1957,F.Zerilli-1970-PRL,F.Zerilli-1970-PRD,H.Nakano-2019,V.Moncrief-1974a,V.Moncrief-1974b,C.T.Cunningham-R.H.Price-V.Moncrief-1978,Chandrasekhar-1983,Gerlach_Sengupta-1979a,Gerlach_Sengupta-1979b,Gerlach_Sengupta-1980,Gerlach_Sengupta-1979c,Gundlach-Martine-Garcia-2000,Gundlach-Martine-Garcia-2001,A.Nagar-L.Rezzolla-2005-2006,K.Martel-E.Poisson-2005}.
This is due to the fact that the set of harmonic functions
\begin{eqnarray}
  \label{eq:harmonic-fucntions-set}
  \left\{
  S,
  \hat{D}_{p}S,
  \epsilon_{pq}\hat{D}^{q}S,
  \frac{1}{2}\gamma_{pq}S,
  \left(\hat{D}_{p}\hat{D}_{q}-\frac{1}{2}\gamma_{pq}\hat{\Delta}\right)S,
  2\epsilon_{r(p}\hat{D}_{q)}\hat{D}^{r}S
  \right\}
\end{eqnarray}
loses its linear independence in $l=0,1$ cases.
To clarify this situation, we consider the inverse relation of the
decomposition formula (\ref{eq:hAB-fourier})--(\ref{eq:hpq-fourier}),
later.
Furthermore, we see that the inverse-relation of the decomposition
formulae (\ref{eq:hAB-fourier})--(\ref{eq:hpq-fourier}) requires the
Green functions of the derivative operators
$\hat{\Delta}:=\hat{D}^{r}\hat{D}_{r}$ and
$\hat{\Delta}+2:=\hat{D}^{r}\hat{D}_{r}+2$, respectively.
The eigen mode of these operators are $l=0$ and $l=1$, respectively.
Actually, for $l=0$ modes, the basis in
(\ref{eq:harmonic-fucntions-set}) vanish except for
$\{S,\frac{1}{2}\gamma_{pq}S\}$.
For $l=1$ modes, we have
$\left(\hat{D}_{p}\hat{D}_{q}-\frac{1}{2}\gamma_{pq}\hat{\Delta}\right)S
=2\epsilon_{r(p}\hat{D}_{q)}\hat{D}^{r}S=0$.
These are explicitly shown in
Appendix~\ref{sec:Explict_form_of_Tlm_for_Ylm}.

%*********************************************************************

Note that the decomposition formulae
(\ref{eq:hAB-fourier})--(\ref{eq:hpq-fourier}) with the spherical
harmonic function $Y_{lm}$ carry out two decompositions.
The first one is the decomposition of the function space through the
spherical harmonic function $Y_{lm}$ as the bases of $L^{2}$-space on
$S^{2}$.
This corresponds to the imposition of the regular boundary conditions
for the perturbations at the starting point.
The second one is the decomposition of the tangent space on $S^{2}$
through the derivative of the scalar harmonic function $S=Y_{lm}$.
The imposition of the boundary conditions at the starting point leads
to the vanishing of vector and tensor harmonics in
(\ref{eq:harmonic-fucntions-set}) for $l=0$ modes and tensor harmonics
in (\ref{eq:harmonic-fucntions-set}) for $l=1$ modes.
These vanishing vector and tensor harmonics leads to the failure of
the decomposition of the tangent space for $l=0,1$ modes.
This is the reason why the special treatments for these modes are
required in many literatures.
At the same time, these vanishing mode functions are an essential
reason for the fact that the proof of
Conjecture~\ref{conjecture:decomposition-conjecture} for perturbations
on the Schwarzschild background spacetime including $l=0,1$ modes is
difficult.

%*********************************************************************

Now, we consider the derivation of the inverse relation of the
decomposition (\ref{eq:hAB-fourier})--(\ref{eq:hpq-fourier}).
In this derivation, we use the orthogonality
\begin{eqnarray}
  \label{eq:spherical_harmonic_orthogonality}
  \int_{S^{2}} d\Omega Y_{lm}^{*} Y_{l'm'} = \delta_{ll'}\delta_{mm'}
\end{eqnarray}
of the spherical harmonic function $S=Y_{lm}$, where
$d\Omega = \sin\theta d\theta d\phi$.
Therefore, we do not show the final expressions as the results of the
application of Eq.~(\ref{eq:spherical_harmonic_orthogonality}).

%*********************************************************************

First, we consider the inverse relation of the decomposition
(\ref{eq:hAp-fourier}).
Taking the divergence of Eq.~(\ref{eq:hAp-fourier}), we obtain
\begin{eqnarray}
  \label{eq:hAp-fourier-div}
  \hat{D}^{p}h_{Ap}
  &=&
      r \sum_{l,m}
      \tilde{h}_{(e1)A} \hat{D}^{p}\hat{D}_{p}S
      =
      r \sum_{l,m,(l\neq 0)}
      \tilde{h}_{(e1)A} \hat{\Delta}S
      .
\end{eqnarray}
Thus, we should regard that the mode coefficient $\tilde{h}_{(e1)A}$
in Eq.~(\ref{eq:hAp-fourier}) does not include $l=0$ mode.
Using the Green function $\hat{\Delta}^{-1}$, we obtain
\begin{eqnarray}
  \label{eq:hAp-fourier-div-Green}
  \sum_{l,m,(l\neq 0)}
  \tilde{h}_{(e1)A} S
  =
  \frac{1}{r} \hat{\Delta}^{-1}\hat{D}^{p}h_{Ap}
  .
\end{eqnarray}
Furthermore, using the orthogonal property
(\ref{eq:spherical_harmonic_orthogonality}) of the $S=Y_{lm}$ with
$l\neq 0$, we obtain the mode coefficient $\tilde{h}_{(e1)A}$ for each
mode, except for $l=0$ mode.
Similarly, taking the rotation of Eq.~(\ref{eq:hAp-fourier}), we
obtain
\begin{eqnarray}
  \label{eq:hAp-fourier-inv-2}
  \sum_{l,m,(l\neq 0)} \tilde{h}_{(o1)A} S
  &=&
  \frac{1}{r}
  \hat{\Delta}^{-1}\hat{D}_{r}\left(
    \epsilon^{rq} h_{Aq}
  \right)
\end{eqnarray}
and the mode coefficient $\tilde{h}_{(o1)A}$ for each
mode, except for $l=0$ mode, through the orthogonal property
(\ref{eq:spherical_harmonic_orthogonality}) of the $S=Y_{lm}$ with
$l\neq 0$.

%*********************************************************************

The explicit form of the Green function is given by
Refs.~\cite{R.Szmytkowski-2006a,R.Szmytkowski-2007}.
The expressions (\ref{eq:hAp-fourier-div-Green}) and
(\ref{eq:hAp-fourier-inv-2}) indicates that the decomposition
(\ref{eq:hAp-fourier}) is meaningless for the modes which belongs to
the kernel $\hat{\Delta}:=\hat{D}^{r}\hat{D}_{r}$, i.e., $l=0$ mode.

%*********************************************************************

Next, we consider the inverse relation of
(\ref{eq:hpq-fourier}).
First, we note that the trace of Eq.~(\ref{eq:hpq-fourier}) yields
\begin{eqnarray}
  \label{eq:hpq-fourier-trace}
  \sum_{l,m} \tilde{h}_{(e0)} S
  &=&
  \frac{1}{r^{2}} \gamma^{pq}h_{pq}
  ,
\end{eqnarray}
and the traceless part of Eq.~(\ref{eq:hpq-fourier}) yields
\begin{eqnarray}
  \label{eq:traceless-calHpq-def}
  \HF_{pq}[h_{tu}]
  &:=&
       h_{pq}
       - \frac{1}{2} \gamma_{pq} \gamma^{rs}h_{rs}
       ,
  \\
  \label{eq:hpq-fourier-traceless}
  &=&
      r^{2} \sum_{l,m} \left[
      \tilde{h}_{(e2)} \left(
      \hat{D}_{p}\hat{D}_{q} - \frac{1}{2} \gamma_{pq} \hat{\Delta}
      \right) S
      +
      2 \tilde{h}_{(o2)} \epsilon_{r(p} \hat{D}_{q)}\hat{D}^{r} S
      \right].
\end{eqnarray}
The mode coefficient $\tilde{h}_{(e0)}$ for each mode is obtained
through the orthogonal property
(\ref{eq:spherical_harmonic_orthogonality}) of the spherical harmonics
$S=Y_{lm}$ from the trace part (\ref{eq:hpq-fourier-trace}) of
$h_{pq}$.
Therefore, we may concentrate on the traceless part
(\ref{eq:hpq-fourier-traceless}) of $h_{pq}$.
Taking the divergence of Eq.~(\ref{eq:hpq-fourier-traceless}),
we obtain
\begin{eqnarray}
  \hat{D}^{p}\HF_{pq}[h_{tu}]
  &=&
      r^{2} \sum_{l,m} \left[
      \tilde{h}_{(e2)}
      \frac{1}{2} \hat{D}_{q}
      \left(
      \hat{\Delta}
      +
      2
      \right) S
      +
      \tilde{h}_{(o2)}
      \epsilon_{rq} \hat{D}^{r}
      \left(
      \hat{\Delta}
      +
      2
      \right)
      S
      \right]
      \label{eq:hpq-fourier-traceless-div}
  \\
  &=&
      r^{2} \sum_{l,m,(l\neq 1)} \left[
      \tilde{h}_{(e2)}
      \frac{1}{2} \hat{D}_{q}
      \left(
      \hat{\Delta}
      +
      2
      \right) S
      +
      \tilde{h}_{(o2)}
      \epsilon_{rq} \hat{D}^{r}
      \left(
      \hat{\Delta}
      +
      2
      \right)
      S
      \right]
      .
      \label{eq:hpq-fourier-traceless-div-exclude-l=1}
\end{eqnarray}
Equation (\ref{eq:hpq-fourier-traceless-div-exclude-l=1}) indicates
that the mode coefficients $\tilde{h}_{(e2)}$ and $\tilde{h}_{(o2)}$
do not include $l=1$ mode if $S=Y_{lm}$ because the $l=1$ spherical
harmonic function $Y_{1m}$ is in the kernel of the derivative operator
$\hat{\Delta}+2$.
Furthermore, we take the divergence of
Eq.~(\ref{eq:hpq-fourier-traceless-div-exclude-l=1}), and obtain
\begin{eqnarray}
  \hat{D}^{q}\hat{D}^{p}
  \HF_{pq}[h_{tu}]
  =
  r^{2} \sum_{l,m,(l\neq 1)} \left[
  \frac{1}{2} \tilde{h}_{(e2)} \hat{\Delta}\left(
  \hat{\Delta} + 2
  \right) S
  \right]
  =
  r^{2} \sum_{l,m,l\geq 2} \left[
  \frac{1}{2} \tilde{h}_{(e2)} \hat{\Delta}\left(
  \hat{\Delta} + 2
  \right) S
  \right]
  .
  \label{eq:hpq-fourier-traceless-div-div}
\end{eqnarray}
Equation (\ref{eq:hpq-fourier-traceless-div-div}) indicates that, in
addition to the $l=1$ mode, the mode coefficient $\tilde{h}_{(e2)}$
does not include the $l=0$ mode which is the kernel mode of the
derivative operator $\hat{\Delta}$.
Then, through the Green functions of the derivative operators
$\hat{\Delta}$ and $(\hat{\Delta}+2)$, we obtain the solution to
Eq.~(\ref{eq:hpq-fourier-traceless-div-div}) as
\begin{eqnarray}
  \sum_{l,m,l\geq 2} \tilde{h}_{(e2)} S
  =
  \frac{2}{r^{2}}
  \left[
  \hat{\Delta} + 2
  \right]^{-1}
  \hat{\Delta}^{-1}
  \hat{D}^{q}\hat{D}^{p}
  \HF_{pq}[h_{tu}]
  .
  \label{eq:hpq-fourier-traceless-div-div-e1-inv}
\end{eqnarray}
From the orthogonal property
(\ref{eq:spherical_harmonic_orthogonality}) of the spherical harmonic
function $S=Y_{lm}$ with $l\geq 2$, we obtain the mode coefficient
$\tilde{h}_{(e2)}$.

%*********************************************************************

On the other hand, multiplying $\epsilon^{qs}$ to
Eq.~(\ref{eq:hpq-fourier-traceless-div-exclude-l=1}), we obtain
\begin{eqnarray}
  \epsilon^{qs}\hat{D}^{p}\HF_{pq}[h_{tu}]
  =
  r^{2} \sum_{l,m,(l\neq 1)} \left[
  \frac{1}{2}
  \tilde{h}_{(e2)}
  \epsilon^{qs}\hat{D}_{q}
  \left(
  \hat{\Delta} + 2
  \right) S
  +
  \tilde{h}_{(o2)} \hat{D}^{s}\left(
  \hat{\Delta} + 2
  \right) S
  \right]
  ,
  \label{eq:hpq-fourier-traceless-div-eps-exclude-l=1}
\end{eqnarray}
and then, taking the divergence of
Eq.~(\ref{eq:hpq-fourier-traceless-div-eps-exclude-l=1}), we obtain
\begin{eqnarray}
  \epsilon^{qs}\hat{D}_{s}\hat{D}^{p}
  \HF_{pq}[h_{tu}]
  =
  r^{2} \hat{\Delta}\left(
  \hat{\Delta} + 2
  \right)
  \sum_{l,m,(l\neq 1)} \tilde{h}_{(o2)} S
  =
  r^{2} \hat{\Delta}\left(
  \hat{\Delta} + 2
  \right)
  \sum_{l,m,(l\geq 2)} \tilde{h}_{(o2)} S
  .
  \label{eq:hpq-fourier-traceless-div-eps-2}
\end{eqnarray}
Equation (\ref{eq:hpq-fourier-traceless-div-eps-2}) indicates that, in
addition to the $l=1$ mode, the mode coefficient $\tilde{h}_{(o2)}$
does not include the $l=0$ mode, which is the kernel mode of the
derivative operator $\hat{\Delta}$.
Through the Green functions of the derivative operators
$\hat{\Delta}$ and $\hat{\Delta}+2$, we can solve
Eq.~(\ref{eq:hpq-fourier-traceless-div-eps-2}) as
\begin{eqnarray}
  \sum_{l,m,l\geq 2} \tilde{h}_{(o2)} S
  =
  \frac{1}{r^{2}}
  \left[\hat{\Delta}+2\right]^{-1}
  \hat{\Delta}^{-1}
  \epsilon^{qs}\hat{D}_{s}\hat{D}^{p}
  \HF_{pq}[h_{tu}]
  .
  \label{eq:hpq-fourier-traceless-div-eps-inv}
\end{eqnarray}
From the orthogonality property
(\ref{eq:spherical_harmonic_orthogonality}) of the spherical harmonic
function $S=Y_{lm}$ with $l\neq 0,1$, we obtain the mode coefficient
$\tilde{h}_{(o2)}$.

%*********************************************************************

Since the eigenvalue of the Laplacian operator $\hat{\Delta}$ on
$S^{2}$ is $-l(l+1)$ with the non-negative integer $l$, the fact that
we have to use the Green function of the operators $\Delta$ and
$(\Delta+2)$ implies that the one-to-one correspondence between the
set of variables $\{h_{pq}\}$ and the set of the variables
$\{\tilde{h}_{(e0)},\tilde{h}_{(e2)},\tilde{h}_{(o2)}\}$ is not
guaranteed for the kernel modes $l=0$ and  $l=1$.

%*********************************************************************

Finally, we also note that the operators
$\hat{\Delta}^{-1}\hat{\Delta}$ and
$\left[\hat{\Delta}+2\right]^{-1}\left[\hat{\Delta}+2\right]$ are not
identity operators but should be regarded as the projection
operators.
We regard that the domain of the operators
$\hat{\Delta}^{-1}\hat{\Delta}$ and
$\left[\hat{\Delta}+2\right]^{-1}\left[\hat{\Delta}+2\right]$ is the
$L^{2}$-space which is spanned by the spherical harmonics $\{Y_{lm}\}$.
Since the operator $\hat{\Delta}$ eliminates the kernel
\begin{eqnarray}
  \label{eq:calK-Delta-def}
  \ScrK_{\hat{\Delta}}:=\{f\in\ScrF|\hat{\Delta}f=0\},
\end{eqnarray}
where $\ScrF$ is the function algebra, the range of the
operator $\hat{\Delta}^{-1}\hat{\Delta}$ is the $L^{2}$-space which is
spanned by the spherical harmonics $\{Y_{lm}|l\neq 0\}$, i.e.,
\begin{eqnarray}
  \{Y_{lm}|l\neq 0\} = L^{2}\backslash\ScrK_{(\hat{\Delta})}.
\end{eqnarray}
Similarly, the domain of the operator
$\left[\hat{\Delta}+2\right]^{-1}\left[\hat{\Delta}+2\right]$ is
the $L^{2}$-space which is spanned by the spherical harmonics
$\{Y_{lm}|l\geq 0\}$, while the kernel
\begin{eqnarray}
  \label{eq:calK-Delta+2-def}
  \ScrK_{\hat{\Delta}+2}:=\{f\in\ScrF|(\hat{\Delta}+2)f=0\}
\end{eqnarray}
is excluded in the range of the operator
$\left[\hat{\Delta}+2\right]^{-1}\left[\hat{\Delta}+2\right]$, i.e.,
the range of this operator is
\begin{eqnarray}
  \{Y_{lm}|l\neq 1\} = L^{2}\backslash\ScrK_{(\hat{\Delta}+2)}.
\end{eqnarray}
Namely, the operators $\hat{\Delta}^{-1}\hat{\Delta}$ and
$[\hat{\Delta}+2]^{-1}[\Delta+2]$ are regarded as the projection
operators as
\begin{eqnarray}
  \label{eq:map_of_DeltainvDelta}
  \hat{\Delta}^{-1}\hat{\Delta}
  \quad &:&
            \quad L^{2}
            \quad \mapsto
            \quad L^{2}\backslash \ScrK_{(\hat{\Delta})}
  \\
  \label{eq:map_of_Delta+2invDelta+2}
  [\hat{\Delta}+2]^{-1}[\Delta+2]
  \quad &:&
            \quad L^{2}
            \quad \mapsto
            \quad L^{2}\backslash \ScrK_{(\hat{\Delta}+2)}
            .
\end{eqnarray}
From Eqs.~(\ref{eq:map_of_DeltainvDelta}) and
(\ref{eq:map_of_Delta+2invDelta+2}), we obtain the projection operator
\begin{eqnarray}
  \label{eq:map_of_Delta+2invDeltainvDelta+2Delta}
  \hat{\Delta}^{-1}[\hat{\Delta}+2]^{-1}\left[\Delta+2\right]\hat{\Delta}
  \quad &:&
            \quad L^{2}
            \quad \mapsto
            \quad L^{2}\backslash \left(\ScrK_{(\hat{\Delta})}\oplus\ScrK_{(\hat{\Delta}+2)}\right)
  .
\end{eqnarray}
This is a reason why we should discuss the treatments of the modes
$l=0$ and $l=1$, separately, if we choose $S=Y_{lm}$.

%*********************************************************************

%%%%%%%%%%%%%%%%%%%%%%%%%%%%%%%%%%%%%%%%%%%
%%%%%%%%%%%%%%%%%%%%%%%%%%%%%%%%%%%%%%%%%%%
\subsection{Treatments of the kernel modes}
\label{sec:2+2-formulation-Addition-of-the-kernel-modes-1}
%%%%%%%%%%%%%%%%%%%%%%%%%%%%%%%%%%%%%%%%%%%
%%%%%%%%%%%%%%%%%%%%%%%%%%%%%%%%%%%%%%%%%%%

%*********************************************************************

As seen in Sec.~\ref{sec:2+2-formulation-Perturbation-decomposition},
the decomposition formulae (\ref{eq:hAB-fourier})--(\ref{eq:hpq-fourier})
with $S=Y_{lm}$ does not include the $l=0,1$ modes of the
perturbations.
In the general-relativistic gauge-invariant perturbation theory
proposed in Refs.~\cite{K.Nakamura-2003,K.Nakamura-2005}, we assumed
the separation of the linear-order metric perturbation into its
gauge-invariant and gauge-variant parts, i.e.,
Conjecture~\ref{conjecture:decomposition-conjecture}.
In Refs.~\cite{K.Nakamura-2011,K.Nakamura-2013}, we discuss a scenario
of the proof of Conjecture~\ref{conjecture:decomposition-conjecture}
on the generic background spacetime.
In this scenario of the proof, we had to use the Green functions of
some elliptic differential operators.
In other words, we ignored the kernel modes of these elliptic
differential operators in the scenario of the proof of
Conjecture~\ref{conjecture:decomposition-conjecture} in
Refs.~\cite{K.Nakamura-2011,K.Nakamura-2013}.
The treatment of these kernel modes was unclear at that time.
We call these kernel modes as zero modes.
Furthermore, we call the problem to find the treatment of these zero
modes as the zero-mode problem.
In the case of the perturbations on the spherically symmetric
background spacetimes, the $l=0,1$ modes correspond to the above zero
mode in Refs.~\cite{K.Nakamura-2011,K.Nakamura-2013}.
This is also the well-known problem as ``$l=0,1$ mode problem'' in the
treatments of perturbations on spherically symmetric background
spacetimes.

%*********************************************************************

Here, we consider the resolution of this $l=0,1$ mode problem.
To carry out this, we re-examine the derivation of the inverse
relations of the decomposition formulae
(\ref{eq:hAB-fourier})--(\ref{eq:hpq-fourier}), again.
In this re-examination, we use the harmonic function $S=Y_{lm}$ for
$l\geq 2$ model, because the set of the harmonic functions
(\ref{eq:harmonic-fucntions-set}) has the linear independence
at least for $l\geq 2$ mode.
For $l=0,1$ mode, we change the harmonic function $S$ from the
spherical harmonic function $Y_{00}$ and $Y_{1m}$ to
$k_{(\hat{\Delta})}$ and $k_{(\hat{\Delta}+2)}$, respectively, i.e.,
we use the harmonic functions $S$ which are given by
\begin{eqnarray}
  \label{eq:harmonic-delta-S-def}
  S
  =
  S_{\delta}
  :=
  \left\{
  \begin{array}{lcl}
    Y_{lm} & \quad & (l\geq 2); \\
    k_{(\hat{\Delta}+2)} & \quad & (l=1); \\
    k_{(\hat{\Delta})} & \quad & (l=0).
  \end{array}
  \right.
\end{eqnarray}
In this paper, we look for the explicit form of functions
$k_{(\hat{\Delta})}$ and $k_{(\hat{\Delta}+2)}$ within the constraints
\begin{eqnarray}
  \label{eq:kernel-mode-def-l=0-l=1}
  k_{(\hat{\Delta})} \in \ScrK_{(\hat{\Delta})}, \quad
  k_{(\hat{\Delta}+2)} \in \ScrK_{(\hat{\Delta}+2)},
\end{eqnarray}
respectively.
Within these domain (\ref{eq:kernel-mode-def-l=0-l=1}) of the kernel
modes, we specify the conditions for the functions
$k_{(\hat{\Delta})}$ and $k_{(\hat{\Delta}+2)}$ to realize the
independence of the set of the harmonic functions
(\ref{eq:harmonic-fucntions-set}).
These introductions of $k_{(\hat{\Delta})}$ and $k_{(\hat{\Delta}+2)}$
correspond to the fact that we do not impose the regular boundary
conditions as the function on $S^{2}$ before the construction of
gauge-invariant variables, which was imposed in the conventional
approach at the starting point.

%*********************************************************************

%%%%%%%%%%%%%%%%%%%%%%%%%%%%%%%%%%%%%%%%%%%
\subsubsection{$h_{pq}$}
\label{sec:hpq}
%%%%%%%%%%%%%%%%%%%%%%%%%%%%%%%%%%%%%%%%%%%

%*********************************************************************

Here, we first consider the decomposition of the component $h_{pq}$.
Previously, we considered the decomposition of the component $h_{pq}$
as Eq.~(\ref{eq:hpq-fourier}):
\begin{eqnarray}
  \label{eq:hpq-fourier-2}
  h_{pq}
  &=&
      r^{2} \sum_{l,m} \left[
      \frac{1}{2} \gamma_{pq} \tilde{h}_{(e0)} S
      +
      \tilde{h}_{(e2)} \left(
      \hat{D}_{p}\hat{D}_{q} - \frac{1}{2} \gamma_{pq} \hat{D}^{r}\hat{D}_{r}
      \right) S
      +
      2 \tilde{h}_{(o2)} \epsilon_{r(p} \hat{D}_{q)}\hat{D}^{r} S
      \right]
      .
\end{eqnarray}
As shown in Eq.~(\ref{eq:hpq-fourier-trace}), we can separate the
component $h_{pq}$ into the trace part and the traceless part.
The trace part of $h_{pq}$ is given by
Eq.~(\ref{eq:hpq-fourier-trace}), which is also given by
\begin{eqnarray}
  \label{eq:hpq-fourier-trace-new-expansion}
  \sum_{l,m,l\geq 2} \tilde{h}_{(e0,l\geq 2)} Y_{lm}
  + \sum_{m=-1,0,1} \tilde{h}_{(e0,l=1)} k_{(\hat{\Delta}+2)}
  + \tilde{h}_{(e0,l=0)} k_{(\hat{\Delta})}
  =
  \frac{1}{r^{2}} \gamma^{pq}h_{pq}
  .
\end{eqnarray}
Here, we note the effects (\ref{eq:map_of_DeltainvDelta}) and
(\ref{eq:map_of_Delta+2invDelta+2}) of the operators
$\hat{\Delta}^{-1}\hat{\Delta}$ and
$[\hat{\Delta}+2]^{-1}[\hat{\Delta}+2]$ as projection operators.
If we apply the derivative operator $[\hat{\Delta}+2]$ to
Eq.~(\ref{eq:hpq-fourier-trace-new-expansion}), we obtain
\begin{eqnarray}
  \label{eq:hpq-fourier-trace-new-expansion-Delta+2}
  \sum_{l,m,l\geq 2} \tilde{h}_{(e0,l\geq 2)} [\hat{\Delta}+2] Y_{lm}
  + 2 \tilde{h}_{(e0,l=0)} k_{(\hat{\Delta})}
  =
  \frac{1}{r^{2}} [\hat{\Delta}+2] \gamma^{pq}h_{pq}
  ,
\end{eqnarray}
since we chose the functions $k_{(\hat{\Delta})}$ and
$k_{(\hat{\Delta}+2)}$ are eigen-functions through
Eqs.~(\ref{eq:kernel-mode-def-l=0-l=1}).
Furthermore, applying the derivative operator $\hat{\Delta}$ to
Eq.~(\ref{eq:hpq-fourier-trace-new-expansion-Delta+2}) as
\begin{eqnarray}
  \label{eq:hpq-fourier-trace-new-expansion-Delta+2-Dleta}
  \sum_{l,m,l\geq 2} \tilde{h}_{(e0,l\geq 2)} \hat{\Delta} [\hat{\Delta}+2] Y_{lm}
  =
  \frac{1}{r^{2}} \hat{\Delta} [\hat{\Delta}+2] \gamma^{pq}h_{pq}
  .
\end{eqnarray}
The left- and right-hand sides of
Eq.~(\ref{eq:hpq-fourier-trace-new-expansion-Delta+2-Dleta}) are in
the domain of the Green functions $[\hat{\Delta}]^{-1}$ and
$[\hat{\Delta}+2]^{-1}$.
Therefore, we may apply the Green functions $[\hat{\Delta}]^{-1}$ and
$[\hat{\Delta}+2]^{-1}$ to
Eq.~(\ref{eq:hpq-fourier-trace-new-expansion-Delta+2-Dleta}) and
obtain
\begin{eqnarray}
  \label{eq:hpq-fourier-trace-new-expansion-Delta-Delta+2-Delta+2-inv-Delta-inv}
  \sum_{l,m,l\geq 2} \tilde{h}_{(e0,l\geq 2)} Y_{lm}
  =
  \frac{1}{r^{2}} [\hat{\Delta}+2]^{-1}\hat{\Delta}^{-1}\hat{\Delta} [\hat{\Delta}+2] \gamma^{pq}h_{pq}
  .
\end{eqnarray}
Through the orthogonal property
(\ref{eq:spherical_harmonic_orthogonality}) of the spherical harmonic
function, we obtain
\begin{eqnarray}
  \label{eq:tildehe0-explicit-result-lgeq2}
  \tilde{h}_{(e0,l\geq 2)}
  =
  \frac{1}{r^{2}}
  \int_{S^{2}} d\Omega Y_{lm}^{*}
  [\hat{\Delta}+2]^{-1} \hat{\Delta}^{-1} \hat{\Delta} [\hat{\Delta}+2] \gamma^{pq}h_{pq}
  =:
  \tilde{h}_{(e0,l\geq 2)}[\![h_{pq}]\!]
  .
\end{eqnarray}
Thus, for $l\geq 2$, the mode coefficients $\tilde{h}_{(e0,l\geq 2)}$
is given by the functional of the original metric component $h_{pq}$.

%*********************************************************************

Substituting Eq.~(\ref{eq:tildehe0-explicit-result-lgeq2}) into
Eq.~(\ref{eq:hpq-fourier-trace-new-expansion-Delta+2}), we obtain
\begin{eqnarray}
  2 \tilde{h}_{(e0,l=0)} k_{(\hat{\Delta})}
  &=&
      \frac{1}{r^{2}} [\hat{\Delta}+2] \gamma^{pq}h_{pq}
      -
      \sum_{l,m,l\geq 2} \tilde{h}_{(e0,l\geq 2)}[\![h_{pq}]\!] [\hat{\Delta}+2] Y_{lm}
      \nonumber\\
  &=:&
       2 \tilde{h}_{(e0,l=0)}[\![h_{pq}]\!] k_{(\hat{\Delta})}
       .
       \label{eq:hpq-fourier-trace-new-expansion-Delta+2-l=0-is}
\end{eqnarray}
Then, the mode coefficient $\tilde{h}_{(e0,l=0)}$ is obtained as a
functional of the original metric perturbation $h_{pq}$ if
$k_{(\hat{\Delta})}\neq 0$.
Furthermore, from Eqs.~(\ref{eq:hpq-fourier-trace-new-expansion}),
(\ref{eq:tildehe0-explicit-result-lgeq2}), and
(\ref{eq:hpq-fourier-trace-new-expansion-Delta+2-l=0-is}), we obtain
\begin{eqnarray}
  \label{eq:hpq-fourier-trace-new-expansion-l=1-is}
  \sum_{m=-1,0,1} \tilde{h}_{(e0.l=1)} k_{(\hat{\Delta}+2)}
  =
  \frac{1}{r^{2}} \gamma^{pq}h_{pq}
  -
  \sum_{l,m,l\geq 2} \tilde{h}_{(e0,l\geq 2)}[\![h_{pq}]\!] Y_{lm}
  -
  \tilde{h}_{(e0,l=0)}[\![h_{pq}]\!] k_{(\hat{\Delta})}
  .
\end{eqnarray}
To resolve the degeneracy of the modes with $m=0,\pm 1$ in
Eq.~(\ref{eq:hpq-fourier-trace-new-expansion-l=1-is}), we choose
$k_{(\hat{\Delta}+2)}$ as
\begin{eqnarray}
  \label{eq:kDelta+2-phi-dependence-requirement}
  k_{(\hat{\Delta}+2)}
  =
  k_{(\hat{\Delta}+2)m}
  =
  \Theta_{1m}(\theta) e^{im\phi}
  .
\end{eqnarray}
Through the orthogonality condition
\begin{eqnarray}
  \frac{1}{2\pi} \int_{0}^{2\pi} d\phi e^{i(m-m')\phi} = \delta_{mm'},
\end{eqnarray}
we obtain
\begin{eqnarray}
  \label{eq:m-degeneracy-resolve-l=1}
  e^{+im\phi} \frac{1}{2\pi} \int_{0}^{2\pi} d\phi e^{-im'\phi}
  k_{(\hat{\Delta}+2)m}
  =
  k_{(\hat{\Delta}+2)m} \delta_{mm'}
  .
\end{eqnarray}
Applying the property (\ref{eq:m-degeneracy-resolve-l=1}) to
Eq.~(\ref{eq:hpq-fourier-trace-new-expansion-l=1-is}), we obtain
\begin{eqnarray}
  \tilde{h}_{(e0.l=1)} k_{(\hat{\Delta}+2)m}
  &=&
      e^{+im\phi}
      \frac{1}{2\pi} \int_{0}^{2\pi} d\phi e^{-im\phi}
      \nonumber\\
  && \quad
     \times
     \left[
     \frac{1}{r^{2}} \gamma^{pq}h_{pq}
     -
     \sum_{l,m',(l\neq 0,1)} \tilde{h}_{(e0,l\geq 2)}[h_{pq}] Y_{lm'}
     -
     \tilde{h}_{(e0,l=0)}[h_{pq}] k_{(\hat{\Delta})}
     \right]
     \nonumber\\
  &=:&
       \tilde{h}_{(e0.l=1)}[\![h_{pq}]\!] k_{(\hat{\Delta}+2)m}
       .
       \label{eq:hpq-fourier-trace-new-expansion-l=1-each-mode}
\end{eqnarray}
Then, if $\Theta_{1m}(\theta)\neq 0$, i.e.,
$k_{(\hat{\Delta}+2)}\neq 0$, the mode coefficient
$\tilde{h}_{(e0,l=1)}$ is given in the functional form of the original
metric perturbation $h_{pq}$.

%*********************************************************************

Thus, the mode decomposition of the trace part
(\ref{eq:hpq-fourier-trace-new-expansion}) of the metric perturbation
$h_{pq}$ is invertible.
In this argument, we essentially used the equations
(\ref{eq:kernel-mode-def-l=0-l=1}) for the eigen functions and the
$\phi$-dependence (\ref{eq:kDelta+2-phi-dependence-requirement}) of
the function $k_{(\hat{\Delta}+2)}$.

%*********************************************************************

Next, we consider the traceless part (\ref{eq:traceless-calHpq-def})
of $h_{pq}$ as Eq.~(\ref{eq:hpq-fourier-traceless}).
Taking the divergence of Eq.~(\ref{eq:hpq-fourier-traceless}), we
obtain
\begin{eqnarray}
  \frac{1}{r^{2}}
  \hat{D}^{p}
  \HF_{pq}[h_{tu}]
  &=&
      \sum_{l,m} \left[
      \frac{1}{2}
      \tilde{h}_{(e2)}
      \hat{D}_{q}
      \left(
      \hat{\Delta}
      +
      2
      \right) Y_{lm}
      -
      \tilde{h}_{(o2)}
      \epsilon_{qr}
      \hat{D}^{r}
      \left(
      \hat{\Delta}
      +
      2
      \right) Y_{lm}
      \right]
      \nonumber\\
  &=&
      \sum_{l,m,l\geq 2} \left[
      \frac{1}{2}
      \tilde{h}_{(e2,l\geq 2)}
      \hat{D}_{q}
      \left(
      \hat{\Delta}
      +
      2
      \right) Y_{lm}
      -
      \tilde{h}_{(o2,l\geq 2)}
      \epsilon_{qr}
      \hat{D}^{r}
      \left(
      \hat{\Delta}
      +
      2
      \right) Y_{lm}
      \right]
      \nonumber\\
  &&
     +
     \tilde{h}_{(e2,l=0)}
     \hat{D}_{q}
     k_{(\hat{\Delta})}
     -
     \tilde{h}_{(o2.l=0)}
     2
     \epsilon_{qr}
     \hat{D}^{r}
     k_{(\hat{\Delta})}
     ,
     \label{eq:hpq-fourier-2-traceless-divergence}
\end{eqnarray}
where we used Eqs.~(\ref{eq:unit-sphere-curvatures}) and
(\ref{eq:kernel-mode-def-l=0-l=1}).
We have to emphasize that the $l=1$ mode does not appear in the
expression (\ref{eq:hpq-fourier-2-traceless-divergence}).
Taking the divergence of
Eq.~(\ref{eq:hpq-fourier-2-traceless-divergence}), again, we have
\begin{eqnarray}
  \frac{1}{r^{2}}
  \hat{D}^{q}\hat{D}^{p}
  \HF_{pq}[h_{tu}]
  &=&
      \frac{1}{2}
      \sum_{l,m,l\geq 2}
      \tilde{h}_{(e2,l\geq 2)}
      \left(
      \hat{\Delta}
      +
      2
      \right)
      \hat{\Delta}
      Y_{lm}
      ,
      \label{eq:hpq-fourier-2-traceless-divergence-divergence}
\end{eqnarray}
where we used the property of the eigen equation for
$k_{(\hat{\Delta})}$ in Eqs.~(\ref{eq:kernel-mode-def-l=0-l=1}).
Through the Green functions $\hat{\Delta}^{-1}$ and
$[\hat{\Delta}+2]^{-1}$ and the orthogonal property
(\ref{eq:spherical_harmonic_orthogonality}) of the spherical harmonics
$Y_{lm}$, we obtain the same result as
Eq.~(\ref{eq:hpq-fourier-traceless-div-eps-inv}) and the mode
coefficient $\tilde{h}_{(e2,l\geq 2)}$ of each mode is given in a
functional form of the original metric perturbation $h_{tu}$ as
\begin{eqnarray}
  \label{eq:he2lgeq2-functional-hpq}
  \tilde{h}_{(e2,l\geq 2)}
  =
  \frac{2}{r^{2}}
  \int_{S^{2}} d\Omega Y_{lm}^{*}
  [\hat{\Delta}]^{-1}
  [\hat{\Delta}+2]^{-1}
  \hat{D}^{q}\hat{D}^{p}
  \HF_{pq}[h_{tu}]
  =:
  \tilde{h}_{(e2,l\geq 2)}[\![h_{tu}]\!]
  .
\end{eqnarray}
On the other hand, taking the rotation of
Eq.~(\ref{eq:hpq-fourier-2-traceless-divergence}) and use the eigen
equation for $k_{(\hat{\Delta})}$ in
Eqs.~(\ref{eq:kernel-mode-def-l=0-l=1}), Green functions
$[\hat{\Delta}]^{-1}$ and $[\hat{\Delta}+2]$, and the orthogonal
properties (\ref{eq:spherical_harmonic_orthogonality}) of the
spherical harmonics $Y_{lm}$, we obtain the mode coefficient
$\tilde{h}_{(o2,l\geq 2)}$ in the functional form of the original metric
perturbation $h_{tu}$ as
\begin{eqnarray}
  \tilde{h}_{(o2,l\geq 2)}
  =
  \frac{1}{r^{2}}
  \int_{S^{2}} d\Omega Y_{lm}^{*}
  [\hat{\Delta}]^{-1}
  [\hat{\Delta}+2]^{-1}
  \epsilon^{ps}
  \hat{D}_{s}
  \hat{D}^{q}
  \HF_{pq}[h_{tu}]
  =:
  \tilde{h}_{(o2,l\geq 2)}[\![h_{tu}]\!]
  .
  \label{eq:ho2lgeq2-functional-hpq}
\end{eqnarray}

%*********************************************************************

Substituting Eqs.~(\ref{eq:he2lgeq2-functional-hpq}) and
(\ref{eq:ho2lgeq2-functional-hpq}) into
Eq.~(\ref{eq:hpq-fourier-2-traceless-divergence}), we obtain
\begin{eqnarray}
  &&
%     \!\!\!\!\!\!\!\!\!\!\!\!\!\!\!\!\!\!\!\!\!\!\!\!\!\!\!\!\!\!\!\!\!\!\!\!\!\!\!\!
     \!\!\!\!\!\!\!\!\!\!\!\!\!\!\!\!\!\!\!\!\!\!\!
     \tilde{h}_{(e2,l=0)}
     \hat{D}_{q}
     k_{(\hat{\Delta})}
     -
     \tilde{h}_{(o2.l=0)}
     2
     \epsilon_{qr}
     \hat{D}^{r}
     k_{(\hat{\Delta})}
     \nonumber\\
  &=&
      -
      \sum_{l,m,l\geq 2} \left[
      \frac{1}{2}
      \tilde{h}_{(e2,l\geq 2)}[\![h_{tu}]\!]
      \hat{D}_{q}
      \left(
      \hat{\Delta}
      +
      2
      \right) Y_{lm}
      -
      \tilde{h}_{(o2,l\geq 2)}[\![h_{tu}]\!]
      \epsilon_{qr}
      \hat{D}^{r}
      \left(
      \hat{\Delta}
      +
      2
      \right) Y_{lm}
      \right]
      \nonumber\\
  &&
     +
     \frac{1}{r^{2}}
     \hat{D}^{p}
     \HF_{pq}[h_{tu}]
     .
     \label{eq:hpq-fourier-2-traceless-divergence-l=0-is}
\end{eqnarray}
If $\hat{D}_{q}k_{(\hat{\Delta})}\neq 0$, the vectors
$\hat{D}_{q}k_{(\hat{\Delta})}$ and
$\epsilon_{qr}\hat{D}^{r}k_{(\hat{\Delta})}$ are orthogonal to each
other.
Then, we have
\begin{eqnarray}
  &&
     \!\!\!\!\!\!\!\!\!\!\!\!\!\!\!\!\!\!\!\!\!\!\!\!\!\!\!\!\!\!\!\!\!\!\!\!
     \tilde{h}_{(e2,l=0)}
     =
     \tilde{h}_{(e2,l=0)}[\![h_{ut}]\!]
     \nonumber\\
  \!\!\!\!\!\!\!\!\!\!\!\!\!\!\!\!\!\!\!\!\!\!\!\!\!\!\!\!\!\!\!\!\!\!\!\!
  &:=&
  \!\!\!\!\!
       \frac{
       \left(
       \hat{D}^{q}
       k_{(\hat{\Delta})}
       \right)
       }{
       \left(
       \hat{D}_{s}
       k_{(\hat{\Delta})}
       \right)
       \left(
       \hat{D}^{s}
       k_{(\hat{\Delta})}
       \right)
       }
       \left[
       \frac{1}{r^{2}}
       \hat{D}^{p}
       \HF_{pq}[h_{tu}]
       \right.
       \nonumber\\
  && \quad\quad\quad\quad\quad\quad\quad\quad\quad\quad
     \left.
       -
       \sum_{l,m,l\geq 2} \left\{
       \frac{1}{2}
       \tilde{h}_{(e2,l\geq 2)}[\![h_{tu}]\!]
       \hat{D}_{q}
       \left(
       \hat{\Delta}
       +
       2
       \right) Y_{lm}
       \right.
       \right.
       \nonumber\\
  && \quad\quad\quad\quad\quad\quad\quad\quad\quad\quad\quad\quad\quad\quad\quad
     \left.
     \left.
     -
     \tilde{h}_{(o2,l\geq 2)}[\![h_{tu}]\!]
     \epsilon_{qr}
     \hat{D}^{r}
     \left(
     \hat{\Delta}
     +
     2
     \right) Y_{lm}
     \right\}
     \right]
     \label{eq:tildehe2l=0-functional-of-hpq}
\end{eqnarray}
and
\begin{eqnarray}
  &&
     \!\!\!\!\!\!\!\!\!\!\!\!\!\!\!\!\!\!\!\!\!\!\!\!\!\!\!\!\!\!\!\!\!\!\!\!
     \tilde{h}_{(o2,l=0)}
     =
     \tilde{h}_{(o2,l=0)}[\![h_{ut}]\!]
     \nonumber\\
  \!\!\!\!\!\!\!\!\!\!\!\!\!\!\!\!\!\!\!\!\!\!\!\!\!\!\!\!\!\!\!\!\!\!\!\!
  &:=&
      \!\!\!\!\!
      \frac{
      \left(
      \epsilon^{qr}
      \hat{D}_{r}
      k_{(\hat{\Delta})}
      \right)
      }{
      \left(
      \hat{D}_{s}
      k_{(\hat{\Delta})}
      \right)
      \left(
      \hat{D}^{s}
      k_{(\hat{\Delta})}
      \right)
      }
      \left[
      \frac{1}{r^{2}}
      \hat{D}^{p}
      \HF_{pq}[h_{tu}]
       \right.
       \nonumber\\
  && \quad\quad\quad\quad\quad\quad\quad\quad\quad\quad
     \left.
     -
     \sum_{l,m,l\geq 2} \left\{
     \frac{1}{2}
     \tilde{h}_{(e2,l\geq 2)}[\![h_{tu}]\!]
     \hat{D}_{q}
     \left(
     \hat{\Delta}
     +
     2
     \right) Y_{lm}
     \right.
     \right.
     \nonumber\\
  && \quad\quad\quad\quad\quad\quad\quad\quad\quad\quad\quad\quad\quad\quad
     \left.
     \left.
     -
     \tilde{h}_{(o2,l\geq 2)}[\![h_{tu}]\!]
     \epsilon_{qr}
     \hat{D}^{r}
     \left(
     \hat{\Delta}
     +
     2
     \right) Y_{lm}
     \right\}
     \right]
     .
     \label{eq:tildeho2l=0-functional-of-hpq}
\end{eqnarray}

%*********************************************************************

Now, we return to the original definition
(\ref{eq:hpq-fourier-traceless}) of the traceless part $\HF_{pq}$.
From Eqs.~(\ref{eq:traceless-calHpq-def}),
(\ref{eq:he2lgeq2-functional-hpq}),
(\ref{eq:ho2lgeq2-functional-hpq}),
(\ref{eq:tildehe2l=0-functional-of-hpq}), and
(\ref{eq:tildeho2l=0-functional-of-hpq}), we obtain
\begin{eqnarray}
  &&
     \!\!\!\!\!\!\!\!\!\!\!\!\!\!\!\!\!\!\!\!\!
     \sum_{m=-1,0,1} \left[
     \tilde{h}_{(e2,l=1,m)} \left(
     \hat{D}_{p}\hat{D}_{q} - \frac{1}{2} \gamma_{pq} \hat{\Delta}
     \right) k_{(\hat{\Delta}+2)}
     +
     2 \tilde{h}_{(o2,l=1,m)} \epsilon_{r(p} \hat{D}_{q)}\hat{D}^{r} k_{(\hat{\Delta}+2)}
     \right]
     \nonumber\\
  &=&
      \frac{1}{r^{2}} \HF_{pq}[h_{tu}]
      -
      \left[
      \sum_{l,m,l\geq 2} \left\{
      \tilde{h}_{(e2)}[\![h_{tu}]\!] \left(
      \hat{D}_{p}\hat{D}_{q} - \frac{1}{2} \gamma_{pq} \hat{\Delta}
      \right) Y_{lm}
      +
      2 \tilde{h}_{(o2)}[\![h_{tu}]\!] \epsilon_{r(p} \hat{D}_{q)}\hat{D}^{r} Y_{lm}
      \right\}
      \right.
      \nonumber\\
  && \quad\quad\quad\quad\quad\quad
     \left.
     +
     \tilde{h}_{(e2,l=0)}[\![h_{tu}]\!] \left(
     \hat{D}_{p}\hat{D}_{q} - \frac{1}{2} \gamma_{pq} \hat{\Delta}
     \right) k_{(\hat{\Delta})}
     +
     2 \tilde{h}_{(o2,l=0)}[\![h_{tu}]\!] \epsilon_{r(p} \hat{D}_{q)}\hat{D}^{r} k_{(\hat{\Delta})}
     \right]
     \nonumber\\
     &=:&
          H_{(\hat{\Delta}+2)pq}[\![h_{tu}]\!]
          .
          \label{eq:m-sum-l=1-is-given-by-htu-functional}
\end{eqnarray}

%*********************************************************************

To simplify the notation, we define
\begin{eqnarray}
  \label{eq:DpDq-halfgammapqDeltakhatDelta+2m-notation}
  K_{(m)pq}
  :=
  \left(
  \hat{D}_{p}\hat{D}_{q} - \frac{1}{2} \gamma_{pq} \hat{D}^{r}\hat{D}_{r}
  \right)  k_{(\hat{\Delta}+2)m}
  ,
  \quad
  J_{(m)pq}
  :=
  2 \epsilon_{r(p} \hat{D}_{q)}\hat{D}^{r} k_{(\hat{\Delta}+2)m}
  ,
\end{eqnarray}
and we evaluate $K_{(m)pq}K_{(m')}^{pq}$, $J_{(m)pq}K_{(m')}^{pq}$,
and $J_{(m)pq}J_{(m')}^{pq}$, which are given by
\begin{eqnarray}
  K_{(m)pq}
  K_{(m')}^{pq}
  &=&
      \left(
      \hat{D}^{p}\hat{D}^{q}
      k_{(\hat{\Delta}+2)m}
      \right)
      \left(
      \hat{D}^{p}\hat{D}^{q}
      k_{(\hat{\Delta}+2)m'}
      \right)
      \nonumber\\
  && \quad
     -
     2
     \left(
     k_{(\hat{\Delta}+2)m}
     \right)
     \left(
     k_{(\hat{\Delta}+2)m'}
     \right)
     ,
     \label{eq:KmpqKm'pq-norm}
  \\
  J_{(m)pq}K_{(m')}^{pq}
  &=&
      2 \epsilon^{rp}
      \hat{D}_{r}\hat{D}_{q}k_{(\hat{\Delta}+2)m}
      \hat{D}_{p}\hat{D}^{q}k_{(\hat{\Delta}+2)m'}
      ,
      \label{eq:JmpqKm'pq-norm}
  \\
  J_{(m)pq}
  J_{(m')}^{pq}
  &=&
      4
      \left[
      \left(
      \hat{D}^{p}\hat{D}^{q} k_{(\hat{\Delta}+2)m}
      \right)
      \left(
      \hat{D}_{p}\hat{D}_{q} k_{(\hat{\Delta}+2)m'}
      \right)
      \right.
      \nonumber\\
  && \quad\quad
     \left.
     -
     2
     \left(k_{(\hat{\Delta}+2)m}\right)
     \left(k_{(\hat{\Delta}+2)m'}\right)
     \right]
     .
     \label{eq:JmpqJm'pq-norm}
\end{eqnarray}

%*********************************************************************

To carry out the resolution of the degeneracy in
Eq.~(\ref{eq:m-sum-l=1-is-given-by-htu-functional}), we use the
property (\ref{eq:kDelta+2-phi-dependence-requirement}) of the
function $k_{(\hat{\Delta}+2)}$.
From the property (\ref{eq:kDelta+2-phi-dependence-requirement}), we
have
\begin{eqnarray}
  \hat{D}_{p}k_{(\hat{\Delta}+2)m}
  &=&
  \left(\frac{d}{d\theta}\Theta_{m}(\theta)\right) e^{im\phi} \theta_{p}
  +
  \frac{i m}{\sin\theta} \Theta_{m}(\theta) e^{im\phi} \phi_{p}
  \label{eq:DpkhatDelta+2m-intermediate-express}
\end{eqnarray}
and
\begin{eqnarray}
  \hat{D}_{p}\hat{D}_{q}k_{(\hat{\Delta}+2)m}
  &=&
      \left(\frac{d^{2}}{d\theta^{2}}\Theta_{m}(\theta)\right) e^{im\phi} \theta_{p} \theta_{q}
      \nonumber\\
  &&
      + \left[
      \left(\frac{d}{d\theta}\Theta_{m}(\theta)\right) \cot\theta
      -  m^{2} \frac{1}{\sin^{2}\theta} \Theta_{m}(\theta)
      \right] e^{im\phi} \phi_{p} \phi_{q}
      \nonumber\\
  &&
      + i m \frac{1}{\sin\theta} \left[
      \frac{d}{d\theta}\Theta_{m}(\theta)
      -  \cot\theta \Theta_{m}(\theta)
      \right] e^{im\phi} 2 \theta_{(p} \phi_{q)}
      .
     \label{eq:DpDqkhatDelta+2m-intermediate-express}
\end{eqnarray}
From Eq.~(\ref{eq:DpDqkhatDelta+2m-intermediate-express}), we obtain
\begin{eqnarray}
  K_{(m)pq}
  &:=&
       \left(
       \hat{D}_{p}\hat{D}_{q}
       - \frac{1}{2} \gamma_{pq}\hat{\Delta}
       \right) k_{(\hat{\Delta}+2)m}
       \nonumber\\
  &=&
      -
      \left(
      \theta_{p}\theta_{q}
      -
      \phi_{p}\phi_{q}
      \right)
      \left[
      \cot\theta \frac{d}{d\theta}\Theta_{m}(\theta)
      + \left( 1 - \frac{m^{2}}{\sin^{2}\theta} \right) \Theta_{m}(\theta)
      \right]
      e^{im\phi}
      \nonumber\\
  &&
     +
     2 \theta_{(p} \phi_{q)}
     \frac{i m}{\sin\theta}
     \left(
     \frac{d}{d\theta}\Theta_{m}(\theta)
     -  \cot\theta \Theta_{m}(\theta)
     \right) e^{im\phi}
     ,
  \label{eq:tracelessDpDqkhatDelta+2m-intermediate-express-final-sum}
\end{eqnarray}
where we used $(\hat{\Delta}+2)k_{(\hat{\Delta}+2)m}=0$,
i.e.,
\begin{eqnarray}
  \label{eq:DeltakhatDelta+2m-intermediate-express-kernel-eq}
  \frac{d^{2}}{d\theta^{2}}\Theta_{m}(\theta)
  +
  \cot\theta
  \frac{d}{d\theta}\Theta_{m}(\theta)
  +
  \left(
  2
  -
  \frac{m^{2}}{\sin^{2}\theta}
  \right) \Theta_{m}(\theta)
  =
  0
  .
\end{eqnarray}
From the expression of the components $K_{(m')pq}$, $J_{(m')pq}$,
$\theta_{p}$, and $\phi_{p}$, we can confirm
\begin{eqnarray}
  \frac{1}{2\pi}\int_{0}^{2\pi} d\phi e^{-im\phi} K_{(m')pq}
  &=&
      K_{(m)pq}
      e^{-im\phi}
      \delta_{mm'}
      ,
      \label{eq:tracelessDpDqkhatDelta+2m-intermediate-express-final-thetatheta-Fourier-sum}
  \\
  \frac{1}{2\pi}\int_{0}^{2\pi} d\phi e^{-im\phi} J_{(m')pq}
  &=&
      J_{(m)pq}
      e^{-im\phi}
      \delta_{mm'}
     .
     \label{eq:epsilonDpDqkhatDelta+2m-intermediate-express-final-thetatheta-Fourier-sum}
\end{eqnarray}
Furthermore, straightforward calculations yield
\begin{eqnarray}
  \label{eq:KmpqKmpq-product}
  K_{(m)pq} K_{(m)}^{pq}
  &=&
      \left(
      \hat{D}_{p}\hat{D}_{q}k_{(\hat{\Delta}+2)m}
      \right)
      \left(
      \hat{D}^{p}\hat{D}^{q}k_{(\hat{\Delta}+2)m}
      \right)
      -
      2
      \left(k_{(\hat{\Delta}+2)m}\right)^{2}
      ,
  \\
  \label{eq:JmpqKmpq-product}
  J_{(m)pq} K_{(m)}^{pq}
  &=&
      0
      ,
  \\
  \label{eq:JmpqJmpq-product}
  J_{(m)pq} J_{(m)}^{pq}
  &=&
      4
      K_{(m)pq} K_{(m)}^{pq}
      .
\end{eqnarray}

%*********************************************************************

Through
Eqs.~(\ref{eq:tracelessDpDqkhatDelta+2m-intermediate-express-final-thetatheta-Fourier-sum})
and
(\ref{eq:epsilonDpDqkhatDelta+2m-intermediate-express-final-thetatheta-Fourier-sum}),
we can consider the resolution of the $m$-degeneracy of $l=1$ mode in
Eq.~(\ref{eq:m-sum-l=1-is-given-by-htu-functional}) as follows:
\begin{eqnarray}
  &&
     \frac{1}{2\pi}  e^{+im\phi} \int_{0}^{2\pi} d\phi e^{-im\phi} H_{(\hat{\Delta}+2)pq}[\![h_{tu}]\!]
     \nonumber\\
  &=&
      \sum_{m'=-1,0,1} \left[
      \tilde{h}_{(e2,l=1,m)} e^{+im\phi} \frac{1}{2\pi} \int_{0}^{2\pi} d\phi e^{-im\phi} K_{(m')pq}
      \right.
      \nonumber\\
  && \quad\quad\quad\quad\quad
     \left.
      +
      \tilde{h}_{(o2,l=1,m)} e^{+im\phi} \frac{1}{2\pi} \int_{0}^{2\pi} d\phi e^{-im\phi} J_{(m')pq}
      \right]
      \nonumber\\
  &=&
      \sum_{m'=-1,0,1} \left[
      \tilde{h}_{(e2,l=1,m)}
      K_{(m)pq}
      \delta_{mm'}
      +
      \tilde{h}_{(o2,l=1,m)}
      J_{(m)pq} \delta_{mm'}
      \right]
      \nonumber\\
  &=&
      \tilde{h}_{(e2,l=1,m)}
      K_{(m)pq}
      +
      \tilde{h}_{(o2,l=1,m)}
      J_{(m)pq}
      .
\end{eqnarray}
Furthermore, from
Eqs.~(\ref{eq:KmpqKmpq-product})--(\ref{eq:JmpqJmpq-product}), we
obtain
\begin{eqnarray}
  \tilde{h}_{(e2,l=1,m)}
  &=&
      [K_{(m)pq}K_{(m)}^{pq}]^{-1}
      K_{(m)}^{pq} \frac{1}{2\pi}  e^{+im\phi} \int_{0}^{2\pi} d\phi e^{-im\phi} H_{(\hat{\Delta}+2)pq}[\![h_{tu}]\!]
      \nonumber\\
  &=:&
       \tilde{h}_{(e2,l=1,m)}[\![h_{tu}]\!]
       ,
       \label{eq:tildehe2l=1m-is}
\end{eqnarray}
and
\begin{eqnarray}
  \tilde{h}_{(o2,l=1,m)}
  &=&
      \frac{1}{4} [K_{(m)pq}K_{(m)}^{pq}]^{-1}
      J_{(m)}^{pq} \frac{1}{2\pi}  e^{+im\phi} \int_{0}^{2\pi} d\phi e^{-im\phi} H_{(\hat{\Delta}+2)pq}[\![h_{tu}]\!]
      \nonumber\\
  &=:&
       \tilde{h}_{(o2,l=1,m)}[\![h_{tu}]\!]
       .
       \label{eq:tildeho2l=1m-is}
\end{eqnarray}
Thus, we have obtained the mode coefficients $\tilde{h}_{(e2,l=1,m)}$
and $\tilde{h}_{(o2,l=1,m)}$ in the functional forms of the original
metric $h_{tu}$.

%*********************************************************************

Here, we summarize the conditions for the eigen functions
$k_{(\hat{\Delta})}$ and $k_{(\hat{\Delta}+2)}$ to obtain the inverse
relation of the metric decomposition (\ref{eq:hpq-fourier-2}).
To obtain the inverse relations of the mode decomposition of
the trace and the traceless-part of Eq.~(\ref{eq:hpq-fourier-2}), we
use the conditions
\begin{eqnarray}
  \label{eq:trace-decomp-inv-cond}
  &&
     k_{(\hat{\Delta})} \in \ScrK_{(\hat{\Delta})}, \quad
     k_{(\hat{\Delta}+2)} \in \ScrK_{(\hat{\Delta}+2)}, \quad
     k_{(\hat{\Delta}+2)} = k_{(\hat{\Delta}+2)m}
     = \Theta_{1m}(\theta) e^{im\phi}
     ,
  \\
  &&
     \left(\hat{D}_{p}k_{(\hat{\Delta})}\right)\left(\hat{D}^{p}k_{(\hat{\Delta})}\right)
     \neq 0
     ,
     \label{eq:traceless-decomp-inv-cond-2}
  \\
  &&
     K_{(m)pq} K_{(m)}^{pq}
     =
     \left(
     \hat{D}_{p}\hat{D}_{q}k_{(\hat{\Delta}+2)m}
     \right)
     \left(
     \hat{D}^{p}\hat{D}^{q}k_{(\hat{\Delta}+2)m}
     \right)
     -
     2
     \left(k_{(\hat{\Delta}+2)m}\right)^{2}
     \neq
     0
     .
  \label{eq:traceless-decomp-inv-cond-3}
\end{eqnarray}
The condition (\ref{eq:traceless-decomp-inv-cond-3}) implies the
nonvanishing $K_{(m)pq}$ and $J_{(m)pq}$.

%*********************************************************************

%%%%%%%%%%%%%%%%%%%%%%%%%%%%%%%%%%%%%%%%%%%
\subsubsection{$h_{Ap}$}
\label{sec:hAp}
%%%%%%%%%%%%%%%%%%%%%%%%%%%%%%%%%%%%%%%%%%%

%*********************************************************************

Next, we consider the inversion relation of the decomposition
(\ref{eq:hAp-fourier}) taking account of the kernel modes
$k_{(\hat{\Delta})}$ and $k_{(\hat{\Delta}+2)}$.
\begin{eqnarray}
  \label{eq:hAp-fourier-2}
  h_{Ap}
  &=&
      r \sum_{l,m} \left[
      \tilde{h}_{(e1)A} \hat{D}_{p}S
      +
      \tilde{h}_{(o1)A} \epsilon_{pq} \hat{D}^{q}S
      \right]
  \\
  &=&
      r \sum_{l,m,l\geq 2} \left[
      \tilde{h}_{(e1,l\geq 2)A} \hat{D}_{p}Y_{lm}
      +
      \tilde{h}_{(o1,l\geq 2)A} \epsilon_{pq} \hat{D}^{q}Y_{lm}
      \right]
      \nonumber\\
  &&
      +
      r \sum_{m} \left[
      \tilde{h}_{(e1,l=1)A} \hat{D}_{p}k_{(\hat{\Delta}+2)}
      +
      \tilde{h}_{(o1,l=1)A} \epsilon_{pq} \hat{D}^{q}k_{(\hat{\Delta}+2)}
      \right]
      \nonumber\\
  &&
      +
      r \left[
      \tilde{h}_{(e1,l=0)A} \hat{D}_{p}k_{(\hat{\Delta})}
      +
      \tilde{h}_{(o1,l=0)A} \epsilon_{pq} \hat{D}^{q}k_{(\hat{\Delta})}
      \right]
      .
      \label{eq:hAp-fourier-2-l=01-separation}
\end{eqnarray}
Taking the divergence of Eq.~(\ref{eq:hAp-fourier-2-l=01-separation}),
we obtain
\begin{eqnarray}
  \hat{D}^{p}h_{Ap}
  &=&
      r \sum_{l,m,l\geq 2}  \tilde{h}_{(e1,l\geq 2)A} \hat{\Delta}Y_{lm}
      - 2 r \sum_{m} \tilde{h}_{(e1,l=1,m)A} k_{(\hat{\Delta}+2)}
      .
      \label{eq:hAp-fourier-2-l=01-separation-div}
\end{eqnarray}
Applying the derivative operator $\hat{\Delta}+2$ to
Eq.~(\ref{eq:hAp-fourier-2-l=01-separation-div}), we obtain
\begin{eqnarray}
  [\hat{\Delta}+2]\hat{D}^{p}h_{Ap}
  &=&
      r \sum_{l,m,l\geq 2}  \tilde{h}_{(e1,l\geq 2)A} [\hat{\Delta}+2]\hat{\Delta}Y_{lm}
      .
      \label{eq:hAp-fourier-2-l=01-separation-div+Delta+2}
\end{eqnarray}
Using the Green functions $[\hat{\Delta}+2]^{-1}$, $\hat{\Delta}^{-1}$,
and the orthogonal property
(\ref{eq:spherical_harmonic_orthogonality}) of the spherical harmonics
$Y_{lm}$, we obtain
\begin{eqnarray}
  \tilde{h}_{(e1,l\geq 2)A}
  &=&
      \frac{1}{r} \int_{S^{2}} Y_{lm}^{*} \hat{\Delta}^{-1}
      [\hat{\Delta}+2]^{-1}[\hat{\Delta}+2]\hat{D}^{p}h_{Ap}
      =:
       \tilde{h}_{(e1,l\geq 2)A}[\![h_{Bs}]\!]
      .
      \label{eq:tildehe1A-as-the-functioal-of-hAp-lgeq2}
\end{eqnarray}
Thus, the mode coefficient $\tilde{h}_{(e1)A}$ is given in the form of
the functional of the original metric component $h_{Ap}$.
Through Eq.~(\ref{eq:tildehe1A-as-the-functioal-of-hAp-lgeq2}),
Eq.~(\ref{eq:hAp-fourier-2-l=01-separation-div}) is expressed as
\begin{eqnarray}
  \sum_{m} \tilde{h}_{(e1,l=1,m)A} k_{(\hat{\Delta}+2)}
  &=&
      \frac{1}{2} \sum_{l,m,l\geq 2}  \tilde{h}_{(e1,l\geq 2)A}[\![h_{Br}]\!] \hat{\Delta}Y_{lm}
      -
      \frac{1}{2r} \hat{D}^{p}h_{Ap}
      .
      \label{eq:hAp-fourier-2-l=01-separation-div-2}
\end{eqnarray}
To resolve the $m$-degeneracy of
Eq.~(\ref{eq:hAp-fourier-2-l=01-separation-div-2}), we use
Eq.~(\ref{eq:trace-decomp-inv-cond}) and
(\ref{eq:m-degeneracy-resolve-l=1}).
Then, we have
\begin{eqnarray}
  \tilde{h}_{(e1,l=1,m)A}
  &=&
      \frac{e^{im\phi}}{k_{(\hat{\Delta}+2)m}}
      \frac{1}{4\pi}
      \int_{0}^{2\pi} d\phi
      e^{-im'\phi}
      \left[
      \sum_{l,m',l\geq 2}  \tilde{h}_{(e1,l\geq 2)A}[\![h_{Br}]\!] \hat{\Delta}Y_{lm'}
      -
      \frac{1}{r} \hat{D}^{p}h_{Ap}
      \right]
      \nonumber\\
  &=:&
      \tilde{h}_{(e1,l=1,m)A}[\![h_{Bs}]\!]
      \label{eq:hAp-fourier-2-l=01-separation-div-3}
\end{eqnarray}

%*********************************************************************

On the other hand, taking the rotation of
Eq.~(\ref{eq:hAp-fourier-2-l=01-separation}), we have
\begin{eqnarray}
  \epsilon^{pq}\hat{D}_{q}h_{Ap}
  &=&
      r \sum_{l,m,l\geq 2} \left[
      \tilde{h}_{(o1,l\geq 2)A} \hat{\Delta}Y_{lm}
      \right]
      - 2 r \sum_{m} \left[
      \tilde{h}_{(o1,l=1)A} k_{(\hat{\Delta}+2)}
      \right]
      .
      \label{eq:hAp-fourier-2-l=01-separation-rotation}
\end{eqnarray}
As in the case of
Eq.~(\ref{eq:tildehe1A-as-the-functioal-of-hAp-lgeq2}), we have
\begin{eqnarray}
  \tilde{h}_{(o1,l\geq 2)A}
  &=&
      \frac{1}{r}\int_{S^{2}} Y_{lm}^{*}
      \hat{\Delta}^{-1}[\hat{\Delta}+2]^{-1}[\hat{\Delta}+2]\epsilon^{pq}\hat{D}_{q}h_{Ap}
      =:
      \tilde{h}_{(o1,l\geq 2)A}[\![h_{Br}]\!]
      ,
      \label{eq:tildeho1A-as-the-functioal-of-hBr-lgeq2}
  \\
  \tilde{h}_{(o1,l=1)A}
  &=&
      \frac{e^{im\phi}}{k_{(\hat{\Delta}+2)m}}
      \frac{1}{4\pi}
      \int_{0}^{2\pi}
      d\phi
      e^{-im'\phi}
      \left[
      \sum_{l,m',l\geq 2} \left\{
      \tilde{h}_{(o1,l\geq 2)A}[\![h_{Br}]\!] \hat{\Delta}Y_{lm'}
      \right\}
      -
      \frac{1}{r} \epsilon^{pq}\hat{D}_{q}h_{Ap}
      \right]
      \nonumber\\
  &=:&
       \tilde{h}_{(o1,l=1)A}[\![h_{Bs}]\!]
       .
       \label{eq:hAp-fourier-2-l=01-separation-rotation-3}
\end{eqnarray}

%*********************************************************************

Through Eqs.~(\ref{eq:tildehe1A-as-the-functioal-of-hAp-lgeq2}),
(\ref{eq:hAp-fourier-2-l=01-separation-div-3}),
(\ref{eq:tildeho1A-as-the-functioal-of-hBr-lgeq2}), and
(\ref{eq:hAp-fourier-2-l=01-separation-rotation-3}), we obtain
\begin{eqnarray}
  &&
     \!\!\!\!\!\!\!\!\!\!\!\!\!\!\!\!\!\!\!\!\!\!\!\!\!\!\!\!\!\!\!\!\!\!\!\!\!\!\!\!
     \tilde{h}_{(e1,l=0)A} \hat{D}_{p}k_{(\hat{\Delta})}
     +
     \tilde{h}_{(o1,l=0)A} \epsilon_{pq} \hat{D}^{q}k_{(\hat{\Delta})}
     \nonumber\\
  &=&
      \frac{1}{r}
      h_{Ap}
      -
      \sum_{l,m,l\geq 2} \left[
      \tilde{h}_{(e1,l\geq 2)A}[\![h_{Bs}]\!] \hat{D}_{p}Y_{lm}
      +
      \tilde{h}_{(o1,l\geq 2)A}[\![h_{Bs}]\!] \epsilon_{pq} \hat{D}^{q}Y_{lm}
      \right]
      \nonumber\\
  &&
     -
     \sum_{m} \left[
     \tilde{h}_{(e1,l=1)A}[\![h_{Bs}]\!] \hat{D}_{p}k_{(\hat{\Delta}+2)}
     +
     \tilde{h}_{(o1,l=1)A}[\![h_{Bs}]\!] \epsilon_{pq} \hat{D}^{q}k_{(\hat{\Delta}+2)}
     \right]
     \label{eq:hAp-fourier-2-l=01-separation-l=0-respect}
     \\
  &=:&
       \HF_{Ap}[\![h_{Bs}]\!]
       .
     \label{eq:H-hAp-def}
\end{eqnarray}
Here, we use the condition (\ref{eq:traceless-decomp-inv-cond-2}).
Then, we have
\begin{eqnarray}
  \tilde{h}_{(e1,l=0)A}
  &=&
      \left[
      \left(\hat{D}^{q}k_{(\hat{\Delta})}\right)
      \left(\hat{D}_{q}k_{(\hat{\Delta})}\right)
      \right]^{-1}
      \hat{D}^{p}k_{(\hat{\Delta})}
      \HF_{Ap}[\![h_{Bs}]\!]
      =:
      \tilde{h}_{(e1,l=0)A}[\![h_{Bs}]\!]
      ,
      \label{eq:tildehe1l=0A-express}
  \\
  \tilde{h}_{(o1,l=0)A}
  &=&
      \left[
      \left(\hat{D}^{r}k_{(\hat{\Delta})}\right)
      \left(\hat{D}_{r}k_{(\hat{\Delta})}\right)
      \right]^{-1}
      \epsilon^{pq}
      \hat{D}_{q}k_{(\hat{\Delta})}
      \HF_{Ap}[\![h_{Bs}]\!]
      =:
      \tilde{h}_{(o1,l=0)A}[\![h_{Bs}]\!]
      .
      \label{eq:tildeho1l=0A-express}
\end{eqnarray}

%*********************************************************************

Thus, we have shown that the mode coefficients $\tilde{h}_{(e1)A}$ and
$\tilde{h}_{(o1)A}$ for all $l\geq 0$ modes are given in the
functional forms (\ref{eq:tildehe1A-as-the-functioal-of-hAp-lgeq2}),
(\ref{eq:hAp-fourier-2-l=01-separation-div-3}),
(\ref{eq:tildeho1A-as-the-functioal-of-hBr-lgeq2}),
(\ref{eq:hAp-fourier-2-l=01-separation-rotation-3}),
(\ref{eq:tildehe1l=0A-express}), and (\ref{eq:tildeho1l=0A-express})
of the original metric $h_{Ap}$ under the conditions
(\ref{eq:trace-decomp-inv-cond})--(\ref{eq:traceless-decomp-inv-cond-3}).

%*********************************************************************

%%%%%%%%%%%%%%%%%%%%%%%%%%%%%%%%%%%%%%%%%%%
\subsubsection{$h_{AB}$}
\label{sec:hAB}
%%%%%%%%%%%%%%%%%%%%%%%%%%%%%%%%%%%%%%%%%%%

%*********************************************************************

Through the harmonic functions $Y_{lm}$ ($l\geq 2$),
$k_{(\hat{\Delta}+2)m}$, and $k_{(\hat{\Delta})}$, the component
$h_{AB}$ of the metric perturbation $h_{ab}$ is decomposed as
\begin{eqnarray}
  \label{eq:new_representation_of_hAB}
  h_{AB}
  =
  \sum_{l,m(l\geq 2)} \tilde{h}_{(l\geq 2)AB} S
  +
  \sum_{m=-1,0,1} \tilde{h}_{(l=1,m)AB} k_{(\hat{\Delta}+2)m}
  +
  \tilde{h}_{(l=0)AB} k_{(\hat{\Delta})}
  .
\end{eqnarray}
This decomposition has the same form as
Eq.~(\ref{eq:hpq-fourier-trace-new-expansion}) for the trace part of
the component $h_{pq}$.
Then, we obtain the inverse relations
\begin{eqnarray}
  \tilde{h}_{(l\geq 2)AB}
  &=&
      \int_{S^{2}} d\Omega Y_{lm}^{*}
      [\hat{\Delta}+2]^{-1} \hat{\Delta}^{-1} \hat{\Delta} [\hat{\Delta}+2] h_{AB}
      =:
      \tilde{h}_{(l\geq 2)AB}[\![h_{AB}]\!]
      ,
      \quad
      l\geq 2
      ,
      \label{eq:tildehAB-explicit-result-lgeq2}
  \\
  \tilde{h}_{(l=0)AB}
  &=&
      \frac{1}{2k_{(\hat{\Delta})}}
      \left[
      [\hat{\Delta}+2] h_{AB}
      -
      \sum_{l,m,(l\neq 0,1)} \tilde{h}_{(l\geq 2)AB}[\![h_{AB}]\!] [\hat{\Delta}+2] Y_{lm}
      \right]
      \nonumber\\
  &=:&
       \tilde{h}_{(l=0)AB}[\![h_{AB}]\!]
       ,
       \label{eq:hAB-fourier-Delta+2-l=0-is}
  \\
  \tilde{h}_{(l=1,m)AB}
  &=&
      \frac{1}{k_{(\hat{\Delta}+2)m}}
      e^{+im\phi}
      \frac{1}{2\pi} \int_{0}^{2\pi} d\phi e^{-im\phi}
      \nonumber\\
  && \quad\quad\quad\quad
     \times
     \left[
     \frac{1}{r^{2}} h_{AB}
     -
     \sum_{l,m',(l\neq 0,1)} \tilde{h}_{(l\geq 2)AB}[\![h_{AB}]\!] Y_{lm'}
     -
     \tilde{h}_{(l=0)AB}[\![h_{AB}]\!] k_{(\hat{\Delta})}
     \right]
     \nonumber\\
  &=:&
       \tilde{h}_{(l=1)AB}[\![h_{AB}]\!]
       .
       \label{eq:hAB-fourier-expansion-l=1-each-mode}
\end{eqnarray}
which correspond to Eqs.~(\ref{eq:hpq-fourier-trace-new-expansion}),
(\ref{eq:hpq-fourier-trace-new-expansion-Delta+2-l=0-is}), and
(\ref{eq:hpq-fourier-trace-new-expansion-l=1-each-mode}),
respectively.

%*********************************************************************

%%%%%%%%%%%%%%%%%%%%%%%%%%%%%%%%%%%%%%%%%%%
\subsubsection{Summary of the mode decomposition including $l=0,1$ modes}
\label{sec:hab-decomp-summary}
%%%%%%%%%%%%%%%%%%%%%%%%%%%%%%%%%%%%%%%%%%%

%*********************************************************************

Here, we summarize the mode decomposition by harmonic functions
$Y_{lm}$ ($l\geq 2$), $k_{(\hat{\Delta}+2)m}$, and
$k_{(\hat{\Delta})}$.
We decompose the components $\{h_{AB}, $ $h_{Ap}$, $h_{pq}\}$ of the
metric perturbation $h_{ab}$ as
Eqs.~(\ref{eq:hAB-fourier})--(\ref{eq:hpq-fourier}) with
\begin{eqnarray}
  \label{eq:harmonics-extended-choice-sum}
  S
  =
  \left\{
  \begin{array}{ccccc}
    Y_{lm} & \quad & \mbox{for} & \quad & l\geq 2; \\
    k_{(\hat{\Delta}+2)m} & \quad & \mbox{for} & \quad & l=1; \\
    k_{(\hat{\Delta})} & \quad & \mbox{for} & \quad & l=0.
  \end{array}
  \right.
\end{eqnarray}
This decomposition is invertible for any $l$, $m$ modes including
$l=0,1$ if the conditions
(\ref{eq:trace-decomp-inv-cond})--(\ref{eq:traceless-decomp-inv-cond-3}),
i.e.,
\begin{eqnarray}
  \label{eq:trace-decomp-inv-cond-sum}
  &&
     k_{(\hat{\Delta})} \in \ScrK_{(\hat{\Delta})}, \quad
     k_{(\hat{\Delta}+2)} \in \ScrK_{(\hat{\Delta}+2)}, \quad
     k_{(\hat{\Delta}+2)} = k_{(\hat{\Delta}+2)m}
     = \Theta_{1m}(\theta) e^{im\phi}
     ,
  \\
  &&
     \left(\hat{D}_{p}k_{(\hat{\Delta})}\right)\left(\hat{D}^{p}k_{(\hat{\Delta})}\right)
     \neq 0
     ,
     \label{eq:traceless-decomp-inv-cond-2-sum}
  \\
  &&
     K_{(m)pq} K_{(m)}^{pq}
     =
     \left(
     \hat{D}_{p}\hat{D}_{q}k_{(\hat{\Delta}+2)m}
     \right)
     \left(
     \hat{D}^{p}\hat{D}^{q}k_{(\hat{\Delta}+2)m}
     \right)
     -
     2
     \left(k_{(\hat{\Delta}+2)m}\right)^{2}
     \neq
     0
  \label{eq:traceless-decomp-inv-cond-3-sum}
\end{eqnarray}
are satisfied.
As the inverse relation of
Eqs.~(\ref{eq:hAB-fourier})--(\ref{eq:hpq-fourier}),
the mode coefficients of these decomposition are given in the
functional form of the metric components $h_{AB}$, $h_{Ap}$, and
$h_{pq}$ as Eqs.~(\ref{eq:tildehe0-explicit-result-lgeq2}),
(\ref{eq:hpq-fourier-trace-new-expansion-Delta+2-l=0-is}),
(\ref{eq:hpq-fourier-trace-new-expansion-l=1-each-mode}),
(\ref{eq:he2lgeq2-functional-hpq}),
(\ref{eq:ho2lgeq2-functional-hpq}),
(\ref{eq:tildehe2l=0-functional-of-hpq}),
(\ref{eq:tildeho2l=0-functional-of-hpq}), (\ref{eq:tildehe2l=1m-is}),
(\ref{eq:tildeho2l=1m-is}),
(\ref{eq:tildehe1A-as-the-functioal-of-hAp-lgeq2}),
(\ref{eq:hAp-fourier-2-l=01-separation-div-3}),
(\ref{eq:tildeho1A-as-the-functioal-of-hBr-lgeq2}),
(\ref{eq:hAp-fourier-2-l=01-separation-rotation-3}),
(\ref{eq:tildehe1l=0A-express}), (\ref{eq:tildeho1l=0A-express}), and
(\ref{eq:tildehAB-explicit-result-lgeq2})--(\ref{eq:hAB-fourier-expansion-l=1-each-mode}).
From
Eqs.~(\ref{eq:hAB-fourier})--(\ref{eq:hpq-fourier}),
the components $\{h_{AB},$ $h_{Ap},$ $h_{pq}\}$ vanish if all mode
coefficients $\{\tilde{h}_{AB},$ $\tilde{h}_{(e1)A},$
$\tilde{h}_{(o1)A},$ $\tilde{h}_{(e0)},$ $\tilde{h}_{(e2)},$
$\tilde{h}_{(o2)}\}$ vanish.
On the contrary, from the obtained functional forms, all mode
coefficients $\{\tilde{h}_{AB},$ $\tilde{h}_{(e1)A},$
$\tilde{h}_{(o1)A},$ $\tilde{h}_{(e0)},$ $\tilde{h}_{(e2)},$
$\tilde{h}_{(o2)}\}$ vanish if the components $\{h_{AB},$ $h_{Ap},$
$h_{pq}\}$ vanish.
This indicates the linear independence of the set of the harmonic
functions (\ref{eq:harmonic-fucntions-set}).
Therefore, the conditions
(\ref{eq:trace-decomp-inv-cond-sum})--(\ref{eq:traceless-decomp-inv-cond-3-sum})
guarantee the linear independence of the set of these harmonic
functions (\ref{eq:harmonic-fucntions-set}).

%*********************************************************************

We also note that the Green functions $\hat{\Delta}^{-1}$ and
$[\hat{\Delta}+2]^{-1}$ which used above do not directly operate to
the functions $k_{(\hat{\Delta})}$, nor $k_{(\hat{\Delta}+2)m}$.
Therefore, the domain of these Green function $\hat{\Delta}^{-1}$ and
$[\hat{\Delta}+2]^{-1}$ may be regarded as the $L^{2}$-space spanned
by $\{Y_{lm}| l\neq 0\}$ and $\{Y_{lm}| l\neq 1\}$, respectively.
The explicit form of these Green functions are given in
Ref.~\cite{R.Szmytkowski-2006a,R.Szmytkowski-2007}.

%*********************************************************************

%%%%%%%%%%%%%%%%%%%%%%%%%%%%%%%%%%%%%%%%%%%
%%%%%%%%%%%%%%%%%%%%%%%%%%%%%%%%%%%%%%%%%%%
\subsection{Explicit form of the mode functions}
\label{sec:Explict_form_of_the_mode_functions}
%%%%%%%%%%%%%%%%%%%%%%%%%%%%%%%%%%%%%%%%%%%
%%%%%%%%%%%%%%%%%%%%%%%%%%%%%%%%%%%%%%%%%%%

%*********************************************************************

Here, we consider the explicit expression of the mode functions
$k_{\hat{\Delta}}$ and $k_{(\hat{\Delta}+2)}$ which satisfy the
conditions
(\ref{eq:trace-decomp-inv-cond-sum})--(\ref{eq:traceless-decomp-inv-cond-3-sum}).
In Appendix~\ref{sec:Explict_form_of_Tlm_for_Ylm}, we explicitly see
that the choice $S=Y_{lm}$ for $l\geq 0$ does not satisfy these
conditions and what is happen in this choice.
As the result of this appendix, in the choice $S=Y_{lm}$, any vector and
tensor harmonics does not have their values for $l=0$ mode.
On the other hand, for $l=1$ modes, the vector harmonics have their
vector value and the trace parts of the second-rank tensor of each
modes have their tensor values, while all traceless even and odd mode
harmonics identically vanish.
Therefore, in the choice $S=Y_{lm}$, the set of harmonics
(\ref{eq:harmonic-fucntions-set}) does not play the role of
basis of tangent space on $S^{2}$ for $l=0,1$ mode.
This situation already appeared in terms of the Green function
$\hat{\Delta}^{-1}$ and $(\hat{\Delta}+2)^{-1}$ in the inverse
relations in
Sec.~\ref{sec:2+2-formulation-Perturbation-decomposition}.
For this reason, we seek an alternative choice of $S$ which satisfy
the conditions
(\ref{eq:trace-decomp-inv-cond-sum})--(\ref{eq:traceless-decomp-inv-cond-3-sum}).

%*********************************************************************

%%%%%%%%%%%%%%%%%%%%%%%%%%%%%%%%%%%%%%%%%%%
\subsubsection{Explicit form of $k_{(\hat{\Delta})}$}
\label{sec:Explict_form_of_kernel_Delta_mode_functions}
%%%%%%%%%%%%%%%%%%%%%%%%%%%%%%%%%%%%%%%%%%%

%*********************************************************************

Here, we treat the modes which belong to the kernel of the derivative
operator $\hat{\Delta}$, i.e.,
\begin{eqnarray}
  \label{eq:l=0-mode-equation}
  \hat{\Delta}k_{(\hat{\Delta})}
  =
  \frac{1}{\sqrt{\gamma}}
  \partial_{p}\left(
  \sqrt{\gamma}\gamma^{pq}\partial_{q}k_{(\hat{\Delta})}
  \right)
  = 0.
\end{eqnarray}
We look for the function which satisfies the conditions
(\ref{eq:trace-decomp-inv-cond-sum}) and
(\ref{eq:traceless-decomp-inv-cond-2-sum}).
We emphasize that we do not impose the regularity on the
function $k_{(\hat{\Delta})}$ on $S^{2}$ itself in this selection of
$k_{(\hat{\Delta})}$.
Since the regularity is a kind of boundary conditions for
perturbations, this regularity may be imposed on the solutions when we
solve the Einstein equations.

%*********************************************************************

Our guiding principle to look for the solution to
Eq.~(\ref{eq:l=0-mode-equation}) with a simple modification from the
conventional spherical harmonic functions.
Although the conditions
(\ref{eq:trace-decomp-inv-cond-sum}) and
(\ref{eq:traceless-decomp-inv-cond-2-sum})
do not restrict the $\phi$-dependence for $k_{(\hat{\Delta})}$, we
look for the solution to Eq.~(\ref{eq:l=0-mode-equation}) which is
independent of $\phi$ as the original $Y_{00}$ in the conventional
spherical harmonics is so.
Then,  in terms of the coordinate system where $\gamma_{ab}$ is given
by Eq.~(\ref{eq:background-metric-2+2-gamma-comp-Schwarzschild}), Eq.~(\ref{eq:l=0-mode-equation}) yields
\begin{eqnarray}
  \label{eq:l=0-mode-equation-explicit-3}
  \frac{d^{2}}{dy^{2}}k_{(\hat{\Delta})} = 0,
\end{eqnarray}
where we introduced an independent variable $y$ by
\begin{eqnarray}
  \label{eq:l=0-mode-y-def-2}
  y = \ln\left(\frac{1-\cos\theta}{1+\cos\theta}\right)^{1/2}.
\end{eqnarray}
As the solution to Eq.~(\ref{eq:l=0-mode-equation-explicit-3}), we
choose
\begin{eqnarray}
  k_{(\hat{\Delta})}
  =
  1 + \delta y
  =
  1 + \delta \ln\left(\frac{1-\cos\theta}{1+\cos\theta}\right)^{1/2},
  \quad \delta \in\RF
  .
  \label{eq:l=0-general-mode-func-specific}
\end{eqnarray}
If $\delta\neq 0$, we see that
\begin{eqnarray}
  \hat{D}_{p}k_{(\hat{\Delta})} (dx^{p})_{a}
  =
  \delta (dy)_{a}
  =
  \frac{\delta}{\sin\theta} (d\theta)_{a}
  \neq 0,
  \label{eq:l=0-extended-mode-func-Dp}
\end{eqnarray}
and
\begin{eqnarray}
  \left(
  \hat{D}_{p}k_{(\Delta)}
  \right)
  \left(
  \hat{D}^{p}k_{(\Delta)}
  \right)
  =
  \frac{\delta^{2}}{\sin^{2}\theta} \neq 0.
  \label{eq:l=0-extended-mode-func-Dp-norm}
\end{eqnarray}
Thus, $\hat{D}_{p}k_{(\hat{\Delta})}$ given by
Eq.~(\ref{eq:l=0-extended-mode-func-Dp}) and
$\epsilon_{pq}\hat{D}^{q}k_{(\hat{\Delta})}$ spans the vector space
though their norm is singular at $\theta=0,\pi$.
The solution (\ref{eq:l=0-general-mode-func-specific}) to
Eq.~(\ref{eq:l=0-mode-equation-explicit-3}) also yields
\begin{eqnarray}
  \left(
  \hat{D}_{p}\hat{D}_{q}
  -
  \frac{1}{2} \gamma_{pq} \hat{\Delta}
  \right)k_{(\hat{\Delta})}
  &=&
  \hat{D}_{p}\hat{D}_{q}k_{(\hat{\Delta})}
      =
      \delta
      \frac{\cos\theta}{\sin^{2}\theta}
      \left(
      -
      \theta_{p} \theta_{q}
      +
      \phi_{p} \phi_{q}
      \right)
      \neq 0
      ,
  \label{eq:l=0-extended-mode-func-DpDq}
  \\
  \epsilon_{r(p} \hat{D}_{q)}\hat{D}^{r}k_{(\hat{\Delta})}
  &=&
      -
      2
      \delta
      \frac{\cos\theta}{\sin^{2}\theta}
      \theta_{(p} \phi_{q)}
      \neq 0
      .
  \label{eq:l=0-extended-mode-func-epsilonDpDq}
\end{eqnarray}
Together with the trace part
\begin{eqnarray}
  \frac{1}{2} \gamma_{pq} k_{(\hat{\Delta})}
  =
  \frac{1}{2} \left(
  1 + \delta \ln\left(\frac{1-\cos\theta}{1+\cos\theta}\right)^{1/2}
  \right)
  (\theta_{p}\theta_{q}+\phi_{p}\phi_{q})
  ,
  \label{eq:l=0-extended-mode-func-gammapq}
\end{eqnarray}
the tensor (\ref{eq:l=0-extended-mode-func-DpDq}) and
(\ref{eq:l=0-extended-mode-func-epsilonDpDq}) span the basis of the
space of the second-rank tensor field though these are singular at
$\theta=0,\pi$.

%*********************************************************************

%%%%%%%%%%%%%%%%%%%%%%%%%%%%%%%%%%%%%%%%%%%
\subsubsection{Explicit form of $k_{(\hat{\Delta}+2)}$}
\label{sec:Explict_form_of_kernel_Delta+2_mode_functions}
%%%%%%%%%%%%%%%%%%%%%%%%%%%%%%%%%%%%%%%%%%%

%*********************************************************************

Here, we consider the kernel mode $k_{(\hat{\Delta}+2)}$ for the
operator $\hat{\Delta}+2$.
The condition (\ref{eq:trace-decomp-inv-cond-sum}) for
$k_{(\hat{\Delta}+2)}$ is given by
\begin{eqnarray}
  \label{eq:l=1-mode-equation}
  \left(\hat{\Delta}+2\right)k_{(\hat{\Delta}+2)}
  =
  \frac{1}{\sqrt{\gamma}}
  \partial_{p}\left(
  \sqrt{\gamma}\gamma^{pq}\partial_{q}k_{(\hat{\Delta}+2)}
  \right)
  +
  2 k_{(\hat{\Delta}+2)}
  = 0.
\end{eqnarray}
We look for the function which satisfies the conditions
(\ref{eq:trace-decomp-inv-cond-sum}) and
(\ref{eq:traceless-decomp-inv-cond-3-sum}).
We emphasize that we do not impose the regularity on the function
$k_{(\hat{\Delta}+2)}$ on $S^{2}$ itself as in the case of
$k_{(\hat{\Delta})}$.
To obtain the solution to Eq.~(\ref{eq:l=1-mode-equation}) which
satisfies the conditions (\ref{eq:trace-decomp-inv-cond-sum}) and
(\ref{eq:traceless-decomp-inv-cond-3-sum}),
we first consider the $\phi$-dependence from the condition
(\ref{eq:trace-decomp-inv-cond-sum}).
Then, Eq.~(\ref{eq:l=1-mode-equation}) is given by
\begin{eqnarray}
  \label{eq:l=1-mode-explicit-equation}
  \sin\theta
  \partial_{\theta}\left(
  \sin\theta \partial_{\theta}\Theta_{1m}(\theta)
  \right)
  - m^{2} \Theta_{1m}(\theta)
  +
  2 \sin^{2}\theta \Theta_{1m}(\theta)
  = 0.
\end{eqnarray}

%*********************************************************************

To solve Eq.~(\ref{eq:l=1-mode-explicit-equation}), we introduce the
independent variable
\begin{eqnarray}
  \label{eq:l=1-mode-z-def}
  z = \cos\theta, \quad dz = - \sin\theta d\theta.
\end{eqnarray}
In terms of the independent variable $z$, we obtain
\begin{eqnarray}
  \sin\theta \frac{d}{d\theta}
  &=&
      - (1-z^{2}) \frac{d}{dz}
      .
      \label{eq:l=1-mode-dzdtheta}
\end{eqnarray}
Then, Eq.~(\ref{eq:l=1-mode-explicit-equation}) is given by
\begin{eqnarray}
  \label{eq:l=1-mode-explicit-equation-2}
  (1-z^{2}) \frac{d^{2}}{dz^{2}}\Theta_{1m}(\theta)
  - 2 z \frac{d}{dz}\Theta_{1m}(\theta)
  +
  \left(
  1(1+1) - \frac{m^{2}}{1-z^{2}}
  \right)
  \Theta_{1m}(\theta)
  = 0.
\end{eqnarray}

%*********************************************************************

Suppose that we have obtained the solution to
Eq.~(\ref{eq:l=1-mode-explicit-equation-2}) as
\begin{eqnarray}
  k_{(\hat{\Delta}+2)m} = \Theta_{1m}(\theta) e^{im\phi}.
\end{eqnarray}
Here, we introduce the ladder operator
$\hat{L}_{\pm}$~\cite{J.J.Sakurai-1994} as
\begin{eqnarray}
  \hat{L}_{\pm}
  :=
  - i e^{\pm i\phi} \left(
  \pm i \partial_{\theta} - \cot\theta \partial_{\phi}
  \right)
  \label{eq:Ladder-operator-defs}
\end{eqnarray}
and examine the function defined by
\begin{eqnarray}
  \hat{L}_{+}k_{(\hat{\Delta}+2,m)}
  &=&
      - i e^{+i\phi} \left(
      + i \partial_{\theta} - \cot\theta \partial_{\phi}
      \right)
      \Theta_{1m}(\theta) e^{im\phi}
      \nonumber\\
  &=&
      \left(
      \partial_{\theta} - m \cot\theta
      \right)
      \Theta_{1m}(\theta) e^{i(m+1)\phi}
      .
      \label{eq:+Ladder-operator-kDelta+2m}
\end{eqnarray}
Evidently, the function given by
Eq.~(\ref{eq:+Ladder-operator-kDelta+2m}) is the eigenfunction of
the operator $-i\partial_{\phi}$ with the eigenvalue $m+1$:
\begin{eqnarray}
  - i \partial_{\phi}\hat{L}_{+}k_{(\hat{\Delta}+2,m)} = (m+1) \hat{L}_{+}k_{(\hat{\Delta}+2,m)}.
  \label{eq:+Ladder-operator-kDelta+2m_is_m+1_eigen}
\end{eqnarray}
Now, we consider the variable $\Phi_{+}$ defined by
\begin{eqnarray}
  \Phi_{+}
  &:=&
       \left(
       \partial_{\theta} - m \cot\theta
       \right)
       \Theta_{1m}
       \nonumber\\
  &=&
      -
      (1-z^{2})^{-1/2}
      \left[
      (1-z^{2}) \frac{d}{dz} \Theta_{1,m}
      +
      m z \Theta_{1m}
      \right]
      ,
\end{eqnarray}
and straightforward calculations using
Eq.~(\ref{eq:l=1-mode-explicit-equation-2}) yields
\begin{eqnarray}
  (1-z^{2}) \frac{d^{2}}{dz^{2}}\Phi_{+}
  - 2 z \frac{d}{dz}\Phi_{+}
  +
  \left(
  1(1+1)  -  \frac{(m+1)^{2}}{1-z^{2}}
  \right)
  \Phi_{+}
  =
  0.
\end{eqnarray}
This indicates
\begin{eqnarray}
  \Phi_{+} = \Theta_{1,m+1}(\theta).
\end{eqnarray}
Therefore, we conclude that
\begin{eqnarray}
  \hat{L}_{+}k_{(\hat{\Delta}+2,m)} = k_{(\hat{\Delta}+2,m+1)}.
\end{eqnarray}

%*********************************************************************

On the other hand, we consider the operator $\hat{L}_{-}$ defined by
\begin{eqnarray}
  \hat{L}_{-}k_{(\hat{\Delta}+2,m)}
  &=&
      - i e^{-i\phi} \left( - i \partial_{\theta} -
      \cot\theta \partial_{\phi} \right)
      \Theta_{1,m} e^{im\phi}
      \nonumber\\
  &=&
      \left( - \partial_{\theta} - m \cot\theta \right)
      \Theta_{1,m} e^{i(m-1)\phi}
      .
      \label{eq:hatL-kDelta+2m}
\end{eqnarray}
Evidently, the function given by Eq.~(\ref{eq:hatL-kDelta+2m}) is an
eigenfunction of the operator $-i\partial_{\phi}$ with the eigenvalue
$m-1$:
\begin{eqnarray}
  - i \partial_{\phi}\hat{L}_{-}k_{(\hat{\Delta}+2,m)} = (m-1) \hat{L}_{-}k_{(\hat{\Delta}+2,m)}.
  \label{eq:-Ladder-operator-kDelta+2m_is_m-1_eigen}
\end{eqnarray}
Now, we consider
\begin{eqnarray}
  \Phi_{-}
  :=
  \left(
  - \partial_{\theta} - m \cot\theta
  \right)
  \Theta_{1,m}
\end{eqnarray}
and straightforward calculations using
Eq.~(\ref{eq:l=1-mode-explicit-equation-2}) yields
\begin{eqnarray}
  (1-z^{2}) \frac{d^{2}}{dz^{2}}\Phi_{-}
  - 2 z \frac{d}{dz}\Phi_{-}
  +
  \left[
  1(1+1)
  -  \frac{(m-1)^{2}}{1-z^{2}}
  \right]
  \Phi_{-}
  =
  0
  .
\end{eqnarray}
This indicates
\begin{eqnarray}
  \Phi_{-} = \Theta_{1,m-1}(\theta).
\end{eqnarray}
Therefore, we conclude that
\begin{eqnarray}
  \hat{L}_{-}k_{(\hat{\Delta}+2)m} = k_{(\hat{\Delta}+2)m-1}.
\end{eqnarray}

%*********************************************************************

From the above operator $\hat{L}_{\pm}$ and
\begin{eqnarray}
  \label{eq:Ladder-generation-of-mpm1}
  \hat{L}_{\pm}k_{(\hat{\Delta}+2)m}=k_{(\hat{\Delta}+2)m\pm 1},
\end{eqnarray}
we may concentrate only to solve $m=0$ case.
Corresponding $m=\pm 1$ modes with $l=1$ can be derived from
Eq.~(\ref{eq:Ladder-generation-of-mpm1}).
Since $k_{(\hat{\Delta}+2)m=0}=\Theta_{10}(\theta)$, the equation for
$\Theta_{10}(\theta)$ is given by
\begin{eqnarray}
  \label{eq:l=1-mode-explicit-equation-m=0}
  (1-z^{2}) \frac{d^{2}}{dz^{2}}\Theta_{10}(\theta)
  - 2 z \frac{d}{dz}\Theta_{10}(\theta)
  + 1(1+1) \Theta_{10}(\theta)
  = 0.
\end{eqnarray}
Here, we note that $\Theta_{10}=z\propto Y_{10}$ should be a solution
to Eq.~(\ref{eq:l=1-mode-explicit-equation-m=0}).
To obtain the other independent solution, we consider the solution in
the form $\Theta_{10}=\Psi(z)z$.
Substituting this into Eq.~(\ref{eq:l=1-mode-explicit-equation-m=0})
we can solve Eq.~(\ref{eq:l=1-mode-explicit-equation-m=0}) as
\begin{eqnarray}
  \label{eq:l=1-mode-explicit-equation-m=0-sol}
  \Theta_{10}
  = z + \delta \left(\frac{1}{2}z\ln\frac{1+z}{1-z}-1\right),
\end{eqnarray}
where we choose one of constant of integration as 1 and $\delta$ is
another integration constant.
Then, we obtain
\begin{eqnarray}
  \label{eq:kDelta+2m=0-sol}
  k_{(\hat{\Delta}+2)m=0} =  z + \delta \left(\frac{1}{2}z\ln\frac{1+z}{1-z}-1\right)
  = P_{1}(z) + \delta Q_{1}(z),
\end{eqnarray}
where $P_{1}(z)$ is the Legendre polynomial and $Q_{1}(z)$ is the
first order and the second kind Legendre function.

%*********************************************************************

Since we have the explicit form (\ref{eq:kDelta+2m=0-sol}) of
$k_{(\hat{\Delta}+2)m=0}$ as
\begin{eqnarray}
  k_{(\hat{\Delta}+2)m} = \Theta_{1m}(\theta) e^{im\phi},
\end{eqnarray}
we can derive the $m=\pm 1$ modes by applying the ladder operators
$\hat{L}_{\pm}$ defined by Eq.~(\ref{eq:Ladder-operator-defs}) as
\begin{eqnarray}
  k_{(\hat{\Delta}+2)m=\pm 1}
  &=&
      \hat{L}_{\pm}k_{(\hat{\Delta}+2)m=0}
      \nonumber\\
  &=&
      \left[
      \sqrt{1-z^{2}}
      +
      \delta \left(
      \frac{1}{2} \sqrt{1-z^{2}} \ln \frac{1+z}{1-z} + \frac{z}{\sqrt{1-z^{2}}}
      \right)
      \right]
      e^{\pm i \phi}
      .
      \label{eq:l=1-mode-mpm1-mode-sol}
\end{eqnarray}
Equations (\ref{eq:kDelta+2m=0-sol}) and
(\ref{eq:l=1-mode-mpm1-mode-sol}) are summarized as
\begin{eqnarray}
  k_{(\hat{\Delta}+2,m=0)}
  &=&
      \cos\theta
      +
      \delta \left(\frac{1}{2} \cos\theta \ln\frac{1+\cos\theta}{1-\cos\theta} -1\right)
      ,
     \quad \delta \in \RF
     ,
     \label{eq:l=1-m=0-mode-func-explicit}
     \\
  k_{(\hat{\Delta}+2,m=\pm1)}
  &=&
      \left[
      \sin\theta
      +
      \delta \left(
      + \frac{1}{2} \sin\theta \ln\frac{1+\cos\theta}{1-\cos\theta}
      + \cot\theta
      \right)
      \right]
      e^{\pm i\phi}
      .
      \label{eq:l=1-m=pm1-mode-func-explicit}
\end{eqnarray}

%*********************************************************************

Here, we check the non-vanishing properties of
$\hat{D}_{p}k_{(\hat{\Delta}+2)}$ and
$\hat{D}_{p}\hat{D}_{q}k_{(\hat{\Delta}+2)}$.
For $m=0$ modes, the vector $\hat{D}_{p}k_{(\hat{\Delta}+2,m=0)}$ is
given by
\begin{eqnarray}
  \hat{D}_{p}k_{(\hat{\Delta}+2,m=0)}
  =
  -
  \left[
  1
  +
  \frac{1}{2} \delta \left(
  \ln\frac{1+\cos\theta}{1-\cos\theta}
  +
  \frac{2\cos\theta}{\sin^{2}\theta}
  \right)
  \right]
  \sin\theta \theta_{p}
  .
  \label{eq:Delta+2m=0gradient}
\end{eqnarray}
Then $\hat{D}_{p}k_{(\hat{\Delta}+2,m=0)}$ and
$\epsilon_{pq}\hat{D}^{q}k_{(\hat{\Delta}+2),m=0}$ span the basis of
the tangent space on $S^{2}$.

%*********************************************************************

Next, we consider the tensor
$\hat{D}_{q}\hat{D}_{p}k_{(\hat{\Delta}+2,m=0)}$ as
\begin{eqnarray}
  &&
     \hat{D}_{q}\hat{D}_{p}k_{(\hat{\Delta}+2,m=0)}
     \nonumber\\
  &=&
      -
      \left[
      \cos\theta
      +
      \frac{1}{2} \delta \left(
      + \cos\theta \ln\frac{1+\cos\theta}{1-\cos\theta}
      -  4
      -  2 \cot^{2}\theta
      \right)
      \right]
      \theta_{p}
      \theta_{q}
      \nonumber\\
  &&
     -
     \left[
     \cos\theta
     +
     \frac{1}{2} \delta \cos\theta \left(
     + \ln\frac{1+\cos\theta}{1-\cos\theta}
     + \frac{2\cos\theta}{\sin^{2}\theta}
     \right)
     \right]
     \phi_{p}
     \phi_{q}
     .
     \label{eq:Delta+2m=0Hesse}
\end{eqnarray}
This does not proportional to $\gamma_{ab}$.
Therefore, we should have nonvanishing $K_{(m)pq}$ and $J_{(m)pq}$
defined by
Eqs.~(\ref{eq:DpDq-halfgammapqDeltakhatDelta+2m-notation}).
To confirm this, we evaluate the condition
(\ref{eq:traceless-decomp-inv-cond-3}) as
\begin{eqnarray}
  \left(\hat{D}_{p}\hat{D}_{q}k_{(\hat{\Delta}+2)}\right)
  \left(\hat{D}^{p}\hat{D}^{q}k_{(\hat{\Delta}+2)}\right)
  -
  2 \left(k_{(\hat{\Delta}+2)}\right)^{2}
  =
  \frac{2 \delta}{\sin^{4}\theta}
  .
  \label{eq:condition-kernel-mode-of-Delta+2-cond-3-check-m=0}
\end{eqnarray}
This indicates that we have nonvanishing $K_{(m)pq}$ and $J_{(m)pq}$
if $\delta\neq 0$.
However, we should note that these tensor singular at $\theta=0,\pi$.

%*********************************************************************

For $m=\pm 1$ modes, the vector
$\hat{D}_{p}k_{(\hat{\Delta}+2,m=\pm 1)}$ is given by
\begin{eqnarray}
  \hat{D}_{p}k_{(\hat{\Delta}+2,m=\pm1)}
  &=&
      \left[
      \cos\theta
      +
      \delta \left(
      + \frac{1}{2} \cos\theta \ln\frac{1+\cos\theta}{1-\cos\theta}
      - 1
      -  \frac{1}{\sin^{2}\theta}
      \right)
      \right]
      e^{\pm i\phi}
      \theta_{p}
      \nonumber\\
  &&
     +
     (\pm i)
     \left[
     1
     +
     \delta \left(
     + \frac{1}{2} \sin\theta \ln\frac{1+\cos\theta}{1-\cos\theta}
     + \frac{\cos\theta}{\sin^{2}\theta}
     \right)
     \right]
     e^{\pm i\phi}
     \phi_{p}
     .
     \label{eq:Delta+2-m=pm1-gradient}
\end{eqnarray}
Finally, we evaluate the condition
(\ref{eq:traceless-decomp-inv-cond-3}) as
\begin{eqnarray}
  \hat{D}_{q}\hat{D}_{p}k_{(\hat{\Delta}+2,m=\pm1)}
  &=&
      \left[
      - \sin\theta
      +
      \delta \left(
      - \frac{1}{2} \sin\theta \ln\frac{1+\cos\theta}{1-\cos\theta}
      - \frac{\cos\theta}{\sin\theta}
      + \frac{2 \cos\theta}{\sin^{3}\theta}
      \right)
      \right]
      e^{\pm i\phi}
      \theta_{p}
      \theta_{q}
      \nonumber\\
  &&
     +
     \left[
     - \sin\theta
     + \delta \left(
     -  \frac{1}{2} \sin\theta \ln\frac{1+\cos\theta}{1-\cos\theta}
     - \frac{\cos\theta}{\sin\theta}
     - \frac{2\cos\theta}{\sin^{3}\theta}
     \right)
     \right]
     e^{\pm i\phi}
     \phi_{p} \phi_{q}
     \nonumber\\
  &&
     \mp
     \frac{4 i \delta}{\sin^{3}\theta}
     e^{\pm i\phi}
     \theta_{(p} \phi_{q)}
     .
     \label{eq:Delta+2m=pm1Hesse}
\end{eqnarray}
This does not proportional to $\gamma_{pq}$.
Therefore, we should have nonvanishing $K_{(m)pq}$ and $J_{(m)pq}$
defined by
Eqs.~(\ref{eq:DpDq-halfgammapqDeltakhatDelta+2m-notation}).
This is confirmed by the check of the condition
(\ref{eq:traceless-decomp-inv-cond-3}) as
\begin{eqnarray}
  \left(\hat{D}_{p}\hat{D}_{q}k_{(\hat{\Delta}+2)}\right)
  \left(\hat{D}^{p}\hat{D}^{q}k_{(\hat{\Delta}+2)}\right)
  -
  2 \left(k_{(\hat{\Delta}+2)}\right)^{2}
  =
  - \frac{8 \delta^{2}}{\sin^{4}\theta}  e^{\pm 2i\phi}
  .
\end{eqnarray}
Then, we have seen that if $\delta\neq 0$, the condition
(\ref{eq:traceless-decomp-inv-cond-3}) is satisfied, though this
norm is singular at $\theta=0,\pi$.
We also note that $K_{(m)pq}$ is orthogonal to $J_{(m)pq}$ as shown in
Eqs.~(\ref{eq:KmpqKmpq-product})--(\ref{eq:JmpqJmpq-product}).
Therefore, $\gamma_{pq}$, $K_{(m)pq}$, and $J_{(m)pq}$ span the basis
of the second-rank tensor field on $S^{2}$.

%*********************************************************************

%%%%%%%%%%%%%%%%%%%%%%%%%%%%%%%%%%%%%%%%%%%
%%%%%%%%%%%%%%%%%%%%%%%%%%%%%%%%%%%%%%%%%%%
\subsection{Proposal of the treatment of $l=0,1$-mode perturbations}
\label{sec:Proposal}
%%%%%%%%%%%%%%%%%%%%%%%%%%%%%%%%%%%%%%%%%%%
%%%%%%%%%%%%%%%%%%%%%%%%%%%%%%%%%%%%%%%%%%%

%*********************************************************************

As shown in above, it is shown that the harmonic decomposition
(\ref{eq:hAB-fourier})--(\ref{eq:hpq-fourier}) have the one-to-one
correspondence between the original metric perturbations
$\{h_{AB},h_{Ap},h_{pq}\}$ and the mode coefficients
$\{\tilde{h}_{AB},$ $\tilde{h}_{(e1)A},$
$\tilde{h}_{(o1)A},$ $\tilde{h}_{(e0)},$ $\tilde{h}_{(e2)},$
$\tilde{h}_{(o2)}\}$ for any modes $l\geq 0$
through the employment of the scalar harmonic functions
\begin{eqnarray}
  \label{eq:harmonics-extended-choice-sum-delta}
  S_{\delta}
  =
  \left\{
  \begin{array}{ccccc}
    Y_{lm} & \quad & \mbox{for} & \quad & l\geq 2; \\
    k_{(\hat{\Delta}+2)m} & \quad & \mbox{for} & \quad & l=1; \\
    k_{(\hat{\Delta})} & \quad & \mbox{for} & \quad & l=0,
  \end{array}
  \right.
\end{eqnarray}
where $k_{(\hat{\Delta})}$ is given by
Eq.~(\ref{eq:l=0-general-mode-func-specific}), i.e.,
\begin{eqnarray}
  k_{(\hat{\Delta})}
  =
  1 + \delta \ln\left(\frac{1-\cos\theta}{1+\cos\theta}\right)^{1/2},
  \quad \delta \in\RF
  \label{eq:l=0-general-mode-func-specific-sum}
\end{eqnarray}
and $k_{(\hat{\Delta}+2)m}$ are given by
Eqs.~(\ref{eq:kDelta+2m=0-sol}) and (\ref{eq:l=1-mode-mpm1-mode-sol}),
i.e.,
\begin{eqnarray}
  \label{eq:kDelta+2m=0-sol-sum}
  k_{(\hat{\Delta}+2)m=0}
  &=&
      \cos\theta + \delta \left(\frac{1}{2}\cos\theta\ln\frac{1+\cos\theta}{1-\cos\theta}-1\right)
      ,
  \\
  k_{(\hat{\Delta}+2)m=\pm 1}
  &=&
  \left[
  \sin\theta
  +
  \delta \left(
  \frac{1}{2} \sin\theta \ln \frac{1+\cos\theta}{1-\cos\theta} + \cot\theta
  \right)
  \right]
  e^{\pm i \phi}
  .
  \label{eq:l=1-mode-mpm1-mode-sol-sum}
\end{eqnarray}
These mode functions $k_{(\hat{\Delta}+2)m}$ and $k_{(\hat{\Delta})}$
are parametrized by the single parameter $\delta$.
This choice satisfies the conditions
(\ref{eq:trace-decomp-inv-cond-sum})--(\ref{eq:traceless-decomp-inv-cond-3-sum})
but singular at $\theta=0,\pi$ if $\delta\neq 0$.
When $\delta=0$, we have $k_{(\hat{\Delta})}\propto Y_{00}$ and
$k_{(\hat{\Delta}+2)m}\propto Y_{1m}$.
In this decomposition, for each mode of any $l\geq 0$, the set of
harmonic functions (\ref{eq:harmonic-fucntions-set}) are a
linear-independent set in the both senses of the second-rank tensor
field and the function on $S^{2}$.

%*********************************************************************

Using the above harmonics functions $S_{\delta}$ in
Eq.~(\ref{eq:harmonics-extended-choice-sum-delta}), we propose the
following strategy~\footnote{
  This statement of the
  proposal~\ref{proposal:treatment-proposal-on-pert-on-spherical-BG}
  actually indicates that at once we ignore the neighborhood of the
  region where the harmonic function $S_{\delta}$ diverges but we
  apply the analytic extension of the linearized solution to these
  regions when we choose $\delta=0$.
}:
\begin{proposal}
  \label{proposal:treatment-proposal-on-pert-on-spherical-BG}
  We decompose the metric perturbation $h_{ab}$ on the background
  spacetime with the metric
  (\ref{eq:background-metric-2+2})--(\ref{eq:background-metric-2+2-gamma-comp-Schwarzschild})
  through Eqs.~(\ref{eq:hAB-fourier})--(\ref{eq:hpq-fourier}) with the
  harmonic function $S_{\delta}$ given by
  Eq.~(\ref{eq:harmonics-extended-choice-sum-delta}).
  Then, Eqs.~(\ref{eq:hAB-fourier})--(\ref{eq:hpq-fourier}) become
  invertible including $l=0,1$ modes.
  After deriving the mode-by-mode field equations such as linearized
  Einstein equations by using the harmonic functions $S_{\delta}$, we
  choose $\delta=0$ as regular boundary condition for solutions when
  we solve these field equations.
\end{proposal}

%*********************************************************************

Since the set of the mode functions (\ref{eq:harmonic-fucntions-set})
with $S=S_{\delta}$ have the linear-independence including $l=0,1$
modes, we can construct gauge-invariant variables and evaluate the
field equations through the mode-by-mode analyses including $l=0,1$
modes through the choice of these mode functions.

%*********************************************************************

%%%%%%%%%%%%%%%%%%%%%%%%%%%%%%%%%%%%%%%%%%%
%%%%%%%%%%%%%%%%%%%%%%%%%%%%%%%%%%%%%%%%%%%
%%%%%%%%%%%%%%%%%%%%%%%%%%%%%%%%%%%%%%%%%%%
\section{Construction of gauge-invariant variables}
\label{sec:gauge-inv-pertur-sphe-sym-spacetime}
%%%%%%%%%%%%%%%%%%%%%%%%%%%%%%%%%%%%%%%%%%%
%%%%%%%%%%%%%%%%%%%%%%%%%%%%%%%%%%%%%%%%%%%
%%%%%%%%%%%%%%%%%%%%%%%%%%%%%%%%%%%%%%%%%%%

%*********************************************************************

In this section, we construct gauge-invariant variables for
perturbations on spherically symmetric background with the metric
(\ref{eq:background-metric-2+2}) through
Proposal~\ref{proposal:treatment-proposal-on-pert-on-spherical-BG}.
To construct gauge-invariant variables, we first discuss the
gauge-transformation rule for the metric perturbation $h_{ab}$.
In the derivation of the gauge-transformation rules for the mode
coefficient in the decomposition
(\ref{eq:hAB-fourier})--(\ref{eq:hpq-fourier}) with the harmonic
function $S=S_{\delta}$ given by
Eq.~(\ref{eq:harmonics-extended-choice-sum-delta}).
In this section, we use the relations of the covariant derivatives
associated with the metrices $g_{ab}$, $y_{ab}$, and $\gamma_{ab}$,
which are summarized in Appendix~\ref{sec:Cov_deriv_2+2-formulation}.
In Sec.~\ref{sec:2+2-formulation-gauge-trans}, we derive the
gauge-transformation rules for the mode coefficients of the metric
perturbation in the decomposition
(\ref{eq:hAB-fourier})--(\ref{eq:hpq-fourier}) with the harmonic
function $S=S_{\delta}$.
In Sec.~\ref{sec:2+2-formulation-gauge-invariants}, we explicitly
construct gauge-invariant variables for the metric perturbations
through the mode-by-mode analyses.
In Sec.~\ref{sec:2+2-formulation-gauge-invariants-summary}, we
summarize gauge-invariant and gauge-variant variables in the
four-dimensional form.

%*********************************************************************

%%%%%%%%%%%%%%%%%%%%%%%%%%%%%%%%%%%%%%%%%%%
%%%%%%%%%%%%%%%%%%%%%%%%%%%%%%%%%%%%%%%%%%%
\subsection{Gauge-transformation rules}
\label{sec:2+2-formulation-gauge-trans}
%%%%%%%%%%%%%%%%%%%%%%%%%%%%%%%%%%%%%%%%%%%
%%%%%%%%%%%%%%%%%%%%%%%%%%%%%%%%%%%%%%%%%%%

%*********************************************************************

Here, we consider the gauge-transformation rules for the
linear-order metric perturbation $h_{ab}$ following to
Proposal~\ref{proposal:treatment-proposal-on-pert-on-spherical-BG}.
The gauge-transformation rule for linear-order metric
perturbation is given by
\begin{eqnarray}
  \label{eq:first-order-gauge-trans-of-metric-AIA2010-00100}
  {}_{\ScrY}\!h_{ab}
  -
  {}_{\ScrX}\!h_{ab}
  =
  {\pounds}_{\xi}g_{ab}
  =
  2 \nabla_{(a}\xi_{b)}.
\end{eqnarray}
We rewrite this gauge-transformation rule in terms of 2+2 formulation.
To do this, the generator of gauge-transformation rules is
decomposed as
\begin{eqnarray}
  \label{eq:generator-components}
  \xi_{a} &=& \xi_{A} (dx^{A})_{a} + \xi_{p} (dx^{p})_{a}.
\end{eqnarray}
Through the component-representations
(\ref{eq:metric-perturbation-components}) and
(\ref{eq:generator-components}), the gauge-transformation rules
(\ref{eq:first-order-gauge-trans-of-metric-AIA2010-00100}) are given by
\begin{eqnarray}
  \label{eq:hAB-gauge-trans}
  {}_{\ScrY}\!h_{AB}
  -
  {}_{\ScrX}\!h_{AB}
  &=&
  \nabla_{A}\xi_{B}
  +
  \nabla_{B}\xi_{A}
  =
  \bar{D}_{A}\xi_{B}
  +
  \bar{D}_{B}\xi_{A}
  , \\
  \label{eq:hAp-gauge-trans}
  {}_{\ScrY}\!h_{Ap}
  -
  {}_{\ScrX}\!h_{Ap}
  &=&
  \nabla_{A}\xi_{p}
  +
  \nabla_{p}\xi_{A}
  =
  \bar{D}_{A}\xi_{p}
  +
  \hat{D}_{p}\xi_{A}
  -
  \frac{2}{r} \bar{D}_{A}r \xi_{p}
  , \\
  \label{eq:hpq-gauge-trans}
  {}_{\ScrY}\!h_{pq}
  -
  {}_{\ScrX}\!h_{pq}
  &=&
  \nabla_{p}\xi_{q}
  +
  \nabla_{q}\xi_{p}
  =
  \hat{D}_{p}\xi_{q}
  +
  \hat{D}_{q}\xi_{p}
  +
  2 r \bar{D}^{A}r \gamma_{pq} \xi_{A}
  .
\end{eqnarray}
Furthermore, through the mode-decomposition
(\ref{eq:hAB-fourier})--(\ref{eq:hpq-fourier}) and
\begin{eqnarray}
  \label{eq:xiA-fourier}
  \xi_{A}
  &=:&
       \sum_{l,m} \zeta_{A} S_{\delta}
       ,
  \\
  \label{eq:xip-fourier}
  \xi_{p}
  &=:&
       r \sum_{l,m} \left[
       \zeta_{(e)} \hat{D}_{p}S_{\delta}
       +
       \zeta_{(o)} \epsilon_{pq}\hat{D}^{q}S_{\delta}
       \right]
\end{eqnarray}
with the harmonic function $S_{\delta}$, we can carry out the
mode-by-mode analyses, since the set of the harmonic functions
(\ref{eq:harmonic-fucntions-set}) has the linear-independence due to
the choice $S=S_{\delta}$.
From Eq.~(\ref{eq:hAB-gauge-trans}), we obtain
\begin{eqnarray}
  {}_{\ScrY}\tilde{h}_{AB}
  -
  {}_{\ScrX}\tilde{h}_{AB}
  =
  2 \bar{D}_{(A}\zeta_{B)}
  .
  \label{eq:gauge-trans-hAB-even}
\end{eqnarray}
From Eq.~(\ref{eq:hAp-gauge-trans}), we obtain
\begin{eqnarray}
  {}_{\ScrY}\!\tilde{h}_{(e1)A}
  -
  {}_{\ScrX}\!\tilde{h}_{(e1)A}
  &=&
      \frac{1}{r}
      \zeta_{A}
      +
      \bar{D}_{A}\zeta_{(e)}
      -
      \frac{1}{r}
      \bar{D}_{A}r \zeta_{(e)}
      ,
      \label{eq:gauge-trans-hAp-even}
  \\
  {}_{\ScrY}\!\tilde{h}_{(o1)A}
  -
  {}_{\ScrX}\!\tilde{h}_{(o1)A}
  &=&
      \bar{D}_{A}\zeta_{(o)}
      -
      \frac{1}{r} \bar{D}_{A}r \zeta_{(o)}
      .
      \label{eq:gauge-trans-hAp-odd}
\end{eqnarray}
Finally, the gauge-transformation rules (\ref{eq:hpq-gauge-trans}) yield
\begin{eqnarray}
  \label{eq:gauge-trans-hpq-trace-even}
  {}_{\ScrY}\!\tilde{h}_{(e0)}
  -
  {}_{\ScrX}\!\tilde{h}_{(e0)}
  &=&
      \frac{4}{r}
      \left(
      - \frac{1}{2} l(l+1) \zeta_{(e)}
      +
      \bar{D}^{A}r \zeta_{A}
      \right)
      ,
  \\
  \label{eq:gauge-trans-hpq-even}
  {}_{\ScrY}\!\tilde{h}_{(e2)}
  -
  {}_{\ScrX}\!\tilde{h}_{(e2)}
  &=&
      \frac{2}{r}
      \zeta_{(e)}
      ,
  \\
  \label{eq:gauge-trans-hpq-odd}
  {}_{\ScrY}\!\tilde{h}_{(o2)}
  -
  {}_{\ScrX}\!\tilde{h}_{(o2)}
  &=&
      -
      \frac{1}{r}
      \zeta_{(o)}
      .
\end{eqnarray}
We note that these gauge-transformation rules
(\ref{eq:gauge-trans-hAB-even})--(\ref{eq:gauge-trans-hpq-odd}) are
not only that for $l\geq 2$ modes but also $l=0,1$ modes.

%*********************************************************************

When we use the usual spherical harmonics $Y_{lm}$ as the scalar
harmonics, i.e., $\delta=0$ from the starting point,  we only have
Eqs.~(\ref{eq:gauge-trans-hAB-even}) and
(\ref{eq:gauge-trans-hpq-trace-even}) with $l=0$ for $l=0$ mode
perturbations and the other gauge-transformation rules
(\ref{eq:gauge-trans-hAp-even}), (\ref{eq:gauge-trans-hAp-odd}),
(\ref{eq:gauge-trans-hpq-even}), and (\ref{eq:gauge-trans-hpq-odd}) do
not appear.
In this case, it is difficult to construct gauge-invariant variables
for $l=0$-mode perturbations through the similar procedure to the
$l\geq 2$-mode case.
For this reason, we usually use the gauge-fixing procedure for $l=0$
mode perturbations from the old paper by
Zerilli~\cite{F.Zerilli-1970-PRD}.
Of course, the construction of gauge-invariant variables might be
possible if we use the integral representations of the original metric
perturbations.
However, such gauge-invariant variables does not match to the
statement of Conjecture~\ref{conjecture:decomposition-conjecture}.
For this reason, we do not consider such integral representations,
here.

%*********************************************************************

Furthermore, for $l=1$ modes with $\delta=0$ from the starting point,
we do not have Eqs.~(\ref{eq:gauge-trans-hpq-even}) nor
(\ref{eq:gauge-trans-hpq-odd}) but we have Eqs.~(\ref{eq:gauge-trans-hAB-even})-(\ref{eq:gauge-trans-hpq-trace-even})
with nonvanishing $\zeta_{(e)}$ and $\zeta_{(o)}$.
For $l=1$ odd-mode perturbations, it is well-known that the variable
defined by
\begin{eqnarray}
  \label{eq:l=1-odd-KIF-gauge-inv}
  \Phi_{KIF}
  &:=&
       \epsilon^{AB}
       \bar{D}_{A}
       \left(
       \frac{1}{r}
       \tilde{h}_{(o1)B}
       \right)
       =
       \frac{1}{r}
       \partial_{t}
       \tilde{h}_{(o1)r}
       -
       \partial_{r}
       \left(
       \frac{1}{r}
       \tilde{h}_{(o1)t}
       \right)
\end{eqnarray}
is gauge invariant under the gauge transformation rule
(\ref{eq:gauge-trans-hAp-odd})~\cite{H.Kodama-H.Ishihara-Y.Fujiwara-1994},
where $\epsilon^{AB}=2(\partial_{t})^{[A}(\partial_{r})^{B]}$ in the
coordinate system (\ref{eq:background-metric-2+2-y-comp-Schwarzschild}).
However, when we reconstruct the original metric perturbations from
this gauge-invariant variables for $l=1$ odd-mode perturbation, we
have to integrate this gauge-invariant variables and we have to carry
out delicate arguments for the problem that the integration constants
are gauge-degree of freedom or not.
On the other hand, such arguments are not necessary for the
gauge-invariant variables given by the statement of
Conjecture~\ref{conjecture:decomposition-conjecture}.
In this sense, the above gauge-invariant variables $\Phi_{KIF}$ for
$l=1$ odd-mode perturbations does not match to the statement
of Conjecture~\ref{conjecture:decomposition-conjecture}.

%*********************************************************************

Moreover, for $l=1$ even-mode perturbations, it is difficult to
eliminate $\zeta_{(e)}$ and $\zeta_{A}$ from the gauge-transformation
of even-mode perturbations through the similar procedure to the
$l\geq 2$-mode case as in the case of $l=0$ modes.
In conventional approach, we use the gauge-fixing procedure for $l=1$
mode perturbations from the old paper by
Zerilli~\cite{F.Zerilli-1970-PRD} due to this reason.
Of course, the construction of gauge-invariant variables for $l=1$
even-modes might be possible if we use the integral representations of
the original metric perturbations.
However, such gauge-invariant variables does not match to the
statement of Conjecture~\ref{conjecture:decomposition-conjecture},
again.
For this reason, we do not consider such integral representation as in
the case of $l=0$ mode perturbation, again.

%*********************************************************************

These situations for $l=0,1$ mode perturbations are the essential
reason for our proposal of the introduction of the singular harmonics
$S=S_{\delta\neq 0}$.
As shown in below, we can construct the gauge-invariant variables
through the similar procedure to $l\geq 2$-mode case if we accept the
introduction of the singular harmonics $S=S_{\delta\neq 0}$ at the
starting point and
Proposal~\ref{proposal:treatment-proposal-on-pert-on-spherical-BG}.

%*********************************************************************

%%%%%%%%%%%%%%%%%%%%%%%%%%%%%%%%%%%%%%%%%%%
%%%%%%%%%%%%%%%%%%%%%%%%%%%%%%%%%%%%%%%%%%%
\subsection{Gauge-invariant and gauge-variant variables}
\label{sec:2+2-formulation-gauge-invariants}
%%%%%%%%%%%%%%%%%%%%%%%%%%%%%%%%%%%%%%%%%%%
%%%%%%%%%%%%%%%%%%%%%%%%%%%%%%%%%%%%%%%%%%%

%*********************************************************************

Inspecting gauge-transformation rules
(\ref{eq:gauge-trans-hAB-even})--(\ref{eq:gauge-trans-hpq-odd}),
we can define gauge-invariant variables.

%*********************************************************************

%%%%%%%%%%%%%%%%%%%%%%%%%%%%%%%%%%%%%%%%%%%
\subsubsection{Odd modes}
\label{sec:2+2-formulation-gauge-invariants-odd}
%%%%%%%%%%%%%%%%%%%%%%%%%%%%%%%%%%%%%%%%%%%

%*********************************************************************

From gauge-transformation rules
(\ref{eq:gauge-trans-hAp-odd}) and
(\ref{eq:gauge-trans-hpq-odd}), we easily find that the
following combination is gauge-invariant:
\begin{eqnarray}
  \tilde{h}_{(o1)A}
  - \bar{D}_{A}\left(
    - r \tilde{h}_{(o2)}
  \right)
  + \frac{1}{r} \bar{D}_{A}r \left(
    - r \tilde{h}_{(o2)}
  \right)
  =
  \tilde{h}_{(o1)A}
  + r \bar{D}_{A}\tilde{h}_{(o2)}
  =: \tilde{F}_{A}
  .
  \label{eq:2+2-gauge-inv-odd-lgeq2-def-tildeFA}
\end{eqnarray}
We also note that the gauge-transformation rule
(\ref{eq:gauge-trans-hpq-odd}) implies that
\begin{eqnarray}
  \label{eq:gauge-trans-tildeho1-sum-2}
  - r^{2} {}_{\ScrY}\tilde{h}_{(o2)}
  + r^{2} {}_{\ScrX}\tilde{h}_{(o2)}
  &=&
  r \zeta_{(o)}
  .
\end{eqnarray}

%*********************************************************************

%%%%%%%%%%%%%%%%%%%%%%%%%%%%%%%%%%%%%%%%%%%
\subsubsection{Even modes}
\label{sec:2+2-formulation-gauge-invariants-even}
%%%%%%%%%%%%%%%%%%%%%%%%%%%%%%%%%%%%%%%%%%%

%*********************************************************************

Now, we note that the gauge-transformation rule
(\ref{eq:gauge-trans-hpq-even}) implies that
\begin{eqnarray}
  \label{eq:gauge-trans-tildehe1-sum-2}
  \frac{r^{2}}{2} {}_{\ScrY}\!\tilde{h}_{(e2)}
  -
  \frac{r^{2}}{2} {}_{\ScrX}\!\tilde{h}_{(e2)}
  &=&
  r \zeta_{(e)}
  .
\end{eqnarray}
Inspecting gauge-transformation rules (\ref{eq:gauge-trans-hAp-even})
and (\ref{eq:gauge-trans-hpq-even}), we define the variable
$\tilde{Y}_{A}$ as
\begin{eqnarray}
  \tilde{Y}_{A}
  &:=&
  r \tilde{h}_{(e1)A}
  - r \bar{D}_{A}\left(\frac{r}{2} \tilde{h}_{(e2)}\right)
  + \bar{D}_{A}r \left(\frac{r}{2} \tilde{h}_{(e2)}\right)
  \nonumber\\
  &=&
  r \tilde{h}_{(e1)A}
  - \frac{r^{2}}{2} \bar{D}_{A}\tilde{h}_{(e2)}
  .
  \label{eq:2+2-gauge-trans-tildeYA-def}
\end{eqnarray}
We easily check that the gauge-transformation rules for the
variable $\tilde{Y}_{A}$ is given by
\begin{eqnarray}
  \label{eq:2+2-tildeYA-def-gauge-trans}
  {}_{\ScrY}\!\tilde{Y}_{A}
  -
  {}_{\ScrX}\!\tilde{Y}_{A}
  =
  \zeta_{A}.
\end{eqnarray}

%*********************************************************************

From the gauge-transformation rules
(\ref{eq:gauge-trans-tildehe1-sum-2}) and
(\ref{eq:2+2-tildeYA-def-gauge-trans}), we easily define the
gauge-invariant variables as follows:
First, from the gauge-transformation rules
(\ref{eq:gauge-trans-hAB-even}) and
(\ref{eq:2+2-tildeYA-def-gauge-trans}), the following
combination is gauge-invariant:
\begin{eqnarray}
  \label{eq:gauge-inv-lgeq2-tildeFAB-def}
  \tilde{F}_{AB}
  :=
  \tilde{h}_{AB}
  - 2 \bar{D}_{(A}\tilde{Y}_{B)}
  .
\end{eqnarray}
Second, from the gauge-transformation rules
(\ref{eq:gauge-trans-hpq-trace-even}),
(\ref{eq:gauge-trans-tildehe1-sum-2}), and
(\ref{eq:2+2-tildeYA-def-gauge-trans}), we can define the
gauge-invariant variables $F$ as follows:
\begin{eqnarray}
  \tilde{F}
  &:=&
  \tilde{h}_{(e0)}
  - \frac{4}{r} \tilde{Y}_{A} \bar{D}^{A}r
  + \frac{2}{r} \frac{r}{2} \tilde{h}_{(e2)} l(l+1)
  \nonumber\\
  &=&
  \tilde{h}_{(e0)}
  - \frac{4}{r} \tilde{Y}_{A} \bar{D}^{A}r
  + \tilde{h}_{(e2)} l(l+1)
  .
  \label{eq:2+2-gauge-inv-tildeF-def}
\end{eqnarray}

%*********************************************************************

%%%%%%%%%%%%%%%%%%%%%%%%%%%%%%%%%%%%%%%%%%%
%%%%%%%%%%%%%%%%%%%%%%%%%%%%%%%%%%%%%%%%%%%
\subsubsection{Summary of gauge-invariant and gauge-variant variables}
\label{sec:2+2-formulation-gauge-invariants-summary}
%%%%%%%%%%%%%%%%%%%%%%%%%%%%%%%%%%%%%%%%%%%
%%%%%%%%%%%%%%%%%%%%%%%%%%%%%%%%%%%%%%%%%%%

%*********************************************************************

In summary, we have defined gauge-invariant variables as
follows:
\begin{eqnarray}
  \label{eq:2+2-gauge-inv-def-tildeFA-sum}
  \tilde{F}_{A}
  &:=&
       \tilde{h}_{(o1)A}
       + r \bar{D}_{A}\tilde{h}_{(o2)}
       ,
  \\
  \label{eq:2+2-gauge-inv-tildeF-def-sum}
  \tilde{F}
  &:=&
       \tilde{h}_{(e0)}
       - \frac{4}{r} \tilde{Y}_{A} \bar{D}^{A}r
       + \tilde{h}_{(e2)} l(l+1)
       ,
  \\
  \label{eq:gauge-inv-tildeFAB-def-sum}
  \tilde{F}_{AB}
  &:=&
       \tilde{h}_{AB}
       - 2 \bar{D}_{(A}\tilde{Y}_{B)}
       ,
\end{eqnarray}
where we defined the variable $\tilde{Y}_{A}$ by
\begin{eqnarray}
  \tilde{Y}_{A}
  &:=&
       r \tilde{h}_{(e1)A}
       - \frac{r^{2}}{2} \bar{D}_{A}\tilde{h}_{(e2)}
       .
       \label{eq:2+2-gauge-trans-tildeYA-def-sum}
\end{eqnarray}
The gauge-transformation rules for the variable $\tilde{Y}_{A}$
is given by
\begin{eqnarray}
  \label{eq:2+2-tildeYA-def-gauge-trans-sum}
  {}_{\ScrY}\!\tilde{Y}_{A}
  -
  {}_{\ScrX}\!\tilde{Y}_{A}
  =
  \zeta_{A}
  .
\end{eqnarray}

%*********************************************************************

We also note that the gauge-transformation rules
(\ref{eq:gauge-trans-tildeho1-sum-2}) and
(\ref{eq:gauge-trans-tildehe1-sum-2}), i.e.,
\begin{eqnarray}
  \label{eq:gauge-trans-tildeho1-sum-3}
  - r^{2} {}_{\ScrY}\tilde{h}_{(o2)}
  + r^{2} {}_{\ScrX}\tilde{h}_{(o2)}
  &=&
  r \zeta_{(o)}
  .
  \\
  \label{eq:gauge-trans-tildehe1-sum-3}
  \frac{r^{2}}{2} {}_{\ScrY}\!\tilde{h}_{(e2)}
  -
  \frac{r^{2}}{2} {}_{\ScrX}\!\tilde{h}_{(e2)}
  &=&
  r \zeta_{(e)}
  .
\end{eqnarray}
Therefore, it is reasonable to define the variables
$\tilde{Y}_{(o)}$ and $\tilde{Y}_{(e)}$ as follows:
\begin{eqnarray}
  \label{eq:tildeYo-def}
  \tilde{Y}_{(o1)}
  &:=&
  - r^{2} \tilde{h}_{(o2)}
  ,
  \\
  \label{eq:tildeYe-def}
  \tilde{Y}_{(e1)}
  &:=&
  \frac{r^{2}}{2} \tilde{h}_{(e2)}
\end{eqnarray}
so that their gauge-transformation rules are given by
\begin{eqnarray}
  &&
  {}_{\ScrY}\!\tilde{Y}_{(o1)}
  -
  {}_{\ScrX}\!\tilde{Y}_{(o1)}
  =
  r \zeta_{(o)}
  , \\
  &&
  {}_{\ScrY}\!\tilde{Y}_{(e1)}
  -
  {}_{\ScrX}\!\tilde{Y}_{(e1)}
  =
  r \zeta_{(e)}
  .
\end{eqnarray}

%*********************************************************************

Furthermore, we define the variable
\begin{eqnarray}
  Y_{a}
  &:=&
       \sum_{l,m} \tilde{Y}_{A} S_{\delta} (dx^{A})_{a}
       +
       \sum_{l,m} \left(
       \tilde{Y}_{(e1)} \hat{D}_{p}S_{\delta}
       +
       \tilde{Y}_{(o1)} \epsilon_{pq}\hat{D}^{q}S_{\delta}
       \right)
       (dx^{p})_{a}
     .
  \label{eq:2+2-Ya-def}
\end{eqnarray}
The gauge transformation rule for the variable $Y_{a}$
is given by
\begin{eqnarray}
  {}_{\ScrY}\!Y_{a}
  -
  {}_{\ScrX}\!Y_{a}
  &=&
      \sum_{l,m} \left(
      {}_{\ScrY}\!\tilde{Y}_{A}
      -
      {}_{\ScrX}\!\tilde{Y}_{A}
      \right) S_{\delta} (dx^{A})_{a}
      \nonumber\\
  &&
     +
     \sum_{l,m} \left(
     \left(
     {}_{\ScrY}\!\tilde{Y}_{(e)}
     -
     {}_{\ScrX}\!\tilde{Y}_{(e)}
     \right) \hat{D}_{p}S_{\delta}
     +
     \left(
     {}_{\ScrY}\!\tilde{Y}_{(o)}
     -
     {}_{\ScrX}\!\tilde{Y}_{(o)}
     \right) \epsilon_{pq}\hat{D}^{q}S_{\delta}
     \right)
     (dx^{p})_{a}
     \nonumber
  \\
  &=&
      \sum_{l,m} \zeta_{A} S_{\delta} (dx^{A})_{a}
      +
      \sum_{l,m} \left(
      r \zeta_{(e)} \hat{D}_{p}S_{\delta}
      +
      r \zeta_{(o)} \epsilon_{pq}\hat{D}^{q}S_{\delta}
      \right)
      (dx^{p})_{a}
      \nonumber
  \\
  &=&
  \xi_{A} (dx^{A})_{a}
  +
  \xi_{p} (dx^{p})_{a}
  \nonumber\\
  &=&
  \xi_{a}
  ,
\end{eqnarray}
where we used Eqs.~(\ref{eq:xiA-fourier}) and (\ref{eq:xip-fourier}).

%*********************************************************************

In terms of the gauge-invariant variables
$\{\tilde{F}_{A},\tilde{F},\tilde{F}_{AB}\}$ defined by
Eqs.~(\ref{eq:2+2-gauge-inv-def-tildeFA-sum})--(\ref{eq:gauge-inv-tildeFAB-def-sum})
and gauge-variant variables $Y_{a}$ defined by (\ref{eq:2+2-Ya-def}),
we can express the original components $\{h_{AB},h_{Ap},h_{pq}\}$.
First, we consider the component $h_{AB}$:
\begin{eqnarray}
  h_{AB}
  &=&
      \sum_{l,m} \left( \tilde{h}_{AB} \right) S_{\delta}
      =
      \sum_{l,m} \left(
      \tilde{F}_{AB} + 2 \bar{D}_{(A}\tilde{Y}_{B)}
      \right) S_{\delta},
      \nonumber
  \\
  &=&
  F_{AB}
  +
  2 \bar{D}_{(A} Y_{B)}
  ,
  \label{eq:hAB-re-expresion}
\end{eqnarray}
where we defined the gauge-invariant variable $F_{AB}$ by
\begin{eqnarray}
  \label{eq:FAB-def}
  F_{AB} := \sum_{l,m} \tilde{F}_{AB} S_{\delta}.
\end{eqnarray}
Next, we consider the component $h_{Ap}$:
\begin{eqnarray}
  h_{Ap}
  &=&
      r \sum_{l,m} \left[
      \left(
      \tilde{h}_{(e1)A}
      \right) \hat{D}_{p}S_{\delta}
      +
      \left(
      \tilde{h}_{(o1)A}
      \right) \epsilon_{pq} \hat{D}^{q}S_{\delta}
      \right]
      \nonumber\\
  &=&
      r F_{Ap}
      + \hat{D}_{p} Y_{A}
      + \bar{D}_{A} Y_{p}
      - \frac{2}{r} \bar{D}_{A}r Y_{p}
      ,
      \label{eq:hAp-re-expresion}
\end{eqnarray}
where we defined
\begin{eqnarray}
  F_{Ap}
  :=
  \sum_{l,m} \left[
    \tilde{F}_{A} \epsilon_{pq} \hat{D}^{q}S_{\delta}
  \right]
  ,
  \quad
  \hat{D}^{p}F_{Ap} = 0
  .
\end{eqnarray}
Finally, we consider the component $h_{pq}$:
\begin{eqnarray}
  h_{pq}
  &=&
      r^{2} \sum_{l,m} \left[
      \tilde{h}_{(e0)}
      \frac{1}{2} \gamma_{pq} S_{\delta}
      +
      \left(
      \tilde{h}_{(e2)}
      \right) \left(
      \hat{D}_{p}\hat{D}_{q} - \frac{1}{2} \gamma_{pq} \hat{D}^{r}\hat{D}_{r}
      \right) S_{\delta}
      \right.
      \nonumber\\
  && \quad\quad\quad
     \left.
      +
      2 \left(
      \tilde{h}_{(o2)}
      \right) \epsilon_{r(p} \hat{D}_{q)}\hat{D}^{r} S_{\delta}
      \right]
      \nonumber\\
  &=&
      \frac{1}{2} \gamma_{pq} r^{2} F
      + 2r \gamma_{pq} \bar{D}^{A}r Y_{A}
      + \hat{D}_{p}Y_{q}
      + \hat{D}_{q}Y_{p}
      ,
      \label{eq:hpq-re-expresion}
\end{eqnarray}
where we have defined
\begin{eqnarray}
  \label{eq:F-def}
  F := \sum_{l,m} \tilde{F} S_{\delta}.
\end{eqnarray}

%*********************************************************************

Then, we have obtained
\begin{eqnarray}
  h_{AB}
  &=&
  F_{AB}
  +
  2 \bar{D}_{(A} Y_{B)}
  ,
  \label{eq:hAB-re-expresion-sum}
  \\
  h_{Ap}
  &=&
  r F_{Ap}
  + \hat{D}_{p} Y_{A}
  + \bar{D}_{A} Y_{p}
  - \frac{2}{r} \bar{D}_{A}r Y_{p}
  ,
  \label{eq:hAp-re-expresion-sum}
  \\
  h_{pq}
  &=&
  \frac{1}{2} \gamma_{pq} r^{2} F
  + 2r \gamma_{pq} \bar{D}^{A}r Y_{A}
  + \hat{D}_{p}Y_{q}
  + \hat{D}_{q}Y_{p}
  .
  \label{eq:hpq-re-expresion-sum}
\end{eqnarray}
Comparing with the gauge-transformation rules
(\ref{eq:hAB-gauge-trans})--(\ref{eq:hpq-gauge-trans}), the
expression
(\ref{eq:hAB-re-expresion-sum})--(\ref{eq:hpq-re-expresion-sum})
are summarized as
\begin{eqnarray}
  \label{eq:2+2-decomposition-of-hab}
  h_{ab} =: \ScrF_{ab} + {\pounds}_{Y}g_{ab},
\end{eqnarray}
where $\ScrF_{ab}$ is the gauge-invariant part in the 2+2
formulation.
The components of $\ScrF_{ab}$ is given by
\begin{eqnarray}
  \label{eq:2+2-gauge-invariant-variables-calFAB}
  \ScrF_{AB}
  &=&
      F_{AB} = \sum_{l,m} \tilde{F}_{AB} S_{\delta}
      ,
  \\
  \label{eq:2+2-gauge-invariant-variables-calFAp}
  \ScrF_{Ap}
  &=&
      r F_{Ap} = r \sum_{l,m} \tilde{F}_{A} \epsilon_{pq}
      \hat{D}^{q}S_{\delta}, \quad
      \hat{D}^{p}\ScrF_{Ap} = 0
      ,
  \\
  \label{eq:2+2-gauge-invariant-variables-calFpq}
  \ScrF_{pq}
  &=&
      \frac{1}{2} \gamma_{pq} r^{2} F
      = \frac{1}{2} \gamma_{pq} r^{2} \sum_{l,m} \tilde{F} S_{\delta}
      .
\end{eqnarray}
Here, we note that the above arguments include not only $l\geq 2$
modes but also $l=0,1$ modes of metric perturbations.
Equations
(\ref{eq:2+2-decomposition-of-hab})--(\ref{eq:2+2-gauge-invariant-variables-calFpq})
is complete proof of the
Conjecture~\ref{conjecture:decomposition-conjecture} for the
perturbations on the spherically symmetric background spacetime and valid even in the case of $\delta=0$.
Therefore, our general arguments on the gauge-invariant perturbation
theory reviewed in
Sec.~\ref{sec:review-of-general-framework-GI-perturbation-theroy} are
applicable to perturbations on the Schwarzschild background spacetime
without special treatment of $l=0,1$ modes.
Thus, we have resolved the zero-mode problem in the perturbations on
the Schwarzschild background spacetime.

%*********************************************************************

We also note that we only used the forms
(\ref{eq:background-metric-2+2}) and
(\ref{eq:background-metric-2+2-gamma-comp-Schwarzschild}) of the
background metric and did not used the specific forms of the
Schwarzschild metric (\ref{eq:background-metric-2+2-y-comp-Schwarzschild}).
Therefore, our construction of the gauge-invariant and gauge-variant
part of the metric perturbation is also valid for the metric
perturbations on any spherically symmetric spacetime.
Thus, if we accept
Proposal~\ref{proposal:treatment-proposal-on-pert-on-spherical-BG}, we
reached to the following statement:
\begin{theorem}
  \label{theorem:decomposition-theorem-spherical}
  If the gauge-transformation rule for a perturbative pulled-back
  tensor field $h_{ab}$ to the background spacetime $\ScrM$ is
  given by ${}_{\ScrY}\!h_{ab}$ $-$ ${}_{\ScrX}\!h_{ab}$ $=$
  ${\pounds}_{\xi_{(1)}}g_{ab}$ with the background metric $g_{ab}$
  with spherically symmetry, there then exist a tensor field
  $\ScrF_{ab}$ and a vector  field $Y^{a}$ such that $h_{ab}$ is
  decomposed as $h_{ab}$ $=:$ $\ScrF_{ab}$ $+$
  ${\pounds}_{Y}g_{ab}$, where $\ScrF_{ab}$ and $Y^{a}$ are
  transformed into ${}_{\ScrY}\!\ScrF_{ab}$ $-$
  ${}_{\ScrX}\!\ScrF_{ab}$ $=$ $0$ and ${}_{\ScrY}\!Y^{a}$
  $-$ ${}_{\ScrX}\!Y^{a}$ $=$ $\xi^{a}_{(1)}$ under the gauge
  transformation, respectively.
\end{theorem}

%*********************************************************************

%%%%%%%%%%%%%%%%%%%%%%%%%%%%%%%%%%%%%%%%%%%
%%%%%%%%%%%%%%%%%%%%%%%%%%%%%%%%%%%%%%%%%%%
%%%%%%%%%%%%%%%%%%%%%%%%%%%%%%%%%%%%%%%%%%%
\section{Einstein equations}
\label{sec:Einstein_equations}
%%%%%%%%%%%%%%%%%%%%%%%%%%%%%%%%%%%%%%%%%%%
%%%%%%%%%%%%%%%%%%%%%%%%%%%%%%%%%%%%%%%%%%%
%%%%%%%%%%%%%%%%%%%%%%%%%%%%%%%%%%%%%%%%%%%

%*********************************************************************

Here, we consider the linearized Einstein equations
(\ref{eq:einstein-equation-gauge-inv}) on the spherically
symmetric background spacetime with the metric
(\ref{eq:background-metric-2+2}).
The gauge-invariant part of the linearized Einstein tensor
${}^{(1)}\ScrG_{a}^{\;\;b}\left[\ScrF\right]$ is given by
Eqs.~(\ref{eq:linear-Einstein-AIA2010-2}) and
(\ref{eq:(1)Sigma-def-linear}).
The components of the tensor fields $H_{abc}[\ScrF]$,
$H_{ab}^{\;\;\;\;\;\;c}[\ScrF]$, and $H_{a}^{\;\;\;bc}[\ScrF]$
in terms of the variables $F_{AB}$, $F_{Ap}$ and $F$ in
Eqs.~(\ref{eq:2+2-gauge-invariant-variables-calFAB})--(\ref{eq:2+2-gauge-invariant-variables-calFpq})
are summarized in Appendix~\ref{sec:2+2-representation-of-H}.
Through these formulae and the mode decomposition in
Eqs.~(\ref{eq:2+2-gauge-invariant-variables-calFAB})--(\ref{eq:2+2-gauge-invariant-variables-calFpq})
with the harmonic functions $S_{\delta}$ defined by
Eq.~(\ref{eq:harmonics-extended-choice-sum-delta}),
the components of the tensor
${}^{(1)}\ScrG_{a}^{\;\;b}\left[\ScrF\right]$ are given by
\begin{eqnarray}
  \!\!\!\!\!\!\!\!\!\!\!\!\!\!\!\!\!\!\!\!
  {}^{(1)}\ScrG_{A}^{\;\;B}
  \!\!\!\!&=&\!\!\!\!
      \frac{1}{2} \sum_{l,m} \left[
      \left(
      -  \bar{D}_{D}\bar{D}^{D}
      + \frac{l(l+1)}{r^{2}}
      -  \frac{2}{r} (\bar{D}_{D}r) \bar{D}^{D}
      \right) \tilde{F}_{A}^{\;\;\;B}
      +
      \left(
      \bar{D}_{D}\bar{D}_{A}
      + \frac{2}{r} (\bar{D}_{D}r) \bar{D}_{A}
      \right) \tilde{F}^{BD}
      \right.
      \nonumber\\
  && \quad\quad
     \left.
     +
     \left(
     \bar{D}^{D}\bar{D}^{B}
     + \frac{2}{r} (\bar{D}^{D}r) \bar{D}^{B}
     \right) \tilde{F}_{AD}
     -  \bar{D}_{A}\bar{D}^{B}\tilde{F}_{D}^{\;\;\;D}
     \right.
     \nonumber\\
  && \quad\quad
     \left.
     -
     \left(
     \bar{D}_{A}\bar{D}^{B}
     + \frac{1}{r} (\bar{D}_{A}r) \bar{D}^{B}
     + \frac{1}{r} (\bar{D}^{B}r) \bar{D}_{A}
     \right) \tilde{F}
     \right] S_{\delta}
     \nonumber\\
  && \!\!\!\!
     + \frac{1}{2} \delta_{A}^{\;\;B} \sum_{l,m} \left[
     \left(
     \bar{D}_{E}\bar{D}^{E}
     -  \frac{l(l+1)+1}{r^{2}}
     + \frac{2}{r} (\bar{D}_{E}r) \bar{D}^{E}
     + \frac{1}{r^{2}} (\bar{D}^{E}r) (\bar{D}_{E}r)
     \right) \tilde{F}_{D}^{\;\;\;D}
     \right.
     \nonumber\\
  && \quad\quad\quad\quad
     \left.
     -
     \left(
     \bar{D}_{D}\bar{D}_{E}
     + \frac{4}{r} (\bar{D}_{D}r) \bar{D}_{E}
     + \frac{2}{r^{2}} (\bar{D}_{E}r) (\bar{D}_{D}r)
     \right) \tilde{F}^{ED}
     \right.
     \nonumber\\
  && \quad\quad\quad\quad
     \left.
     +
     \left(
     \bar{D}_{D}\bar{D}^{D}
     -  \frac{l(l+1)-2}{2r^{2}}
     + \frac{3}{r} (\bar{D}^{D}r) \bar{D}_{D}
     \right) \tilde{F}
     \right] S_{\delta}
     ,
     \label{eq:cal-G-def-linear-AIA2010-4-AB-mode-dec-sum}
\end{eqnarray}
\begin{eqnarray}
  \!\!\!\!\!\!\!\!\!\!\!\!\!\!\!\!\!\!\!\!
  {}^{(1)}\ScrG_{A}^{\;\;q}
  \!\!\!\!&=&\!\!\!\!
      \frac{1}{2r^{2}}
      \sum_{l,m} \left[
      \left(
      -  \bar{D}_{A}
      + \frac{1}{r} (\bar{D}_{A}r)
      \right) \tilde{F}_{D}^{\;\;\;D}
      + \bar{D}^{D}\tilde{F}_{AD}
      - \frac{1}{2} \bar{D}_{A}\tilde{F}
      \right] \hat{D}^{q}S_{\delta}
      \nonumber\\
  && \!\!\!\!
     + \frac{1}{2r} \sum_{l,m} \left[
     \left(
     -  \bar{D}^{D}\bar{D}_{D}
     + \frac{l(l+1)}{r^{2}}
     -  \frac{2}{r} (\bar{D}^{D}r) \bar{D}_{D}
     + \frac{3}{2r^{2}} \left\{
     (\bar{D}^{D}r) (\bar{D}_{D}r) - 1
     \right\}
     \right) \tilde{F}_{A}
     \right.
     \nonumber\\
  && \quad
     \left.
     +
     \left(
     \bar{D}^{D}\bar{D}_{A}
     + \frac{3}{r} (\bar{D}^{D}r) \bar{D}_{A}
     -  \frac{1}{r} (\bar{D}_{A}r) \bar{D}^{D}
     -  \frac{2}{r^{2}} (\bar{D}_{A}r) (\bar{D}^{D}r)
     \right) \tilde{F}_{D}
     \right] \epsilon^{qt} \hat{D}_{t}S_{\delta}
     ,
     \label{eq:cal-G-def-linear-AIA2010-4-Aq-mode-dec-sum}
\end{eqnarray}
\begin{eqnarray}
  \!\!\!\!\!\!\!\!\!\!\!\!\!\!\!\!\!\!\!\!
  {}^{(1)}\ScrG_{p}^{\;\;B}
  \!\!\!\!&=&\!\!\!\!
      \frac{1}{2} \sum_{l,m} \left[
      \left(
      -                       \bar{D}^{B}
      + \frac{1}{r} (\bar{D}^{B}r)
      \right) \tilde{F}_{D}^{\;\;\;D}
      +                      \bar{D}_{D}\tilde{F}^{BD}
      -  \frac{1}{2} \bar{D}^{B}\tilde{F}
      \right] \hat{D}_{p}S_{\delta}
      \nonumber\\
  && \!\!\!\!
     + \frac{r}{2} \sum_{l,m} \left[
     \left(
     - \bar{D}_{D}\bar{D}^{D}
     + \frac{l(l+1)}{r^{2}}
     -  \frac{2}{r} (\bar{D}_{D}r) \bar{D}^{D}
     + \frac{3}{2r^{2}}  \left\{ (\bar{D}^{D}r) (\bar{D}_{D}r) - 1 \right\}
     \right) \tilde{F}^{B}
     \right.
     \nonumber\\
  && \quad
     \left.
     +
     \left(
     \bar{D}_{D}\bar{D}^{B}
     + \frac{3}{r} (\bar{D}_{D}r) \bar{D}^{B}
     -  \frac{1}{r} (\bar{D}^{B}r) \bar{D}_{D}
     -  \frac{2}{r^{2}} (\bar{D}^{B}r) (\bar{D}_{D}r)
     \right) \tilde{F}^{D}
     \right] \epsilon_{pq} \hat{D}^{q}S_{\delta}
     ,
     \label{eq:cal-G-def-linear-AIA2010-4-pB-mode-dec-sum}
\end{eqnarray}
\begin{eqnarray}
  \!\!\!\!\!\!\!\!\!\!\!\!\!\!\!\!\!\!\!\!
  {}^{(1)}\ScrG_{p}^{\;\;q}
  \!\!\!\!&=&\!\!\!\!
      \sum_{l,m}
      \left[
      \frac{1}{2r^{2}}
      \bar{D}_{D}\left(r^{2} \bar{D}^{D}\tilde{F}\right)
      -
      \frac{1}{r^{2}}
      \bar{D}_{D}\left(r^{2} \bar{D}_{E}\tilde{F}^{ED}\right)
     \right.
      \nonumber\\
  && \quad
     \left.
      +
      \left(
      \bar{D}_{E}\bar{D}^{E}
      + \frac{1}{r} (\bar{D}_{E}r) \bar{D}^{E}
      -  \frac{l(l+1)}{2r^{2}}
      \right) \tilde{F}_{D}^{\;\;\;D}
     \right]
      \frac{1}{2} \gamma_{p}^{\;\;q} S_{\delta}
      \nonumber\\
  && \!\!\!\!
     + \frac{1}{2r^{2}} \sum_{l,m} \left[
     - \tilde{F}_{D}^{\;\;\;D}
     \left(
     \hat{D}_{p}\hat{D}^{q}
     -
     \frac{1}{2}
     \gamma_{p}^{\;\;q} \hat{D}^{s}\hat{D}_{s}
     \right) S_{\delta}
     \right.
     % \right]
     \nonumber\\
          && \quad\quad\quad\quad
      -
      % \frac{1}{2r^{2}} \left[
      \left.
      \bar{D}_{D}\left(r\tilde{F}^{D}\right)
      \left(
      \epsilon^{sq}\hat{D}_{p}\hat{D}_{s}  + \epsilon_{sp}\hat{D}^{q}\hat{D}^{s}
      \right) S_{\delta}
      \right]
      ,
     \label{eq:cal-G-def-linear-AIA2010-4-pq-mode-dec-sum}
\end{eqnarray}
where we used the fact that the background Ricci curvature vanishes
and the background Einstein equations
(\ref{eq:BG-Einstein-tensor-AB-trace-2}) and
(\ref{eq:BG-Einstein-tensor-AB-traceless-2}).

%*********************************************************************

We also decompose the components of the linearized energy-momentum
tensor ${}^{(1)}\!\ScrT_{a}^{\;\;b}$ as follows:
\begin{eqnarray}
  {}^{(1)}\!\ScrT_{A}^{\;\;\;B}
  &=&
      \sum_{l,m}
      \tilde{T}_{A}^{\;\;\;B}
      S_{\delta}
      ,
      \label{eq:1st-pert-calTab-du-decomp-2-AB}
  \\
  {}^{(1)}\!\ScrT_{A}^{\;\;q}
  &=&
      \frac{1}{r}
      \sum_{l,m} \left\{
      \tilde{T}_{(e1)A} \hat{D}^{q}S_{\delta}
      +
      \tilde{T}_{(o1)A} \epsilon^{qr}\hat{D}_{r}S_{\delta}
      \right\}
      ,
      \label{eq:1st-pert-calTab-du-decomp-2-Aq}
  \\
  {}^{(1)}\!\ScrT_{p}^{\;\;B}
  &=&
      r
      \sum_{l,m} \left\{
      \tilde{T}_{(e1)}^{B} \hat{D}_{p}S_{\delta}
      +
      \tilde{T}_{(o1)}^{B} \epsilon_{pr}\hat{D}^{r}S_{\delta}
      \right\}
      ,
      \label{eq:1st-pert-calTab-du-decomp-2-pB}
  \\
  {}^{(1)}\!\ScrT_{p}^{\;\;q}
  &=&
     \sum_{l,m} \left\{
     \tilde{T}_{(e0)} \frac{1}{2} \gamma_{p}^{\;\;q} S_{\delta}
     +
     \tilde{T}_{(e2)} \left(
     \hat{D}_{p}\hat{D}^{q}S_{\delta}
     -
     \frac{1}{2} \gamma_{p}^{\;\;q} \hat{D}_{r}\hat{D}^{r}S_{\delta}
     \right)
     \right.
     \nonumber\\
  && \quad\quad\quad\quad\quad
     \left.
     + \tilde{T}_{(o2)} \left(
     \epsilon_{sp}\hat{D}^{q}\hat{D}^{s}S_{\delta}
     +
     \epsilon^{sq}\hat{D}_{p}\hat{D}_{s}S_{\delta}
     \right)
     \right\}
     .
     \label{eq:1st-pert-calTab-du-decomp-2-pq}
\end{eqnarray}

%*********************************************************************

The linearized continuity equation
(\ref{eq:divergence-barTab-linear-vac-back-u}) for the energy-momentum tensor
$\ScrT_{a}^{\;\;b}$ is summarized as
\begin{eqnarray}
  &&
     \bar{D}^{C}\tilde{T}_{C}^{\;\;B}
     + \frac{2}{r} (\bar{D}^{D}r)\tilde{T}_{D}^{\;\;\;B}
     -  \frac{1}{r} l(l+1) \tilde{T}_{(e1)}^{B}
     -  \frac{1}{r} (\bar{D}^{B}r) \tilde{T}_{(e0)}
     =
     0
     ,
     \label{eq:div-barTab-linear-vac-back-u-A-mode-dec-sum}
  \\
  &&
     \bar{D}^{C}\tilde{T}_{(e1)C}
     + \frac{3}{r} (\bar{D}^{C}r) \tilde{T}_{(e1)C}
     + \frac{1}{2r} \tilde{T}_{(e0)}
     -  \frac{1}{2r} (l-1)(l+2) \tilde{T}_{(e2)}
     =
     0
     ,
     \label{eq:div-barTab-linear-vac-back-u-p-mode-dec-even-sum}
  \\
  &&
     \bar{D}^{C}\tilde{T}_{(o1)C}
     + \frac{3}{r} (\bar{D}^{D}r) \tilde{T}_{(o1)D}
     + \frac{1}{r} (l-1)(l+2) \tilde{T}_{(o2)}
     =
     0
     .
     \label{eq:div-barTab-linear-vac-back-u-p-mode-dec-odd-sum}
\end{eqnarray}

%*********************************************************************

Through the components (\ref{eq:cal-G-def-linear-AIA2010-4-AB-mode-dec-sum})--(\ref{eq:cal-G-def-linear-AIA2010-4-pq-mode-dec-sum})
for the linearized Einstein tensor and the components
(\ref{eq:1st-pert-calTab-du-decomp-2-AB})--(\ref{eq:1st-pert-calTab-du-decomp-2-pq})
for the linearized energy-momentum tensor, we evaluate the linearized Einstein equation
(\ref{eq:einstein-equation-gauge-inv}).
Due to the linear-independence of the set of harmonics
(\ref{eq:harmonic-fucntions-set}), we can carry out the mode-by-mode
analyses including $l=0,1$ modes.
Since the odd-mode perturbations and the even-mode perturbations are
decoupled with each other, we consider these perturbations, separately.

%*********************************************************************

%%%%%%%%%%%%%%%%%%%%%%%%%%%%%%%%%%%%%%%%%%%
%%%%%%%%%%%%%%%%%%%%%%%%%%%%%%%%%%%%%%%%%%%
\subsection{Odd mode perturbation equations}
\label{sec:Schwarzschild_Background-non-vaccum-2-odd-eq}
%%%%%%%%%%%%%%%%%%%%%%%%%%%%%%%%%%%%%%%%%%%
%%%%%%%%%%%%%%%%%%%%%%%%%%%%%%%%%%%%%%%%%%%

%*********************************************************************

From the linearized Einstein equation
(\ref{eq:einstein-equation-gauge-inv}) through Eqs.~(\ref{eq:cal-G-def-linear-AIA2010-4-AB-mode-dec-sum})--(\ref{eq:cal-G-def-linear-AIA2010-4-pq-mode-dec-sum})
and
Eq.~(\ref{eq:1st-pert-calTab-du-decomp-2-AB})--(\ref{eq:1st-pert-calTab-du-decomp-2-pq}),
the odd-mode part in the linearized Einstein equations are simplified
as the constraint equation
\begin{eqnarray}
  \bar{D}_{D}(r\tilde{F}^{D})
  =
  - 16 \pi r^{2} \tilde{T}_{(o2)}
  ,
  \label{eq:1st-p-Ein-non-vac-pq-traceless-odd-sum-3}
\end{eqnarray}
and the evolution equation
\begin{eqnarray}
  &&
     - \left[ \bar{D}^{D}\bar{D}_{D} - \frac{l(l+1)}{r^{2}} \right] (r\tilde{F}_{A})
     - \frac{2}{r^{2}} (\bar{D}^{D}r) (\bar{D}_{A}r) (r\tilde{F}_{D})
     + \frac{2}{r} (\bar{D}^{D}r) \bar{D}_{A}(r\tilde{F}_{D})
     \nonumber\\
  &=&
  16 \pi r \left(
  \tilde{T}_{(o1)A}
  + r \bar{D}_{A}\tilde{T}_{(o2)}
  \right)
  .
  \label{eq:1st-p-Ein-non-vac-Aq-odd-sum-2-red-3}
\end{eqnarray}
Furthermore, we have the continuity equation
(\ref{eq:div-barTab-linear-vac-back-u-p-mode-dec-odd-sum}) for the
odd-mode matter perturbation which is derived from the divergence of
the first-order perturbation of the energy-momentum tensor.
The explicit strategy to solve these odd-mode perturbations and
$l=0,1$ mode solutions will be discussed in
Sec.~\ref{sec:Schwarzschild_Background-non-vaccum-odd-Nakano-treatment}
in this paper.

%*********************************************************************

%%%%%%%%%%%%%%%%%%%%%%%%%%%%%%%%%%%%%%%%%%%
%%%%%%%%%%%%%%%%%%%%%%%%%%%%%%%%%%%%%%%%%%%
\subsection{Even mode perturbation equations}
\label{sec:Schwarzschild_Background-non-vaccum-2-even-eq}
%%%%%%%%%%%%%%%%%%%%%%%%%%%%%%%%%%%%%%%%%%%
%%%%%%%%%%%%%%%%%%%%%%%%%%%%%%%%%%%%%%%%%%%

%*********************************************************************

Here, we consider the even-mode perturbations from
Eqs.~(\ref{eq:cal-G-def-linear-AIA2010-4-AB-mode-dec-sum})--(\ref{eq:cal-G-def-linear-AIA2010-4-pq-mode-dec-sum})
and
(\ref{eq:1st-pert-calTab-du-decomp-2-AB})--(\ref{eq:1st-pert-calTab-du-decomp-2-pq}).
The traceless even part of the $(p,q)$-component of the linearized Einstein
equation (\ref{eq:einstein-equation-gauge-inv}) is given by
\begin{eqnarray}
  \label{eq:linearized-Einstein-pq-traceless-even}
  \tilde{F}_{D}^{\;\;\;D}
  =
  -
  16 \pi r^{2}
  \tilde{T}_{(e2)}
  .
\end{eqnarray}
Using this equation, the even part of $(A,q)$-component, equivalently
$(p,B)$-component, of the linearized Einstein equation
(\ref{eq:einstein-equation-gauge-inv}) yields
\begin{eqnarray}
  \bar{D}^{D}\tilde{\FF}_{AD}
  - \frac{1}{2} \bar{D}_{A}\tilde{F}
  =
  16 \pi
  \left[
  r \tilde{T}_{(e1)A}
  - \frac{1}{2} r^{2} \bar{D}_{A}\tilde{T}_{(e2)}
  \right]
  =:
  16 \pi S_{(ec)A}
  \label{eq:even-FAB-divergence-3}
\end{eqnarray}
through the definition of the traceless part $\tilde{\FF}_{AB}$ of
the variable $\tilde{F}_{AB}$ defined by
\begin{eqnarray}
  \label{eq:FF-def}
  \tilde{\FF}_{AB} := \tilde{F}_{AB} - \frac{1}{2} y_{AB} \tilde{F}_{C}^{\;\;C}.
\end{eqnarray}
Using Eqs.~(\ref{eq:linearized-Einstein-pq-traceless-even}),
(\ref{eq:even-FAB-divergence-3}), and the component
(\ref{eq:BG-Einstein-tensor-AB-trace-2}) of background Einstein
equation, the trace part of $(p,q)$-component of
the linearized Einstein equation
(\ref{eq:einstein-equation-gauge-inv}) is given by
\begin{eqnarray}
  \bar{D}^{D}\tilde{T}_{(e1)D}
  + \frac{3}{r} (\bar{D}^{D}r) \tilde{T}_{(e1)D}
  + \frac{1}{2r} \tilde{T}_{(e0)}
  -  \frac{(l-1)(l+2)}{2r} \tilde{T}_{(e2)}
  =
  0
  .
\end{eqnarray}
This coincides with the component
(\ref{eq:div-barTab-linear-vac-back-u-p-mode-dec-even-sum}) of the
continuity equation for the linearized energy-momentum tensor.
Next, we consider the $(A,B)$-components of the linearized Einstein
equation (\ref{eq:einstein-equation-gauge-inv}).

%*********************************************************************

Through Eqs.~(\ref{eq:linearized-Einstein-pq-traceless-even}) and
(\ref{eq:even-FAB-divergence-3}), the trace part of the
$(A,B)$-component of the linearized Einstein equation
(\ref{eq:einstein-equation-gauge-inv}) is given by
\begin{eqnarray}
  &&
     \!\!\!\!\!\!\!\!\!\!\!\!\!\!\!\!\!\!\!\!\!\!\!\!
     \left(
     \bar{D}_{D}\bar{D}^{D}
     + \frac{2}{r} (\bar{D}^{D}r) \bar{D}_{D}
     -  \frac{(l-1)(l+2)}{r^{2}}
     \right) \tilde{F}
     - \frac{4}{r^{2}} (\bar{D}_{C}r) (\bar{D}_{D}r) \tilde{\FF}^{CD}
  =
      16 \pi S_{(F)}
      ,
     \label{eq:even-mode-tildeF-master-eq-mod-3}
  \\
  S_{(F)}
  \!\!\!\!&:=&\!\!\!\!
       \tilde{T}_{C}^{\;\;\;C}
       + 4 (\bar{D}_{D}r) \tilde{T}_{(e1)}^{D}
       -  2 r (\bar{D}_{D}r) \bar{D}^{D}\tilde{T}_{(e2)}
       -  (l(l+1)+2) \tilde{T}_{(e2)}
       .
               \label{eq:sourcee0-def}
\end{eqnarray}

%*********************************************************************

On the other hand, the traceless part of the $(A,B)$-component of the
linearized Einstein equation (\ref{eq:einstein-equation-gauge-inv}) is
given by
\begin{eqnarray}
  &&
%     \!\!\!\!\!\!\!\!\!\!\!\!\!\!\!\!\!\!\!\!\!\!\!\!
     \!\!\!\!\!\!\!\!\!\!
     \left[
     -  \bar{D}_{D}\bar{D}^{D}
     -  \frac{2}{r} (\bar{D}_{D}r) \bar{D}^{D}
     + \frac{4}{r} (\bar{D}^{D}\bar{D}_{D}r)
     + \frac{l(l+1)}{r^{2}}
     \right]
     \tilde{\FF}_{AB}
     \nonumber\\
  && \quad\quad
     + \frac{4}{r} (\bar{D}^{D}r) \bar{D}_{(A}\tilde{\FF}_{B)D}
     -  \frac{2}{r} (\bar{D}_{(A}r) \bar{D}_{B)}\tilde{F}
     \nonumber\\
  \!\!\!\!&=&\!\!\!\!
  16 \pi S_{(\FF)AB}
              ,
              \label{eq:1st-pert-Einstein-non-vac-AB-traceless-final-3}
  \\
  S_{(\FF)AB}
  \!\!\!\!&:=&\!\!\!\!
      T_{AB} - \frac{1}{2} y_{AB} T_{C}^{\;\;\;C}
      - 2 \left( \bar{D}_{(A}(r \tilde{T}_{(e1)B)}) - \frac{1}{2} y_{AB} \bar{D}^{D}(r \tilde{T}_{(e1)D}) \right)
      \nonumber\\
  &&\!\!\!\!
      + 2 \left( (\bar{D}_{(A}r) \bar{D}_{B)} - \frac{1}{2} y_{AB} (\bar{D}^{D}r) \bar{D}_{D}  \right) ( r \tilde{T}_{(e2)} )
      \nonumber\\
  &&\!\!\!\!
      + r \left( \bar{D}_{A}\bar{D}_{B} - \frac{1}{2} y_{AB} \bar{D}^{D}\bar{D}_{D} \right)(r \tilde{T}_{(e2)})
      \nonumber\\
  &&\!\!\!\!
      + 2 \left( (\bar{D}_{A}r) (\bar{D}_{B}r) - \frac{1}{2} y_{AB} (\bar{D}^{C}r) (\bar{D}_{C}r) \right) \tilde{T}_{(e2)}
      \nonumber\\
  &&\!\!\!\!
      + 2 y_{AB} (\bar{D}^{C}r) \tilde{T}_{(e1)C}
      -  r y_{AB} (\bar{D}^{C}r) \bar{D}_{C}\tilde{T}_{(e2)}
     ,
     \nonumber\\
  \label{eq:souce(FF)-def}
\end{eqnarray}
where we used the background Einstein equation
(\ref{eq:BG-Einstein-tensor-AB-traceless-2}).

%*********************************************************************

Equations (\ref{eq:linearized-Einstein-pq-traceless-even}),
(\ref{eq:even-FAB-divergence-3}),
(\ref{eq:even-mode-tildeF-master-eq-mod-3}), and
(\ref{eq:1st-pert-Einstein-non-vac-AB-traceless-final-3}) are all
independent equations of the linearized Einstein equation for
even-mode perturbations.
These equations are coupled equations for the variables
$\tilde{F}_{C}^{\;\;\;C}$, $F$, and $\tilde{\FF}_{AB}$ and the
energy-momentum tensor for the matter field.
When we solve these equations, we have to take into account of the
continuity equations
(\ref{eq:div-barTab-linear-vac-back-u-A-mode-dec-sum}) and
(\ref{eq:div-barTab-linear-vac-back-u-p-mode-dec-even-sum}) for the
matter fields.
We note that these equations are valid not only for $l\geq 2$ modes
but also $l=0,1$ modes in our formulation.

%*********************************************************************

The explicit strategy to solve these Einstein equations for
even modes, and the explicit solution for $l=0,1$ mode perturbations
are discussed in the Part II paper~\cite{K.Nakamura-2021c}.

%*********************************************************************

%%%%%%%%%%%%%%%%%%%%%%%%%%%%%%%%%%%%%%%%%%%
%%%%%%%%%%%%%%%%%%%%%%%%%%%%%%%%%%%%%%%%%%%
%%%%%%%%%%%%%%%%%%%%%%%%%%%%%%%%%%%%%%%%%%%
\section{Component treatment for the odd-mode perturbations of the
  Einstein equations}
\label{sec:Schwarzschild_Background-non-vaccum-odd-Nakano-treatment}
%%%%%%%%%%%%%%%%%%%%%%%%%%%%%%%%%%%%%%%%%%%
%%%%%%%%%%%%%%%%%%%%%%%%%%%%%%%%%%%%%%%%%%%
%%%%%%%%%%%%%%%%%%%%%%%%%%%%%%%%%%%%%%%%%%%

%*********************************************************************

%%%%%%%%%%%%%%%%%%%%%%%%%%%%%%%%%%%%%%%%%%%
%%%%%%%%%%%%%%%%%%%%%%%%%%%%%%%%%%%%%%%%%%%
\subsection{Strategy to solve odd-mode perturbations}
\label{sec:Schwarzschild_Background_odd_sol_strategy}
%%%%%%%%%%%%%%%%%%%%%%%%%%%%%%%%%%%%%%%%%%%
%%%%%%%%%%%%%%%%%%%%%%%%%%%%%%%%%%%%%%%%%%%

%*********************************************************************

Here, we consider the component treatment for the odd-mode
perturbations based on the old paper by Regge and
Wheeler~\cite{T.Regge-J.A.Wheeler-1957}, and
Zerilli~\cite{F.Zerilli-1970-PRL,F.Zerilli-1970-PRD}
and re-derivation by Nakano~\cite{H.Nakano-2019}.
We introduce the component of $r\tilde{F}^{D}$ as
\begin{eqnarray}
  \label{eq:component-odd-rtildeFD}
  r \tilde{F}_{D} =: X_{(o)}(dt)_{D} + Y_{(o)}(dr)_{D}, \quad
  r \tilde{F}^{D} = - f^{-1} X_{(o)} (\partial_{t})^{D} + f Y_{(o)} (\partial_{r})^{D},
\end{eqnarray}
where the background metric is given by
Eqs.~(\ref{eq:background-metric-2+2})--(\ref{eq:background-metric-2+2-gamma-comp-Schwarzschild}).
In terms of the components (\ref{eq:component-odd-rtildeFD}),
Eq.~(\ref{eq:1st-p-Ein-non-vac-pq-traceless-odd-sum-3}) is given by
\begin{eqnarray}
  -  \partial_{t}X_{(o)}
  + f f' Y_{(o)}
  + f^{2} \partial_{r}Y_{(o)}
  =
  - 16 \pi r^{2} f \tilde{T}_{(o2)}
  ,
  \label{eq:divrFD=To2-2-sum}
\end{eqnarray}
where $f'=\partial_{r}f$.
The components of Eq.~(\ref{eq:1st-p-Ein-non-vac-Aq-odd-sum-2-red-3})
are summarized as follows:
\begin{eqnarray}
  &&
     \!\!\!\!\!\!\!\!\!\!\!\!\!\!\!\!\!\!\!\!\!\!\!\!
     \frac{1}{f} \partial_{t}^{2}X_{(o)}
     -  f \partial_{r}^{2}X_{(o)}
     -  \frac{2(1-f)}{r^{2}} X_{(o)}
     + \frac{l(l+1)}{r^{2}} X_{(o)}
     -  \frac{1-3f}{r} \partial_{t}Y_{(o)}
     \nonumber\\
  &=&
      16 \pi r \left(
      \tilde{T}_{(o1)t}
      + r \partial_{t}\tilde{T}_{(o2)}
      \right)
      ,
      \label{eq:EvolofFA=source-t-component-sum}
  \\
  &&
     \!\!\!\!\!\!\!\!\!\!\!\!\!\!\!\!\!\!\!\!\!\!\!\!
      \partial_{t}^{2}Y_{(o)}
      -  f \partial_{r}(f \partial_{r}Y_{(o)})
      + \frac{2(2f-1)f}{r} \partial_{r}Y_{(o)}
      + \frac{(l-1)(l+2)}{r^{2}} f Y_{(o)}
      + \frac{(1-f)(5f-1)}{r^{2}} Y_{(o)}
      \nonumber\\
  &=&
      + 16 \pi r \left(
      f \tilde{T}_{(o1)r}
      + r f \partial_{r}\tilde{T}_{(o2)}
      + (1-f) \tilde{T}_{(o2)}
      \right)
      .
      \label{eq:EvolofFA=source-r-component-sum}
\end{eqnarray}
In addition to these equations, the odd-mode perturbation
(\ref{eq:div-barTab-linear-vac-back-u-p-mode-dec-odd-sum}) of the
divergence of the energy-momentum tensor.

%*********************************************************************

Here, we consider Eqs.~(\ref{eq:EvolofFA=source-r-component-sum}).
We define the dependent variable $Z_{(o)}$ by
\begin{eqnarray}
  \label{eq:Regge-Wheeler-variable}
  Y_{(o)} =: \frac{r}{f} Z_{(o)}
\end{eqnarray}
and we have obtained the famous equation which is called Regge-Wheeler
equation
\begin{eqnarray}
  \partial_{t}^{2}Z_{(o)}
  -  f \partial_{r}( f \partial_{r}Z_{(o)})
  + \frac{1}{r^{2}} f \left[ l(l+1) - 3 (1-f) \right] Z_{(o)}
  =
  16 \pi f \left[
  f \tilde{T}_{(o1)r}
  +
  r \partial_{r}\left( f \tilde{T}_{(o2)}\right)
  \right]
  .
  \label{eq:odd-master-equation-Regge-Wheeler}
\end{eqnarray}
We can solve Eq.~(\ref{eq:odd-master-equation-Regge-Wheeler}) with
appropriate boundary conditions and obtain the variable
$Y_{(o)}$ through Eq.~(\ref{eq:Regge-Wheeler-variable}).
For the $l\geq 2$ case, the analytic solutions to
Eq.~(\ref{eq:odd-master-equation-Regge-Wheeler}) are constructed by
the formulation proposed by Mano, Suzuki, and
Takasugi~\cite{S.Mano-H.Suzuki-E.Takasugi-1996a,S.Mano-H.Suzuki-E.Takasugi-1996b,S.Mano-E.Takasugi-1997,H.Tagoshi-S.Mano-E.Takasugi-1997}
(MST formulation).
However, this is a partial solution to the odd-mode Einstein
equations.
We cannot regard such solutions as the solution to the total Einstein
equation for odd-mode perturbations, because we have other two
equations of the Einstein equation
(\ref{eq:EvolofFA=source-t-component-sum}) and the constraint equation
(\ref{eq:divrFD=To2-2-sum}).
To obtain the solution to the total Einstein equations for odd-mode
perturbations, we have to discuss Eqs.~(\ref{eq:divrFD=To2-2-sum}),
(\ref{eq:EvolofFA=source-t-component-sum}), and
(\ref{eq:div-barTab-linear-vac-back-u-p-mode-dec-odd-sum}), i.e.,
\begin{eqnarray}
  - \frac{1}{f} \partial_{t}\tilde{T}_{(o1)t}
  + f \partial_{r}\tilde{T}_{(o1)r}
  + f' \tilde{T}_{(o1)r}
  + \frac{3}{r} f \tilde{T}_{(o1)r}
  + \frac{1}{r} (l-1)(l+2) \tilde{T}_{(o2)}
  =
  0
  .
  \label{eq:div-barTab-linear-odd-component}
\end{eqnarray}
in addition to Eq.~(\ref{eq:odd-master-equation-Regge-Wheeler}).

%*********************************************************************

To obtain the solution to the total Einstein equations for odd-mode
perturbations, it is convenient to introduce the
Cunningham-Price-Moncrief variable
$\Phi_{(o)}$~\cite{C.T.Cunningham-R.H.Price-V.Moncrief-1978} by
\begin{eqnarray}
  \label{eq:Cunningham-variable}
  \Phi_{(o)}
  &:=&
       2r \left[
       r^{2} \partial_{r}\left(\frac{X_{(o)}}{r^{2}}\right) - \partial_{t}Y_{(o)}
       \right]
  \\
  &=&
      2 r \partial_{r}X_{(o)}
      -  4 X_{(o)}
      -  2 r \partial_{t}Y_{(o)}
      .
  \label{eq:Cunningham-variable-2}
\end{eqnarray}
Here, we consider the time derivative of $\Phi_{(o)}$ and use
Eqs.~(\ref{eq:divrFD=To2-2-sum}),
(\ref{eq:EvolofFA=source-r-component-sum}), and the background
Einstein equation (\ref{eq:BG-Einstein-tensor-AB-2-sol-2}) as
\begin{eqnarray}
  \partial_{t}\Phi_{(o)}
  &=&
      2 \frac{(l-1)(l+2)}{r} f Y_{(o)}
      - 32 \pi r^{2} f \tilde{T}_{(o1)r}
      \nonumber\\
  &=&
      2 (l-1)(l+2) Z_{(o)}
      - 32 \pi r^{2} f \tilde{T}_{(o1)r}
      .
      \label{eq:Psio-Zo-relation}
\end{eqnarray}
The relation (\ref{eq:Psio-Zo-relation}) indicates that the variable
$Z_{(o)}$ is related to $\Phi_{(o)}$ for $l\neq 1$ modes, while the
time derivative of $\Phi_{(o)}$ is just the matter degree of freedom
$\tilde{T}_{(o1)r}$ for the $l=1$ mode.
This relation also gives the relation with the metric perturbation
$Y_{(o)}$ as
\begin{eqnarray}
  (l-1) (l+2) Y_{(o)}
  =
  \frac{r}{2f} \partial_{t}\Phi_{(o)}
  + 16 \pi r^{3} \tilde{T}_{(o1)r}
  .
  \label{eq:Yo-Psio-relation}
\end{eqnarray}
On the other hand, using
Eqs.~(\ref{eq:divrFD=To2-2-sum}) and
(\ref{eq:EvolofFA=source-t-component-sum}), the $r$-derivative of
$\Phi_{(o)}$ through Eq.~(\ref{eq:Cunningham-variable-2}) is given by
\begin{eqnarray}
  \partial_{r}\Phi_{(o)}
  =
  - \frac{1}{r} \Phi_{(o)}
  + \frac{2}{rf} (l-1)(l+2) X_{(o)}
  - 32 \pi \frac{r^{2}}{f} \tilde{T}_{(o1)t}
  .
  \label{eq:Xo-Psio-relation-pre}
\end{eqnarray}
Then, we obtain the relation
\begin{eqnarray}
  (l-1)(l+2) X_{(o)}
  =
  \frac{f}{2}
  \left(
  r \partial_{r}\Phi_{(o)}
  + \Phi_{(o)}
  \right)
  + 16 \pi r^{3} \tilde{T}_{(o1)t}
  .
  \label{eq:Xo-Psio-relation}
\end{eqnarray}
From Eqs.~(\ref{eq:Psio-Zo-relation}) and
(\ref{eq:Xo-Psio-relation-pre}) and the constraint
(\ref{eq:divrFD=To2-2-sum}), we obtain
\begin{eqnarray}
     \partial_{r}\partial_{t}\Phi_{(o)}
     -
     \partial_{t}\partial_{r}\Phi_{(o)}
  &=&
      \partial_{r}\left[
      + 2 \frac{(l-1)(l+2)}{r} f Y_{(o)}
      - 32 \pi r^{2} f \tilde{T}_{(o1)r}
      \right]
      \nonumber\\
  &&
      -
      \partial_{t}\left[
      - \frac{1}{r} \Phi_{(o)}
      + \frac{1}{r} 2 (l-1)(l+2) \frac{1}{f} X_{(o)}
      - 32 \pi r^{2} \frac{1}{f} \tilde{T}_{(o1)t}
      \right]
      \nonumber\\
  &=&
      - 32 \pi r^{2}
      \left[
      - \frac{1}{f} \partial_{t}\tilde{T}_{(o1)t}
      + f' \tilde{T}_{(o1)r}
      +  f \partial_{r}\tilde{T}_{(o1)r}
      \right.
      \nonumber\\
  && \quad\quad\quad\quad\quad\quad
     \left.
      + \frac{3}{r} f \tilde{T}_{(o1)r}
      + \frac{1}{r} (l-1)(l+2) \tilde{T}_{(o2)}
      \right]
      \nonumber\\
  &=&
      0
      .
\end{eqnarray}
The final equality comes from the odd-mode perturbation
(\ref{eq:div-barTab-linear-odd-component}) of the divergence of the
energy-momentum tensor.
Thus, Eqs.~(\ref{eq:Psio-Zo-relation}) and
(\ref{eq:Xo-Psio-relation-pre}) are integrable under the constraint
(\ref{eq:divrFD=To2-2-sum}) and the continuity equation
(\ref{eq:div-barTab-linear-odd-component}).

%*********************************************************************

We emphasize that the relations (\ref{eq:Yo-Psio-relation}) and
(\ref{eq:Xo-Psio-relation}) gives the relations of the metric
components ($X_{(o)}$,$Y_{(o)}$) and the master variable
$\Phi_{(o)}$ only for $l\neq 1$ mode.
In the case of the $l=1$ mode, these equations give the constraint of
the master variable $\Phi_{(o)}$ and the matter degree of freedom.
Furthermore, in the derivation of the relation
(\ref{eq:Xo-Psio-relation}), we used
Eq.~(\ref{eq:EvolofFA=source-t-component-sum}) and
(\ref{eq:divrFD=To2-2-sum}), which means that the relation
(\ref{eq:Xo-Psio-relation}) carries the information of
Eq.~(\ref{eq:EvolofFA=source-t-component-sum}).

%*********************************************************************

From Eq.~(\ref{eq:Psio-Zo-relation}), we evaluate the second
time-derivative of the master variable $\Phi_{(o)}$.
On the other hand, from Eq.~(\ref{eq:Xo-Psio-relation-pre}) we also
evaluate the second derivative of $\Phi_{(o)}$ with respect to the
tortoise coordinate $f\partial_{r}$.
Furthermore, using Eqs.~(\ref{eq:Cunningham-variable-2}) and
(\ref{eq:Xo-Psio-relation}), we obtain
\begin{eqnarray}
  &&
     \partial_{t}^{2}\Phi_{(o)}
     - f \partial_{r}\left[ f \partial_{r}\Phi_{(o)} \right]
     + \frac{1}{r^{2}} f \left[ l(l+1) - 3(1-f) \right] \Phi_{(o)}
     \nonumber\\
  &=&
      32 \pi r f \left[
      \partial_{r}(r \tilde{T}_{(o1)t})
      -  r \partial_{t}\tilde{T}_{(o1)r}
      \right]
      .
      \label{eq:odd-master-equation-full-information}
\end{eqnarray}
This has the same form as
Eq.~(\ref{eq:odd-master-equation-Regge-Wheeler}) but we have different
source terms from Eq.~(\ref{eq:odd-master-equation-Regge-Wheeler}).
For the $l\geq 2$ case, the analytic solutions to
Eq.~(\ref{eq:odd-master-equation-full-information}) is also
constructed by the MST formulation~\cite{S.Mano-H.Suzuki-E.Takasugi-1996a,S.Mano-H.Suzuki-E.Takasugi-1996b,S.Mano-E.Takasugi-1997,H.Tagoshi-S.Mano-E.Takasugi-1997}.
In the vacuum case, Eq.~(\ref{eq:Psio-Zo-relation}) with $l\neq 1$ implies
that the component $Y_{(o)}$ of the metric perturbation corresponds to the
time-derivative of the variable $\Phi_{(o)}$.
This indicates that
Eq.~(\ref{eq:odd-master-equation-full-information}) corresponds to the
time-integration of Eq.~(\ref{eq:odd-master-equation-Regge-Wheeler})
in the vacuum case.
However, there is no degree of freedom of the integration constant in
Eq.~(\ref{eq:odd-master-equation-full-information}).
Therefore, we may say that the initial conditions for
Eq.~(\ref{eq:odd-master-equation-full-information}) is restricted more
than that of Eq.~(\ref{eq:odd-master-equation-Regge-Wheeler}).

%*********************************************************************

Here, we note that Eq.~(\ref{eq:Yo-Psio-relation}) is derived from
Eqs.~(\ref{eq:divrFD=To2-2-sum}) and
(\ref{eq:odd-master-equation-Regge-Wheeler}).
This means that the relation (\ref{eq:Yo-Psio-relation}) does not includes the information
(\ref{eq:EvolofFA=source-t-component-sum}).
On the other hand, the relation (\ref{eq:Xo-Psio-relation}) is derived from
Eq.~(\ref{eq:divrFD=To2-2-sum}) and (\ref{eq:EvolofFA=source-t-component-sum}).
This means that the relation (\ref{eq:Xo-Psio-relation}) does not includes the information
of Eq.~(\ref{eq:odd-master-equation-Regge-Wheeler}).
In other words, we may regard the relation (\ref{eq:Yo-Psio-relation}) as a result of
Eq.~(\ref{eq:odd-master-equation-Regge-Wheeler}), while Eq.~(\ref{eq:Xo-Psio-relation-pre})
as a result of Eq.~(\ref{eq:EvolofFA=source-t-component-sum}).
Therefore, we obtain the two equations (\ref{eq:Yo-Psio-relation})
and (\ref{eq:Xo-Psio-relation}) from the three equations (\ref{eq:divrFD=To2-2-sum}),
(\ref{eq:odd-master-equation-Regge-Wheeler}),
and (\ref{eq:EvolofFA=source-t-component-sum}).
On the other hand, we have derived
Eq.~(\ref{eq:odd-master-equation-full-information}) from
Eqs.~(\ref{eq:divrFD=To2-2-sum}), (\ref{eq:odd-master-equation-Regge-Wheeler}),
and (\ref{eq:EvolofFA=source-t-component-sum}), which is independent
of Eqs.~(\ref{eq:Yo-Psio-relation}) and (\ref{eq:Xo-Psio-relation}).
Thus, we may regard that all information of the set of three equations
(\ref{eq:divrFD=To2-2-sum}), (\ref{eq:odd-master-equation-Regge-Wheeler}), and
(\ref{eq:EvolofFA=source-t-component-sum}) is included in the set of three equations
(\ref{eq:Yo-Psio-relation}), (\ref{eq:Xo-Psio-relation}), and
(\ref{eq:odd-master-equation-full-information}).
In addition to these equations, we have to take into account of the
continuity equation (\ref{eq:div-barTab-linear-odd-component}) for the
odd-mode perturbations of the matter field.

%*********************************************************************

However, as emphasized above, these arguments are not valid for $l=1$ mode.
Therefore, we have to reconsider the derivation of equations in the case of $l=1$
mode, separately.
Here, we examine the $l=1$ modes.
In this case, Eq.~(\ref{eq:Psio-Zo-relation}) is still valid, though
this equation does not give the component $Y_{(o)}$ of the metric
perturbations.
In this case, the time-derivative of the variable $\Phi_{(o)}$ is given by
\begin{eqnarray}
  \partial_{t}\Phi_{(o)}
  =
  - 32 \pi r^{2} f \tilde{T}_{(o1)r}
  ,
  \label{eq:Psio-Zo-relation-l=1}
\end{eqnarray}
which indicates that $\partial_{t}\Phi_{(o)}$ is determined by the
matter degree of freedom.
Similarly, Equation (\ref{eq:Xo-Psio-relation-pre}) is also valid even
in the case of $l=1$ mode, though this equation does not give the
component $X_{(o)}$ of the metric perturbations.
In this case, we obtain
\begin{eqnarray}
  f \partial_{r}\Phi_{(o)}
  =
  - \frac{1}{r} f \Phi_{(o)}
  - 32 \pi r^{2} \tilde{T}_{(o1)t}
  .
  \label{eq:Xo-Psio-relation-pre-l=1}
\end{eqnarray}
This equation indicates that the $\partial_{r}\Phi_{(o)}$ is also
determined by the matter degree of freedom.
From Eqs.~(\ref{eq:Psio-Zo-relation-l=1}) and
(\ref{eq:Xo-Psio-relation-pre-l=1}), we can confirm that the variable
$\Phi_{(o)}$ satisfies
Eq.~(\ref{eq:odd-master-equation-full-information}) with $l=1$.
However, we do not have to solve
Eq.~(\ref{eq:odd-master-equation-full-information}) with $l=1$ in this
case, because we can directly integrate
Eqs.~(\ref{eq:Psio-Zo-relation-l=1}) and
(\ref{eq:Xo-Psio-relation-pre-l=1}).
Actually, the integrability condition
$\partial_{t}\partial_{r}\Phi_{(o)}=\partial_{r}\partial_{t}\Phi_{(o)}$
of Eqs.~(\ref{eq:Psio-Zo-relation-l=1}) and
(\ref{eq:Xo-Psio-relation-pre-l=1}) can be checked through the
continuity equation (\ref{eq:div-barTab-linear-odd-component}) with
$l=1$.

%*********************************************************************

Since we obtain the variable $\Phi_{(o)}$ by the direct integration of
Eqs.~(\ref{eq:Psio-Zo-relation-l=1}) and
(\ref{eq:Xo-Psio-relation-pre-l=1}), we can obtain the relation
between the components $X_{(o)}$ and $Y_{(o)}$ of the metric
perturbations through the definition (\ref{eq:Cunningham-variable-2}).
In addition to the solution $\Phi_{(o)}$, if we have a solution to
$Z_{(o)}=\frac{f}{r}Y_{(o)}$, independently, we obtain the components
$X_{(o)}$ and $Y_{(o)}$ of the metric perturbations through the above
relation between $X_{(o)}$ and $Y_{(o)}$.
Note that $Z_{(o)}=\frac{f}{r}Y_{(o)}$ can be determined through the
integration of Eq.~(\ref{eq:odd-master-equation-Regge-Wheeler}) with
$l=1$ with appropriate boundary conditions.
In this case, the continuity equation
(\ref{eq:div-barTab-linear-odd-component}) for odd-mode matter
perturbations is used as consistency check of the solutions.

%*********************************************************************

%%%%%%%%%%%%%%%%%%%%%%%%%%%%%%%%%%%%%%%%%%
%%%%%%%%%%%%%%%%%%%%%%%%%%%%%%%%%%%%%%%%%%%
\subsection{Odd-mode solutions}
\label{sec:Schwarzschild_Background_odd_vacuum_sol}
%%%%%%%%%%%%%%%%%%%%%%%%%%%%%%%%%%%%%%%%%%%
%%%%%%%%%%%%%%%%%%%%%%%%%%%%%%%%%%%%%%%%%%%

%*********************************************************************

Since the construction of solutions for $l\geq 2$ mode is accomplished
by the MST
formulation~\cite{S.Mano-H.Suzuki-E.Takasugi-1996a,S.Mano-H.Suzuki-E.Takasugi-1996b,S.Mano-E.Takasugi-1997,H.Tagoshi-S.Mano-E.Takasugi-1997},
we discuss the $l=0,1$-mode solutions for odd-mode perturbations along
Proposal~\ref{proposal:treatment-proposal-on-pert-on-spherical-BG} and
the strategy discussed in
Sec.~\ref{sec:Schwarzschild_Background_odd_sol_strategy}.

%*********************************************************************

%%%%%%%%%%%%%%%%%%%%%%%%%%%%%%%%%%%%%%%%%%%
\subsubsection{$l=0$ odd mode}
\label{sec:Schwarzschild_Background_odd_l=1_vacuum_sol_l=0}
%%%%%%%%%%%%%%%%%%%%%%%%%%%%%%%%%%%%%%%%%%%

%*********************************************************************

We choose Eq.~(\ref{eq:l=0-general-mode-func-specific}) as the
harmonic function $k_{(\hat{\Delta})}$ and used the set
$\{\hat{D}_{p}k_{(\hat{\Delta})},$
$\epsilon_{pr}\hat{D}^{r}k_{(\hat{\Delta})},$
$\hat{D}_{p}\hat{D}_{q}k_{(\hat{\Delta})},$
$2\epsilon_{r(p}\hat{D}_{q)}\hat{D}^{r}k_{(\hat{\Delta})}\}$ as the
basis of the vector and tensors on $S^{2}$.
The bases of the odd-mode perturbations are
$\epsilon_{pr}\hat{D}^{r}k_{(\hat{\Delta})}$ and
$2\epsilon_{r(p}\hat{D}_{q)}\hat{D}^{r}k_{(\hat{\Delta})}$.
Following
Proposal~\ref{proposal:treatment-proposal-on-pert-on-spherical-BG}, we
choose $\delta=0$ as the regularity of solutions when we solve the
linearized Einstein equations.
As shown in Eqs.~(\ref{eq:l=0-extended-mode-func-Dp-norm}) and
(\ref{eq:l=0-extended-mode-func-epsilonDpDq}),
$\epsilon_{pr}\hat{D}^{r}k_{(\hat{\Delta})}=0=2\epsilon_{r(p}\hat{D}_{q)}\hat{D}^{r}k_{(\hat{\Delta})}$.
Then, we conclude that there is no nontrivial solution for odd-mode
perturbations with $l=0$.

%*********************************************************************

%%%%%%%%%%%%%%%%%%%%%%%%%%%%%%%%%%%%%%%%%%%
\subsubsection{$l=1$ odd-mode vacuum solution}
\label{sec:Schwarzschild_Background_odd_l=1_vacuum_sol_l=1}
%%%%%%%%%%%%%%%%%%%%%%%%%%%%%%%%%%%%%%%%%%%

%*********************************************************************

Following the strategy to solve the $l=1$ odd-mode perturbation given
in Sec.~\ref{sec:Schwarzschild_Background_odd_sol_strategy},
we consider the equations
(\ref{eq:odd-master-equation-Regge-Wheeler}),
(\ref{eq:div-barTab-linear-odd-component}) with $l=1$,
(\ref{eq:Cunningham-variable-2}), (\ref{eq:Psio-Zo-relation-l=1}),
and (\ref{eq:Xo-Psio-relation-pre-l=1}).
To derive the non-vacuum solution to the linearized Einstein equations
for $l=1$ odd-mode perturbations, it is instructive to consider the
vacuum case in which
$\tilde{T}_{(o1)t}=\tilde{T}_{(o1)r}=\tilde{T}_{(o2)}=0$.
From Eqs.~(\ref{eq:Psio-Zo-relation-l=1}) and
(\ref{eq:Xo-Psio-relation-pre-l=1}), we obtain the solution to these
equations as~\footnote{
  From Eq.~(\ref{eq:vacuum-l=1-Psio}) and the descriptions in
  Ref.~\cite{H.Kodama-H.Ishihara-Y.Fujiwara-1994}, readers might regard
  that the extension to $l=1$ mode case of the
  Cunningham-Price-Moncrief variable $\Phi_{(o)}$ is the same variable
  as the gauge-invariant variable $\Phi_{KIF}$ defined by
  Eq.~(\ref{eq:l=1-odd-KIF-gauge-inv}).
  Actually, if we can identify $\tilde{h}_{(o1)A}$ with
  $\tilde{F}_{A}$, the extension to $l=1$ mode case of the
  Cunningham-Price-Moncrief variable $\Phi_{(o)}$ coincides with the
  definition of $\Phi_{KIF}$ and there is the description in
  Ref.~\cite{H.Kodama-H.Ishihara-Y.Fujiwara-1994} which is similar to
  Eq.~(\ref{eq:vacuum-l=1-Psio}).
  However, this identification is not appropriate, since
  $\tilde{F}_{A}$ is gauge invariant in the sense of the second-kind
  but $\tilde{h}_{(o1)A}$ is not gauge-invariant.
  We actually take $\delta=0$ in the singular harmonic when we solve
  the mode-by-mode Einstein equations.
  However, this does not mean $\tilde{h}_{(o2)}=0$, nevertheless the
  term $\tilde{h}_{(o2)}$ in the metric perturbation disappear since
  the singular harmonic function vanishes due to the choice $\delta=0$.
  This difference also appears when we obtain the
  gauge-invariant relation between the components of $\tilde{F}_{A}$
  and the extension to $l=1$ mode case of Cunningham-Price-Moncrief
  variable $\Phi_{(o)}$ by integrating the linearized Einstein equations.
  In this integration, the integration constants appear in the
  relation between the components $\tilde{F}_{A}$ and $\Phi_{(o)}$.
  This integration ``constants'' are automatically gauge-invariant in
  the sense of second-kind.
  On the other hand, when we integrate $\Psi_{KIF}$ to obtain the
  explicit relation with $\tilde{h}_{(o1)A}$, there is no guarantee
  that the integration ``constants'' are gauge invariant, because
  $\tilde{h}_{(o1)A}$ is not gauge-invariant.
}
\begin{eqnarray}
  \label{eq:vacuum-l=1-Psio}
  \Phi_{(o)} = \frac{\alpha}{r},
\end{eqnarray}
where $\alpha$ is constant of integration.

%*********************************************************************

On the other hand, $Y_{(o)}$ is obtained as the solution to the $l=1$
version of the Regge-Wheeler equation
(\ref{eq:odd-master-equation-Regge-Wheeler}) without source terms
through Eq.~(\ref{eq:Regge-Wheeler-variable}).
Here, we consider the case $Y_{(o)}=0$, at first.
The derivations of solutions under the assumption $Y_{(o)}=0$ is an
instructive lesson for the derivation of the general solutions of the
$l=1$ odd-mode perturbations.
Through the definition (\ref{eq:Cunningham-variable}) of the variable
$\Phi_{(o)}$ and Eq.~(\ref{eq:vacuum-l=1-Psio}), we obtain
\begin{eqnarray}
  \label{eq:Cunningham-variable-l=1-vacuum-Yo=0}
  \frac{\alpha}{r}
  &=&
      2r \left[
      r^{2} \partial_{r}\left(\frac{X_{(o)}}{r^{2}}\right)
      \right]
      .
\end{eqnarray}
The solution to Eq.~(\ref{eq:Cunningham-variable-l=1-vacuum-Yo=0})
together with the assumption $Y_{(o)}=0$ is a special solution to the
linearized Einstein equations for $l=1$ odd-mode perturbations as follows:
\begin{eqnarray}
  \label{eq:odd-mode-l=1-vacuum-special-so-suml}
  X_{(o)}
  =
  - \frac{\alpha}{6r} + \beta_{1} r^{2}
  ,
  \quad
  Y_{(o)}
  =
  0
  ,
\end{eqnarray}
where $\beta_{1}$ is constant~\footnote{
  Although the simple integration of
  Eq.~(\ref{eq:Cunningham-variable-l=1-vacuum-Yo=0}) yields the
  time-dependence of $\beta_{1}$, this time-dependence is inconsistent with
  Eq.~(\ref{eq:divrFD=To2-2-sum}).
  This inconsistency is due to the fact that we just use the
  constraint (\ref{eq:divrFD=To2-2-sum}) in the form
  $\partial_{r}\left(r^{2}\mbox{(\ref{eq:divrFD=To2-2-sum})}\right)$
  when we derive Eq.~(\ref{eq:Psio-Zo-relation}).
}.
From Eqs.~(\ref{eq:2+2-gauge-invariant-variables-calFAp}) and
(\ref{eq:component-odd-rtildeFD}), we can derive the gauge-invariant
metric perturbation $\ScrF_{Ap}$ which corresponds to the solution
(\ref{eq:odd-mode-l=1-vacuum-special-so-suml}).
In the $l=1$ modes, there are $m=0,\pm 1$ modes.
In this paper, we only consider the $m=0$-mode perturbation, since the
generalization to $m=\pm 1$ modes is straightforward.
If we choose $\delta=0$ in the mode function
(\ref{eq:Delta+2m=0gradient}), we obtain
\begin{eqnarray}
  \label{eq:gauge-inv-metric-pert-FAp-tildeFA-rela-l=1-m=0-def}
  \ScrF_{Ap}
  =
  r F_{Ap}
  =
  r \tilde{F}_{A} \epsilon_{pq} \hat{D}^{q}k_{(\hat{\Delta}+2,m=0)}
  ,
  \quad
  \epsilon_{pq} \hat{D}^{q}k_{(\hat{\Delta}+2,m=0)}
  =
  \sin^{2}\theta (d\phi)_{p}
  .
\end{eqnarray}
Then, we have
\begin{eqnarray}
  \label{eq:gauge-inv-metric-pert-FAp-tildeFA-rela-l=1-m=0}
  2 \ScrF_{Ap} (dx^{A})_{(a}(dx^{p})_{b)}
  &=&
      2 r F_{(Ap)} (dx^{A})_{(a}(dx^{p})_{b)}
      \nonumber\\
  &=&
      \label{eq:gauge-inv-metric-pert-2calFAp-Xo-Yo-l=1-m=0}
      2 X_{(o)} \sin^{2}\theta (dt)_{(a} (d\phi)_{b)}
      +
      2 Y_{(o)} \sin^{2}\theta (dr)_{(a} (d\phi)_{b)}
      \\
  &=&
      \left(
      - \frac{\alpha}{3r} + 2\beta_{1} r^{2}
      \right)
      \sin^{2}\theta (dt)_{(a} (d\phi)_{b)}
      .
      \label{eq:gauge-inv-metric-pert-2calFAp-Xo-Yo-l=1-m=0-special-sol}
\end{eqnarray}
Here, the term $\beta_{1}r^{2}$ is diverge as $r\rightarrow
\infty$.
At this moment, we choose the arbitrary function
$\beta_{1}=0$ to derive a special solution.
Then, we have obtained
\begin{eqnarray}
  2 \ScrF_{Ap} (dx^{A})_{(a}(dx^{p})_{b)}
  =
  - \frac{\alpha}{3r} \sin^{2}\theta (dt)_{(a} (d\phi)_{b)}
  .
  \label{eq:gauge-inv-metric-pert-2calFAp-Xo-Yo-l=1-m=0-special-sol-beta=0}
\end{eqnarray}
Eq.~(\ref{eq:gauge-inv-metric-pert-2calFAp-Xo-Yo-l=1-m=0-special-sol-beta=0})
is the linearized Kerr solution.
Actually, the Kerr solution with the Kerr parameter $a$ is expressed
as~\cite{R.H.Boyer-R.W.Lindquist-1967,Wald-book}
\begin{eqnarray}
  g_{ab}
  &=&
      -
      \left[
      1
      -
      \frac{2Mr}{\Sigma}
      \right]
      (dt)_{a}(dt)_{b}
      -
      \frac{2 a M r \sin^{2}\theta}{\Sigma} (dt)_{(a}(d\phi)_{b)}
      +
      \frac{\Sigma}{\Delta} (dr)_{a}(dr)_{b}
      \nonumber\\
  &&
     +
     \Sigma(d\theta)_{a}(d\theta)_{b}
     +
     \left[
     r^{2} + a^{2} + \frac{2Mr}{\Sigma} a^{2}\sin^{2}\theta
     \right]
     \sin^{2}\theta (d\phi)_{a}(d\phi)_{b}
     ,
     \label{eq:Kerr-solution-from-Wald}
\end{eqnarray}
where
\begin{eqnarray}
  \label{eq:Kerr-Sigma-Delta-def}
  \Sigma := r^{2} + a^{2} \cos^{2}\theta, \quad
  \Delta := r^{2} + a^{2} - 2 Mr
  .
\end{eqnarray}
In the metric (\ref{eq:Kerr-solution-from-Wald}), we replace
$a\rightarrow \epsilon a$, where $\epsilon$ is the parameter for
the perturbative expansion.
Then, when the Kerr metric (\ref{eq:Kerr-solution-from-Wald}) is
expressed as follows:
\begin{eqnarray}
  g_{ab}
  &=&
      y_{ab}
      +
      r^{2}
      \gamma_{ab}
      +
      \epsilon
      \left(
      -
      \frac{2 a M}{r}  \sin^{2}\theta (dt)_{(a}(d\phi)_{b)}
      \right)
      +
      O(\epsilon^{2})
      .
     \label{eq:linearized-Kerr-solution-from-Wald}
\end{eqnarray}
Comparing
Eqs.~(\ref{eq:gauge-inv-metric-pert-2calFAp-Xo-Yo-l=1-m=0-special-sol-beta=0})
and (\ref{eq:linearized-Kerr-solution-from-Wald}), the constant of
integration $\alpha$ in
Eq.~(\ref{eq:gauge-inv-metric-pert-2calFAp-Xo-Yo-l=1-m=0-special-sol-beta=0})
is identified as the angular momentum
perturbation in Kerr solution by choosing
\begin{eqnarray}
  \frac{\alpha}{3} = 2 a M =: 2 a_{10} M.
  \label{eq:Kerr-parameter-a1}
\end{eqnarray}
Thus, we have seen that the solution
(\ref{eq:gauge-inv-metric-pert-2calFAp-Xo-Yo-l=1-m=0-special-sol}) is
given using the Kerr parameter $a_{10}$ as follows:
\begin{eqnarray}
  2 \ScrF_{Ap} (dx^{A})_{(a}(dx^{p})_{b)}
  =
  2 \left(
  - \frac{a_{10}M}{r} + \beta_{1} r^{2}
  \right)
  \sin^{2}\theta (dt)_{(a} (d\phi)_{b)}
  .
  \label{eq:2calFAp-Xo-Yo-l=1-m=0-special-sol-with-Kerr}
\end{eqnarray}

%*********************************************************************

Next, we consider the physical meaning of the constant $\beta_{1}$ in the solution
(\ref{eq:gauge-inv-metric-pert-2calFAp-Xo-Yo-l=1-m=0-special-sol}).
If we consider the frame with the rigid rotation
\begin{eqnarray}
  \label{eq:rigid-rotation}
  t = t', \quad \phi = \varphi + \epsilon \omega t'.
\end{eqnarray}
In terms of $(t',\varphi)$, the background metric
(\ref{eq:background-metric-2+2}) with
Eqs.~(\ref{eq:background-metric-2+2-y-comp-Schwarzschild}) and
(\ref{eq:background-metric-2+2-gamma-comp-Schwarzschild}) is given by
\begin{eqnarray}
  g_{ab}
  &=&
      -  f(dt')_{a} (dt')_{b}
      + f^{-1}(dr)_{a}(dr)_{b}
      + r^{2}\left[
      (d\theta)_{a}(d\theta)_{b}
      +
      \sin^{2}\theta
      (d\varphi)_{a}
      (d\varphi)_{b}
      \right]
      \nonumber\\
  &&
     +
     2
     \epsilon
     \omega
     r^{2}
     \sin^{2}\theta
     (dt')_{(a} (d\varphi)_{b)}
     +
     O(\epsilon^{2})
     .
     \label{eq:flat-metric-spherical-rigid-rotation-linear}
\end{eqnarray}
Comparing Eq.~(\ref{eq:flat-metric-spherical-rigid-rotation-linear})
and Eq.~(\ref{eq:2calFAp-Xo-Yo-l=1-m=0-special-sol-with-Kerr}),
we can see that the arbitrary function $\beta_{1}$ corresponds to
\begin{eqnarray}
  \label{eq:beta-omega-relation}
  \beta_{1} = \omega
  .
\end{eqnarray}
Thus, we may interpret the integration constant $\beta_{1}$
as non-inertia term due to the rigidly rotating frame with the angular
velocity $\omega$.

%*********************************************************************

Finally, we consider the general solution for $l=1$ odd-mode
perturbations which includes the case $Y_{(o)}\neq 0$ through
Eqs.~(\ref{eq:Regge-Wheeler-variable}) and
(\ref{eq:odd-master-equation-Regge-Wheeler}).
Here, we consider the situation $Y_{(o)}\neq 0$ and introduce the
variable $W_{(o)}$ as follows:
\begin{eqnarray}
  \label{eq:Wo-def}
  Y_{o} =: r^{2}\partial_{r}W_{(o)},
  \quad
  Z_{(o)} = \frac{f}{r} Y_{(o)} = r f \partial_{r}W_{(o)}.
\end{eqnarray}
Through the solution (\ref{eq:vacuum-l=1-Psio}) with
Eq.~(\ref{eq:Kerr-parameter-a1}) and the definition
(\ref{eq:Cunningham-variable}) of the variable $\Phi_{(o)}$, we obtain
the equation
\begin{eqnarray}
  \label{eq:Cunningham-variable-l=1-vacuum-perturbation-general}
  \frac{6 a_{10}M}{r}
  =
  2r \left[
  r^{2} \partial_{r}\left(\frac{X_{(o)}}{r^{2}}\right) - r^{2}\partial_{t}\partial_{r}W_{(o)}
  \right]
  .
\end{eqnarray}
Integrating this equation, we obtain
\begin{eqnarray}
  X_{(o)}
  =
  - \frac{a_{10}M}{r}
  + \beta_{1} r^{2}
  + r^{2}\partial_{t}W_{(o)}
  .
  \label{eq:Cunningham-variable-l=1-vacuum-perturbation-general-sol}
\end{eqnarray}
Through Eqs.~(\ref{eq:2+2-gauge-invariant-variables-calFAp}) and
(\ref{eq:component-odd-rtildeFD}), we obtain
\begin{eqnarray}
  2 \ScrF_{Ap} (dx^{A})_{(a}(dx^{p})_{b)}
  &=&
      2 \left(
      - \frac{a_{10}M}{r}
      + r^{2} \beta_{1}
      + r^{2}\partial_{t}W_{(o)}
      \right)
      \sin^{2}\theta
      (dt)_{(a} (d\phi)_{b)}
     \nonumber\\
  &&
      +
      2
      r^{2}\partial_{r}W_{(o)}
      \sin^{2}\theta
      (dr)_{(a} (d\phi)_{b)}
      .
  \label{eq:l=1-odd-mode-propagating-sol}
\end{eqnarray}
Note again that the variable $Z_{(o)} = r f \partial_{r}W_{(o)}$
satisfy the Regge-Wheeler equation
(\ref{eq:odd-master-equation-Regge-Wheeler}) with $l=1$.

%*********************************************************************

The above interpretation of the arbitrary function $\beta_{1}$ as the
inertia force on the rigidly rotation frame is instructive to consider
the interpretation of the odd-mode vacuum solution
(\ref{eq:l=1-odd-mode-propagating-sol}).
To see this, we consider the component expression of
${\pounds}_{V}g_{ab}$, where $V^{a}$ is constructed from
gauge-invariant variables, which is discussed in
Sec.~\ref{sec:review-of-general-framework-GI-perturbation-theroy}.
To obtain the components of ${\pounds}_{V}g_{ab}$, the explicit
components of the Christoffel symbol $\Gamma_{ab}^{\;\;\;\;c}$ for the
background metric (\ref{eq:background-metric-2+2}) with
Eqs.~(\ref{eq:background-metric-2+2-y-comp-Schwarzschild}) and
(\ref{eq:background-metric-2+2-gamma-comp-Schwarzschild}) are
convenient, which are summarized in
Eqs.~(\ref{eq:Background-connection-explicit}).
Here, we assume that $V_{a}=V_{\phi}(d\phi)_{a}$, then the
non-vanishing components of ${\pounds}_{V}g_{ab}$ are given by
\begin{eqnarray}
     {\pounds}_{V}g_{t\phi}
     =
     \partial_{t}V_{\phi}
     ,
  \quad
     {\pounds}_{V}g_{r\phi}
     =
     \partial_{r}V_{\phi} - \frac{2}{r} V_{\phi}
     ,
  \quad
     {\pounds}_{V}g_{\theta\phi}
     =
     \partial_{\theta}V_{\phi} - 2 \cot\theta V_{\phi}
     .
  \label{eq:poundsVgtphi-poundsVrphi-poundsVthetaphi}
\end{eqnarray}
Comparing Eqs.~(\ref{eq:l=1-odd-mode-propagating-sol}) and
(\ref{eq:poundsVgtphi-poundsVrphi-poundsVthetaphi}), we obtain
\begin{eqnarray}
  V_{a}
  &=&
      \left(\beta_{1}t+\beta_{0} + W_{(o)}(t,r)\right) r^{2} \sin^{2}\theta (d\phi)_{a}
      ,
  \\
  {\pounds}_{V}g_{ab}
  &=&
      \partial_{t}\left(\beta_{1}t + \beta_{0} + W_{(o)}(t,r)\right)r^{2}\sin^{2}\theta
      2 (dt)_{(a}(d\phi)_{b)}
      \nonumber\\
  &&
      +
      \left(\partial_{r}W_{(o)}(t,r)\right) r^{2} \sin^{2}\theta
      2 (dr)_{(a}(d\phi)_{b)}
      ,
\end{eqnarray}
where $\beta_{0}$ is constant.
This coincides with the perturbation
(\ref{eq:l=1-odd-mode-propagating-sol}) with the condition of the
vanishing Kerr parameter $a_{10}=0$.
Then, we have
\begin{eqnarray}
  \label{eq:l=1-odd-mode-propagating-sol-ver2}
  2 \ScrF_{Ap} (dx^{A})_{(a}(dx^{p})_{b)}
  &=&
      - \frac{2 a_{10}M}{r}
      \sin^{2}\theta
      (dt)_{(a} (d\phi)_{b)}
      +
      {\pounds}_{V}g_{ab}
      ,
  \\
  \label{eq:l=1-odd-mode-propagating-sol-ver2-Va-def}
  V_{a}
  &=&
      \left(\beta_{1}t + \beta_{0} + W_{(o)}(t,r)\right) r^{2} \sin^{2}\theta (d\phi)_{a}
      .
\end{eqnarray}
Here, we note that the vector field $V_{a}$ and ${\pounds}_{V}g_{ab}$
are gauge-invariant.
The interpretation of this term ${\pounds}_{V}g_{ab}$, which is gauge
invariant in the sense of the second kind, is extensively discussed in
Sec.~\ref{sec:summary_and_discussions}.

%*********************************************************************

%%%%%%%%%%%%%%%%%%%%%%%%%%%%%%%%%%%%%%%%%%
%%%%%%%%%%%%%%%%%%%%%%%%%%%%%%%%%%%%%%%%%%%
\subsection{Odd mode non-vacuum $l=1$ solution}
\label{sec:Schwarzschild_Background_odd_non-vacuum_l=1_sol}
%%%%%%%%%%%%%%%%%%%%%%%%%%%%%%%%%%%%%%%%%%%
%%%%%%%%%%%%%%%%%%%%%%%%%%%%%%%%%%%%%%%%%%%

%*********************************************************************

Inspecting the derivation of the vacuum solution for $l=1$ modes in
Sec.~\ref{sec:Schwarzschild_Background_odd_l=1_vacuum_sol_l=1}, we
consider the non-vacuum solution for $l=1$ modes.
For $l=1$ modes, the linearized Einstein equations for the master
variable $\Phi_{(o)}$ defined by Eq.~(\ref{eq:Cunningham-variable})
are given by Eqs.~(\ref{eq:Psio-Zo-relation-l=1}) and
(\ref{eq:Xo-Psio-relation-pre-l=1}).
As mentioned in
Sec.~\ref{sec:Schwarzschild_Background_odd_sol_strategy}, the
integrability condition for these equations is guaranteed by the
continuity equation (\ref{eq:div-barTab-linear-odd-component}) with
$l=1$.
Inspecting Eqs.~(\ref{eq:vacuum-l=1-Psio}) and
(\ref{eq:Kerr-parameter-a1}), we consider the solution in the form
\begin{eqnarray}
  \label{eq:non-vacuum-l=1-Psio-sol-form}
  \Phi_{(o)} = \frac{6 M a_{1}(t,r)}{r}.
\end{eqnarray}
Substituting Eq.~(\ref{eq:non-vacuum-l=1-Psio-sol-form}) into
Eqs.~(\ref{eq:Psio-Zo-relation-l=1}) and
(\ref{eq:Xo-Psio-relation-pre-l=1}), we obtain
\begin{eqnarray}
  \partial_{t}a_{1}(t,r)
  =
  - \frac{16 \pi}{3M} r^{3} f \tilde{T}_{(o1)r}
  , \quad
  \partial_{r}a_{1}(t,r)
  =
  - \frac{16 \pi}{3M} r^{3} \frac{1}{f} \tilde{T}_{(o1)t}
  .
  \label{eq:Xo-Psio-relation-pre-l=1-sum-3}
\end{eqnarray}
The integrability of Eqs.~(\ref{eq:Xo-Psio-relation-pre-l=1-sum-3}) is
equivalent to the integrability of Eqs.~(\ref{eq:Psio-Zo-relation-l=1}) and
(\ref{eq:Xo-Psio-relation-pre-l=1}) which is guaranteed by the
continuity equation (\ref{eq:div-barTab-linear-odd-component}) with
$l=1$.
Then, we may integrate Eqs.~(\ref{eq:Xo-Psio-relation-pre-l=1-sum-3})
as follows:
\begin{eqnarray}
  a_{1}(t,r)
  &=&
      - \frac{16 \pi}{3M} r^{3} f \int dt \tilde{T}_{(o1)r} + a_{10}
      \nonumber\\
  &=&
      - \frac{16 \pi}{3M} \int dr r^{3} \frac{1}{f} \tilde{T}_{(o1)t} + a_{10}
      ,
      \label{eq:Psio-Zo-relation-l=1-sum-6}
\end{eqnarray}
where $a_{10}$ is the constant which corresponds to the Kerr parameter
$a$ in Eq.~(\ref{eq:Kerr-solution-from-Wald}) as shown in the vacuum
case.

%*********************************************************************

Similar arguments to those in
Sec.~\ref{sec:Schwarzschild_Background_odd_l=1_vacuum_sol_l=1}, which
lead the results (\ref{eq:l=1-odd-mode-propagating-sol-ver2}) and
(\ref{eq:l=1-odd-mode-propagating-sol-ver2-Va-def}), also leads
\begin{eqnarray}
  \label{eq:l=1-odd-mode-propagating-sol-ver2-2}
  &&
     2 \ScrF_{Ap} (dx^{A})_{(a}(dx^{p})_{b)}
     =
     6 M r^{2} \left[\int dr \frac{a_{1}(t,r)}{r^{4}}\right]
     \sin^{2}\theta
     (dt)_{(a} (d\phi)_{b)}
     +
     {\pounds}_{V}g_{ab}
     ,
  \\
  \label{eq:l=1-odd-mode-propagating-sol-ver2-Va-def-2}
  &&
     V_{a}
     =
     \left(\beta_{1}t + \beta_{0} + W_{(o)}(t,r)\right) r^{2} \sin^{2}\theta (d\phi)_{a}
     .
\end{eqnarray}
Here, we note that the vector field $V_{a}$ and ${\pounds}_{V}g_{ab}$
are gauge-invariant in the sense of the second kind.
The term ${\pounds}_{V}g_{ab}$ may always appear due to the symmetry
of the linearized Einstein equation as pointed out through
Eq.~(\ref{eq:einstein-equation-gauge-inv-calH}).
However, it is also true that we can eliminate the term
${\pounds}_{V}g_{ab}$ by an infinitesimal coordinate transformation at
any time.
The interpretation of the term ${\pounds}_{V}g_{ab}$ will be discussed
in Sec.~\ref{sec:summary_and_discussions}.

%*********************************************************************

%%%%%%%%%%%%%%%%%%%%%%%%%%%%%%%%%%%%%%%%%%%
%%%%%%%%%%%%%%%%%%%%%%%%%%%%%%%%%%%%%%%%%%%
%%%%%%%%%%%%%%%%%%%%%%%%%%%%%%%%%%%%%%%%%%%
\section{Summary and Discussions}
\label{sec:summary_and_discussions}
%%%%%%%%%%%%%%%%%%%%%%%%%%%%%%%%%%%%%%%%%%%
%%%%%%%%%%%%%%%%%%%%%%%%%%%%%%%%%%%%%%%%%%%
%%%%%%%%%%%%%%%%%%%%%%%%%%%%%%%%%%%%%%%%%%%

%*********************************************************************

In summary, after reviewing our general framework of the
gauge-invariant perturbation theory, we discussed a resolution of
the ``zero-mode problem'' in perturbations on the Schwarzschild
background spacetime.
The ``zero-mode problem'' in the context of our general framework of
the gauge-invariant perturbation theory corresponds to the $l=0,1$
mode problem in perturbations of the Schwarzschild background
spacetime.
In the review of our general framework of the gauge invariant
perturbation theory, we emphasize the importance of the distinction of
the first- and the second-kind gauge in general relativity.
It should be also emphasized that our general framework for the
gauge-invariant perturbation theory is a formulation to exclude
the second-kind gauge degree of freedom, but we do not exclude
first-kind gauge degree of freedom.

%*********************************************************************

As emphasize in
Sec.~\ref{sec:review-of-general-framework-GI-perturbation-theroy},
Conjecture~\ref{conjecture:decomposition-conjecture} is the
non-trivial and an important premise of our general framework of
gauge-invariant perturbation theories.
If Conjecture~\ref{conjecture:decomposition-conjecture} is actually
true, we can develop gauge-invariant perturbation theory on general
background spacetime and we can also extend this gauge-invariant
perturbation theory to higher-order perturbation theory.
For this reason, the gauge-invariant treatment of the $l=0,1$ modes in
perturbations of the Schwarzschild background spacetime is important
not only for the development of the linear perturbations but also for
the development of the higher-order perturbation theory on the
Schwarzschild background spacetime.

%*********************************************************************

To find the gauge-invariant treatments of the $l=0,1$ mode
perturbations on the Schwarzschild background spacetime, we first
reviewed 2+2 formulation in which the decomposition formulae
(\ref{eq:hAB-fourier})--(\ref{eq:hpq-fourier}) with the spherical
harmonic functions $Y_{lm}$ as the scalar harmonic function $S$ and
explained why $l=0,1$ modes should be separately treated in
conventional perturbation theory on the Schwarzschild background
spacetime.
The special treatment in the conventional formulation caused by the
loss of the linear independence of the set
(\ref{eq:harmonic-fucntions-set}) of the tensor harmonic functions on
$S^{2}$, i.e., vector and/or tensor harmonic functions vanishes in
$l=0,1$ modes and does not play a role of the bases of tangent space
on $S^{2}$.

%*********************************************************************

To recover this situation, instead of the spherical harmonics $Y_{00}$
and $Y_{1m}$ for $l=0,1$ modes, we introduce the mode functions
$k_{(\hat{\Delta})}$ and $k_{(\hat{\Delta}+2)m}$, which belongs to the
kernel of the derivative operator $\hat{\Delta}$ and $\hat{\Delta}+2$,
respectively.
We also derive the sufficient condition for which the decomposition
formulae (\ref{eq:hAB-fourier})--(\ref{eq:hpq-fourier}) with the
harmonic function $S=S_{\delta}$ defined by
Eq.~(\ref{eq:harmonic-delta-S-def}) is invertible not only for
$l\geq 2$ modes but also $l=0,1$ modes.
As the result, we showed that the mode functions
(\ref{eq:l=0-general-mode-func-specific-sum})--(\ref{eq:l=1-mode-mpm1-mode-sol-sum})
with the parameter $\delta$ for $l=0,1$ modes satisfy this
sufficient condition.
These mode functions realize the conventional spherical harmonic
functions $Y_{00}$ and $Y_{1m}$ when $\delta=0$.
However, in this case, the set of harmonic functions
(\ref{eq:harmonic-delta-S-def}) loses the linear independence as the
bases of the tangent space on $S^{2}$ as the conventional case,
nevertheless the set $\{Y_{lm}\}$ of the spherical harmonics is a
complete bases set of the $L^{2}$-space of scalar functions on $S^{2}$.
On the other hand, when $\delta\neq 0$, the set of the mode functions
(\ref{eq:harmonic-delta-S-def}) has the linear-independence as the
bases of the tangent space on $S^{2}$.
However, the mode functions $k_{(\hat{\Delta})}$ and
$k_{(\hat{\Delta}+2)m}$ with $\delta\neq 0$ are singular functions.

%*********************************************************************

Due to this situation, we proposed
Proposal~\ref{proposal:treatment-proposal-on-pert-on-spherical-BG} as
a strategy to define the gauge-invariant variables for $l=0,1$ modes
and to derive and solve the linearized Einstein equation.
Following
Proposal~\ref{proposal:treatment-proposal-on-pert-on-spherical-BG}, we
can construct gauge-invariant and gauge-variant variables for linear
metric perturbation through the similar manner to the case of the
$l\geq 2$ modes.
This construction is a proof of
Conjecture~\ref{conjecture:decomposition-conjecture} for the
perturbations on the spherically symmetric background spacetime.
Then, we reach to the statement
Theorem~\ref{theorem:decomposition-theorem-spherical}.
Owing to Theorem~\ref{theorem:decomposition-theorem-spherical}, we can
develop gauge-invariant perturbation theory on spherically symmetric
background spacetimes including $l=0,1$ modes.
Furthermore, Theorem~\ref{theorem:decomposition-theorem-spherical}
yields that we can develop higher-order gauge-invariant perturbation
theory on any spherically symmetric background spacetimes, although
this development is beyond the current scope of this paper.
A brief discussion of this development to higher-order perturbations
was already given in Ref.~\cite{K.Nakamura-2021b}.

%*********************************************************************

Besides the discussion on the extension to the higher-order
perturbation theory, it is also true that we are proposing different
procedure from the conventional one as
Proposal~\ref{proposal:treatment-proposal-on-pert-on-spherical-BG}.
The difference is in the timing of the imposition of the boundary
conditions on the functions on $S^{2}$ to solve the Einstein equations.
In conventional treatments, we restrict the function on $S^{2}$ to the
$L^{2}$-space through the mode decomposition using the spherical
harmonics $Y_{lm}$ from the starting point.
In Proposal~\ref{proposal:treatment-proposal-on-pert-on-spherical-BG}
in this paper, we do not impose the regular boundary condition on the
functions $S^{2}$ at the starting point, but we impose the regular
boundary condition $\delta=0$ after the construction of the
gauge-invariant variables and the derivation of the mode-by-mode
Einstein equations.
Physically, this different timing of the imposition of the boundary
condition should not affect the physical properties of the solution to
the Einstein equations.
Therefore, we have to confirm that the solutions to the Einstein
equation derived by
Proposal~\ref{proposal:treatment-proposal-on-pert-on-spherical-BG} are
physically reasonable.
To check this, we derived the linearized Einstein equations on the
Schwarzschild background spacetime following
Proposal~\ref{proposal:treatment-proposal-on-pert-on-spherical-BG}.
We consider the mode decomposition of the general expression of the
linearized energy-momentum tensor as the source term of the linearized
Einstein equations.
To solve the derived linearized Einstein equations, the linearized
perturbations of the continuity equation of the energy-momentum tensor
should be taken into account.
The metric perturbations on the Schwarzschild spacetime are classified
into the odd-mode and the even-mode perturbations.
In this Part I paper, we concentrate only on the odd-mode
perturbations and derive the $l=0,1$-mode solutions following
Proposal~\ref{proposal:treatment-proposal-on-pert-on-spherical-BG}.

%*********************************************************************

For odd-mode perturbations, we examined the strategy to solve
the linearized Einstein equations for any $l$ modes following the
Proposal~\ref{proposal:treatment-proposal-on-pert-on-spherical-BG},
through we take care of the structure of equations for $l=1$ mode
perturbations.
As well-known, to solve the odd-mode perturbations, Einstein equations
for the $l\geq 2$ odd-mode perturbations are reduced to the
Regge-Wheeler equation.
Furthermore, the solutions to the Regge-Wheeler equation for $l\geq 2$
modes are constructed through the MST formulation~\cite{S.Mano-H.Suzuki-E.Takasugi-1996a,S.Mano-H.Suzuki-E.Takasugi-1996b,S.Mano-E.Takasugi-1997,H.Tagoshi-S.Mano-E.Takasugi-1997}.
Therefore, we concentrated on the $l=0,1$ mode perturbations.

%*********************************************************************

Following
Proposal~\ref{proposal:treatment-proposal-on-pert-on-spherical-BG},
for $l=0$ odd-mode perturbations, we reached to the conclusion that
there is no non-trivial solution to the linearized Einstein
equation as expected.
Then, we carefully examined the solutions to the Einstein equations
for $l=1$ odd-mode perturbations.
We first consider the vacuum solution to the linearized Einstein
equation in which the linear perturbation of the energy-momentum
tensor vanishes.
Then, we obtain the linearized Kerr parameter perturbation with
the term given in the form of the Lie derivative of the background
metric $g_{ab}$.
Through the variation of constant, we derived the general
solutions to non-vacuum linearized Einstein equations for the $l=1$
odd-mode perturbations.
Since we use the constant Kerr parameter in the variation of constant,
we can expect that the obtained general solution describes
the spin-up or the spin-down of the black hole due to the effect of
the linearized energy-momentum tensor.

%*********************************************************************

In addition to the Kerr parameter perturbations, we obtain the
term which has the form of the Lie derivative of the background
metric $g_{ab}$ in our derived solution.
The appearance of such term is natural consequence due to the
symmetry of the linearized Einstein equations as discussed in
Sec.~\ref{sec:The_general-relativistic_gauge-invariant_linear_perturbation_theory}.
Actually, gauge-invariant variables defined through
Conjecture~\ref{conjecture:decomposition-conjecture} is not unique as
pointed out by Eq.~(\ref{eq:gauge-inv-nonunique}) in
Sec.~\ref{sec:review-of-general-framework-GI-perturbation-theroy}.
It is easy to show that new gauge-invariant variable $\ScrH_{ab}$
defined by Eq.~(\ref{eq:gauge-inv-nonunique}) is also a solution to
the linearized Einstein equation (\ref{eq:einstein-equation-gauge-inv})
through Eqs.~(\ref{eq:Gab-Tab-decomp}) and the background Einstein
equation $G_{a}^{\;\;b}$ $=$ $8\pi T_{a}^{\;\;b}$ if the original
gauge-invariant variable $\ScrF_{ab}$ in
Eq.~(\ref{eq:gauge-inv-nonunique}) is a solution to the linearized
Einstein equations (\ref{eq:einstein-equation-gauge-inv}).
This is a diffeomorphism symmetry of the linearized Einstein
equations.

%*********************************************************************

The appearance of the term which has the form of the Lie derivative of
the background metric $g_{ab}$ in the derived solution is a natural
consequence in the sense of the above diffeomorphism symmetry of the
linearized Einstein equation.
In the case where the conventional expansion through the spherical
harmonics $Y_{lm}$ at the starting point and the gauge-fixing method
are used, the appearance of this type of solutions is well-known as
the residual gauge degree of freedom.
It might be able to regard that the term of the Lie derivative of the
background metric $g_{ab}$ in
Eqs.~(\ref{eq:l=1-odd-mode-propagating-sol-ver2}) and
(\ref{eq:l=1-odd-mode-propagating-sol-ver2-2}) corresponds to these
``residual gauge'' solutions.
On the other hand, we are using the gauge-invariant perturbation
theory in which the gauge degree of freedom of the second kind is
completely excluded.
Therefore, the term which has the form of the Lie derivative is
{\it not} the gauge degree of freedom of the second kind.
On the other hand, in our gauge-invariant perturbation theory, we do
not exclude the gauge degree of freedom of the first kind as carefully
explained in Secs.~\ref{sec:The-first-kind-gauge}
and~\ref{sec:The-second-kind-gauge}.
The term of the Lie derivative of the background metric $g_{ab}$
in Eqs.~(\ref{eq:l=1-odd-mode-propagating-sol-ver2}) and
(\ref{eq:l=1-odd-mode-propagating-sol-ver2-2}) appears even if we
completely excluded the gauge degree of freedom of the second kind.
Therefore, we should regard that the term of the Lie derivative of
the background metric $g_{ab}$ in
Eqs.~(\ref{eq:l=1-odd-mode-propagating-sol-ver2}) and
(\ref{eq:l=1-odd-mode-propagating-sol-ver2-2}) as the gauge degree of
freedom of the first kind which is represented in
Eq.~(\ref{eq:infinitesimal-coordinate-trans-first-kind-metric-pert}).
Actually, we can interpret the term of the Lie derivative of the
background metric $g_{ab}$ can be eliminate by the infinitesimal
coordinate transformation on the background spacetime at any time.
As an example, in
Sec.~\ref{sec:Schwarzschild_Background_odd_l=1_vacuum_sol_l=1}, we
explained that the constant $\beta_{1}$ in
the solution (\ref{eq:2calFAp-Xo-Yo-l=1-m=0-special-sol-with-Kerr})
can be regarded as the degree of freedom of the infinitesimal
coordinate transformation by Eq.~(\ref{eq:rigid-rotation}).

%*********************************************************************

\begin{figure}
  \begin{center}
    \includegraphics[width=0.958\textwidth]{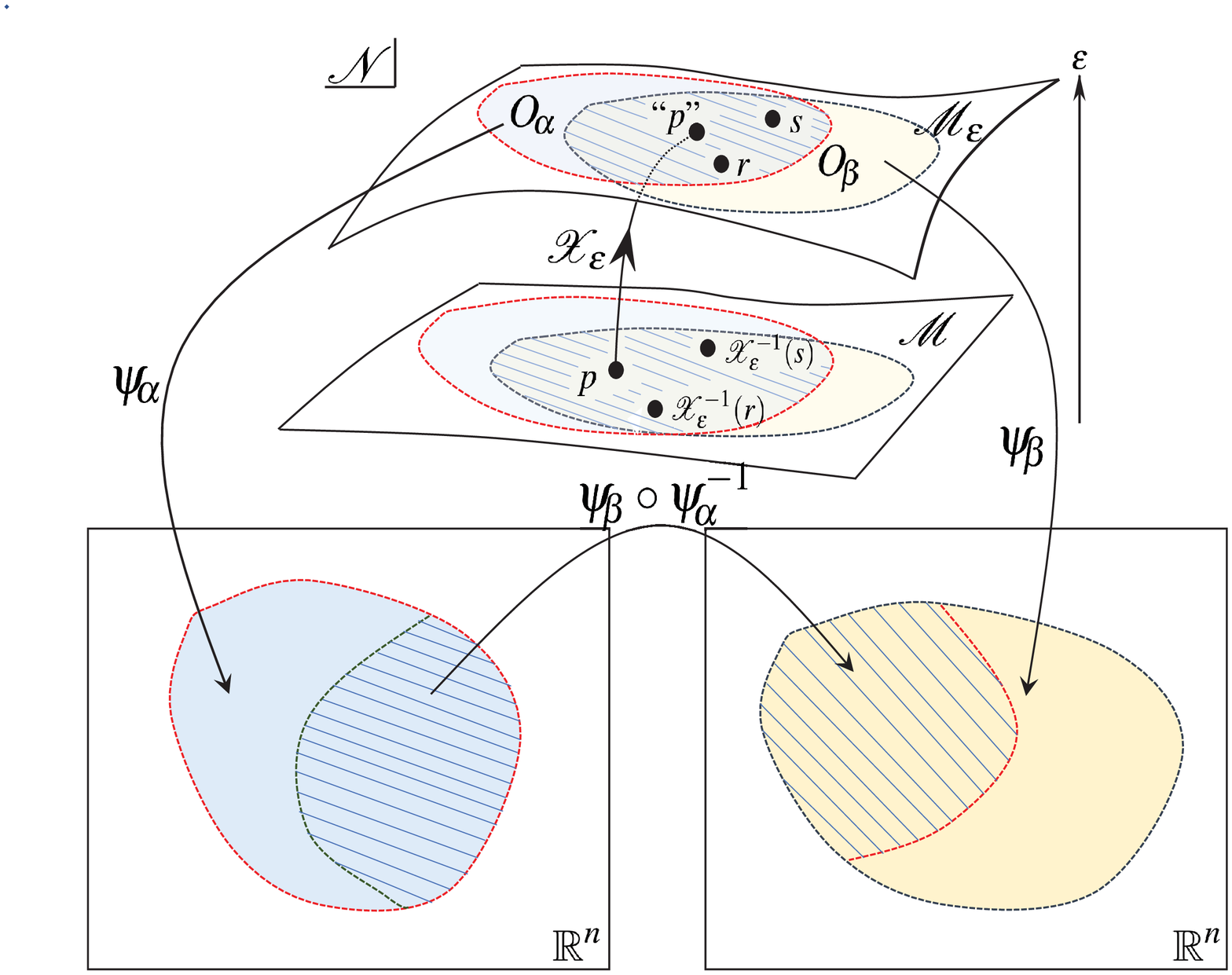}
  \end{center}
  \caption{
    Consider the $n$-dimensional physical manifolds
    $\ScrM_{\epsilon}$ and the background $\ScrM$.
    We may introduce the coordinate transformation on the physical
    spacetime $\ScrM_{\epsilon}$, even if we completely fix the
    second-kind gauge as $\ScrX_{\epsilon}$.
    Actually, we may introduce the diffeomorphism $\psi_{\alpha}$ from
    the open set $O_{\alpha}$ to an open set on $\RF^{n}$ and the
    diffeomorphism $\psi_{\beta}$ from the open set $O_{\beta}$ to an
    open set on the other $\RF^{n}$.
    If $O_{\alpha}\cap O_{\beta}\neq\emptyset$, we can consider the
    coordinate transformation $\psi_{\beta}\circ\psi_{\alpha}^{-1}$ which
    transforms the coordinate system $(O_{\alpha},\psi_{\alpha})$ to
    $(O_{\beta},\psi_{\beta})$.
    If we choose the gauge-choice of the second-kind by
    $\ScrX_{\epsilon}$, this gauge-choice induce the coordinate
    systems
    $\{\ScrX_{\epsilon}^{-1}O_{\alpha},\psi_{\alpha}\circ\ScrX_{\epsilon}\}$
    and
    $\{\ScrX_{\epsilon}^{-1}O_{\beta},\psi_{\beta}\circ\ScrX_{\epsilon}\}$
    on $\ScrM$.
    Furthermore, the coordinate transformation is given by
    $(\psi_{\beta}\circ\ScrX_{\epsilon})\circ(\psi_{\alpha}\circ\ScrX_{\epsilon})^{-1}$
    $=$ $\psi_{\beta}\circ\psi_{\alpha}^{-1}$.
  }
  \label{fig:residual-is-first-kind-gauge}
\end{figure}

%*********************************************************************

Now, we confirm the geometrical meaning of the gauge degree of freedom
of the first kind in the context of the perturbation theory through
Fig.~\ref{fig:residual-is-first-kind-gauge}.
Here, we consider the $n$-dimensional physical manifolds
$\ScrM_{\epsilon}$ and the background manifold $\ScrM$.
As depicted in Fig.~\ref{fig:residual-is-first-kind-gauge}, we show
that we may introduce the coordinate transformation on the physical
spacetime $\ScrM_{\epsilon}$, even if we completely fix the
second-kind gauge as $\ScrX_{\epsilon}$.
Actually, we may introduce the diffeomorphism $\psi_{\alpha}$, i.e.,
a coordinate system on $O_{\alpha}\subset\ScrM_{\epsilon}$, from
the open set $O_{\alpha}$ to an open set on $\RF^{n}$ and the
diffeomorphism $\psi_{\beta}$, i.e., a coordinate system on
$O_{\beta}\subset\ScrM_{\epsilon}$,  from the open set $O_{\beta}$
to an open set on the other $\RF^{n}$.
If $O_{\alpha}\cap O_{\beta}\neq\emptyset$, we can consider the
coordinate transformation $\psi_{\beta}\circ\psi_{\alpha}^{-1}$ which
transforms the coordinate system $(O_{\alpha},\psi_{\alpha})$ to
$(O_{\beta},\psi_{\beta})$.
This is the first-kind gauge on $\ScrM_{\epsilon}$ as shown in
Fig.~\ref{fig:first-kind-gauge}.
If we choose the gauge-choice of the second-kind by
$\ScrX_{\epsilon}$ as depicted in
Fig.~\ref{fig:residual-is-first-kind-gauge}, this gauge-choice induce
the diffeomorphisms
$\ScrX_{\epsilon}^{-1}:O_{\alpha}\rightarrow
\ScrX_{\epsilon}^{-1}O_{\alpha}\subset\ScrM$
and
$\ScrX_{\epsilon}^{-1}:O_{\beta}\rightarrow
\ScrX_{\epsilon}^{-1}O_{\beta}\subset\ScrM$.
Then, the coordinate systems $(O_{\alpha},\psi_{\alpha})$ and
$(O_{\beta},\psi_{\beta})$ on $\ScrM_{\epsilon}$ induce the
coordinate systems
$\{\ScrX_{\epsilon}^{-1}O_{\alpha},\psi_{\alpha}\circ\ScrX_{\epsilon}\}$
and
$\{\ScrX_{\epsilon}^{-1}O_{\beta},\psi_{\beta}\circ\ScrX_{\epsilon}\}$
on $\ScrM$.
Actually, $\psi_{\alpha}\circ\ScrX_{\epsilon}$ is a diffeomorphism
which maps from
$\ScrX_{\epsilon}^{-1}O_{\alpha}\subset\ScrM$ to $\RF^{n}$
and $\psi_{\beta}\circ\ScrX_{\epsilon}$ is a diffeomorphism
which maps from
$\ScrX_{\epsilon}^{-1}O_{\beta}\subset\ScrM$ to $\RF^{n}$.
Furthermore, the coordinate transformation is given by
$(\psi_{\beta}\circ\ScrX_{\epsilon})\circ(\psi_{\alpha}\circ\ScrX_{\epsilon})^{-1}$
$=$ $\psi_{\beta}\circ\ScrX_{\epsilon}\circ\ScrX_{\epsilon}^{-1}\circ\psi_{\alpha}^{-1}$
$=$ $\psi_{\beta}\circ\psi_{\alpha}^{-1}$.
Thus indicates that the first-kind gauge transformation on the
physical spacetime $\ScrM_{\epsilon}$ coincides with that on the
background spacetime $\ScrM$.
Thus, even if we fix the gauge choice $\ScrX_{\epsilon}$ of the
second kind, the gauge degree of freedom of the first kind on the
background spacetime $\ScrM$ is induced by the gauge degree of
freedom of the first kind on the physical spacetime
$\ScrM_{\epsilon}$.
This induced gauge degree of freedom of the first-kind entirely
depends entirely on the gauge choice $\ScrX_{\epsilon}$.
Actually, the gauge choice $\psi_{\alpha}\circ\ScrX_{\epsilon}$ of
the first kind does depend on the gauge choice $\ScrX_{\epsilon}$
of the second kind.
However, the first-kind gauge transformation rule
$(\psi_{\beta}\circ\ScrX_{\epsilon})\circ(\psi_{\alpha}\circ\ScrX_{\epsilon})^{-1}$
$=$ $\psi_{\beta}\circ\psi_{\alpha}^{-1}$ is independent of the gauge
choice $\ScrX_{\epsilon}$ of the second kind.

%*********************************************************************

The above geometrical arguments indicates that even if we completely
exclude the gauge-degree of freedom of the second kind, the
gauge-degree of freedom of the first kind still remains.
This situation support the existence of the term of the Lie derivative
of the background metric $g_{ab}$ in the solution
(\ref{eq:l=1-odd-mode-propagating-sol-ver2-2}) of the linear metric
perturbation.
Actually, we may consider the point replacement $s=\Psi_{\lambda}(r)$
as Eq.~(\ref{eq:psibeta-Psizeta-psialphainv}) on the physical
spacetime $\ScrM_{ph}=\ScrM_{\epsilon}$.
If we express the point replacement $\Psi_{\lambda}$ through the point
identification $\ScrX_{\epsilon}$ to the background spacetime
$\ScrM$, the diffeomorphism $\Psi_{\lambda}$ should be regarded as
$\ScrX_{\epsilon}^{-1}(s)$ $=$
$\ScrX_{\epsilon}^{-1}\circ\Psi_{\lambda}\circ\ScrX_{\epsilon}(\ScrX_{\epsilon}^{-1}(r))$.
This point replacement
$\ScrX_{\epsilon}^{-1}\circ\Psi_{\lambda}\circ\ScrX_{\epsilon}$
$:$ $\ScrX_{\epsilon}^{-1}(r)$ $\mapsto$
$\ScrX_{\epsilon}^{-1}(s)$ on the background spacetime $\ScrM$
is completely depends on the second-kind gauge choice
$\ScrX_{\epsilon}$.
However, if we use the coordinate systems
$\{\ScrX_{\epsilon}^{-1}O_{\alpha},\psi_{\alpha}\circ\ScrX_{\epsilon}\}$
and
$\{\ScrX_{\epsilon}^{-1}O_{\beta},\psi_{\beta}\circ\ScrX_{\epsilon}\}$
on the background spacetime $\ScrM$, which are induced from the
coordinate system on physical spacetime $\ScrM_{\epsilon}$, the
action (\ref{eq:psibeta-Psizeta-psialphainv}) of the diffeomorphism is
given by
\begin{eqnarray}
  &&
     \left(
     \psi_{\beta}
     \circ\ScrX_{\epsilon}
     \right)
     \circ\ScrX_{\epsilon}^{-1}
     \circ\Psi_{\lambda}
     \circ\ScrX_{\epsilon}
     \circ
     \left(
     \psi_{\alpha}\circ\ScrX_{\epsilon}
     \right)^{-1}
     \nonumber\\
  &=&
      \psi_{\beta}
      \circ\ScrX_{\epsilon}
      \circ\ScrX_{\epsilon}^{-1}
      \circ\Psi_{\lambda}
      \circ\ScrX_{\epsilon}
      \circ
      \ScrX_{\epsilon}^{-1}\circ\psi_{\alpha}^{-1}
      \nonumber\\
  &=&
      \psi_{\beta}\circ\Psi_{\lambda}\circ\psi_{\alpha}^{-1}
      .
      \label{eq:psibeta-Psizeta-psialphainv-calX}
\end{eqnarray}
This is just the ``coordinate transformation''
(\ref{eq:psibeta-Psizeta-psialphainv}) and does not depend on the
gauge choice $\ScrX_{\epsilon}$ of the second-kind, i.e., is the
gauge-invariant in the sense of the second-kind.
Therefore, the coordinate transformation
(\ref{eq:psibeta-Psizeta-psialphainv-calX}) may be regarded as
the representation of the coordinate transformation
(\ref{eq:psibeta-Psizeta-psialphainv}), i.e., the replacement of
points $r$ $\mapsto$ $s$ on the physical spacetime
$\ScrM_{\epsilon}$.

%*********************************************************************

The solution (\ref{eq:l=1-odd-mode-propagating-sol-ver2-2}) is gauge
invariant in the sense of the second kind, i.e., the degree of freedom
of the point-identifications between the physical spacetime
$\ScrM_{\epsilon}$ and the background spacetime $\ScrM$ is
completely excluded.
However, in this gauge-invariant solutions in the sense of the second
kind, there still exists the term ${\pounds}_{V}g_{ab}$.
As noted in
Sec.~\ref{sec:The_general-relativistic_gauge-invariant_linear_perturbation_theory},
such terms may be included in the solution to the linearized Einstein
equation due to the symmetry of the linearized Einstein equation as
the gauge-invariant terms in the sense of the second-kind.
Therefore, the term ${\pounds}_{V}g_{ab}$ in
Eq.~(\ref{eq:l=1-odd-mode-propagating-sol-ver2-2}) is no longer
regarded as the gauge degree of the second kind, but we should regard
this term as the gauge degree of freedom of the first kind as
discussed above.
Actually, the coordinate transformation (\ref{eq:rigid-rotation})
should be regarded as the ``coordinate transformation''
(\ref{eq:psibeta-Psizeta-psialphainv-calX}), because $\beta_{1}$ is
gauge invariant in the sense of the second-kind.
Furthermore, we note that the infinitesimal ``coordinate
transformation'' which eliminate the term ${\pounds}_{V}g_{ab}$ in the
solution (\ref{eq:l=1-odd-mode-propagating-sol-ver2-2}) should be
regarded as the ``coordinate transformation''
(\ref{eq:psibeta-Psizeta-psialphainv-calX}) due to the same reason.
As explained in Sec.~\ref{sec:The-first-kind-gauge}, the coordinate
transformation (\ref{eq:psibeta-Psizeta-psialphainv}) is regarded as
the first-kind gauge degree of freedom.
Then, the term ${\pounds}_{V}g_{ab}$ in the
solution (\ref{eq:l=1-odd-mode-propagating-sol-ver2-2}) should be
regarded as the degree of freedom of the first-kind gauge.
As pointed out in Sec.~\ref{sec:The-first-kind-gauge}, the first kind
gauge is often used to predict or to interpret the measurement results
in observations and experiments.
In this sense, this term of the Lie derivative of the background
metric $g_{ab}$ in the solution
(\ref{eq:l=1-odd-mode-propagating-sol-ver2-2}) should have their
physical meaning.
This is the reason why we emphasized the importance of the distinction
of the notions of the first-kind gauge and the second-kind gauge.

%*********************************************************************

We have to emphasize that this conclusion is the consequence of our
complete exclusion of the second-kind gauge degree of freedom which
includes not only $l\geq 2$ modes but also $l=0,1$ modes of
perturbations and our
proposal~\ref{proposal:treatment-proposal-on-pert-on-spherical-BG}.
From the view point of the gauge-invariant perturbation theory
developed in this paper, the conventional gauge-fixing procedure
corresponds to the partial gauge-fixing.
Therefore, it will be difficult to reach the above conclusion through
the conventional gauge-fixing procedure.
Furthermore, in conventional approach, there is no distinction between
the first- and the second-kind gauge and all terms which have the form
${\pounds}_{V}g_{ab}$ may be regarded as the ``gauge-degree of
freedom'' and these are ``unphysical degree of freedom'' because we
can always eliminate these terms through the infinitesimal coordinate
transformation.
If the concept of ``the complete gauge fixing'' corresponds to the
standing point that all terms which have the form
${\pounds}_{V}g_{ab}$ are ``unphysical degree of freedom'', this
concept of ``the complete gauge-fixing'' is stronger restriction of
the metric perturbation than the concept of ``gauge-invariant of the
second kind'' in this paper.
Thus, we may say that these conceptual discussion is an important
result comes from the realization of the gauge-invariant formulation
including $l=0,1$ modes in this paper.
Similar results are also obtained in even-mode perturbations which
will be shown in the Part II paper~\cite{K.Nakamura-2021c}.

%*********************************************************************

Apart from these terms of the Lie derivative of the background
metric $g_{ab}$, in vacuum case, the only non-trivial solution in
$l=1$ odd-mode perturbation is the Kerr parameter perturbations.
This will be related to the uniqueness of the Kerr solution in the
vacuum Einstein equations in the local
sense~\cite{M.Kimura-T.Harada-A.Naruko-K.Toma-2021}, though the
assertion of the uniqueness theorem of the Kerr solution includes
topological statement.
Besides the relation of the uniqueness theorem of Kerr black hole, at
least, we may say that the derived vacuum solution for $l=0,1$
odd-mode perturbations is physically reasonable.
In the paper~\cite{K.Nakamura-2021c}, we derive the $l=0,1$ even-mode
solution to the linearized Einstein equation which also includes the terms of
the Lie derivative of the background metric.
In the Part III paper~\cite{K.Nakamura-2021d}, we show that the
derived solutions in Ref.~\cite{K.Nakamura-2021c} realize the
linearized Lema\^itre-Tolman-Bondi solution and the linearized
non-rotating C-metric.
Due to these facts, we may say that our solutions derived through
Proposal~\ref{proposal:treatment-proposal-on-pert-on-spherical-BG} are
physically reasonable.
In this sense, we may say that
Proposal~\ref{proposal:treatment-proposal-on-pert-on-spherical-BG}
is also physically reasonable.

%*********************************************************************

%%%%%%%%%%%%%%%%%%%%%%%%%%%%%%%%%%%%%%%%%%%
%%%%%%%%%%%%%%%%%%%%%%%%%%%%%%%%%%%%%%%%%%%
%%%%%%%%%%%%%%%%%%%%%%%%%%%%%%%%%%%%%%%%%%%
\section*{Acknowledgements}
%%%%%%%%%%%%%%%%%%%%%%%%%%%%%%%%%%%%%%%%%%%
%%%%%%%%%%%%%%%%%%%%%%%%%%%%%%%%%%%%%%%%%%%
%%%%%%%%%%%%%%%%%%%%%%%%%%%%%%%%%%%%%%%%%%%

%*********************************************************************

The author deeply acknowledged to Professor Hiroyuki Nakano for
various discussions and suggestions.
The author also thanks to Professor Takahiro Tanaka for valuable
discussions.
Finally, he deeply thanks for anonymous referees of Progress of
Theoretical and Experimental Physics for the comments and questions
which were helpful for the improvement of this manuscript.

%*********************************************************************

%%%%%%%%%%%%%%%%%%%%%%%%%%%%%%%%%%%%%%%%%%%
%%%%%%%%%%%%%%%%%%%%%%%%%%%%%%%%%%%%%%%%%%%
%%%%%%%%%%%%%%%%%%%%%%%%%%%%%%%%%%%%%%%%%%%
\appendix
%%%%%%%%%%%%%%%%%%%%%%%%%%%%%%%%%%%%%%%%%%%
%%%%%%%%%%%%%%%%%%%%%%%%%%%%%%%%%%%%%%%%%%%
%%%%%%%%%%%%%%%%%%%%%%%%%%%%%%%%%%%%%%%%%%%

%*********************************************************************

%%%%%%%%%%%%%%%%%%%%%%%%%%%%%%%%%%%%%%%%%%%
%%%%%%%%%%%%%%%%%%%%%%%%%%%%%%%%%%%%%%%%%%%
%%%%%%%%%%%%%%%%%%%%%%%%%%%%%%%%%%%%%%%%%%%
\section*{Appendix}
%%%%%%%%%%%%%%%%%%%%%%%%%%%%%%%%%%%%%%%%%%%
%%%%%%%%%%%%%%%%%%%%%%%%%%%%%%%%%%%%%%%%%%%
%%%%%%%%%%%%%%%%%%%%%%%%%%%%%%%%%%%%%%%%%%%

%*********************************************************************

%%%%%%%%%%%%%%%%%%%%%%%%%%%%%%%%%%%%%%%%%%%
%%%%%%%%%%%%%%%%%%%%%%%%%%%%%%%%%%%%%%%%%%%
%%%%%%%%%%%%%%%%%%%%%%%%%%%%%%%%%%%%%%%%%%%
\section{Explicit form of conventional spherical harmonics on $S^{2}$}
\label{sec:Explict_form_of_Tlm_for_Ylm}
%%%%%%%%%%%%%%%%%%%%%%%%%%%%%%%%%%%%%%%%%%%
%%%%%%%%%%%%%%%%%%%%%%%%%%%%%%%%%%%%%%%%%%%
%%%%%%%%%%%%%%%%%%%%%%%%%%%%%%%%%%%%%%%%%%%

%*********************************************************************

First, we summarize the properties of the conventional spherical
harmonic functions $Y_{lm}$.
The spherical harmonic functions $Y_{lm}(\theta,\phi)$ satisfy the equations
\begin{eqnarray}
  \label{eq:H.Kodama-H.Ishihara-Y.Fujiwara-1994-A4}
  && \left[\hat{\Delta} + l(l+1)\right]Y_{lm} = 0, \\
  \label{eq:H.Kodama-H.Ishihara-Y.Fujiwara-1994-A5}
  && \partial_{\phi}Y_{lm} = imY_{lm}.
\end{eqnarray}
To be explicit, they are expressed in terms of the Legendre functions
as
\begin{eqnarray}
  \label{eq:H.Kodama-H.Ishihara-Y.Fujiwara-1994-A6}
  Y_{lm}(\theta,\phi)
  =
  \sqrt{\frac{(2l+1)(l-m)!}{4\pi(l+m)!}} P_{l}^{m}(\cos\theta) e^{im\phi}.
\end{eqnarray}
For $l=0,1$ modes, the spherical harmonic functions $Y_{lm}=Y_{l,m}$ are
explicitly given by
\begin{eqnarray}
  \label{eq:Y00-explicit}
  Y_{00}
  &=&
  \sqrt{\frac{1}{4\pi}}, \\
  \label{eq:Y1m-explicit}
  Y_{10}
  &=&
  \sqrt{\frac{3}{4\pi}} \cos\theta
  ,
  \quad
  Y_{11}
  =
  \sqrt{\frac{3}{8\pi}} \sin\theta e^{i\phi}
  ,
  \quad
  Y_{1-1}
  =
  - \sqrt{\frac{3}{8\pi}} \sin\theta e^{-i\phi}
  .
\end{eqnarray}
Employing these spherical harmonic functions
(\ref{eq:H.Kodama-H.Ishihara-Y.Fujiwara-1994-A6}) as the scalar
harmonics, we construct the set of the tensor harmonics on $S^{2}$.
Since the dimension of $S^{2}$ is two, we have enough number of
tensor harmonic functions as bases of tangent space on $S^{2}$.

%*********************************************************************

On the unit sphere any vector field $v^{p}$ is written in terms of two
scalar functions $v$ and $w$ as
\begin{eqnarray}
  \label{eq:H.Kodama-H.Ishihara-Y.Fujiwara-1994-A8}
  v^{p} = \hat{D}^{p}v + \epsilon^{pq}\hat{D}_{q}w.
\end{eqnarray}
Here, $\hat{D}^{p}v$ is even part and $\epsilon^{pq}\hat{D}_{q}w$ is
the odd part, which corresponds to $\hat{D}_{p}S$ and
$\epsilon_{pq}\hat{D}^{q}S$ in Eq.~(\ref{eq:hAp-fourier}),
respectively.
If we choose $S=Y_{lm}$, these vectors are given by
\begin{eqnarray}
  \label{eq:even-vector-odd-vectors-lm}
  \hat{D}_{p}Y_{lm}, \quad \epsilon_{pq}\hat{D}^{q}Y_{lm}.
\end{eqnarray}
For $l=0$ modes, the spherical harmonic function $Y_{00}$ is constant
as in Eq.~(\ref{eq:Y00-explicit}) and corresponding vector harmonics
vanish:
\begin{eqnarray}
  \label{eq:even-vector-odd-vectors-00}
  \hat{D}_{p}Y_{00} = 0, \quad \epsilon_{pq}\hat{D}^{q}Y_{00} = 0.
\end{eqnarray}
On the other hand, for $l=1$ modes, vector harmonics has the vector
values as
\begin{eqnarray}
  \label{eq:even-vector-l=1-m=0-explicit}
  \hat{D}_{p}Y_{10}
  &=&
      - \sqrt{\frac{3}{4\pi}} \sin\theta \theta_{p}
      ,
  \\
  \label{eq:even-vector-l=1-m=1-1-explicit}
  \hat{D}_{p}Y_{11}
  &=&
      \sqrt{\frac{3}{8\pi}}
      e^{i\phi}
      \left(
      \cos\theta \theta_{p}
      +
      i \phi_{p}
      \right)
      ,
      \quad
      \hat{D}_{p}Y_{1-1}
      =
      \sqrt{\frac{3}{8\pi}}
      e^{-i\phi}
      \left(
      - \cos\theta \theta_{p}
      + i \phi_{p}
      \right)
\end{eqnarray}
and
\begin{eqnarray}
  \label{eq:odd-vector-l=1-m=0-explicit}
  \epsilon_{pq}\hat{D}^{q}Y_{10}
  &=&
      \sqrt{\frac{3}{4\pi}} \sin\theta \phi_{p}
      ,
  \\
  \label{eq:odd-vector-l=1-m=1-1-explicit}
  \epsilon_{pq}\hat{D}^{q}Y_{11}
  &=&
      \sqrt{\frac{3}{8\pi}}
      e^{i\phi}
      \left(
      -  \cos\theta \phi_{p}
      + i \theta_{p}
      \right)
      ,
      \quad
      \epsilon_{pq}\hat{D}^{q}Y_{1-1}
      =
      \sqrt{\frac{3}{8\pi}}
      e^{-i\phi}
      \left(
      \cos\theta \phi_{p}
      + i \theta_{p}
      \right)
      .
\end{eqnarray}
Thus, vector harmonics has its vector value for $l=1$ modes, while
does not for $l=0$ mode.

%*********************************************************************

Any smooth symmetric second-rank tensor field $t^{pq}$ on the unit
sphere can be expressed in terms of its trace $t=t_{p}^{p}$ and two
scalar fields $v$ and $w$ as
\begin{eqnarray}
  \label{eq:H.Kodama-H.Ishihara-Y.Fujiwara-1994-A16}
  t^{pq}
  =
  \frac{1}{2}t \gamma^{pq}
  +
  \left(
  \hat{D}^{p}\hat{D}^{q} - \frac{1}{2}\gamma^{pq}\hat{\Delta}
  \right)v
  +
  2 \epsilon^{r(q}\hat{D}^{p)}D_{r}w
  .
\end{eqnarray}
These three terms correspond to the terms proportional to
$\displaystyle \frac{1}{2}\gamma_{pq}S$, $\displaystyle
\left(\hat{D}_{p}\hat{D}_{q}-\frac{1}{2}\gamma^{pq}\hat{\Delta}\right)S$,
and $2\epsilon_{r(p}\hat{D}_{q)}\hat{D}^{r}S$ in
Eq.~(\ref{eq:hpq-fourier}).
As in the case of vector harmonics above, for $l=0$ modes, the
spherical harmonic function $Y_{00}$ is constant as in
Eq.~(\ref{eq:Y00-explicit}) and the only non-vanishing harmonics is
its trace part
\begin{eqnarray}
  \label{eq:even-trace-tensor-harmonics}
  \frac{1}{2}\gamma_{pq}Y_{00} = \frac{1}{2} \gamma_{pq} \sqrt{\frac{1}{4\pi}}
\end{eqnarray}
and the other traceless even and odd parts vanish.
For $l=1$ modes, from Eqs.~(\ref{eq:Y1m-explicit}), the trace parts
are trivially given by
\begin{eqnarray}
  \label{eq:tensor-trace-Y10-explicit}
  \frac{1}{2} \gamma_{pq} Y_{10}
  &=&
      \frac{1}{2} \sqrt{\frac{3}{4\pi}} \cos\theta \gamma_{pq}
      ,
  \\
  \label{eq:tensor-trace-Y11-explicit}
  \frac{1}{2} \gamma_{pq} Y_{11}
  &=&
      \frac{1}{2} \sqrt{\frac{3}{8\pi}} \sin\theta e^{i\phi} \gamma_{pq}
      ,
  \\
  \label{eq:tensor-trace-Y1-1-explicit}
  \frac{1}{2} \gamma_{pq} Y_{1-1}
  &=&
      - \frac{1}{2} \sqrt{\frac{3}{8\pi}} \sin\theta e^{-i\phi} \gamma_{pq}
      .
\end{eqnarray}
On the other hand, the traceless even and odd parts for $\displaystyle
\left(\hat{D}_{p}\hat{D}_{q}-\frac{1}{2}\gamma^{pq}\hat{\Delta}\right)Y_{1m}$,
and $2\epsilon_{r(p}\hat{D}_{q)}\hat{D}^{r}Y_{1m}$ identically vanish
for all $m=-1,0,1$.

%*********************************************************************

As a summary of $S=Y_{lm}$ cases, for $l=0$ mode, any vector and
tensor harmonics does not have their values, and these do not play
roles of bases of the tangent space on $S^{2}$.
On the other hand, for $l=1$ modes, the vector harmonics have their
vector value and play roles of bases of the tangent space on $S^{2}$.
The trace parts of the second-rank tensor of each modes have their
tensor values, while all traceless even and odd mode harmonics
identically vanish and does not play roles of bases of the tangent
space on $S^{2}$.

%*********************************************************************

%%%%%%%%%%%%%%%%%%%%%%%%%%%%%%%%%%%%%%%%%%%
%%%%%%%%%%%%%%%%%%%%%%%%%%%%%%%%%%%%%%%%%%%
%%%%%%%%%%%%%%%%%%%%%%%%%%%%%%%%%%%%%%%%%%%
\section{Covariant derivatives in 2+2 formulation and background curvatures}
\label{sec:Cov_deriv_2+2-formulation}
%%%%%%%%%%%%%%%%%%%%%%%%%%%%%%%%%%%%%%%%%%%
%%%%%%%%%%%%%%%%%%%%%%%%%%%%%%%%%%%%%%%%%%%
%%%%%%%%%%%%%%%%%%%%%%%%%%%%%%%%%%%%%%%%%%%

%*********************************************************************

In this Appendix, we summarize the relation between the covariant
derivatives $\nabla_{a}$ associated with the metric $g_{ab}$,
$\bar{D}_{A}$ associated with the metric $y_{ab}$, and $\hat{D}_{p}$
associated with the metric $\gamma_{ab}$.
These formulae are convenient to derive the gauge-transformation
rules, linearized Einstein equations, and so on.
Here, the metrices $g_{ab}$, $y_{ab}$, and $\gamma_{ab}$ are given by
Eq.~(\ref{eq:background-metric-2+2}).
We assume that $y_{ab}$ depends on $\{x^{A}\}$ and $r=r(x^{A})$.
We also assume that $\gamma_{ab}$ depends only on $\{x^{p}\}$.
Under these assumptions, the Christoffel symbol
$\Gamma_{ab}^{\;\;\;\;\;c}$ are given by
\begin{eqnarray}
  \Gamma_{ab}^{\;\;\;\;\;c}
  &=&
  \frac{1}{2} g^{cd} \left(
    \partial_{a}g_{db}
    +
    \partial_{b}g_{da}
    -
    \partial_{d}g_{ab}
  \right)
  ,
\end{eqnarray}
\begin{eqnarray}
  \label{eq:Background-connection-ABC}
  \Gamma_{AB}^{\;\;\;\;\;\;C}
  &=&
  \frac{1}{2} y^{CD} \left(
    \partial_{A}y_{DB}
    +
    \partial_{B}y_{DA}
    -
    \partial_{D}y_{AB}
  \right)
  =:
  \bar{\Gamma}_{AB}^{\;\;\;\;\;\;C}
  , \\
  \label{eq:Background-connection-pBC}
  \Gamma_{pB}^{\;\;\;\;\;\;C}
  &=&
  0
  , \\
  \label{eq:Background-connection-pqC}
  \Gamma_{pq}^{\;\;\;\;\;\;C}
  &=&
  -
  r (\bar{D}^{C}r) \gamma_{pq}
  , \\
  \label{eq:Background-connection-ABp}
  \Gamma_{AB}^{\;\;\;\;\;\;p}
  &=&
  0
  , \\
  \label{eq:Background-connection-qAp}
  \Gamma_{qA}^{\;\;\;\;\;p}
  &=&
  \frac{1}{r} (\bar{D}_{A}r) \gamma_{q}^{\;\;p}
  , \\
  \label{eq:Background-connection-qrp}
  \Gamma_{qr}^{\;\;\;\;p}
  &=&
  \frac{1}{2} \gamma^{pd} \left(
    \partial_{q}\gamma_{dr}
    +
    \partial_{r}\gamma_{dq}
    -
    \partial_{d}\gamma_{qr}
  \right)
  =:
  \hat{\Gamma}_{qr}^{\;\;\;\;p}
  .
\end{eqnarray}

%*********************************************************************

Here, we note that
\begin{eqnarray}
  \hat{D}_{p}\bar{D}_{A}t_{B}
  &=&
      \bar{D}_{A}\hat{D}_{p}t_{B}
      ,
\end{eqnarray}
and
\begin{eqnarray}
  \hat{D}_{p}\bar{D}_{A}t_{q}
  &=&
      \partial_{p}\bar{D}_{A}t_{q} - \hat{\Gamma}_{qp}^{\;\;\;\;\;\;r}\bar{D}_{A}t_{q}
      =
      \bar{D}_{A}\hat{D}_{p}t_{q}
      ,
\end{eqnarray}
since
\begin{eqnarray}
  \partial_{p}\bar{\Gamma}_{AB}^{\;\;\;\;\;\;C} = 0,
  \quad
  \partial_{A}\hat{\Gamma}_{pq}^{\;\;\;\;\;\;r} = 0.
\end{eqnarray}

%*********************************************************************

Then, we obtain the formulae for the covariant derivatives
$\nabla_{a}v_{b}$ and $\nabla_{a}t^{b}$ as
\begin{eqnarray}
  \nabla_{A}v_{B}
  &=&
  \bar{D}_{A}v_{B}
  , \\
  \nabla_{A}v_{p}
  &=&
  \bar{D}_{A}v_{p}
  -
  \frac{1}{r} \bar{D}_{A}r v_{p}
  , \\
  \nabla_{p}v_{A}
  &=&
  \hat{D}_{p}v_{A}
  -
  \frac{1}{r} \bar{D}_{A}r v_{p}
  , \\
  \nabla_{p}v_{q}
  &=&
  \hat{D}_{p}v_{q}
  +
  r \bar{D}^{A}r \gamma_{pq} v_{A}
  , \\
  \nabla_{A}t^{B}
  &=&
  \bar{D}_{A}t^{B}
  , \\
  \nabla_{A}t^{p}
  &=&
  \partial_{A}t^{p}
  +
  \frac{1}{r} \bar{D}_{A}r t^{p}
  , \\
  \nabla_{p}t^{A}
  &=&
  \hat{D}_{p}t^{A}
  -
  r \bar{D}^{A}r \gamma_{pq} t^{q}
  , \\
  \nabla_{p}t^{q}
  &=&
  \hat{D}_{p}t^{q}
  +
  \frac{1}{r} \bar{D}_{A}r \gamma_{p}^{\;\;q} t^{A}
  .
\end{eqnarray}

%*********************************************************************

Here, we also summarize the expression of $\nabla_{a}T_{bc}$ for an
arbitrary tensor $T_{bc}$ in terms of the covariant derivatives
$\bar{D}_{A}$ and $\hat{D}_{p}$ which are associated with the metric
$y_{AB}$ and $\gamma_{pq}$, respectively, from
\begin{eqnarray}
  \nabla_{a}T_{bc}
  &=&
      \partial_{a}T_{bc}
      -
      \Gamma_{ba}^{\;\;\;\;\;d}T_{dc}
      -
      \Gamma_{ca}^{\;\;\;\;\;d}T_{bd}
      .
\end{eqnarray}
These are given by
\begin{eqnarray}
  \nabla_{A}T_{BC}
  &=&
      \bar{D}_{A}T_{BC}
      ,
  \\
  \nabla_{A}T_{Bp}
  &=&
      \bar{D}_{A}T_{Bp}
      -
      \frac{1}{r} \bar{D}_{A}r T_{Bp}
      ,
  \\
  \nabla_{A}T_{pC}
  &=&
      \bar{D}_{A}T_{pC}
      -
      \frac{1}{r} \bar{D}_{A}r T_{pC}
      ,
  \\
  \nabla_{p}T_{BC}
  &=&
      \hat{D}_{p}T_{BC}
      -
      \frac{1}{r} \bar{D}_{B}r T_{pC}
      -
      \frac{1}{r} \bar{D}_{C}r T_{Bp}
      ,
  \\
  \nabla_{p}T_{qC}
  &=&
      \hat{D}_{p}T_{qC}
      +
      r \bar{D}^{D}r \gamma_{qp} T_{DC}
      -
      \frac{1}{r} \bar{D}_{C}r T_{qp}
      ,
  \\
  \nabla_{p}T_{Bq}
  &=&
      \hat{D}_{p}T_{Bq}
      -
      \frac{1}{r} \bar{D}_{B}r T_{pq}
      +
      r \bar{D}^{D}r \gamma_{qp} T_{BD}
      ,
  \\
  \nabla_{A}T_{pq}
  &=&
      \bar{D}_{A}T_{pq}
      -
      \frac{2}{r} \bar{D}_{A}r T_{pq}
      ,
  \\
  \nabla_{p}T_{qr}
  &=&
      \hat{D}_{p}T_{qr}
      +
      r \bar{D}^{D}r \gamma_{qp} T_{Dr}
      +
      r \bar{D}^{D}r \gamma_{rp} T_{qD}
      .
\end{eqnarray}

%*********************************************************************

Furthermore, the derive the linearized Einstein equation, we have to
derive the components of
\begin{eqnarray}
  \label{eq:first-nablaaHdbd-antisym-calF}
  \nabla_{a}^{}H_{c}^{\;\;\;bd}
  &=&
      \partial_{a}^{}H_{c}^{\;\;\;bd}
      -
      \Gamma_{ca}^{\;\;\;\;\;e}H_{e}^{\;\;\;bd}
      +
      \Gamma_{ea}^{\;\;\;\;\;b}H_{c}^{\;\;\;ed}
      +
      \Gamma_{ea}^{\;\;\;\;\;d}H_{c}^{\;\;\;be}
      .
\end{eqnarray}
Then, these are summarized as
\begin{eqnarray}
  &&
     \nabla_{A}^{}H_{C}^{\;\;\;BD}
     =
     \bar{D}_{A}^{}H_{C}^{\;\;\;BD}
     ,
  \\
  &&
     \nabla_{A}^{}H_{C}^{\;\;\;Bs}
     =
     \bar{D}_{A}^{}H_{C}^{\;\;\;Bs}
     +
     \frac{1}{r} \bar{D}_{A}r H_{C}^{\;\;\;Bs}
     ,
  \\
  &&
     \nabla_{A}^{}H_{C}^{\;\;\;qD}
     =
     \bar{D}_{A}^{}H_{C}^{\;\;\;qD}
     +
     \frac{1}{r} \bar{D}_{A}r H_{C}^{\;\;\;qD}
     ,
  \\
  &&
     \nabla_{A}^{}H_{C}^{\;\;\;qs}
     =
     \bar{D}_{A}^{}H_{C}^{\;\;\;qs}
     +
     \frac{2}{r} \bar{D}_{A}r H_{C}^{\;\;\;qs}
     ,
\end{eqnarray}
\begin{eqnarray}
  &&
     \nabla_{A}^{}H_{r}^{\;\;\;BD}
     =
     \bar{D}_{A}^{}H_{r}^{\;\;\;BD}
     -
     \frac{1}{r} \bar{D}_{A}r H_{r}^{\;\;\;BD}
     ,
  \\
  &&
     \nabla_{A}^{}H_{r}^{\;\;\;Bs}
     =
     \bar{D}_{A}^{}H_{r}^{\;\;\;Bs}
     ,
  \\
  &&
     \nabla_{A}^{}H_{r}^{\;\;\;qD}
     =
     \bar{D}_{A}^{}H_{r}^{\;\;\;qD}
     ,
  \\
  &&
     \nabla_{A}^{}H_{r}^{\;\;\;qs}
     =
     \bar{D}_{A}^{}H_{r}^{\;\;\;qs}
     +
     \frac{1}{r} \bar{D}_{A}r H_{r}^{\;\;\;qs}
     ,
\end{eqnarray}
\begin{eqnarray}
  &&
     \nabla_{p}^{}H_{C}^{\;\;\;BD}
     =
     \hat{D}_{p}^{}H_{C}^{\;\;\;BD}
     -
     \frac{1}{r} \bar{D}_{C}r H_{p}^{\;\;\;BD}
     -
     r \bar{D}^{B}r \gamma_{tp} H_{C}^{\;\;\;tD}
     -
     r \bar{D}^{D}r \gamma_{tp} H_{C}^{\;\;\;Bt}
     ,
  \\
  &&
     \nabla_{p}^{}H_{C}^{\;\;\;Bs}
     =
     \hat{D}_{p}^{}H_{C}^{\;\;\;Bs}
     -
     \frac{1}{r} \bar{D}_{C}r H_{p}^{\;\;\;Bs}
     +
     \frac{1}{r} \bar{D}_{E}r \gamma_{p}^{\;\;s} H_{C}^{\;\;\;BE}
     -
     r \bar{D}^{B}r \gamma_{tp} H_{C}^{\;\;\;ts}
     ,
  \\
  &&
     \nabla_{p}^{}H_{C}^{\;\;\;qD}
     =
     \hat{D}_{p}^{}H_{C}^{\;\;\;qD}
     -
     \frac{1}{r} \bar{D}_{C}r H_{p}^{\;\;\;qD}
     +
     \frac{1}{r} \bar{D}_{E}r \gamma_{p}^{\;\;q} H_{C}^{\;\;\;ED}
     -
     r \bar{D}^{D}r \gamma_{tp} H_{C}^{\;\;\;qt}
     ,
  \\
  &&
     \nabla_{p}^{}H_{C}^{\;\;\;qs}
     =
     \hat{D}_{p}^{}H_{C}^{\;\;\;qs}
     -
     \frac{1}{r} \bar{D}_{C}r H_{p}^{\;\;\;qs}
     +
     \frac{1}{r} \bar{D}_{E}r \gamma_{p}^{\;\;q} H_{C}^{\;\;\;Es}
     +
     \frac{1}{r} \bar{D}_{E}r \gamma_{p}^{\;\;s} H_{C}^{\;\;\;qE}
     ,
\end{eqnarray}
\begin{eqnarray}
  &&
     \nabla_{p}^{}H_{r}^{\;\;\;BD}
     =
     \hat{D}_{p}^{}H_{r}^{\;\;\;BD}
     +
     r \bar{D}^{E}r \gamma_{rp} H_{E}^{\;\;\;BD}
     -
     r \bar{D}^{B} \gamma_{tp} H_{r}^{\;\;\;tD}
     -
     r \bar{D}^{D} \gamma_{tp} H_{r}^{\;\;\;Bt}
     ,
  \\
  &&
     \nabla_{p}^{}H_{r}^{\;\;\;Bs}
     =
     \hat{D}_{p}^{}H_{r}^{\;\;\;Bs}
     + r \bar{D}^{E}r \gamma_{rp} H_{E}^{\;\;\;Bs}
     -  r \bar{D}^{B}r \gamma_{tp} H_{r}^{\;\;\;ts}
     + \frac{1}{r} \bar{D}_{E}r \gamma_{p}^{\;\;s}H_{r}^{\;\;\;BE}
     ,
  \\
  &&
     \nabla_{p}^{}H_{r}^{\;\;\;qD}
     =
     \hat{D}_{p}^{}H_{r}^{\;\;\;qD}
     +
     r \bar{D}^{E}r \gamma_{rp} H_{E}^{\;\;\;qD}
     -
     r \bar{D}^{D}r \gamma_{tp} H_{r}^{\;\;qt}
     +
     \frac{1}{r} \bar{D}_{E}r \gamma_{p}^{\;\;q} H_{r}^{\;\;\;ED}
     ,
  \\
  &&
     \nabla_{p}^{}H_{r}^{\;\;\;qs}
     =
     \hat{D}_{p}^{}H_{r}^{\;\;\;qs}
     +
     r \bar{D}^{E}r \gamma_{rp} H_{E}^{\;\;\;qs}
     +
     \frac{1}{r} \bar{D}_{E}r \gamma_{p}^{\;\;q} H_{r}^{\;\;\;Es}
     +
     \frac{1}{r} \bar{D}_{E}r \gamma_{p}^{\;\;s} H_{r}^{\;\;\;qE}
      .
\end{eqnarray}

%*********************************************************************

Next, we summarize the components of the background curvatures induced
by the metric Eq.~(\ref{eq:background-metric-2+2}).
We derive these components through the components of the connection
(\ref{eq:Background-connection-ABC})--(\ref{eq:Background-connection-qrp})
and the formula of the Riemann curvature
\begin{eqnarray}
  R_{abc}^{\;\;\;\;\;\;\;d}
  =
  \partial_{b}\Gamma_{ac}^{\;\;\;\;d}
  -
  \partial_{a}\Gamma_{bc}^{\;\;\;\;d}
  +
  \Gamma_{ac}^{\;\;\;\;\;e}\Gamma_{eb}^{\;\;\;\;\;d}
  -
  \Gamma_{bc}^{\;\;\;\;\;e}\Gamma_{ea}^{\;\;\;\;\;d}
  .
\end{eqnarray}
To derive the components of this curvature, we use
\begin{eqnarray}
  \bar{D}_{A}\gamma_{pq} = 0 = \hat{D}_{p}y_{AB},
  \quad
  \hat{D}_{p}r = 0.
\end{eqnarray}

%*********************************************************************

The components of the non-vanishing Riemann curvature are summarized
as
\begin{eqnarray}
  R_{ABC}^{\;\;\;\;\;\;\;\;\;D}
  &=&
      {}^{(2)}\!\bar{R}_{ABC}^{\;\;\;\;\;\;\;\;\;D}
      ,
      \label{eq:BG-riemman-curvature-ABCD-general}
  \\
  R_{pBr}^{\;\;\;\;\;\;\;D}
  &=&
      -
      r
      (\bar{D}_{B}\bar{D}^{D}r)
      \gamma_{pr}
      ,
      \label{eq:BG-riemman-curvature-pBrD-general}
  \\
  R_{pBC}^{\;\;\;\;\;\;\;\;\;s}
  &=&
      \frac{1}{r} (\bar{D}_{B}\bar{D}_{C}r) \gamma_{p}^{\;\;s}
      ,
      \label{eq:BG-riemman-curvature-pBCs-general}
  \\
  R_{pqr}^{\;\;\;\;\;\;\;s}
  &=&
      {}^{(2)}\!\hat{R}_{pqr}^{\;\;\;\;\;\;\;s}
      -
      2
      (\bar{D}^{E}r)
      (\bar{D}_{E}r)
      \gamma_{r[p}^{} \gamma_{q]}^{\;\;s}
      .
      \label{eq:BG-riemman-curvature-pqrs-general}
\end{eqnarray}

%*********************************************************************

The components of the Ricci curvature are summarized as
\begin{eqnarray}
  R_{AC}
  &=&
      {}^{(2)}\!\bar{R}_{AC}
      - \frac{2}{r} (\bar{D}_{A}\bar{D}_{C}r)
      ,
      \label{eq:background-Ricci-AC}
  \\
  R_{Ar}
  &=&
      0
      ,
      \label{eq:background-Ricci-Ap}
  \\
  R_{pr}
  &=&
      {}^{(2)}\!\hat{R}_{pr}
      -
      \left[
      r (\bar{D}_{E}\bar{D}^{E}r) + (\bar{D}^{E}r) (\bar{D}_{E}r)
      \right]
       \gamma_{pr}
      \label{eq:background-Ricci-pr}
      .
\end{eqnarray}
The Ricci scalar curvature is given by
\begin{eqnarray}
  R
  &=&
      g^{ac}R_{ac}
      =
      {}^{(2)}\!\bar{R}
      +
      \frac{1}{r^{2}} {}^{(2)}\!\hat{R}
      -
      \frac{4}{r} (\bar{D}^{C}\bar{D}_{C}r)
      -
      \frac{2}{r^{2}} (\bar{D}^{E}r) (\bar{D}_{E}r)
      .
      \label{eq:BG-scalar-curvature-general}
\end{eqnarray}
Next, we derive the components of the Einstein tensor
\begin{eqnarray}
  G_{ab} := R_{ab} - \frac{1}{2} g_{ab}R
\end{eqnarray}
and its components are summarized as
\begin{eqnarray}
  G_{AB}
  &=&
      -  \frac{2}{r} (\bar{D}_{A}\bar{D}_{B}r)
      + \frac{1}{r^{2}} y_{AB}
      \left[
      -  1
      + 2r (\bar{D}^{C}\bar{D}_{C}r)
      + (\bar{D}^{E}r) (\bar{D}_{E}r)
      \right]
      ,
      \label{eq:BG-Einstein-tensor-AB}
  \\
  G_{Aq}
  &=&
      0
      ,
      \label{eq:BG-Einstein-tensor-Aq}
  \\
  G_{pq}
  &=&
      \gamma_{pq}
      \left[
      r (\bar{D}^{C}\bar{D}_{C}r)
      - \frac{1}{2} r^{2} {}^{(2)}\!\bar{R}
      \right]
      ,
      \label{eq:BG-Einstein-tensor-pq}
\end{eqnarray}
where we used the two-dimensional Einstein tensors are identically
vanish and the fact that the metric $\gamma_{pq}$ is the maximally
symmetric space with positive curvature, i.e.,
\begin{eqnarray}
  {}^{(2)}\!\hat{R}_{pqrs} = 2 \gamma_{p[r}\gamma_{s]q}, \quad
  {}^{(2)}\!\hat{R}_{pr} = \gamma_{pr}, \quad
  {}^{(2)}\!\hat{R} = 2.
  \label{eq:unit-sphere-condition}
\end{eqnarray}

%*********************************************************************

Here, we consider the static solution whose metric is given by
\begin{eqnarray}
  y_{AB} = - f (dt)_{A} (dt)_{B} + f^{-1} (dr)_{A}(dr)_{B},
  \label{eq:static-yAB-assumption}
\end{eqnarray}
where $f=f(r)$.
Due to the Birkhoff theorem, the vacuum solution with the spherically
symmetric spacetime must be the Schwarzschild spacetime.
We check this fact from
Eqs.~(\ref{eq:BG-Einstein-tensor-AB})--(\ref{eq:BG-Einstein-tensor-pq})
with the substitution (\ref{eq:static-yAB-assumption}).
Actually, we obtain
\begin{eqnarray}
  \bar{D}_{B}r
  =
  (dr)_{B}
  , \quad
  \bar{D}^{B}r = f \left(\frac{\partial}{\partial r}\right)^{B}
  , \quad
  \bar{D}_{A}\bar{D}_{B}r
  =
  \frac{f'}{2} y_{AB}
  \label{eq:gradient-r-hessian-r}
  .
\end{eqnarray}
Then, we have
\begin{eqnarray}
  \label{eq:gradient-r-norm-hessisan-r-2}
  (\bar{D}^{B}r) (\bar{D}_{B}r)
  =
  f
  , \quad
  \bar{D}^{A}\bar{D}_{B}r = \frac{f'}{2} \delta_{B}^{\;\;A}
  , \quad
  \bar{D}^{C}\bar{D}_{C}r
  =
  f'
  .
\end{eqnarray}
From Eq.~(\ref{eq:BG-Einstein-tensor-AB}) as
\begin{eqnarray}
  y^{AB}G_{AB}
  =
  \frac{2}{r}
  \left(
  f'
  - \frac{1-f}{r}
  \right)
  =
  0
  , \quad
  G_{AB} - \frac{1}{2}y_{AB}G_{AB}
  =
  0
  .
  \label{eq:BG-Einstein-tensor-AB-2}
\end{eqnarray}
The solution to Eq.~(\ref{eq:BG-Einstein-tensor-AB-2}) is given by
\begin{eqnarray}
  \label{eq:BG-Einstein-tensor-AB-2-sol-1}
  f &=& 1 - \frac{2M}{r}, \\
  \label{eq:BG-Einstein-tensor-AB-2-sol-2}
  f' &=&  \frac{1-f}{r},
\end{eqnarray}
where $M$ is the constant of integration.
This is the Schwarzschild metric.
We also evaluate the component $G_{pq} = 0$ through
Eq.~(\ref{eq:BG-Einstein-tensor-pq}) using
Eq.~(\ref{eq:BG-Einstein-tensor-AB-2-sol-1}) as
\begin{eqnarray}
  {}^{(2)}\!\bar{R}
  =
  \frac{2}{r} (\bar{D}^{C}\bar{D}_{C}r)
  \label{eq:Background-Einstein-pq-2-org}
\end{eqnarray}

%*********************************************************************

As the summary of the background vacuum Einstein equations, we have
\begin{eqnarray}
  &&
     r (\bar{D}^{C}\bar{D}_{C}r)
     +   (\bar{D}^{E}r) (\bar{D}_{E}r)
     =
     1
     ,
     \label{eq:BG-Einstein-tensor-AB-trace-2}
  \\
  &&
     (\bar{D}_{A}\bar{D}_{B}r)
     =
     \frac{1}{2} y_{AB} (\bar{D}^{C}\bar{D}_{C}r)
     ,
      \label{eq:BG-Einstein-tensor-AB-traceless-2}
  \\
  &&
     {}^{(2)}\!\bar{R}
     =
     \frac{2}{r} (\bar{D}^{C}\bar{D}_{C}r)
     .
     \label{eq:Background-Einstein-pq-2}
\end{eqnarray}
Eq.~(\ref{eq:BG-Einstein-tensor-AB-trace-2}) is equivalent to
Eq.~(\ref{eq:BG-Einstein-tensor-AB-2-sol-2}).
Since the two-dimensional curvature ${}^{(2)}\!\bar{R}_{DAEC}$ has
only one independent component, ${}^{(2)}\!\bar{R}_{DAEC}$ is written as
\begin{eqnarray}
  \label{eq:background-y-Riemman-curv-}
  {}^{(2)}\!\bar{R}_{DAEC}
  =
  \frac{2}{r} (\bar{D}^{F}\bar{D}_{F}r) y_{D[E}y_{C]A}
  ,
  \quad
  {}^{(2)}\!\bar{R}_{DE}
  =
  \frac{1}{r} (\bar{D}^{F}\bar{D}_{F}r) y_{DE}
  ,
  \quad
  {}^{(2)}\!\bar{R}
  =
  \frac{2}{r} (\bar{D}^{F}\bar{D}_{F}r)
  .
\end{eqnarray}

%*********************************************************************

The above formulae are expressed the covariant form of the 2+2
formulation.
However, the explicit components of $\Gamma_{ab}^{\;\;\;\;\;c}$ are also
convenient to leads the results in
Sec.~\ref{sec:Schwarzschild_Background_odd_vacuum_sol}.
From
Eqs.~(\ref{eq:Background-connection-ABC})--(\ref{eq:Background-connection-qrp})
and the background metric (\ref{eq:background-metric-2+2}) with
Eqs.~(\ref{eq:background-metric-2+2-y-comp-Schwarzschild}) and
(\ref{eq:background-metric-2+2-gamma-comp-Schwarzschild}),
non-vanishing components of $\Gamma_{ab}^{\;\;\;\;\;c}$ are summarized
as
\begin{eqnarray}
  && \Gamma_{tr}^{\;\;\;\;t} = \frac{f'}{2f}, \quad
         \Gamma_{tt}^{\;\;\;r} = \frac{1}{2} f f' , \quad
         \Gamma_{rr}^{\;\;\;\;r} = - \frac{f'}{2f}, \quad
         \Gamma_{\theta\theta}^{\;\;\;\;\;\;r} = - r f,
     \nonumber\\
  && \Gamma_{\phi\phi}^{\;\;\;\;\;\;r} = - r f \sin^{2}\theta, \quad
         \Gamma_{r\theta}^{\;\;\;\;\theta} = \frac{1}{r}, \quad
         \Gamma_{\phi\phi}^{\;\;\;\;\;\theta} = - \sin\theta \cos\theta,
     \label{eq:Background-connection-explicit}
  \\
  && \Gamma_{r\phi}^{\;\;\;\;\phi} = \frac{1}{r}, \quad
         \Gamma_{\phi\theta}^{\;\;\;\;\;\phi} = \cot\theta.
     \nonumber
\end{eqnarray}

%*********************************************************************

%%%%%%%%%%%%%%%%%%%%%%%%%%%%%%%%%%%%%%%%%%%
%%%%%%%%%%%%%%%%%%%%%%%%%%%%%%%%%%%%%%%%%%%
%%%%%%%%%%%%%%%%%%%%%%%%%%%%%%%%%%%%%%%%%%%
\section{Summary of the 2+2 representations of the tensor
  $H_{abc}[\ScrF]$, $H_{ab}^{\;\;\;\;c}[\ScrF]$,
  $H_{a}^{\;\;bc}[\ScrF]$}
\label{sec:2+2-representation-of-H}
%%%%%%%%%%%%%%%%%%%%%%%%%%%%%%%%%%%%%%%%%%%
%%%%%%%%%%%%%%%%%%%%%%%%%%%%%%%%%%%%%%%%%%%
%%%%%%%%%%%%%%%%%%%%%%%%%%%%%%%%%%%%%%%%%%%

%*********************************************************************

Here, we summarize the components of $H_{abc}[\ScrF]$ through the
expressions
(\ref{eq:2+2-gauge-invariant-variables-calFAB})--(\ref{eq:2+2-gauge-invariant-variables-calFpq}):
\begin{eqnarray}
  H_{ABC}
  &=&
      \bar{D}_{(A}F_{B)C}
      -
      \frac{1}{2}
      \bar{D}_{C}F_{AB}
      ,
  \\
  H_{pBC}
  &=&
      \frac{1}{2}
      \left(
      \hat{D}_{p}F_{BC}
      + r \bar{D}_{B}F_{Cp}
      -  r \bar{D}_{C}F_{Bp}
      -  (\bar{D}_{B}r) F_{Cp}
      -  (\bar{D}_{C}r) F_{Bp}
      \right)
      ,
  \\
  H_{pqC}
  &=&
      \frac{1}{2} \left(
      2 r \hat{D}_{(p}F_{q)C}
      -  \frac{1}{2} \gamma_{pq} r^{2} \bar{D}_{C}F
%      \right.
%      \nonumber\\
%  && \quad\quad\quad
%     \left.
      -  r (\bar{D}_{C}r) \gamma_{pq} F
      + 2 r (\bar{D}^{D}r) \gamma_{pq} F_{DC}
      \right)
      ,
  \\
  H_{ABr}
  &=&
      r \bar{D}_{(A}F_{B)r}
      + (\bar{D}_{(A}r) F_{B)r}
      -  \frac{1}{2} \hat{D}_{r}F_{AB}
      ,
  \\
  H_{pBr}
  &=&
      \frac{1}{2}
      \left(
      r \hat{D}_{p}F_{rB}
      -  r \hat{D}_{r}F_{pB}
      + \frac{1}{2} r^{2} \gamma_{pr} \bar{D}_{B}F
      \right)
      ,
  \\
  H_{pqr}
  &=&
      \frac{1}{2} r^{2} \gamma_{r(q} \hat{D}_{p)}F
      -  \frac{1}{4} r^{2} \gamma_{pq} \hat{D}_{r}F
      + r^{2} \bar{D}^{D}r \gamma_{pq} F_{Dr}
      .
\end{eqnarray}

%****************************************************************

Next, we summarize the components of $H_{ab}^{\;\;\;\;c}[\ScrF]$ through the
expressions
(\ref{eq:2+2-gauge-invariant-variables-calFAB})--(\ref{eq:2+2-gauge-invariant-variables-calFpq}):
\begin{eqnarray}
  H_{AB}^{\;\;\;\;\;\;C}
  &=&
      \bar{D}_{(A}F_{B)}^{\;\;\;\;C}
      -
      \frac{1}{2}
      \bar{D}^{C}F_{AB}
      ,
  \\
  H_{pB}^{\;\;\;\;\;\;C}
  &=&
      \frac{1}{2}
      \left(
      \hat{D}_{p}F_{B}^{\;\;\;C}
      + r \bar{D}_{B}F_{p}^{\;\;C}
      -  r \bar{D}^{C}F_{Bp}
      -  (\bar{D}_{B}r) F_{p}^{\;\;C}
      -  (\bar{D}^{C}r) F_{Bp}
      \right)
      ,
  \\
  H_{pq}^{\;\;\;\;\;\;C}
  &=&
      \frac{1}{2} \left(
      2 r \hat{D}_{(p}F_{q)}^{\;\;\;C}
      -  \frac{1}{2} \gamma_{pq} r^{2} \bar{D}^{C}F
%      \right.
%      \nonumber\\
%  && \quad\quad
%     \left.
      -  r (\bar{D}^{C}r) \gamma_{pq} F
      + 2 r (\bar{D}^{D}r) \gamma_{pq} F_{D}^{\;\;\;C}
      \right)
      ,
  \\
  H_{AB}^{\;\;\;\;\;\;r}
  &=&
      \frac{1}{r} \bar{D}_{(A}F_{B)}^{\;\;\;\;r}
      + \frac{1}{r^{2}} (\bar{D}_{(A}r) F_{B)}^{\;\;\;\;r}
      -  \frac{1}{2r^{2}} \hat{D}^{r}F_{AB}
      ,
  \\
  H_{pB}^{\;\;\;\;\;\;r}
  &=&
      \frac{1}{2r} \hat{D}_{p}F_{B}^{\;\;\;r}
      -  \frac{1}{2r} \hat{D}^{r}F_{Bp}
      + \frac{1}{4} \gamma_{p}^{\;\;r} \bar{D}_{B}F
      ,
  \\
  H_{pq}^{\;\;\;\;\;\;r}
  &=&
      \frac{1}{2} \gamma_{(q}^{\;\;\;r} \hat{D}_{p)}F
      -  \frac{1}{4} \gamma_{pq} \hat{D}^{r}F
      + (\bar{D}^{D}r) \gamma_{pq} F_{D}^{\;\;\;r}
      .
\end{eqnarray}

%****************************************************************

Finally, we summarize the component $H_{a}^{\;\;bc}[\ScrF]$ through
the expression
(\ref{eq:2+2-gauge-invariant-variables-calFAB})--(\ref{eq:2+2-gauge-invariant-variables-calFpq}):
\begin{eqnarray}
  H_{A}^{\;\;\;BC}
  &=&
      \frac{1}{2}
      \left(
      \bar{D}_{A}F^{BC}
      +
      \bar{D}^{B}F_{A}^{\;\;\;C}
      -
      \bar{D}^{C}F_{A}^{\;\;\;B}
      \right)
      ,
  \\
  H_{A}^{\;\;\;Br}
  &=&
      \frac{1}{2r} \bar{D}_{A}F^{Br}
      + \frac{1}{2r} \bar{D}^{B}F_{A}^{\;\;\;r}
%      \nonumber\\
%  &&
      + \frac{1}{2r^{2}} (\bar{D}_{A}r) F^{Br}
      + \frac{1}{2r^{2}} (\bar{D}^{B}r) F_{A}^{\;\;\;r}
      -  \frac{1}{2r^{2}} \hat{D}^{r}F_{A}^{\;\;\;B}
      ,
  \\
  H_{A}^{\;\;\;qC}
  &=&
      \frac{1}{2r^{2}} \left(
      \hat{D}^{q}F_{A}^{\;\;\;C}
      + r \bar{D}_{A}F^{qC}
      -  r \bar{D}^{C}F_{A}^{\;\;\;q}
      -  (\bar{D}_{A}r) F^{qC}
      -  (\bar{D}^{C}r) F_{A}^{\;\;\;q}
      \right)
      ,
  \\
  H_{A}^{\;\;\;qr}
  &=&
      \frac{1}{2r^{3}} \left[
      \hat{D}^{q}F_{A}^{\;\;\;r}
      -  \hat{D}^{r}F_{A}^{\;\;\;q}
      + \frac{1}{2} r \gamma^{qr} \bar{D}_{A}F
      \right]
      ,
  \\
  H_{p}^{\;\;\;BC}
  &=&
      \frac{1}{2}
      \left(
      \hat{D}_{p}F^{BC}
      + r \bar{D}^{B}F_{p}^{\;\;C}
      -  r \bar{D}^{C}F_{p}^{\;\;B}
      -  (\bar{D}^{B}r) F_{p}^{\;\;C}
      -  (\bar{D}^{C}r) F_{p}^{\;\;B}
      \right)
      ,
  \\
  H_{p}^{\;\;\;Br}
  &=&
      \frac{1}{2r} \hat{D}_{p}F^{Br}
      -  \frac{1}{2r} \hat{D}^{r}F_{p}^{\;\;B}
      + \frac{1}{4} \gamma_{p}^{\;\;r} \bar{D}^{B}F
      ,
  \\
  H_{p}^{\;\;\;qC}
  &=&
      \frac{1}{2r^{2}} \left(
      r \hat{D}_{p}F^{qC}
      + r \hat{D}^{q}F_{p}^{\;\;\;C}
      -  \frac{1}{2} \gamma_{p}^{\;\;q} r^{2} \bar{D}^{C}F
%      \right.
%      \nonumber\\
%  && \quad\quad\quad
%     \left.
      -  r (\bar{D}^{C}r) \gamma_{p}^{\;\;q} F
      + 2 r (\bar{D}^{D}r) \gamma_{p}^{\;\;q} F_{D}^{\;\;\;C}
      \right)
      ,
  \\
  H_{p}^{\;\;\;qr}
  &=&
      \frac{1}{r^{2}} \left(
      \frac{1}{4} \gamma^{qr} \hat{D}_{p}F
      + \frac{1}{4} \gamma_{p}^{\;\;r} \hat{D}^{q}F
      -  \frac{1}{4} \gamma_{p}^{\;\;q} \hat{D}^{r}F
      + (\bar{D}^{D}r) \gamma_{p}^{\;\;q} F_{D}^{\;\;\;r}
      \right)
      .
\end{eqnarray}

%****************************************************************

%%%%%%%%%%%%%%%%%%%%%%%%%%%%%%%%%%%%%%%%%%%
%%%%%%%%%%%%%%%%%%%%%%%%%%%%%%%%%%%%%%%%%%%
%%%%%%%%%%%%%%%%%%%%%%%%%%%%%%%%%%%%%%%%%%%
%\section*{References}
%%%%%%%%%%%%%%%%%%%%%%%%%%%%%%%%%%%%%%%%%%%
%%%%%%%%%%%%%%%%%%%%%%%%%%%%%%%%%%%%%%%%%%%
%%%%%%%%%%%%%%%%%%%%%%%%%%%%%%%%%%%%%%%%%%%

\end{document}